\documentclass[twocolumn]{aastex631}
\usepackage{array}
\usepackage{subfigure}
\usepackage{natbib}
\usepackage{enumitem}

\usepackage{graphicx}	% Including figure files
\usepackage{amsmath}	% Advanced maths commands
\usepackage{amssymb}	% Extra maths symbols
\usepackage{bm}		% Bold maths symbols, including upright Greek
\usepackage{color}

\let\tablenum\relax % Make sure to clear the existing definition
\usepackage{siunitx}
\usepackage{anyfontsize}

\usepackage{threeparttable}

\makeatletter
\let\TPT@hookin\@gobble
\let\TPT@hookarg\@gobble
\makeatother

\def\hi{\textsc{HI~}}

\newcommand{\kms}{\,km\,s$^{-1}$} % kilometres per second
\newcommand{\cm}{\ensuremath{\,{\rm cm}}}
\newcommand{\m}{\ensuremath{\,{\rm m}}}

\renewcommand{\kms}{\ensuremath{\,{\rm km\,s^{-1}}}} % kilometres per second

\newcommand{\K}{\ensuremath{\, {\rm K}}}

\newcommand{\mJy}{\ensuremath{\,{\rm mJy}}}

\shorttitle{\hi absorption from FAST}
\shortauthors{Hu et al.}
\begin{document}

\title[A large 21cm-selected sample of \hi absorption from blind survey in CRAFTS]
{The FAST \hi 21-cm absorption blind survey. \\ II. -- Statistic Exploration for Associated and Intervening Systems}

\correspondingauthor{Wenkai Hu}
\author[0000-0002-3108-5591]{Wenkai Hu}
\email{wkhu@nao.cas.cn}
\affiliation{Department of Physics and Astronomy, University of the Western Cape, Robert Sobukwe Road, Bellville, 7535, South Africa}
\affiliation{ARC Centre of Excellence for All Sky Astrophysics in 3 Dimensions (ASTRO 3D), Australia}

\author[0000-0003-0631-568X]{Yougang Wang}
\affiliation{National Astronomical Observatories, Chinese Academy of Sciences, Beijing 100101, China}
\affiliation{Key Laboratory of Radio Astronomy and Technology, Chinese Academy of Sciences, A20 Datun Road, Chaoyang District, Beijing 100101, China}
\affiliation{School of Astronomy and Space Science, University of Chinese Academy of Sciences, Beijing 100049, China}
\affiliation{Key Laboratory of Cosmology and Astrophysics (Liaoning) \& College of Sciences, Northeastern University, Shenyang 110819, China}
\author[0000-0003-1962-2013]{Yichao Li}
\affiliation{Key Laboratory of Cosmology and Astrophysics (Liaoning) \& College of Sciences, Northeastern University, Shenyang 110819, China}

\author[0009-0006-2521-025X]{Wenxiu Yang}
\affiliation{National Astronomical Observatories, Chinese Academy of Sciences, Beijing 100101, China}
\affiliation{School of Astronomy and Space Science, University of Chinese Academy of Sciences, Beijing 100049, China}

\author[0000-0003-3224-4125]{Yidong Xu}
\affiliation{National Astronomical Observatories, Chinese Academy of Sciences, Beijing 100101, China}
\affiliation{Key Laboratory of Radio Astronomy and Technology, Chinese Academy of Sciences, A20 Datun Road, Chaoyang District, Beijing 100101, China}

\author[0000-0002-6174-8640]{Fengquan Wu}
\affiliation{National Astronomical Observatories, Chinese Academy of Sciences, Beijing 100101, China}
\affiliation{Key Laboratory of Radio Astronomy and Technology, Chinese Academy of Sciences, A20 Datun Road, Chaoyang District, Beijing 100101, China}

\author[0000-0003-2155-9578]{Ue-Li Pen}
\affiliation{Institute of Astronomy and Astrophysics, Academia Sinica, Astronomy-Mathematics Building, No. 1, Sec. 4, Roosevelt Road, Taipei 10617, Taiwan, China}
\affiliation{Canadian Institute for Theoretical Astrophysics, University of Toronto, 60 Saint George Street, Toronto, ON M5S 3H8, Canada}
\affiliation{Canadian Institute for Advanced Research, 180 Dundas St West, Toronto, ON M5G 1Z8, Canada}
\affiliation{Dunlap Institute for Astronomy and Astrophysics, University of Toronto, 50 St George Street, Toronto, ON M5S 3H4, Canada}
\affiliation{Perimeter Institute of Theoretical Physics, 31 Caroline Street North, Waterloo, ON N2L 2Y5, Canada}

\author[0000-0002-9937-2351]{Jie Wang}
\affiliation{National Astronomical Observatories, Chinese Academy of Sciences, Beijing 100101, China}
\affiliation{School of Astronomy and Space Science, University of Chinese Academy of Sciences, Beijing 100049, China}

\author{Yingjie Jing}
\affiliation{National Astronomical Observatories, Chinese Academy of Sciences, Beijing 100101, China}

\author{Chen Xu}
\affiliation{National Astronomical Observatories, Chinese Academy of Sciences, Beijing 100101, China}
\affiliation{School of Astronomy and Space Science, University of Chinese Academy of Sciences, Beijing 100049, China}

\author{Qingze Chen}
\affiliation{National Astronomical Observatories, Chinese Academy of Sciences, Beijing 100101, China}
\affiliation{School of Astronomy and Space Science, University of Chinese Academy of Sciences, Beijing 100049, China}

\author[0009-0005-9546-4573]{Zheng Zheng}
\affiliation{National Astronomical Observatories, Chinese Academy of Sciences, Beijing 100101, China}
\affiliation{Research Center for Intelligent Computing Platforms, Zhejiang Laboratory, Hangzhou 311100, China}

\author[0000-0003-3010-7661]{Di Li}
\affiliation{National Astronomical Observatories, Chinese Academy of Sciences, Beijing 100101, China}
%\affiliation{CAS Key Laboratory of FAST, National Astronomical Observatories, Chinese Academy of Sciences, Beijing 100101, China}
%\affiliation{University of Chinese Academy of Sciences, Beijing 100049, China}
\affiliation{Key Laboratory of Radio Astronomy and Technology, Chinese Academy of Sciences, A20 Datun Road, Chaoyang District, Beijing 100101, China}
%\affiliation{School of Astronomy and Space Science, University of Chinese Academy of Sciences, Beijing 100049, China}
\affiliation{Department of Astronomy, Tsinghua Univerisity, 30 Shuangqing Road, Beijing 100084, People’s Republic of China}

\author{Ming Zhu}
\affiliation{National Astronomical Observatories, Chinese Academy of Sciences, Beijing 100101, China}
\affiliation{Key Laboratory of Radio Astronomy and Technology, Chinese Academy of Sciences, A20 Datun Road, Chaoyang District, Beijing 100101, China}
\affiliation{Guizhou Radio Astronomical Observatory, Guizhou University, Guiyang 550000, China}
%\affiliation{CAS Key Laboratory of FAST, National Astronomical Observatories, Chinese Academy of Sciences, Beijing 100101, China}

\author[0000-0002-6029-1933]{Xin Zhang}
\affiliation{Key Laboratory of Cosmology and Astrophysics (Liaoning) \& College of Sciences, Northeastern University, Shenyang 110819, China}
\affiliation{National Frontiers Science Center for Industrial Intelligence and Systems Optimization, Northeastern University, Shenyang 110819, China}
\affiliation{Key Laboratory of Data Analytics and Optimization for Smart Industry (Ministry of Education), Northeastern University, Shenyang 110819, China}

\author[0000-0001-6475-8863]{Xuelei Chen}
\affiliation{National Astronomical Observatories, Chinese Academy of Sciences, Beijing 100101, China}
\affiliation{Key Laboratory of Radio Astronomy and Technology, Chinese Academy of Sciences, A20 Datun Road, Chaoyang District, Beijing 100101, China}
%\affiliation{University of Chinese Academy of Sciences, Beijing 100049, China}
\affiliation{School of Astronomy and Space Science, University of Chinese Academy of Sciences, Beijing 100049, China}
\affiliation{Key Laboratory of Cosmology and Astrophysics (Liaoning) \& College of Sciences, Northeastern University, Shenyang 110819, China}

\begin{abstract}

We present an extragalactic \hi 21-cm absorption lines catalog from a blind search at z $\leqslant$ 0.35, using drift-scan data collected in 1325.6 hours by the ongoing Commensal Radio Astronomy FasT Survey (CRAFTS) and FAST All Sky \hi Survey (FASHI), which spans a sky area of 6072.0 deg$^{2}$ and covers 84533 radio sources with a flux density greater than 12 mJy. 14 previously identified \hi absorbers and 20 newly discovered \hi absorbers were detected, comprising 15 associated systems, 10 intervening systems, and 9 systems with undetermined classifications. Through spectral stacking, the mean peak optical path, mean velocity-integrated optical path, mean FWHM and mean \hi column density are measured to be 0.47 and 0.30; 27.19 and 4.36 $\kms$; 42.61 and 9.33 $\kms$; 0.49 and 0.08 T$_{s} \times$ 10$^{20}$cm$^{-2}$K$^{-1}$, for the associated and intervening samples, respectively. Statistical analysis also reveals that associated systems tend to be hosted by red ($g-r>$0.7) galaxies at lower redshifts, whereas galaxies hosting intervening \hi absorption are typically found at higher redshifts and are of a bluer ($g-r\leqslant$0.7) type. A noticeable difference is observed in the positions of foregrounds, backgrounds of intervening systems, and high-redshift and low-redshift associated systems on the WISE color-color diagram. All identified foreground sources in our sample have W1-W2 magnitudes below 0.8, suggesting no Active Galactic Nuclei (AGN). In contrast, backgrounds of intervening systems tend to have W1-W2 magnitudes above 0.8, indicating AGN presence. For associated absorption, most low-redshift ($z\leqslant$0.5) systems show W1-W2 values below 0.8, while higher-redshift associated absorption ($z>$0.5) displays a broader range of W1-W2 values.

%Additionally, it has been shown that associated \hi 21-cm absorptions connected to compact radio sources display higher N$_{\hi}$ values compared to those linked with extended radio sources.
%Assuming an \hi spin temperature to source covering fraction ratio of $T_{s}/c_{f}$ = 100 K, the total comoving absorption path length spanned by our data set and sensitive to Damped Lyman $\alpha$ Absorbers (DLAs; $N_{\hi} \geqslant 2\times10^{20} \cm^{-2}$) is $\Delta X^{\mathrm{inv}}$ = 4.07$\times10^4$ ($\Delta z^{\mathrm{inv}} = 3.19\times10^{4}$) for intervening absorption. For associated absorption, the corresponding value is $\Delta X^{\mathrm{asc}}$ = 1.81$\times10^{2}$ ($\Delta z^{\mathrm{asc}} = 1.31\times10^{2}$).

\end{abstract}

%% You will be asked to select these concepts during the submission process


\section{Introduction} \label{sec:intro}

The \hi 21-cm absorption lines are spectral features that arise from foreground gas absorbing the flux of a background bright radio source. Since the detectability of \hi absorption depends only on the column density of foreground gas and the strength of background radio sources, it could be a complement to the observation of \hi emission at higher redshifts, where direct detection of emission becomes challenging. \hi absorption is an excellent tool to measure \hi content and constrain redshift-evolution of cosmic \hi relative density ($\Omega_{\hi}$) at intermediate redshifts $0.2 < z < 2$, probe physical conditions in ISM \citep{2003ApJ...586.1067H,2003ApJ...587..278W,2018A&ARv..26....4M}, trace mergers and interactions of galaxies \citep{2018MNRAS.480..947D,2017A&A...607A..43V}, reveal evolution history of SFR density \citep{2017A&A...606A..56C,2019JApA...40...41D}, place stringent constraints on variation of fundamental constants of physics \citep{2012MNRAS.425..556R} and provide direct proof of cosmic acceleration \citep{2012ApJ...761L..26D,2014PhRvL.113d1303Y,2015aska.confE..27K,2020JCAP...01..054J,2023MNRAS.521.3150L}.

In the study of \hi absorption systems, there are two primary categories. The first is the (source) associated absorption system, where the gas absorbing the light is located in the same extragalactic object (often an AGN) that emits the bright continuum. Research in this area primarily focuses on the AGN itself and its interaction with Interstellar Medium (ISM) in the same extragalactic object. This research includes examining the kinematics and structure of the ISM in galaxies hosting AGNs and exploring how \hi gas might fuel supermassive black holes (SMBH)(see \citep{2018A&ARv..26....4M} and the references therein).

The second category is known as the intervening absorption system, which occurs when gas in a foreground galactic or extragalactic object absorbs light from a bright, unrelated background source. These systems are commonly used to study ISM properties of both Galaxy \citep{2013PASA...30....3D,2022ApJ...926..186D} and distant galaxies \citep{2020MNRAS.499.4293S}. A significant portion of post-reionization universe's neutral hydrogen is found in the Damped Lyman-alpha systems (DLAs, $N_{\hi} \geqslant 2\times10^{20} \cm^{-2}$; \citep{2005ARA&A..43..861W}), observed in absorption lines of optical quasi-stellar objects (QSOs). However, the bright light from the background QSOs blinds the direct study of DLAs themselves. Despite extensive research, the precise physical nature of DLAs remains an unresolved issue (for recent work, see, e.g. \citealt{Bordoloi2022}). Although associated HI absorption can reach DLA-level column densities, it is considered separately due to its proximity to host galaxies, unlike intervening DLAs along random sightlines. DLAs typically refer to intervening systems that reveal the large-scale distribution of neutral hydrogen across cosmic time.

Despite significant advances in this field, current findings are constrained by limited sample sizes and biases resulting from previously chosen targets. Conducting an unbiased large radio survey would directly tackle the DLA issue, helping to surmount these restrictions. Additionally, it would offer a more thorough understanding of the interaction between the AGN and the ISM of the host galaxy. 

Due to the limitation that only \hi gas along the line of sight to the background radio source can be traced, and considering the sensitivity of the telescopes, only a small number of \hi absorption systems have been detected through blind surveys. Recently, there have been promising developments in the search for \hi absorption systems, utilizing the capabilities of advanced radio telescopes from the latest generation, including \citet{2020MNRAS.494.3627A,2022MNRAS.516.2947S,2024MNRAS.527.8511A} from the First Large Absorption Survey in \hi (FLASH; \citealt{2022PASA...39...10A}), \citet{2021ApJ...907...11G,2024A&A...687A..50D} from the MeerKAT Absorption Line Survey (MALS; \citealt{2024ApJS..270...33D}), and the searches described below with FAST.

As the largest single-dish telescope in the world, FAST \citep{2011IJMPD..20..989N,2020RAA....20...64J} is equipped with a multibeam feed system and low-noise cryogenic receivers, ideal for conducting large blind \hi absorption surveys. Research efforts using \hi absorption have been successfully conducted by FAST, including studies on cosmic acceleration \citep{Kang_2024}, constraining the OH-to-\hi relative abundance ([OH]/[\hi]) \citep{2020MNRAS.499.3085Z}, and both targeted \citep{2021MNRAS.503.5385Z,2024ApJ...973...48C,2023ApJ...952..144Y} and blind \citet{2021MNRAS.503.5385Z,2023A&A...675A..40H} \hi absorption searches. These studies highlight that FAST is an excellent telescope for conducting \hi absorption science.

%\citet{2021MNRAS.503.5385Z} observed five \hi absorption systems that were previously identified in the 40$\%$ data release of the ALFALFA survey, with much higher spectral resolution and signal-to-noise ratio ($\mathrm{S/N}$) than those obtained in previous searches, demonstrating the power of FAST in revealing detailed structures of \hi absorption lines. \citet{2020MNRAS.499.3085Z} carried out an OH absorption survey towards eight associated and one intervening \hi absorbers at redshifts of z $\in$ [0.1919, 0.2241] using FAST, and constrained the OH-to-\hi relative abundance ([OH]/[\hi]) to be $\leqslant$ 5.45 $\times 10^{-8}$. Thanks to the high sensitivity of FAST, \citet{2023arXiv230808851K} investigated the \hi absorption spectrum of PKS\,1413+135, starting the first observational experiment to illustrate the cosmic acceleration by the direct measurement of the time evolution of the redshift of \hi 21cm absorption. \citet{2023ApJ...952..144Y} reported the discovery of three \hi absorbers toward low-power AGNs. \citet{2024ApJ...973...48C} reported 5 new \hi absorption detections in a \hi and OH absorption survey FAST towards 40 radio sources of low-intermediate radio luminosity and red mid-infrared color. These studies highlight that FAST is an excellent telescope for conducting \hi absorption science.

In \citet[][hereafter Paper I]{2023A&A...675A..40H}, we carried out a purely blind \hi absorption survey in the 1300-1450 MHz band of 3155\,deg$^{2}$ CRAFTS data. Three known associated absorbers (UGC\,00613, 3C\,293 and 4C\,+27.14) and two new absorbers (NVSS J231240\allowbreak-052547 and NVSS J053118\allowbreak+315412) were detected. The search technique detailed in Paper I was then extended to a broader range of CRAFTS \citep{2018IMMag..19..112L} and FASHI \citep{2024SCPMA..6719511Z} data, leading to a discovery of an increased number of \hi 21cm absorption features.

In this paper, we report a 21cm-selected sample of \hi absorption from a purely blind search in the 1050-1450 MHz band in the sky covered by the CRAFTS and FASHI (1325.6 hours and 6072.0 deg$^{2}$). Another 11 known \hi absorbers and 18 new \hi absorbers are found using our search pipeline. Combined with Paper I, a sample of 34 21cm-selected \hi absorption systems is obtained. We made a comprehensive statistical study for the intervening and associated absorption. 

This paper is organized as follows: Section~\ref{sec:Data} describes the survey data and follow-up observation used in this work. We summarize data processing, candidate selection method and \hi absorption measuring method in Section~\ref{sec:Data_Analysis}. The physical properties of confirmed \hi absorption are presented in Section~\ref{sec:Absorption_Systems}. Section~\ref{sec:Absorption_Properties} statistically studies the characters of associated and intervening absorption systems. We discuss the comoving absorption path and statistic completeness in Section~\ref{sec:Discussion}. In Section~\ref{sec:Summary} a summary of this work is presented. Throughout this paper we use H$_{0} = 70$ km s$^{-1}$ Mpc$^{-1}$, $\Omega_{\rm m} = 0.3$ and $\Omega_{\Lambda} = 0.7$.

\section{Data}
\label{sec:Data}
\subsection{Radio Data}
\subsubsection{CRAFTS}

CRAFTS is a multi-purpose drift-scan survey that aims to observe galactic and extragalactic \hi emissions, measure continuum signals, and search for new pulsars and fast radio bursts (FRB). The survey uses the FAST L-band Array of 19 feed-horns (FLAN, \citet{8105012}) covering the frequency band from 1050 MHz to 1450 MHz. CRAFTS drift scans started in early 2020 with the 19-beam feed rotated by 23.4° to achieve a super-Nyquist sampling while drifting. Two-pass drift scans are planned to be made and over 22000 deg$^{2}$ within a declination (Dec) range between $-14^\circ$ and $66^\circ$. Limited by allocated time, there is only one survey pass at present. 

\subsubsection{FASHI}
FASHI is a comprehensive observational initiative designed to map the sky accessible by FAST, encompassing an area of about 22,000 square degrees within declinations from $-14^\circ$ to $66^\circ$ and frequencies between 1050 and 1450 MHz. 
The primary objective of FASHI is to conduct a detailed survey of \hi gas in the nearby universe, with an additional goal to catalog at least 100,000 \hi galaxies. The survey employed a strategy of fixing declination and conducting drift scans throughout the project. From August 2020 to June 2023, FASHI successfully surveyed over 7,600 square degrees, achieving a median sensitivity of approximately 0.76 mJy beam$^{-1}$ and a spectral line velocity resolution of about 6.4 $\kms$ at 1.4 GHz. In their findings, \citet{2024SCPMA..6719511Z} reported the detection of 41,741 extragalactic \hi sources within the frequency range of 1305.5$-$1419.5 MHz, which corresponds to a redshift limit of z $\leqslant$ 0.09.

\begin{figure*}[hbt!]
    \centering
    \includegraphics[width=0.9\textwidth]{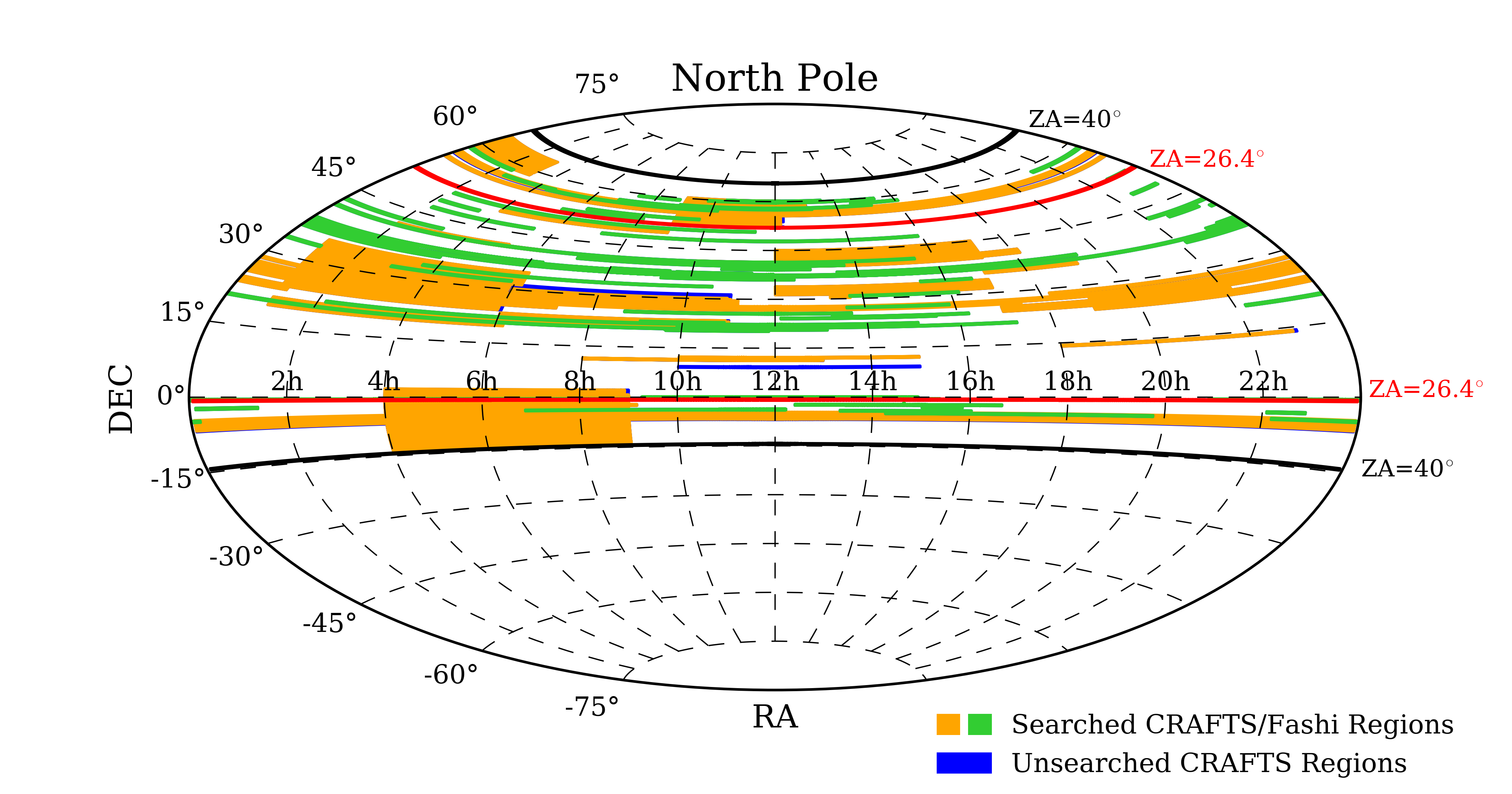}
    \caption{Up-to-date (2023-08-29) CRAFTS sky coverage in Equatorial coordinates. Search areas in CRAFTS and FASHI regions are highlighted by orange and green labels, respectively. Blue regions refer to CRAFTS sky that has not been searched. Zenith angle of 40° (maximum zenith angle for FAST) and 26.4° (zenith angle within which FAST has full gain) are shown as black circles and red circles, respectively.}
    \label{surveyed_sky}
\end{figure*}

As of August 29, 2023, our search has been conducted in part of the CRAFTS and FASHI regions, marked by the orange and green labels in Figure~\ref{surveyed_sky}. \footnote{Further information about the finished scans can be found in \url{http://groups.bao.ac.cn/ism/CRAFTS/CRAFTS/}.} 

\subsubsection{Follow-up Observation}
In the blind searching in CRAFTS data, a total of 34 candidates, including 14 previously known \hi absorption systems and 20 newly discovered ones are detected. All absorption systems are verified by the CRAFTS data of several neighboring beams with a high signal-to-noise ratio. To confirm the signal of these 20 first discovered candidates and obtain a more significant \hi absorption feature for known absorption systems, we made several follow-up observations with FAST from 2021 to 2023.

To suppress the fluctuations in bandpass and remove the sky signal, follow-up observations used ON-OFF tracking mode. The integration time for both the source-on and source-off observations is set as 990s. A cycle of 330s source-on followed by 330s source-off observation is repeated 3 times for each target. The data set of follow-up observations has a time resolution of 1 s, and a frequency resolution of 7.63 kHz(equivalent to 1.6 $\kms$ at the redshift of 0), covering the frequency from 1050 MHz to 1450 MHz. A noise level of 0.338 mJy/beam can be achieved with an integration time of 990 seconds and a frequency resolution of 7.63 kHz.

\subsection{Infrared Data}

We extract the infrared properties of the absorption systems from the Wide-field Infrared Survey Explorer (WISE; \citealt{2010AJ....140.1868W}) data. WISE is an all-sky survey mapping the infrared sky at 3.4, 4.6, 12, and \SI{22}{\micro\metre} (W1, W2, W3, and W4) with angular resolutions of 6.1 arcsec, 6.4 arcsec, 6.5 arcsec and 12.0 arcsec in the four bands respectively. We recognize the WISE counterparts of the \hi absorption systems using the NASA/IPAC Extragalactic Database (NED, \citealt{1991ASSL..171...89H}). WISE continuum map, if available, for each \hi absorption system is extracted from the NASA/IPAC Infrared Science Archive (IRSA) and presented below. 

The W1-W2$>$0.8 \citep{2012ApJ...753...30S} threshold is a widely used and reliable diagnostic for identifying AGN in mid-infrared surveys like WISE. AGNs exhibit a characteristic infrared excess caused by hot dust near the central black hole \citep{2006MNRAS.366..767F}, which emits strongly in the W2 band, leading to a W1-W2 color greater than 0.8. In this work, we use the W1-W2 threshold to infer the origin of the infrared emission and assess the presence of AGN associated with each \hi absorption. Additionally, the WISE color-color diagram \citep{2010AJ....140.1868W} is a valuable tool for identifying and classifying various types of astronomical objects based on their infrared properties. This method is employed in our analysis to classify object types (see Section~\ref{sec:WISE_color_color}).

\subsection{Optical Data}

The optical observation data of our background sources are obtained from the Sloan Digital Sky Survey (SDSS, \citealt{2000AJ....120.1579Y}) Data Release 16. The optical counterparts of the background/foreground sources are selected by cross-matching the coordinates of the sources with the SDSS catalog using a 2-arcsec matching radius. Among the 29 \hi absorption detections presented in this paper, 6 are outside the footprint of SDSS.

\section{Analysis}
\label{sec:Data_Analysis}
\subsection{Searching Algorithms}
\label{sec:algorithms}

Considering both CRAFTS and FASHI surveys have been conducted in the drift-scan mode, we rebinned the data into a time resolution of $\sim$ 12$/\cos\delta_{\rm{dec}}$ s ($\sim$ the transit time in drift scan, where $\delta_{\rm{dec}}$ is the declination of the pointing) and a frequency resolution of $\Delta\nu$ = 15.26 kHz, and applied the same data-processing method to them. The search technique has been described in detail in Paper I. In summary, after the removal of the baseline (estimated using a low-pass filter), \hi absorbers are blindly searched by cross-correlating flux spectra with Gaussian templates (matched-filtering approach, \citep{2007AJ....133.2087S}). We searched for \hi absorption signal in both XX and YY polarizations of each beam. We select candidates with a combined velocity-integrated signal-to-noise ratio ($S/N$) exceeding 5.5 and those that are found at nearly the same frequency ($\delta f < 0.04$ MHz) with an individual $S/N$ exceeding 3.5 in both XX and YY polarizations. Final candidates are selected by use of transit information recorded by the 19-beams of FAST.

\subsection{\hi Absorption Measurement}
\label{sec:theory}

\hi 21-cm absorption lines arise from foreground \hi gas absorbing the flux of the background bright radio source. The \hi column density of the foreground gas can be calculated by integrating observed absorption over velocity \citep{2018A&ARv..26....4M}:
\begin{eqnarray}
    N_{\hi}[\mathrm{cm^{-2}}] = 1.82 \times 10^{18} T_{\rm s}[\mathrm{K}] \int \tau(V)dV[\mathrm{km\,s^{-1}}],
    \label{column_density}
\end{eqnarray}
where $\tau(V)$ and $T_{\rm s}$ refer to optical depth and spin temperature of the \hi source, respectively. Under the assumption that $T_{\rm s} \ll c_{\rm f}T_{\rm c}$, the optical depth $\tau(V)$ can be expressed as:
\begin{eqnarray}
    \tau(V) = -\mathrm{ln}(1+\Delta T(V)/(c_{\rm f}T_{\rm c})),
    \label{tau}
\end{eqnarray}
where $c_{\rm f}$ is covering factor. $T_{\rm c}$ is the brightness temperature of the background continuum source, which was deduced from the line-free parts of the spectrum in this work. $\Delta T(V)$ is the difference between the observed signal of \hi gas and brightness temperature of background continuum source. If $T_{\rm s}>c_{\rm f}T_{\rm c}$, the spectral line will be seen in emission, while if $T_{\rm s} < c_{\rm f}T_{\rm c}$, absorption dominates.

Throughout this paper we take the assumption that $T_{\rm s} \ll c_{\rm f}T_{\rm c}$ and $c_{\rm f} = 1$ \citep{2017A&A...604A..43M} for the calculation of $\tau$ and $N_{\hi}$. The new \hi absorption will be classified as ``associated absorption'' if the velocity difference between the \hi absorption and the background is within 1500 $\kms$.

\section{Verified \hi Absorption Systems}
\label{sec:Absorption_Systems}

From August 2021 to October 2023, we conducted follow-up observations using FAST in the ON-OFF tracking observation mode, with 990 seconds of on-source integration for each absorption system. In total, we verified 34 \hi absorption, comprising 14 previously known absorbers and 20 newly discovered ones. The spectra of these \hi absorption were calibrated using a built-in noise diode that was activated for 1 second every 8 seconds. To remove the bandpass baseline, a polynomial function was fitted to the absorption line-free parts of the spectrum. A Doppler shift correction was then applied to each absorption spectrum. Finally, we employed multi-component Gaussian functions to fit the profiles of authentic \hi absorption.

For each source, we display the absorption spectrum along with its Gaussian fitting. Additionally, we present an image centered on the radio source, created by using radio data at S-band from the Karl G. Jansky Very Large Array Sky Survey (VLASS, \citet{2020PASP..132c5001L}), along with infrared data from WISE and optical data from SDSS (when available).

We document the fundamental physical information for each radio source in Table~\ref{radiosource_table}. Tables~\ref{WISE_counterpart_table} and \ref{SDSS_counterpart_table} display infrared and optical magnitudes corresponding to the WISE and SDSS counterparts of the radio sources, respectively. Table~\ref{known_absorption_table_1} shows the measurements of \hi absorption signals and the corresponding \hi column densities for previously identified absorption. Additionally, Tables~\ref{new_absorption_table_1} provides the same information for the newly detected absorbers.

\subsection{Previously Known Absorbers}

\subsubsection{4C +56.02}

4C\,+56.02 is also a little-studied bright compact radio source. The absence of redshift data for 4C\,+56.02 has left the identification of the foreground counterpart in an uncertain state, emphasizing the need for further high-resolution follow-up observations to attain clarity. %The WISE counterpart to 4C\,+56.02 is WISEA J011057.55+563216.8 as shown in the NASA/IPAC Extragalactic Database. The WISE W1[\SI{3.4}{\micro\metre}], W2[\SI{4.6}{\micro\metre}], W3[\SI{12.1}{\micro\metre}] and W4[\SI{22.2}{\micro\metre}] magnitudes for 4C\,+56.02 are 14.795 $\pm$ 0.033, 14.021 $\pm$ 0.043, 12.190 $\pm$ 0.329 and 8.956, respectively. The W1-W2 color of WISEA J011057.55+563216.8 is 0.774, indicating that the mid-IR emission comes mainly from the stars.
\citet{1982AJ.....87..278D} used the 91m- and 43m- single-dish telescopes as an interferometer to observe its 21-cm line, and reported the parameter of absorption as peak optical depth $\tau_{\rm peak} \sim 0.48$ and $N_{\hi} \sim 22.44\times$10$^{20}$cm$^{-2}$K$^{-1}$. 

\begin{figure*}[hbt!]
    \centering
    \includegraphics[width=0.25\textwidth]{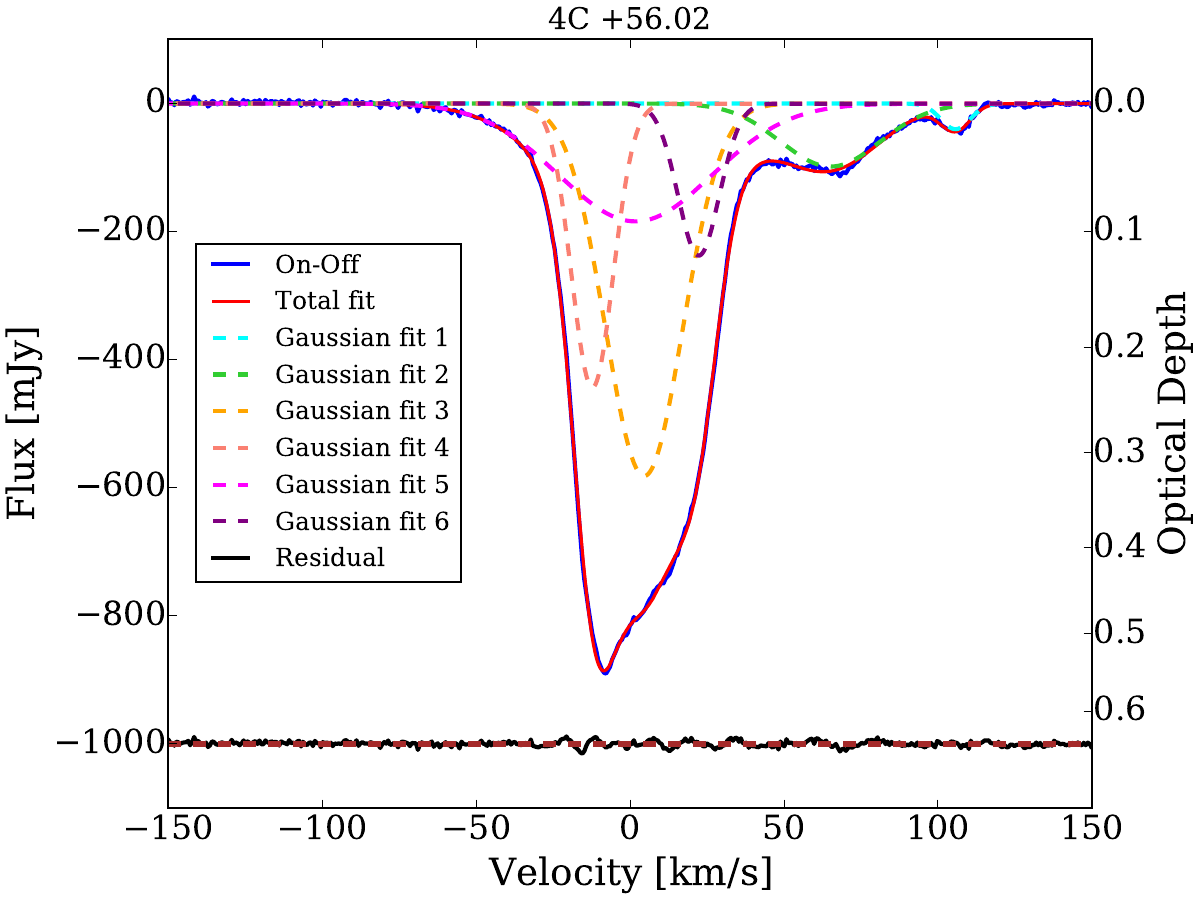}
    \includegraphics[width=0.25\textwidth]{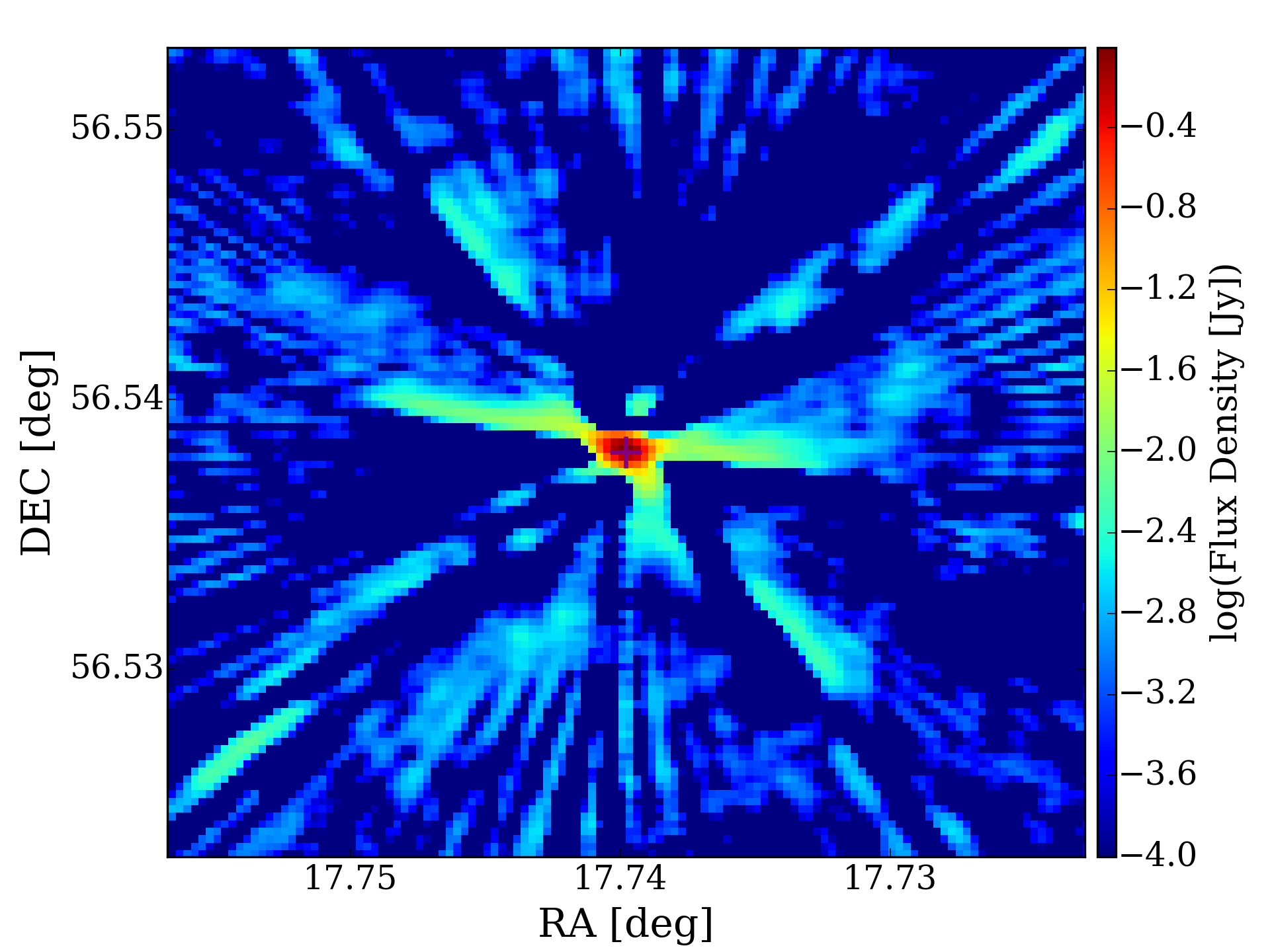}
    \includegraphics[width=0.25\textwidth]{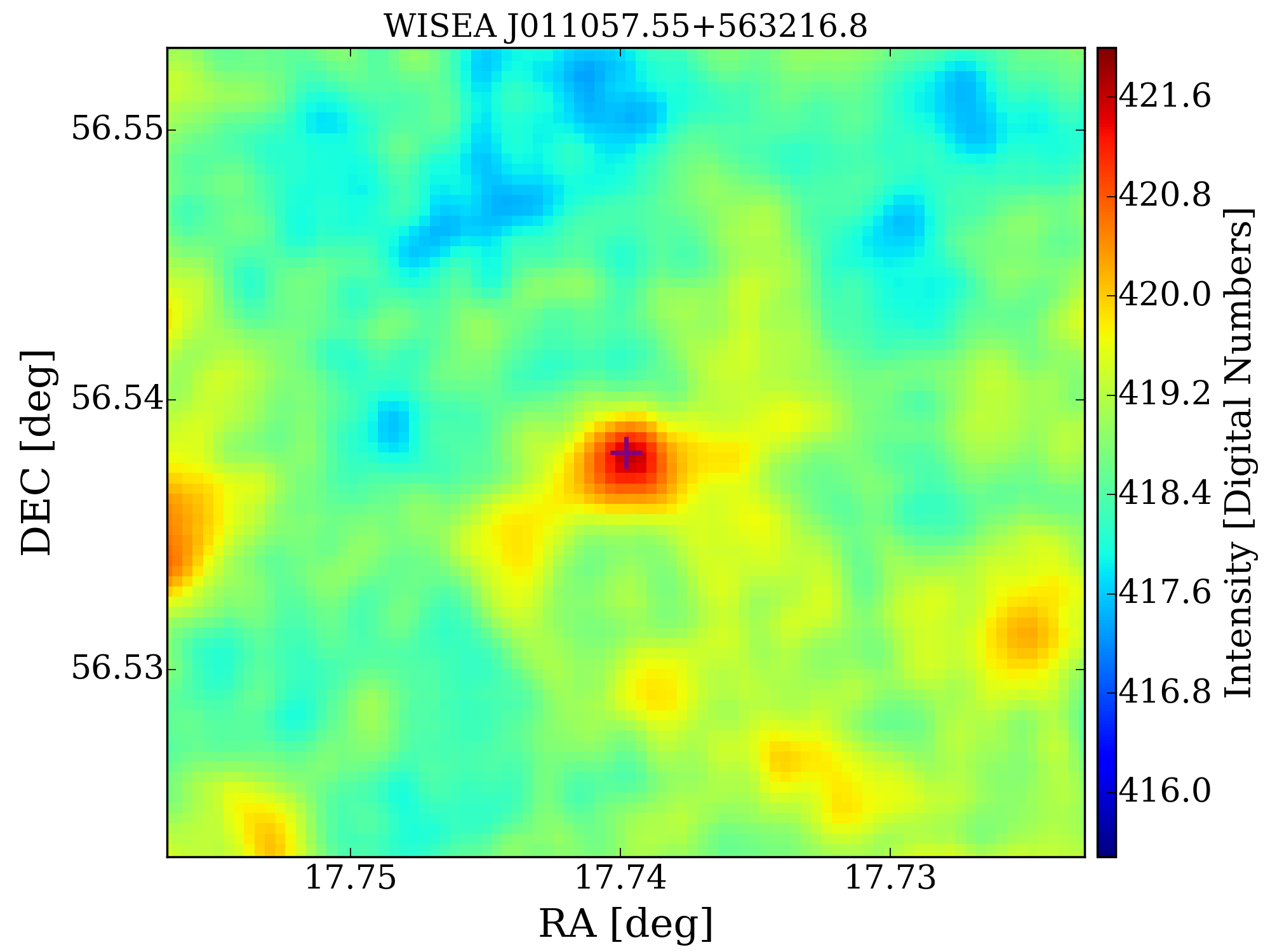}
    \caption{Left: \hi absorption feature of 4C\,+56.02. The blue solid line shows the absorption spectrum and the red solid line shows the fit with a six-component Gaussian model. The fitting residual is shown as a black-solid line at the bottom. The optical depth value for \hi absorption is shown on the right scale. Middle: the radio map from VLASS centered at 4C\,+56.02. Right: the W3 band infrared map centred at 4C\,+56.02.}
    \label{4C+56.02_fit}
\end{figure*}

We blindly re-detected its absorbing profile and presented its absorption spectrum in the left panel of Figure~\ref{4C+56.02_fit}. Its absorption spectrum is well-fitted using the six-components Gaussian function. Our measurements of the \hi absorption towards 4C\,+56.02 give a flux density depth of $S_{\hi,\rm peak} \sim -885.85 \mJy$ and FWHM of $\sim 46.32 \kms$, peak optical depth $\tau_{\rm peak} \sim 0.55$ and $N_{\hi} \sim 0.47 T_{s}\times 10^{20}\cm^{-2}\K^{-1}$.

%The middle and right panel of Figure~\ref{4C+56.02_fit} show the image centered at 4C\,+56.02, constructed using radio data at S-band from VLASS and infrared data at 12.1 microns (W3) from WISE, respectively. The straight cross shows the position of 4C\,+56.02.

\subsubsection{4C +31.04}

\begin{figure*}[hbt!]
    \centering
    \includegraphics[width=0.25\textwidth]{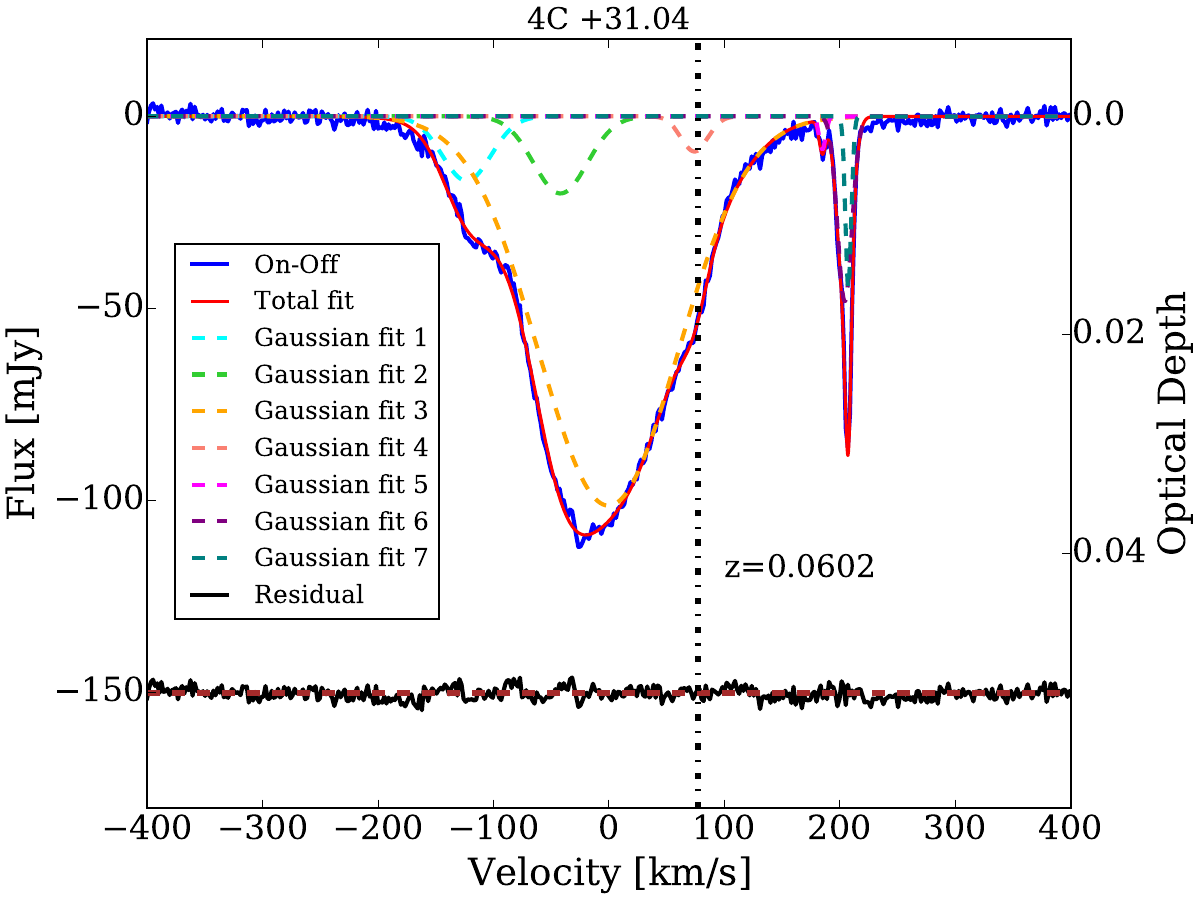}
    \includegraphics[width=0.25\textwidth]{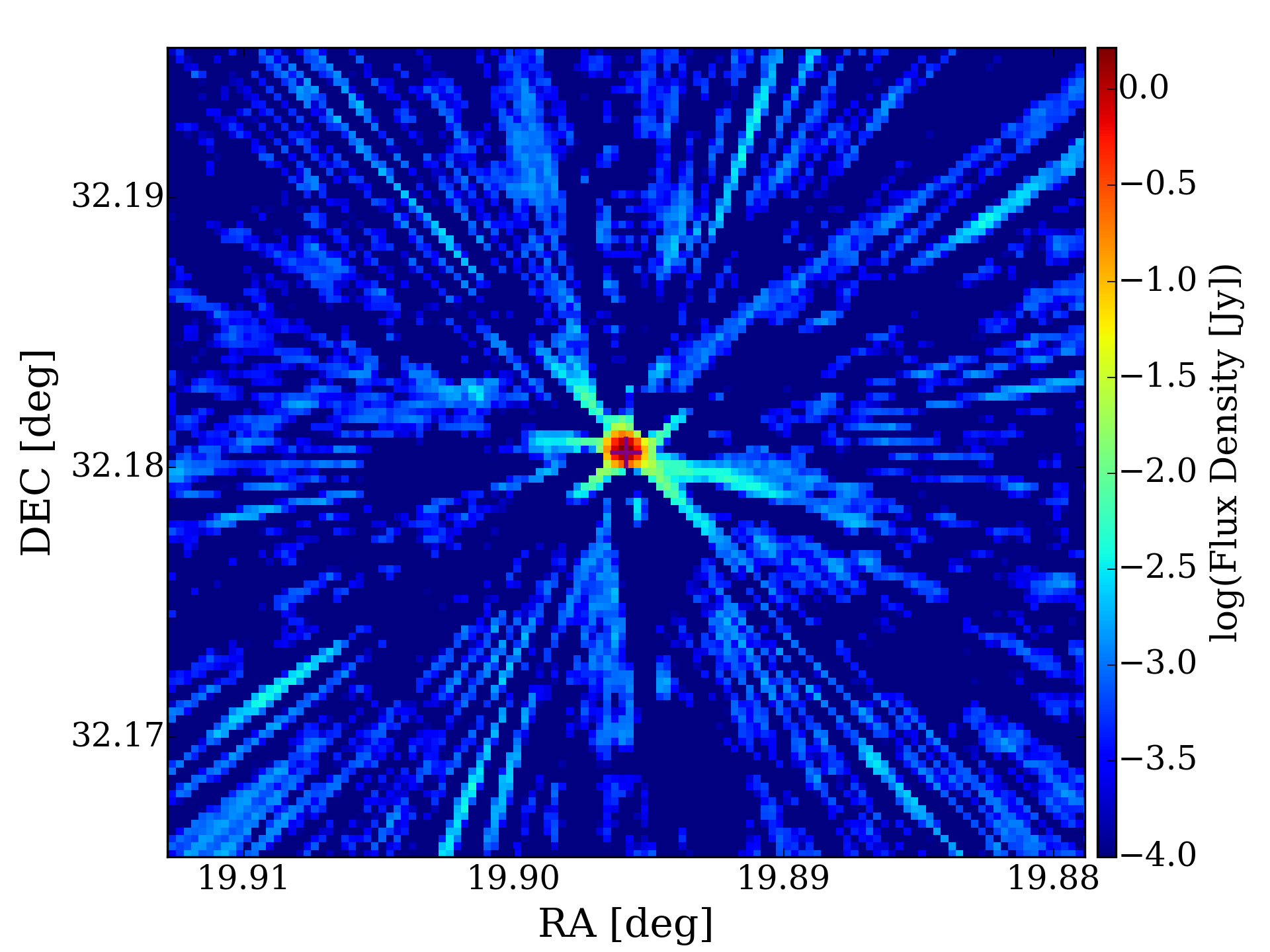}
    \includegraphics[width=0.25\textwidth]{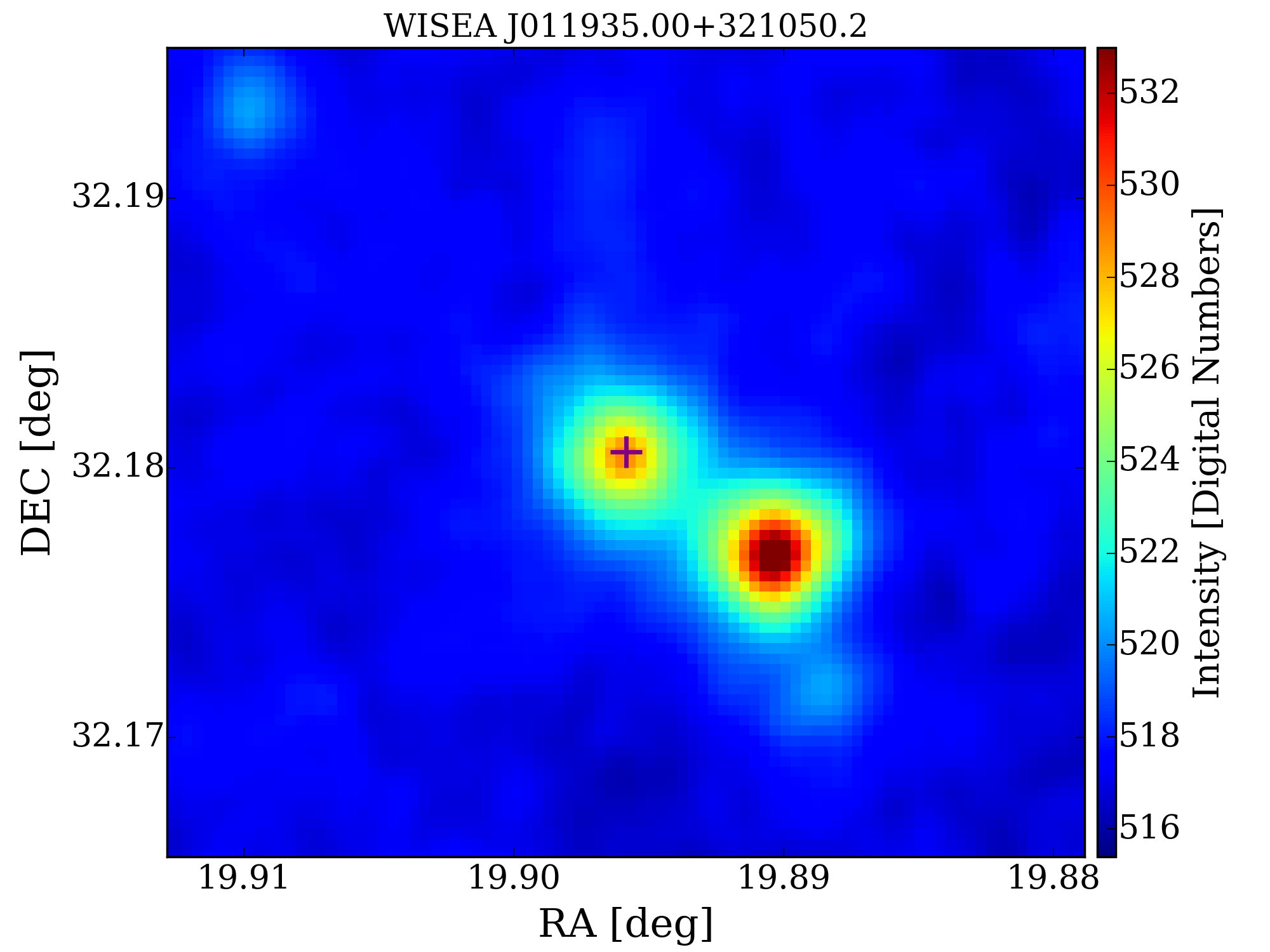}
    \hspace*{0.25cm}
    \includegraphics[width=0.18\textwidth]{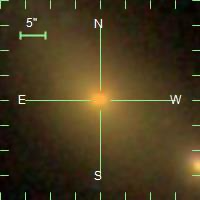}
    \caption{Left: same as Figure~\ref{4C+56.02_fit}, but for 4C\,+31.04. The black dot-dashed vertical line marks the redshift as given in the NED. Middle left: radio map from VLASS centered at 4C\,+31.04. Middle right: W3 band infrared map of WISEA J011935.00\allowbreak+321050.2 from WISE. Right: optical map of 4C\,+31.04 from SDSS.}
    \label{4C+31.04_fit}
\end{figure*}

4C\,+31.04 (Figure~\ref{4C+31.04_fit}) is a GHz-peaked spectrum (GPS) and Compact Steep Spectrum (CSS) source with compact-symmetric-object (CSO) morphology \citep{2003A&A...399..889G,2017ApJ...849...34O}. The host galaxy of 4C\,+31.04 is MCG 5-4-18, a giant elliptical galaxy with a Seyfert 2-like optical spectrum. 4C\,+31.04 has two asymmetric lobes, suggesting strong interactions between the jets and a dense ISM \citep{2001ApJ...552..508G}. Shocked molecular and ionized gas resulting from jet-driven feedback in 4C\,+31.04 was observed and reported by \citet{2019MNRAS.484.3393Z}. The jet-blown bubble pushes a forward shock into ambient ISM, giving rise to [Fe II] emission. The H$_{2}$ emission, arising from the shock-excited molecular gas, is also detected. 
4C\,+31.04 has a close companion, MCG 5-4-17, which is a spiral galaxy at a projected distance of $\sim$20 kpc and a velocity offset of 1560 \kms.

The WISE counterpart to 4C\,+31.04 given by NED is WISEA J011935.00\allowbreak+321050.2. WISEA J011935.00\allowbreak+321050.2 has a W1-W2 of 0.021, indicating that the mid-IR emission comes mainly from stars. 

%The WISE W1[\SI{3.4}{\micro\metre}], W2[\SI{4.6}{\micro\metre}], W3[\SI{12.1}{\micro\metre}] and W4[\SI{22.2}{\micro\metre}] magnitudes for WISEA J011935.00+321050.2 are 11.641 $\pm$ 0.023, 11.620 $\pm$ 0.021, 10.024 $\pm$ 0.039 and 7.578 $\pm$ 0.100, respectively. 
 
The \hi absorption associated with 4C\,+31.04 was firstly reported in \citet{1989AJ.....97..708V} with a velocity of 17391 $\pm$ 3 \kms, a FWHM of 153 $\pm$ 6 \kms, peak optical depth $\tau_{\rm peak} \sim 0.037$ and $N_{\hi} \sim 0.096T_{s}$10$^{20}$cm$^{-2}$K$^{-1}$. \citet{2006MNRAS.373..972G} presented the results of \hi and OH absorption measurements towards 4C\,+31.04 using the Arecibo 305-m telescope, giving the peak optical depth $\tau_{\rm peak} \sim 0.038$ and $N_{\hi} \sim 0.122T_{s}$10$^{20}$cm$^{-2}$K$^{-1}$. \citet{2012A&A...546A..22S} presented Very Long Baseline Array (VLBA) \hi absorption observations of the circumnuclear environments of 4C\,+31.04, and absorption was detected against the core and both lobes. 

We blindly re-detected its associated absorbing profile and updated its parameter by use of the data from the follow-up observation. We present the up-to-date finest structure of its absorption profile in Figure~\ref{4C+31.04_fit}, which is fitted by the seven-components Gaussian function. Our measurements of the \hi absorption associated with 4C\,+31.04 give a flux density depth of $S_{\hi,\rm peak} \sim -108.96 \mJy$, an FWHM of $\sim 151.94 \kms$, $\int\tau dv\sim 6.55 \kms$ and $N_{\hi} \sim 0.119T_{s}$10$^{20}$cm$^{-2}$K$^{-1}$, which is consistent with \citet{1989AJ.....97..708V} and \citet{2006MNRAS.373..972G}.

Besides the broad absorption centered at 17926.39 \kms, a narrow absorption component at 18152.40 \kms ~is also detected in the CRAFTS data. Our detection of the sub-absorption is consistent with the absorption profile presented in \citet{1989AJ.....97..708V} and \citet{2012A&A...546A..22S}. 4C\,+31.04 does not show any sign of recent interaction such as tidal tails \citep{2001AJ....122..536P}. On the other hand, the velocity offset between 4C\,+31.04 and its companion far exceeds the measured velocity difference between broad and narrow absorption ($\sim$ 200 \kms). The narrow absorption may come from another un-discovered optical-faint but gas-rich companion galaxy \citep{2012A&A...546A..22S}. 

%The middle left, middle right panel and right panel of Figure~\ref{4C+31.04_fit} show the image centered at 4C\,+31.04, constructed using radio data at S-band from the Karl G. Jansky Very Large Array Sky Survey (VLASS, \citet{2020PASP..132c5001L}), infrared data at 12.1 microns (W3) from WISE and optical data from SDSS, respectively. The straight cross shows the position of 4C\,+31.04. The position of 4C\,+31.04 matches WISEA\,J011935.00+321050.2 and the source in VLASS very well. 

\subsubsection{3C 84 (NGC 1275)}

\begin{figure*}[hbt!]
    \centering
    \includegraphics[width=0.2\textwidth]{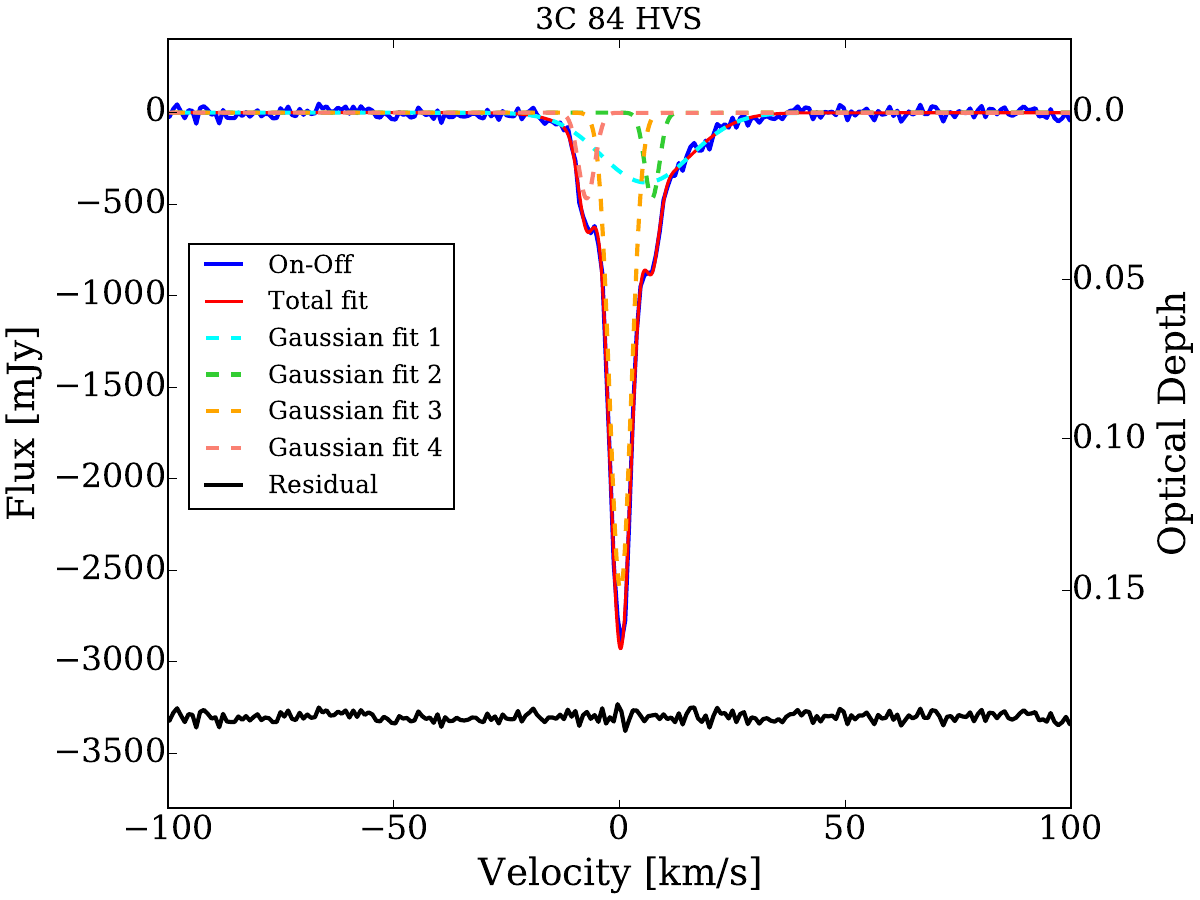}
    \includegraphics[width=0.2\textwidth]{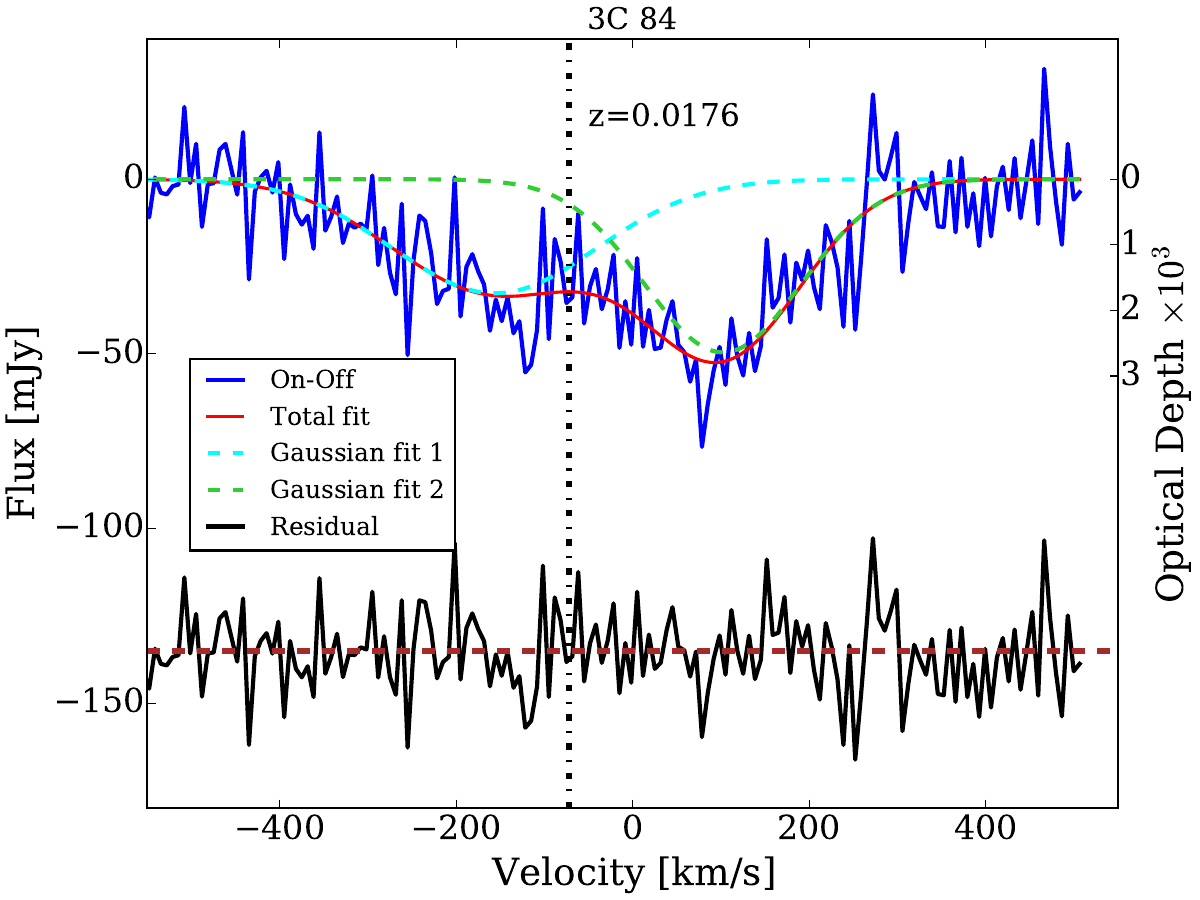}
    \includegraphics[width=0.2\textwidth]{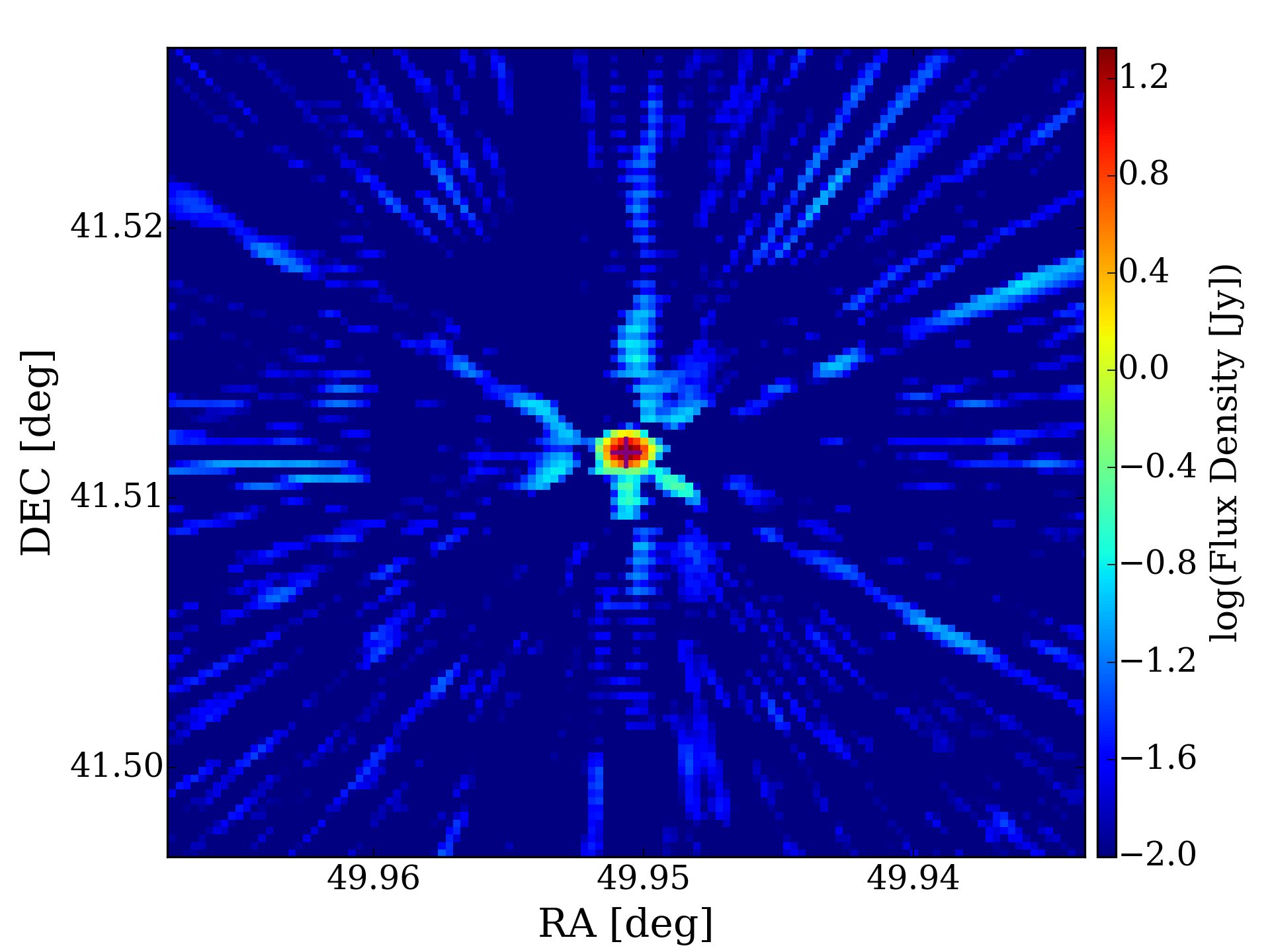}
    \includegraphics[width=0.2\textwidth]{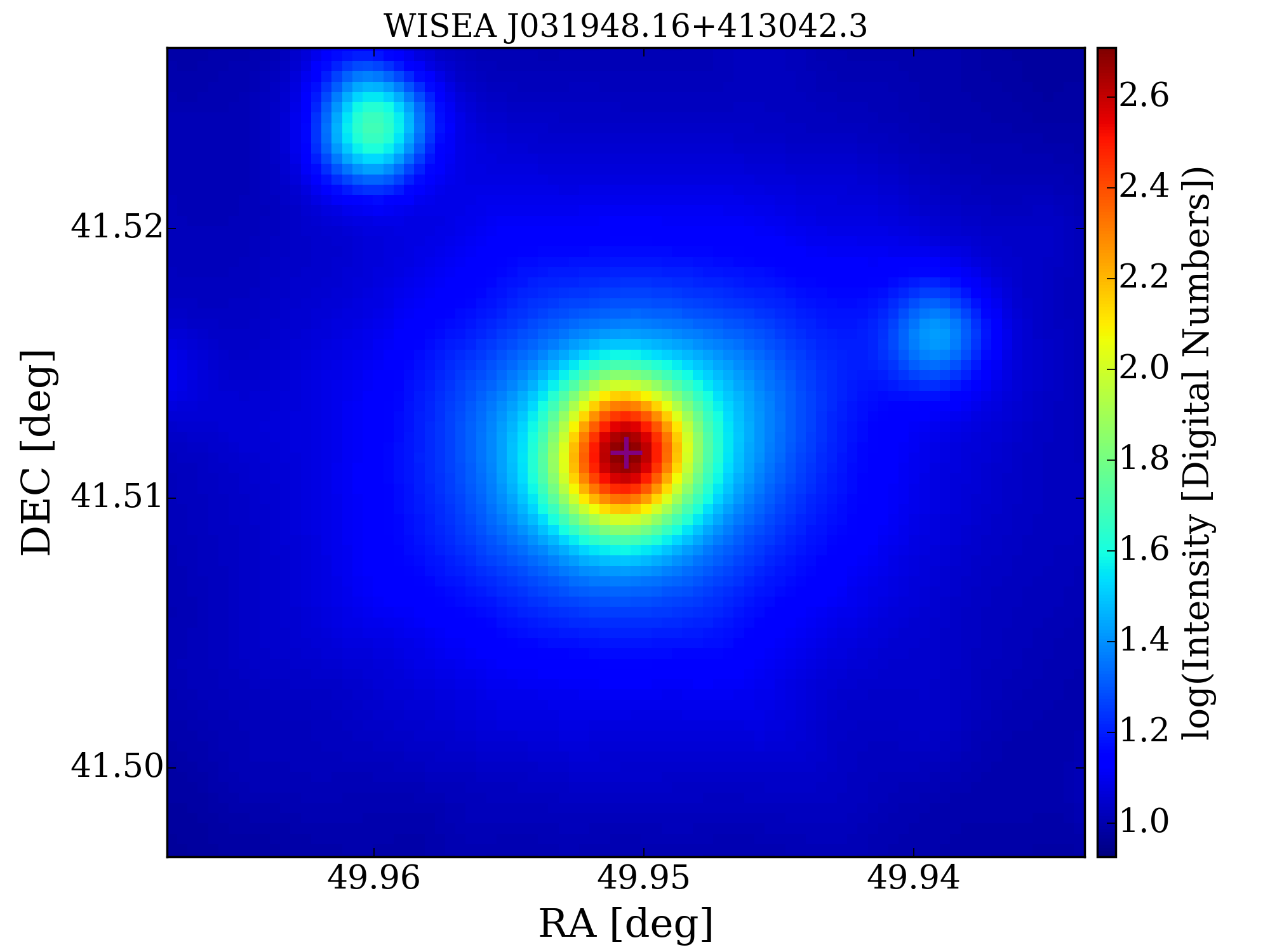}
    \includegraphics[width=0.15\textwidth]{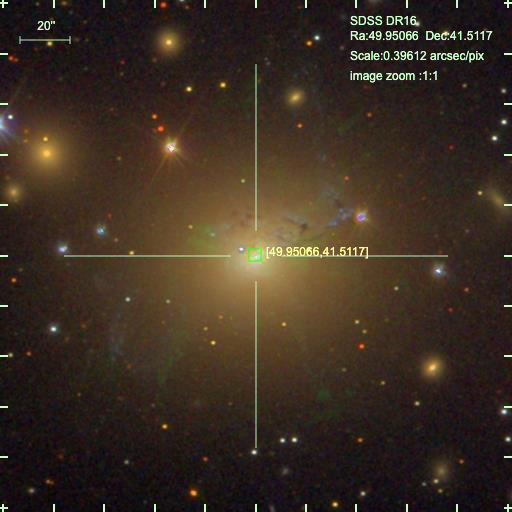}
    \caption{Left and Middle Left: same as Figure~\ref{4C+56.02_fit}, but for 3C\,84 HVS and 3C\,84. Middle: the radio map from VLASS centered at 3C\,84. Middle right: W2 band infrared map of WISEA J031948.16\allowbreak+413042.3 from WISE. Right: SDSS optical map of the optical counterpart of WISEA J031948.16\allowbreak+413042.3.}
    \label{3C_84_fit}
\end{figure*}

NGC\,1275 (Figure~\ref{3C_84_fit}) is a giant cD elliptical galaxy situated close to the center of the Perseus cluster, stands out as the nearest cool-core cluster Brightest Cluster Galaxy (BCG) detected with \hi, showcasing a rich cluster environment \citep{2023MNRAS.519.4128S}. The galaxy hosts an active nucleus, evident through the presence of the powerful compact radio source 3C\,84. Positioned in front of NGC\,1275 is another high-velocity system (HVS) at around 8200 $\kms$, observable in both H$\alpha$ emission \citep{1992ApJ...388..301C} and \hi absorption \citep{1973ApJ...185..809D}. Through comprehensive global VLBI observations, \citet{Momjian_2002} presented a high dynamic range image of this \hi absorption feature, detecting six distinct absorption peaks that suggest the presence of multiple \hi clouds. The likely nature of the foreground object is a gas-rich galaxy.

In addition to the \hi absorption associated with the HVS, a broad \hi absorption feature centered on the systemic velocity of NGC \,1275 (around 5300 $\kms$) was also detected \citep{1990A&A...240..254J}. Recently, \citet{2023A&A...678A..42M}, using JVLA and VLBA observations, detected both narrow and broad \hi absorptions with arcsecond resolution. By comparing the properties of the HI absorption to those of the molecular circumnuclear disc (CND) in NGC\,1275, they concluded that the \hi arises from the fast-rotating CND, and that neutral atomic hydrogen is present as close as ~20 pc from the SMBH. Additionally, with the JVLA, they discovered a new, faint absorbing system redshifted by approximately 2660 $\kms$ relative to NGC 1275, which they identified as gas stripped from a foreground galaxy falling into the Perseus cluster.

In our data, we re-detected the previously known narrow and broad \hi absorptions associated with the HVS and NGC\,1275. As shown in Figure~\ref{3C_84_fit}, our measurements are in close agreement with those from earlier studies.

%The WISE counterpart to 3C\,84 is WISEA\,J031948.16+413042.3 as shown in the NASA/IPAC Extragalactic Database. The WISE W1[\SI{3.4}{\micro\metre}], W2[\SI{4.6}{\micro\metre}], W3[\SI{12.1}{\micro\metre}] and W4[\SI{22.2}{\micro\metre}] magnitudes for WISEA J031948.16+413042.3 are 9.136 $\pm$ 0.022, 8.113 $\pm$ 0.019, 4.062 $\pm$ 0.014 and 1.150 $\pm$ 0.014, respectively. The W1-W2 color of WISEA J031948.16+413042.3 is 1.023, which means that the mid-IR emission comes mainly from the AGN. According to the W2-W3 value of 4.051 mag, WISEA J031948.16+413042.3 locates in the intersection of the Seyferts region and the Starburst region in the WISE color-color diagram. 

\subsubsection{4C -06.18}

\begin{figure*}[hbt!]
    \centering
    \includegraphics[width=0.25\textwidth]{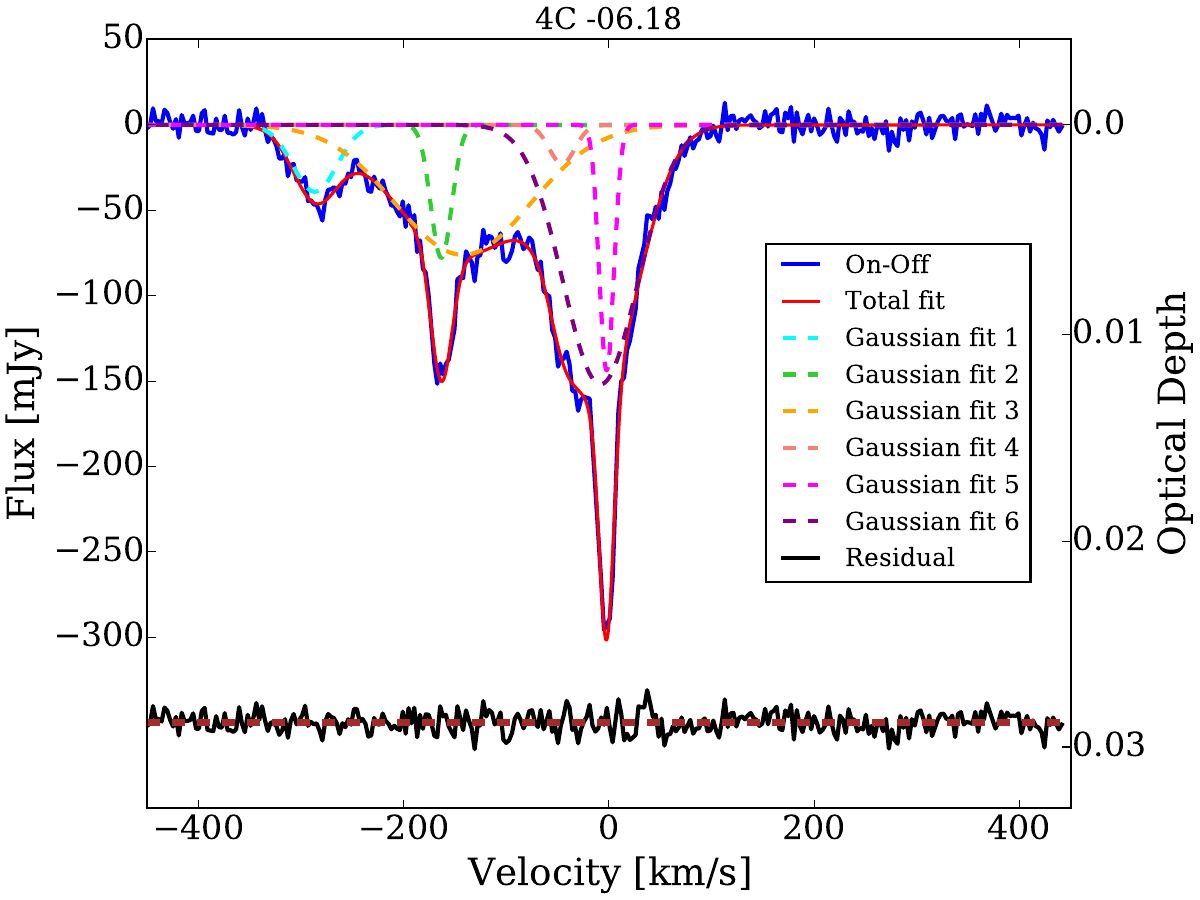}
    \includegraphics[width=0.25\textwidth]{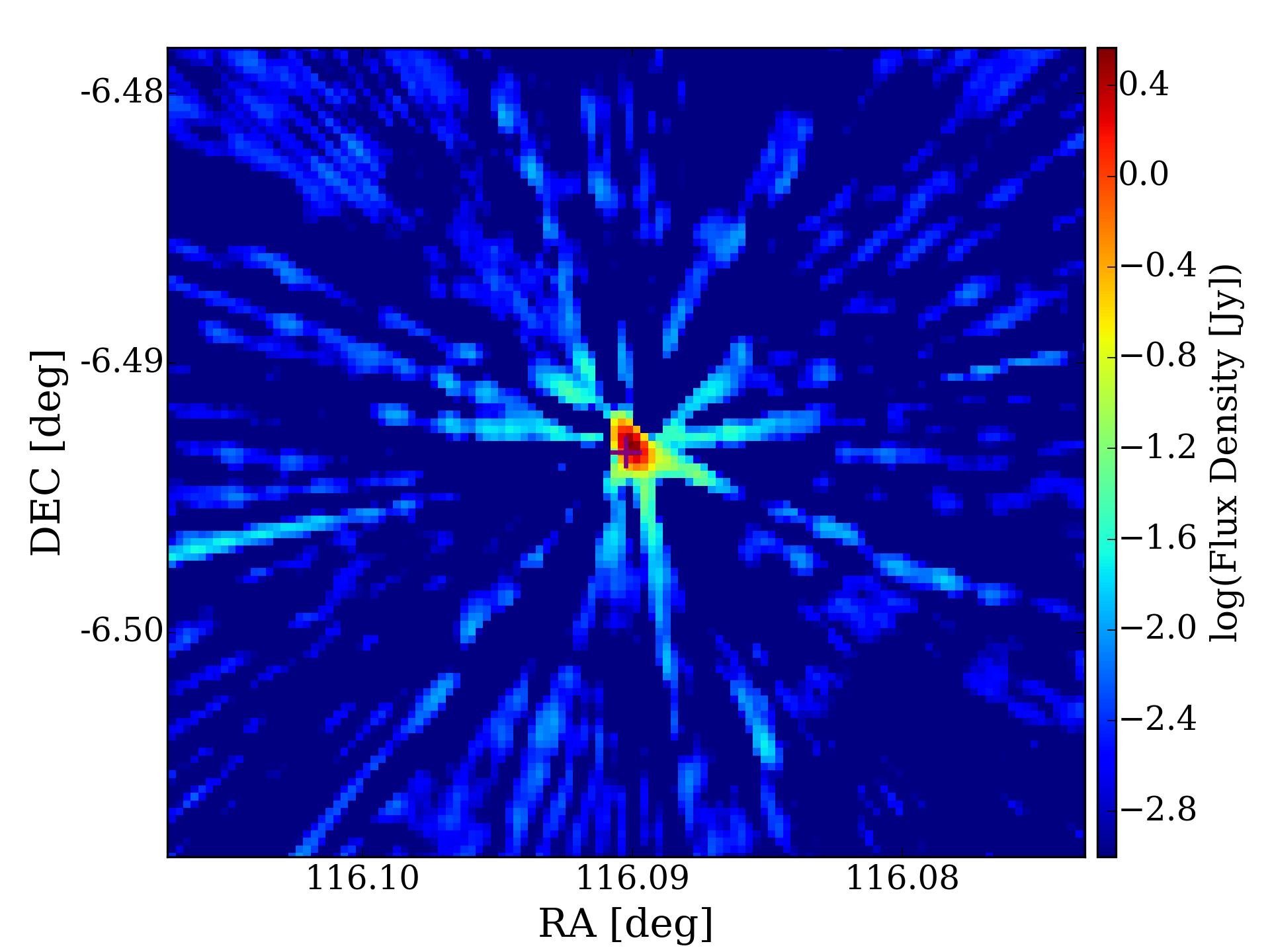}
    \includegraphics[width=0.25\textwidth]{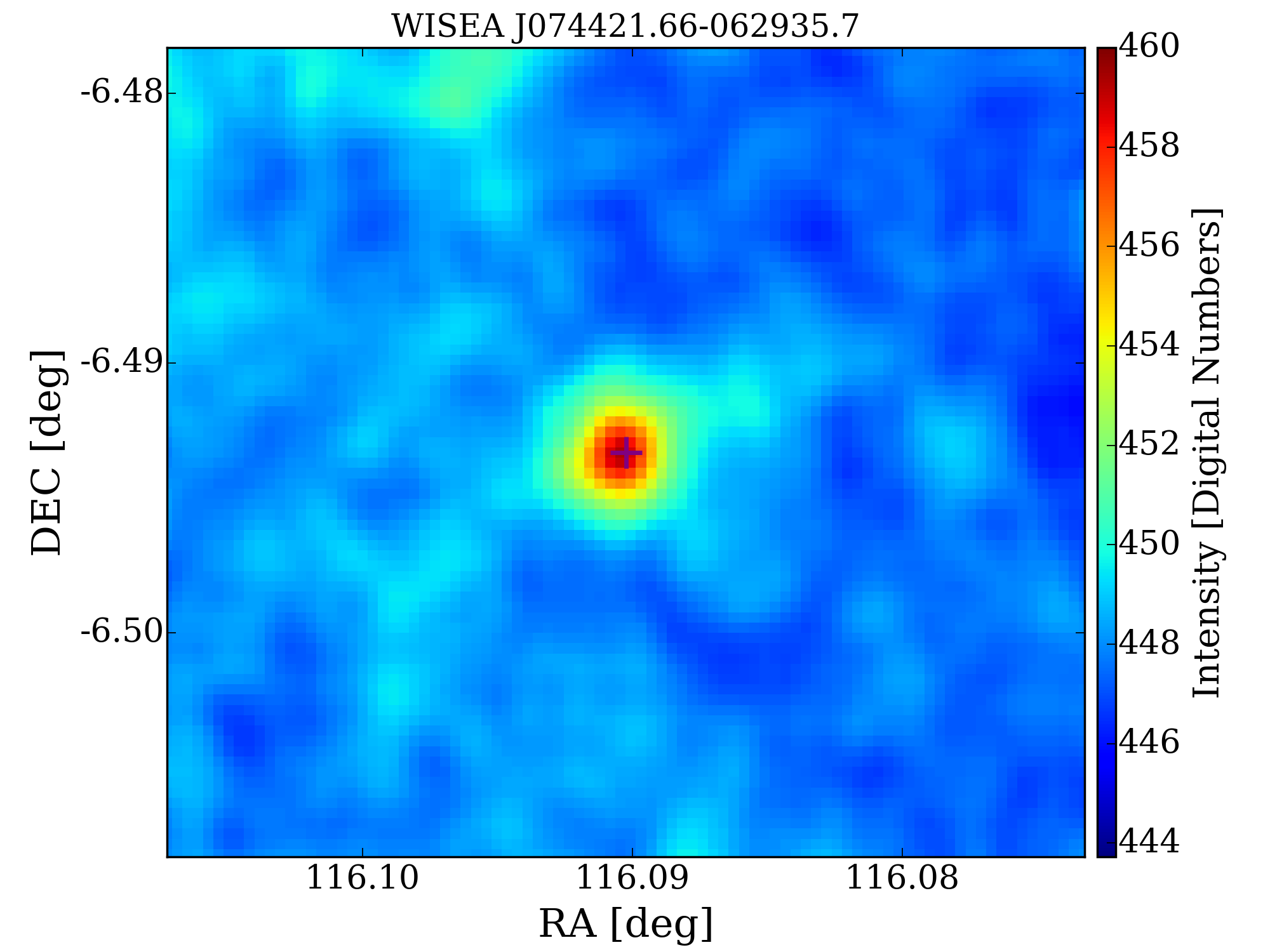}
    \caption{Left: same as Figure~\ref{4C+56.02_fit}, but for 4C\,-06.18. Middle: the radio map from VLASS centered at 4C\,-06.18. Right: the W3 band infrared map from WISE centred at 4C\,-06.18.}
    \label{4C-06.18_fit}
\end{figure*}

4C\,-06.18 (Figure~\ref{4C-06.18_fit}) is a little-studied bright radio source, its redshift information is unavailable, leaving the counterpart to the foreground in an uncertain state. It necessitates additional high-resolution follow-up observations. %The WISE counterpart to 4C\,-06.18 is WISEA J074421.66\allowbreak-062935.7 as shown in the NASA/IPAC Extragalactic Database. The WISE W1[\SI{3.4}{\micro\metre}], W2[\SI{4.6}{\micro\metre}], W3[\SI{12.1}{\micro\metre}] and W4[\SI{22.2}{\micro\metre}] magnitudes for WISEA J074421.66\allowbreak-062935.7 are 15.440 $\pm$ 0.056, 14.327 $\pm$ 0.062, 9.821 $\pm$ 0.043 and 6.232 $\pm$ 0.048, respectively. The W1-W2 colour of WISEA J074421.66\allowbreak-062935.7 is 1.113, indicating that the AGN in WISEA J074421.66-062935.7 contributes a large fraction to total emission. According to the W2-W3 value of 4.506 mag, WISEA J074421.66-062935.7 is located in the Luminous Infrared Galaxy (LIRG) region in the WISE color–color diagram. Taking into consideration the measurement errors WISEA J074421.66\allowbreak-062935.7 probably located in any region of Seyferts, Ultra-luminous Infrared Galaxies (ULIRGs) and Obscured AGN. 
The absorption towards 4C\,-06.18 was firstly and exclusively reported by \citet{1978MNRAS.182..209R}, with the following absorption parameters: peak optical depth $\tau_{\rm peak} \sim 1.4$, $\int\tau dv\sim 4.5 \kms$ and $N_{\hi} \sim 2.9\times$10$^{20}$cm$^{-2}$ (spin temperature $T_{\rm s} \sim 35$K), which are close to our measurements. We blindly re-detected its absorbing profile and obtained its fine structure in the follow-up observation. Its absorption spectrum is well-fitted using the six-components Gaussian function and is shown in the left panel of Figure~\ref{4C-06.18_fit}. The strongest \hi absorption profile exhibits a nearly symmetric and narrow profile, likely originating from a gas disk. Additionally, there is a broader wing with two shallower peaks at lower redshifts, indicating the existence of unsettled gas structures and a potential jet of gas.

\subsubsection{NVSS J085521+575143}

\begin{figure*}[hbt!]
    \centering
    \includegraphics[width=0.25\textwidth]{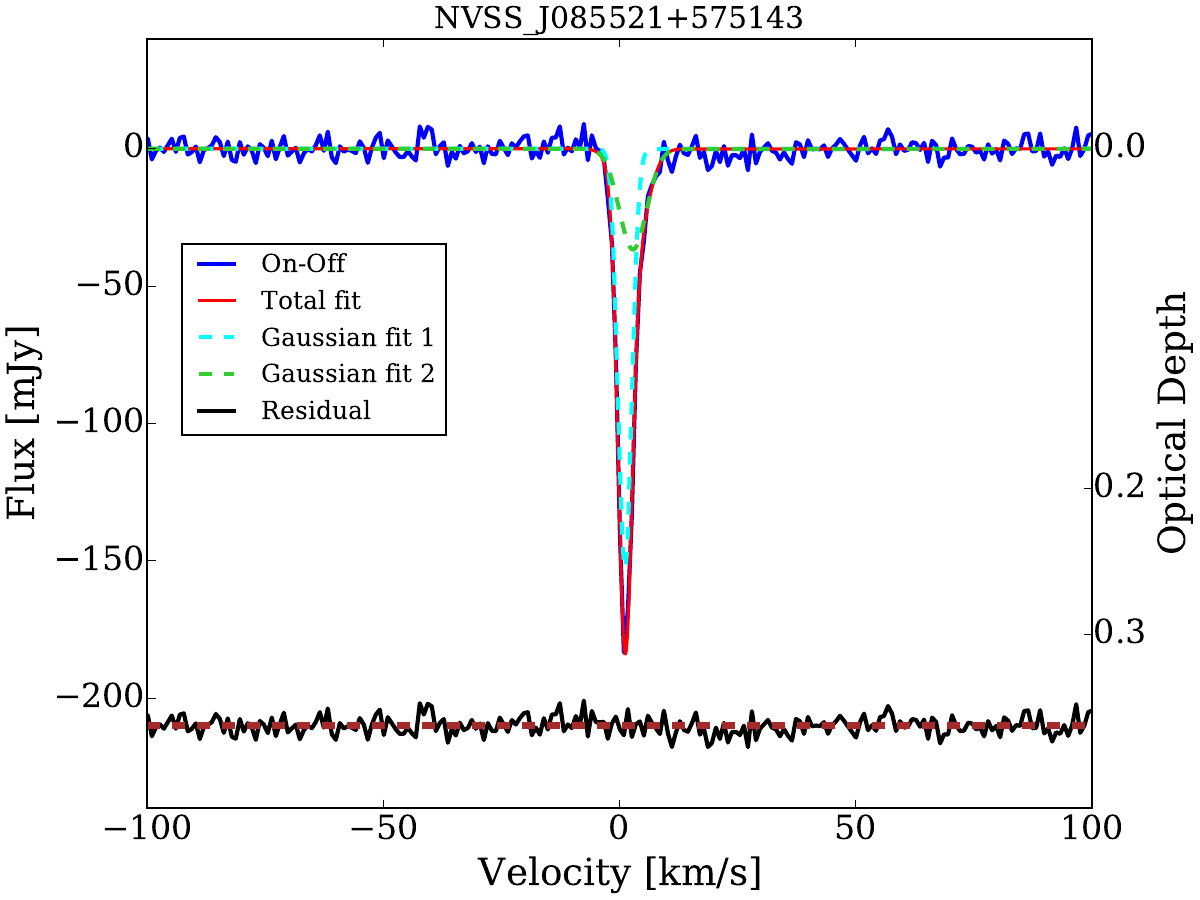}
    \includegraphics[width=0.25\textwidth]{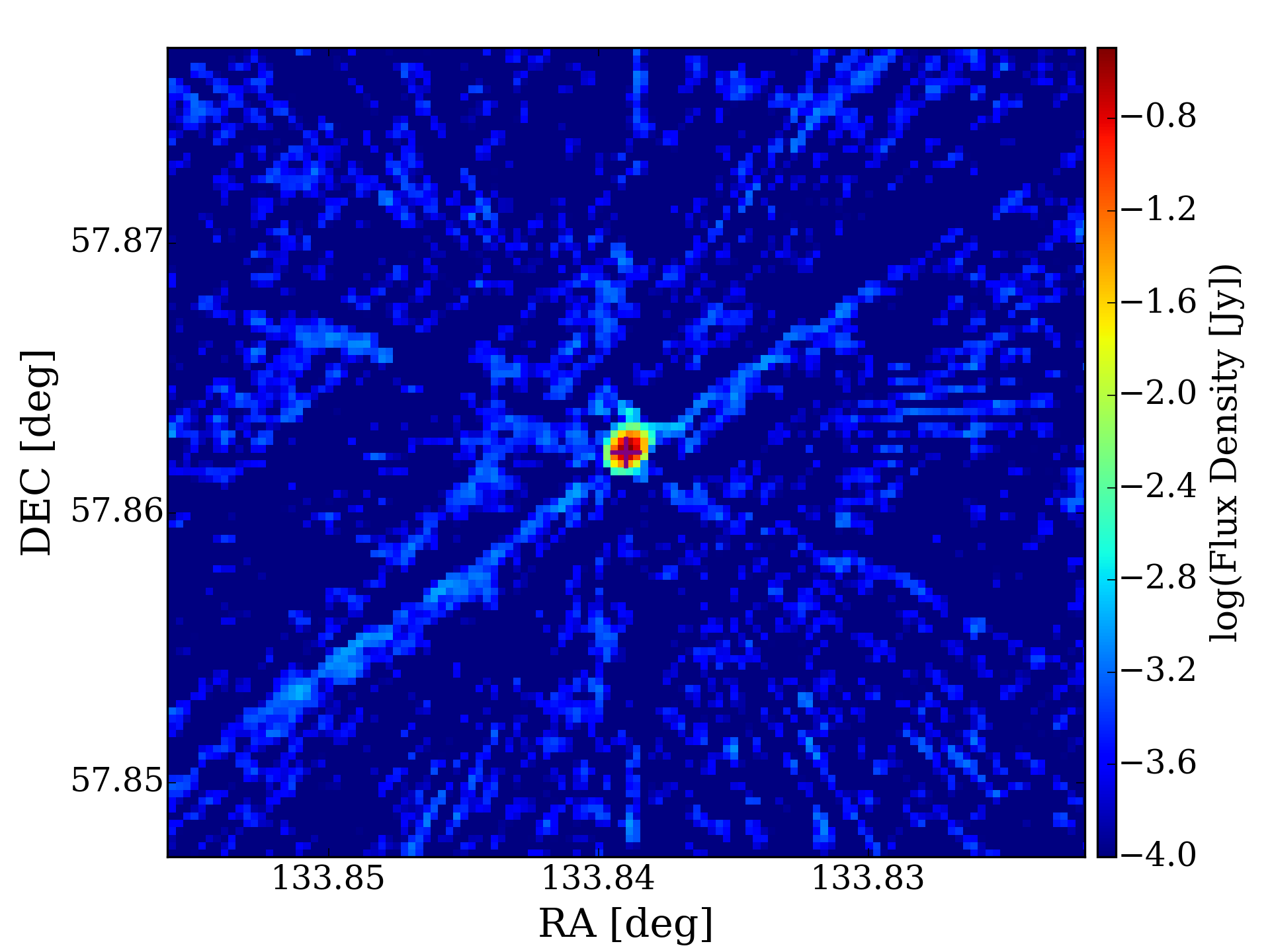}
    \includegraphics[width=0.25\textwidth]{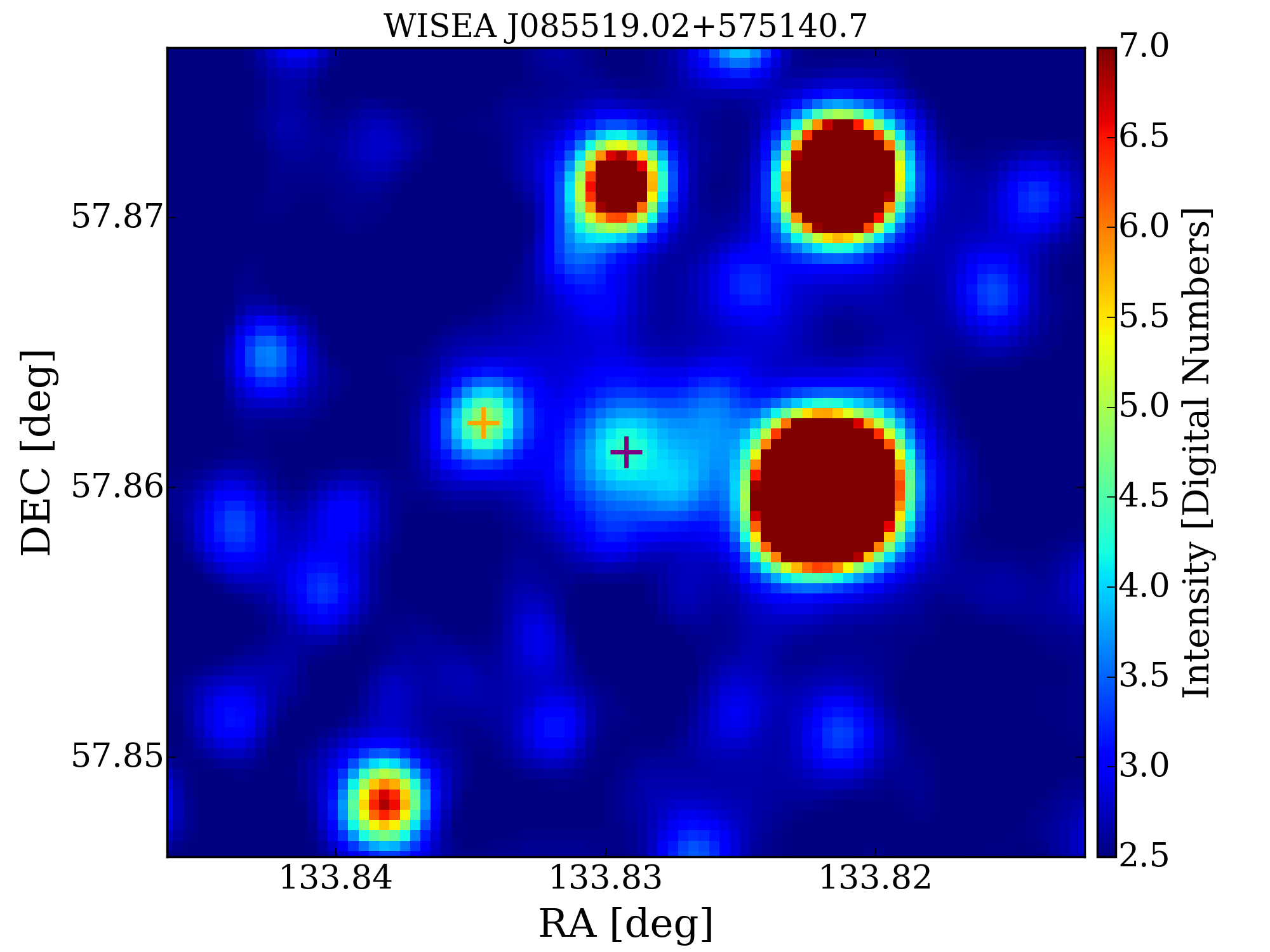}
    \includegraphics[width=0.18\textwidth]{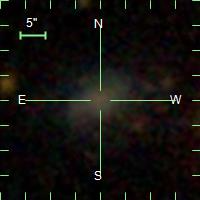}
    \caption{Left: same as Figure~\ref{4C+56.02_fit}, but for NVSS J085521\allowbreak+575143. Middle left: radio map from VLASS centered at NVSS J085521\allowbreak+575143. Middle right: W1 band infrared map of WISEA J085519.02\allowbreak+575140.7 (shown as a purple cross) and WISEA J085519.81\allowbreak+575141.1 (WISE counterpart to background NVSS J085521\allowbreak+575143, shown as an orange cross) from WISE. Right: SDSS optical map of the optical counterpart of WISEA J085519.02\allowbreak+575140.7.}
    \label{NVSS_J085521+575143_fit}
\end{figure*}

Intervening absorption is found towards NVSS J085521\allowbreak+575143, the background source is identified as an FR II radio source \citep{2016MNRAS.462.2819B} and a CSO with two lobes separated by approximately 55 mas \citep{2007ApJ...658..203H} (Figure~\ref{NVSS_J085521+575143_fit}). The distorted appearance of the background radio source suggests a strong jet–cloud interaction in its host galaxy \citep{2016MNRAS.462.2819B}. The absorption towards NVSS J085521\allowbreak+575143 was initially reported by \citet{2015MNRAS.453.1268Z}, the absorption parameters were reported as peak optical depth $\tau_{\rm peak} \sim 0.24$ and $\int\tau dv \sim$ 1.02, which closely aligns with our findings.

The SDSS counterpart to the foreground galaxy is SDSS J085519.05\allowbreak+575140.7, whose magnitudes are u=19.437, g=17.765, r=17.265, i=17.043 and z=16.834. The WISE counterpart to SDSS J085519.05\allowbreak+575140.7 given by NED is WISEA J085519.02\allowbreak+575140.7. The WISE W1[\SI{3.4}{\micro\metre}], W2[\SI{4.6}{\micro\metre}], W3[\SI{12.1}{\micro\metre}] and W4[\SI{22.2}{\micro\metre}] magnitudes for WISEA J085519.02\allowbreak+575140.7 are 15.535 $\pm$ 0.047, 15.288 $\pm$ 0.105, 11.935 $\pm$ 0.214 and 8.706, respectively. The W1-W2 color of WISEA J085519.02\allowbreak+575140.7 is 0.247, which indicates that the mid-IR emission comes mainly from stars. According to the W2-W3 value of 3.353 mag, WISEA J085519.02\allowbreak+575140.7 is located in the intersection of the Spirals region and the LIRGs region in the WISE color–color diagram.
%The WISE counterpart to NVSS\,J085521+575143 is WISEA\,J085521.34+575144.5 as shown in the NASA/IPAC Extragalactic Database. The WISE W1[\SI{3.4}{\micro\metre}], W2[\SI{4.6}{\micro\metre}], W3[\SI{12.1}{\micro\metre}] and W4[\SI{22.2}{\micro\metre}] magnitudes for WISEA\,J085521.34+575144.5 are 15.126 $\pm$ 0.039, 14.971 $\pm$ 0.082, 11.808 $\pm$ 0.193 and 8.772, respectively. The W1-W2 color of WISEA\,J085521.34+575144.5 is 0.155, which means that the mid-IR emission comes mainly from the stars. According to the W2-W3 value of 3.163 mag, WISEA\,J085521.34+575144.5 is located in the intersection of the Spirals region and the LIRGs region in the WISE color-color diagram.

\subsubsection{NVSS J112832+583346 (NGC 3690)}

\begin{figure*}[hbt!]
    \centering
    \includegraphics[width=0.25\textwidth]{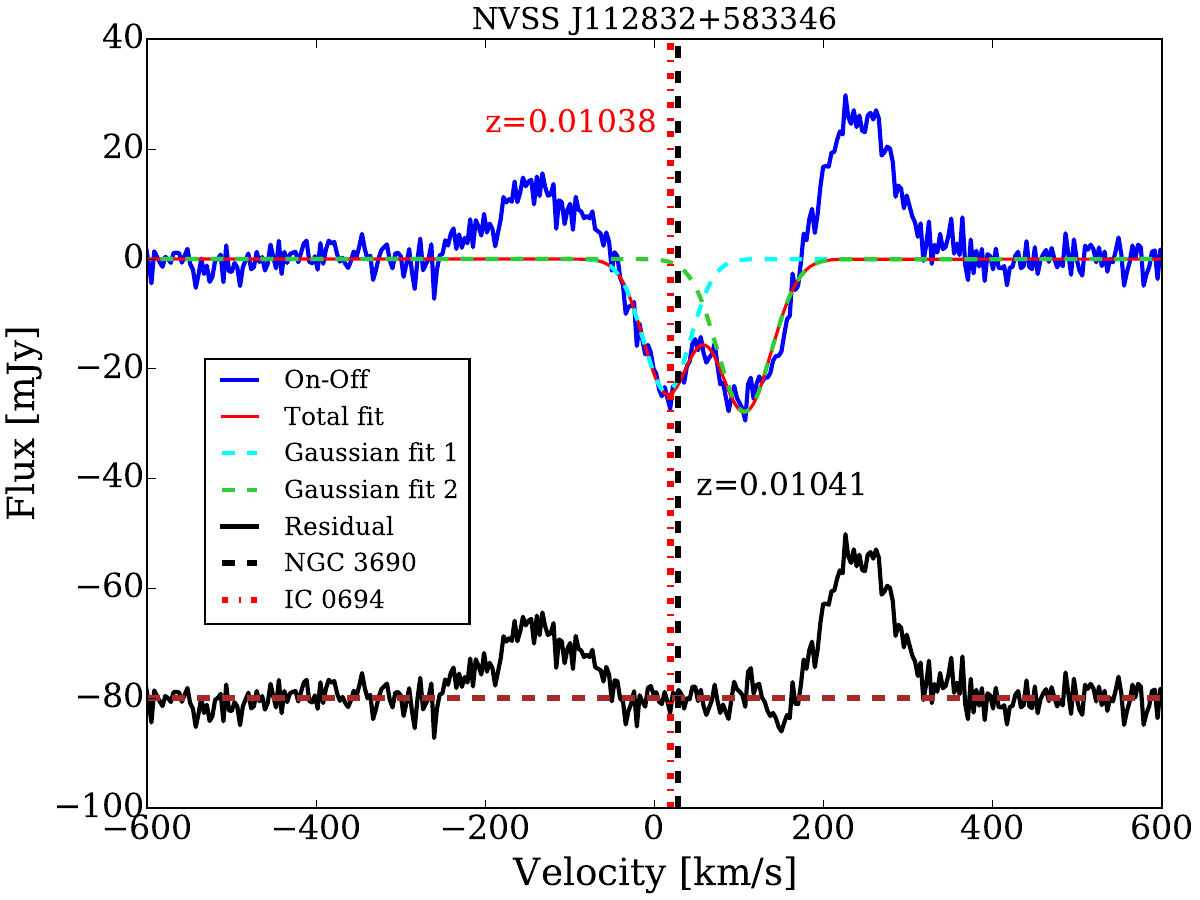}
    \includegraphics[width=0.25\textwidth]{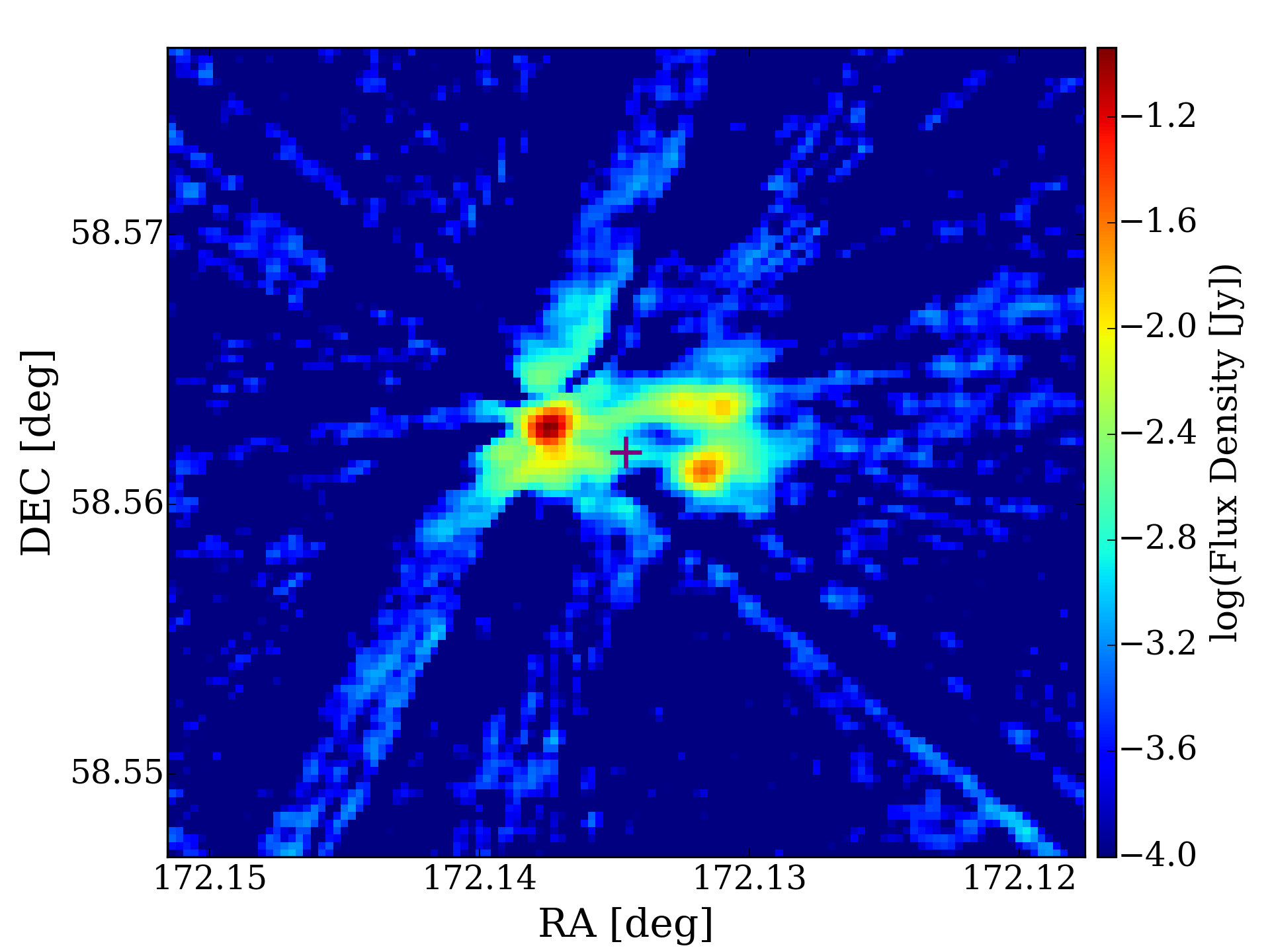}
    \includegraphics[width=0.25\textwidth]{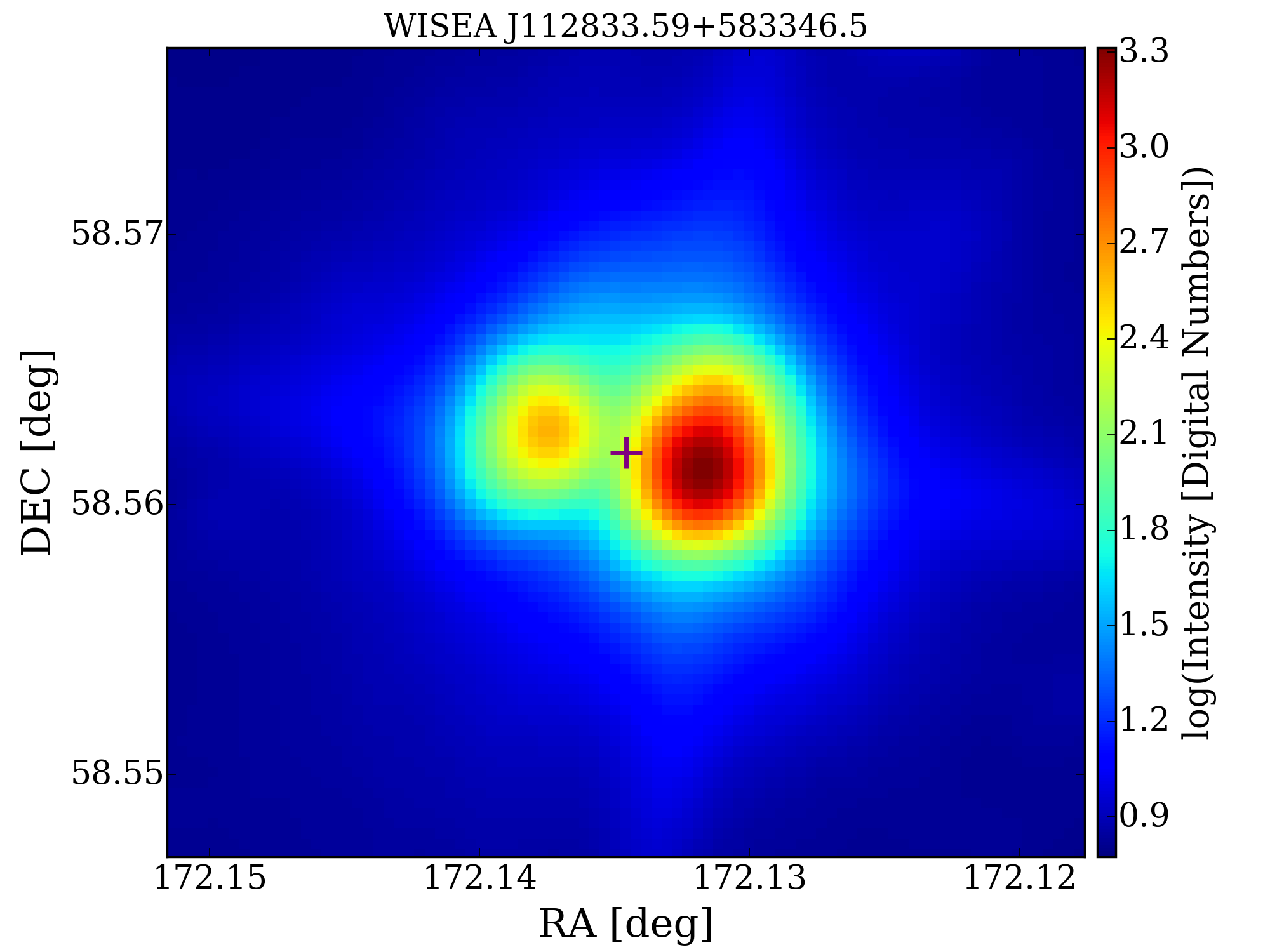}
    \includegraphics[width=0.18\textwidth]{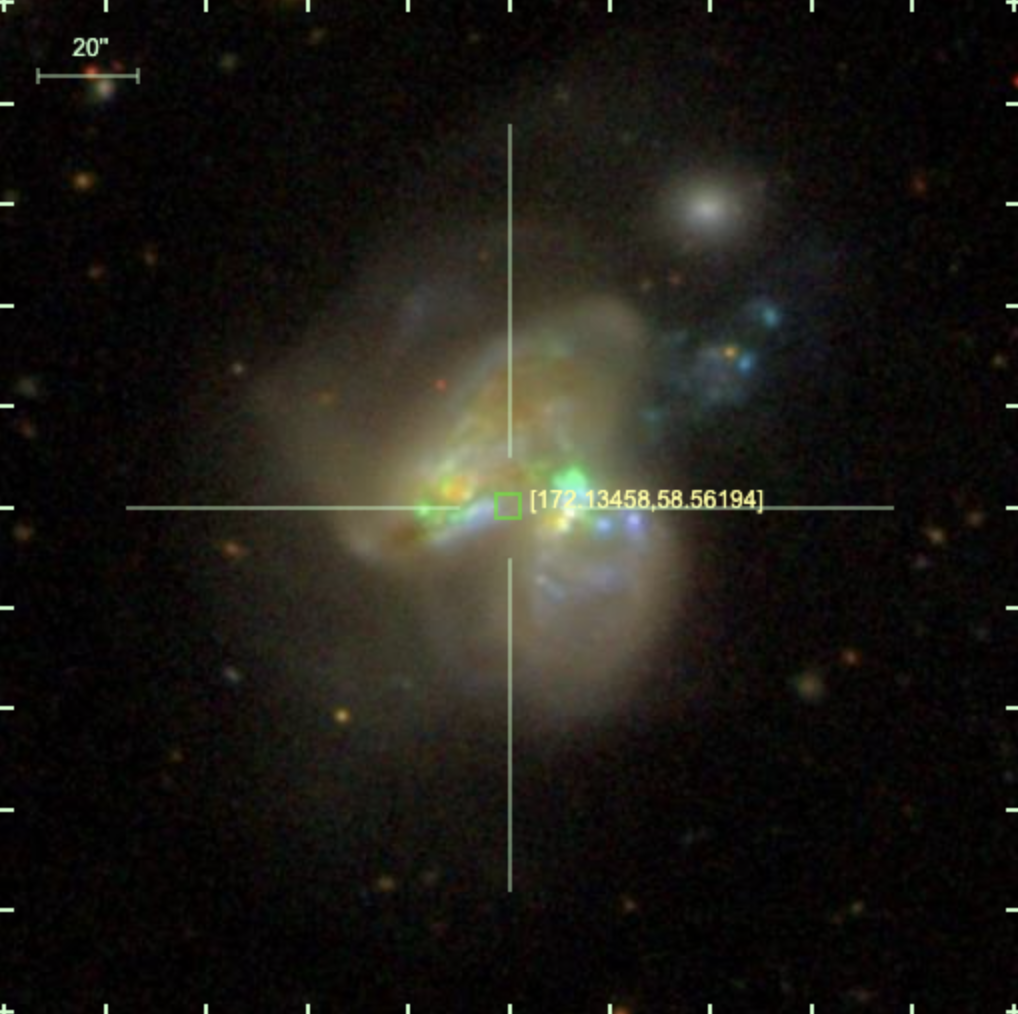}
    \caption{Left: same as Figure~\ref{4C+31.04_fit}, but for NVSS J112832\allowbreak+583346. Middle left: radio map from VLASS centered at NVSS J112832\allowbreak+583346. Middle right: W2 band infrared map of WISEA J112833.59\allowbreak+583346.5 (WISE counterpart to NVSS J112832\allowbreak+583346 ) from WISE. Right: SDSS optical map of the optical counterpart of WISEA J112833.59\allowbreak+583346.5.}
    \label{NVSS_J112832+583346_fit}
\end{figure*}

NGC 3690 (Figure~\ref{NVSS_J112832+583346_fit}) is  within the nearby (cz = 3121 \kms) merging system recognized as Arp\,299 or Mrk\,171. Positioned in the early stages of a merger, this system comprises NGC\,3690, constituting the western portion of the merger, and IC\,694, serving as the eastern component. The exceptional burst of star formation in Arp\,299 is evidenced by strong \SI{10}{\micro\metre} emission \citep{1983ApJ...267..551G}, extended H$_{2}$ features \citep{1983ApJ...273L..27F}, and optical emission fluxes \citep{1972ApJ...171....5W,1985A&A...147..273A}. The total molecular gas mass (M$_{H_{2}}$) derived for Arp 299 is 6 $\times$ 10$^{9}$ M$_{\odot}$, with M$_{\hi}$ exceeding 6 $\times$ 10$^{9}$ M$_{\odot}$ \citep{1989A&A...224...31C}. The system exhibits a high infrared luminosity, L$_{IR}$ = 5 $\times$ 10$^{11}$ L$_{\odot}$, attributed to re-radiation of dust heated by remarkable starburst activity, with no discernible evidence of an AGN within the system.

The detection of \hi absorption linked to NGC 3690 was initially documented by \citet{1982AJ.....87..278D,1986ApJ...300..190D}. Subsequently, employing the VLA, \citet{1990ApJ...364...65B} identified \hi absorption across all locations within Arp\,299, revealing sixteen distinct velocity components. The \hi gas originating from IC\,694 is observed in absorption at the nucleus of NGC\,3690, and vice versa.

We blindly detected \hi absorption in Arp 299 in the FAST survey. The FAST beam, with a size of 3 arcmins, covers the entire merging system and captures emissions and absorption in front of the extended continuum sources (depicted in Figure~\ref{NVSS_J112832+583346_fit}). Our spectra align with those of \citet{1989A&A...224...31C}, who presented \hi spectra obtained using the Nançay telescope (with a beam size of 4 arcminutes $\times$ 20 arcminutes). 

In our observed absorption profile, two components are discernible. Given that NGC 3690 is nearly face-on and IC 694 is almost edge-on, it is reasonable to infer that the velocity component around 3184 \kms corresponds to gas from IC 694, while the component at 3094 \kms belongs to NGC 3690 \citep{1990ApJ...364...65B}. However, due to the limitations of our single-dish observation, which can not differentiate \hi absorption in each object within Arp 299, we cannot provide accurate estimates for optical depth and column density.

%The WISE counterpart to NVSS\,J112832+583346 is WISEA\,J112833.59+583346.5 as shown in the NASA/IPAC Extragalactic Database. The WISE W1[\SI{3.4}{\micro\metre}], W2[\SI{4.6}{\micro\metre}], W3[\SI{12.1}{\micro\metre}] and W4[\SI{22.2}{\micro\metre}] magnitudes for WISEA\,J112833.59+583346.5 are 9.342 $\pm$ 0.023, 8.299 $\pm$ 0.021, 3.667 $\pm$ 0.015 and 0.726 $\pm$ 0.025, respectively. The W1-W2 color of WISEA\,J112833.59+583346.5 is 1.043, which means that the mid-IR emission comes mainly from the AGN. According to the W2-W3 value of 4.632 mag, WISEA\,J112833.59+583346.5 is located in the ULIRGs/LINERs region in the WISE color-color diagram.

\subsubsection{NVSS J134035+444817}

\begin{figure*}[hbt!]
    \centering
    \includegraphics[width=0.25\textwidth]{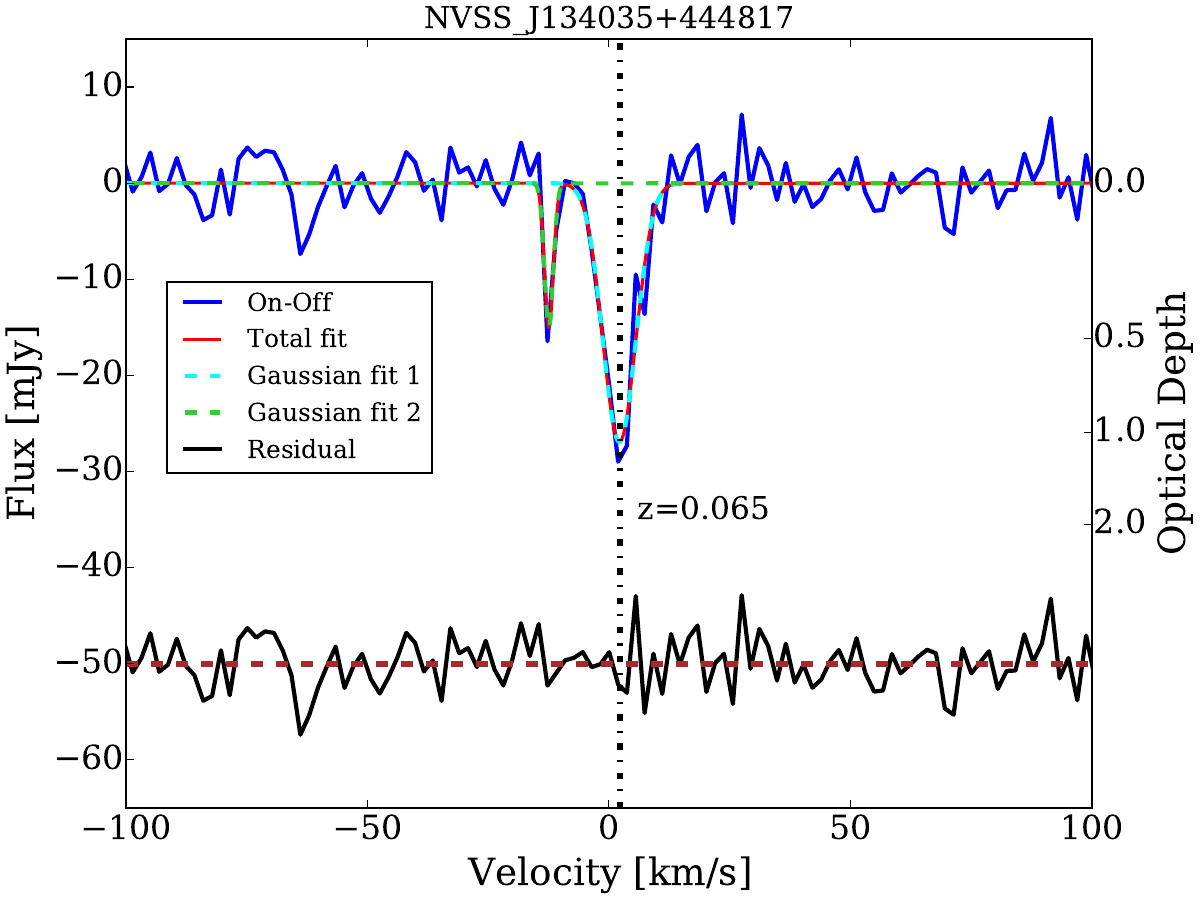}
    \includegraphics[width=0.25\textwidth]{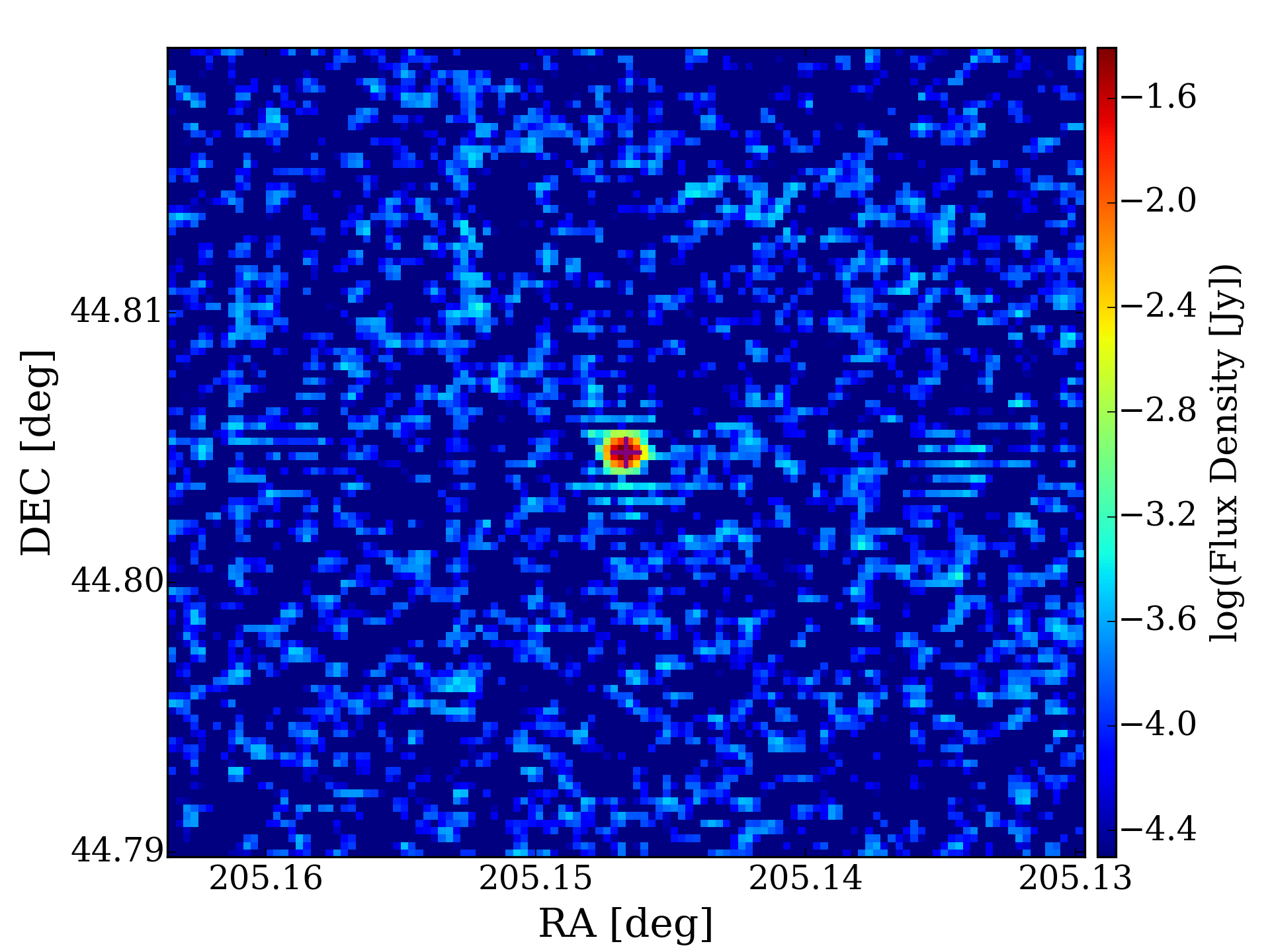}
    \includegraphics[width=0.25\textwidth]{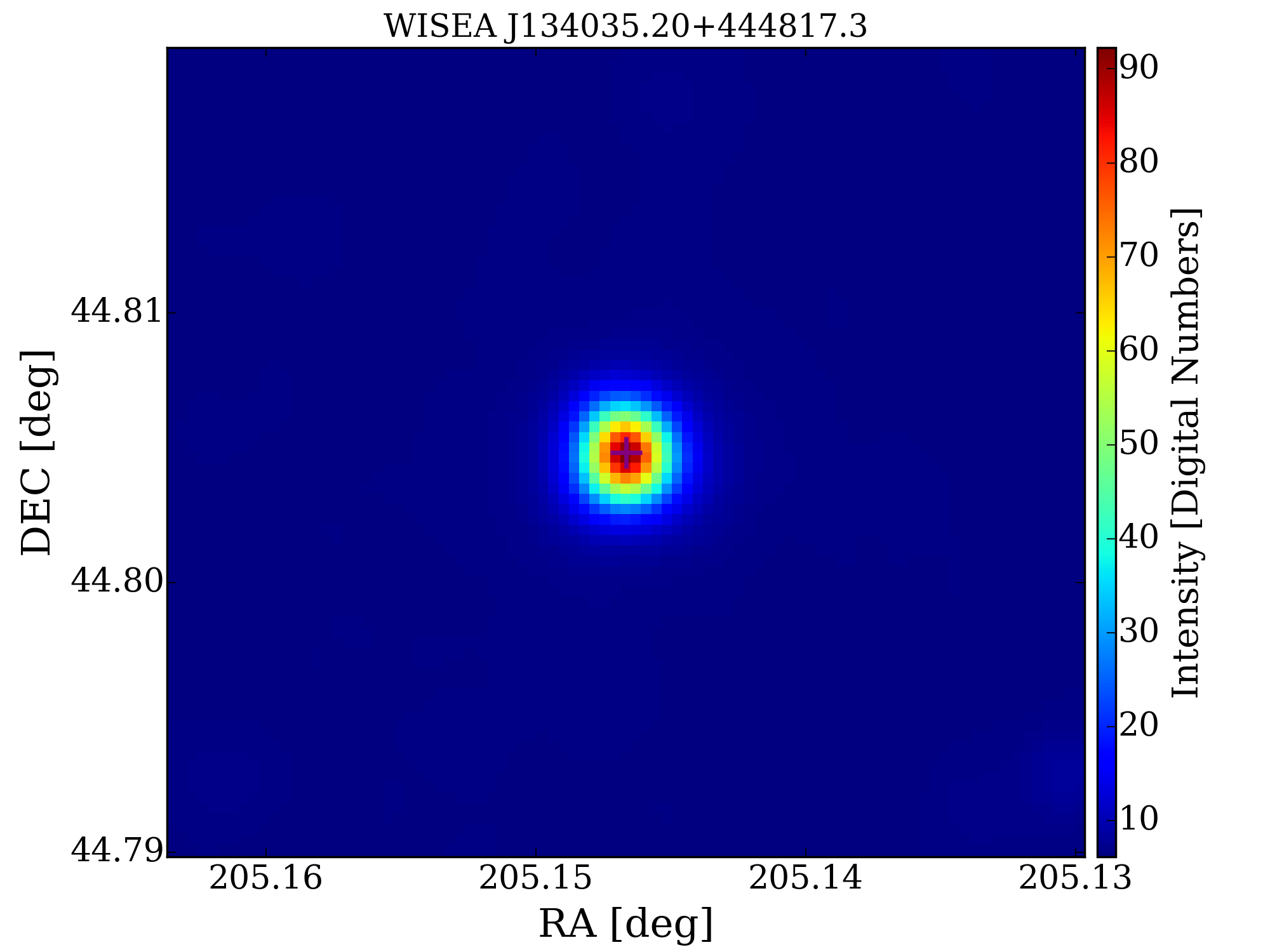}
    \includegraphics[width=0.18\textwidth]{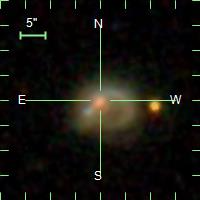}
    \caption{Left: same as Figure~\ref{4C+31.04_fit}, but for NVSS J134035\allowbreak+444817. Middle left: radio map from VLASS centered at NVSS J134035\allowbreak+444817. Middle right: W2 band infrared map of WISEA J134035.20\allowbreak+444817.3 (WISE counterpart to NVSS J134035\allowbreak+44481). from WISE. Right: SDSS optical map of the optical counterpart of WISEA J134035.20\allowbreak+444817.3.}
    \label{NVSS_J134035+444817_fit}
\end{figure*}

NVSS J134035\allowbreak+444817 (Figure~\ref{NVSS_J134035+444817_fit}) is categorized as a Flat-Spectrum Radio Source in \citet{2007ApJS..171...61H}. Within the CORALZ sample, this radio source is identified as a potential compact core-jet (CJ) source \citep{2009A&A...498..641D}. The linked \hi absorption was initially documented by \citet{2015A&A...575A..44G}. A slender \hi absorption profile is observed at the velocity of NVSS J134035\allowbreak+444817, suggesting the presence of a gas disk.

%The WISE counterpart to NVSS\,J134035+444817 is WISEA\,J134035.20+444817.3 as shown in the NASA/IPAC Extragalactic Database. The WISE W1[\SI{3.4}{\micro\metre}], W2[\SI{4.6}{\micro\metre}], W3[\SI{12.1}{\micro\metre}] and W4[\SI{22.2}{\micro\metre}] magnitudes for WISEA\,J134035.20+444817.3 are 11.066 $\pm$ 0.023, 10.050 $\pm$ 0.020, 7.202 $\pm$ 0.016 and 4.728 $\pm$ 0.026, respectively. The W1-W2 color of WISEA\,J134035.20+444817.3 is 1.016, which means that the mid-IR emission comes mainly from the AGN. According to the W2-W3 value of 2.848 mag, WISEA\,J134035.20+444817.3 is located in the QSO region in the WISE color-color diagram.

\subsubsection{4C +57.23}

\begin{figure*}[hbt!]
    \centering
    \includegraphics[width=0.25\textwidth]{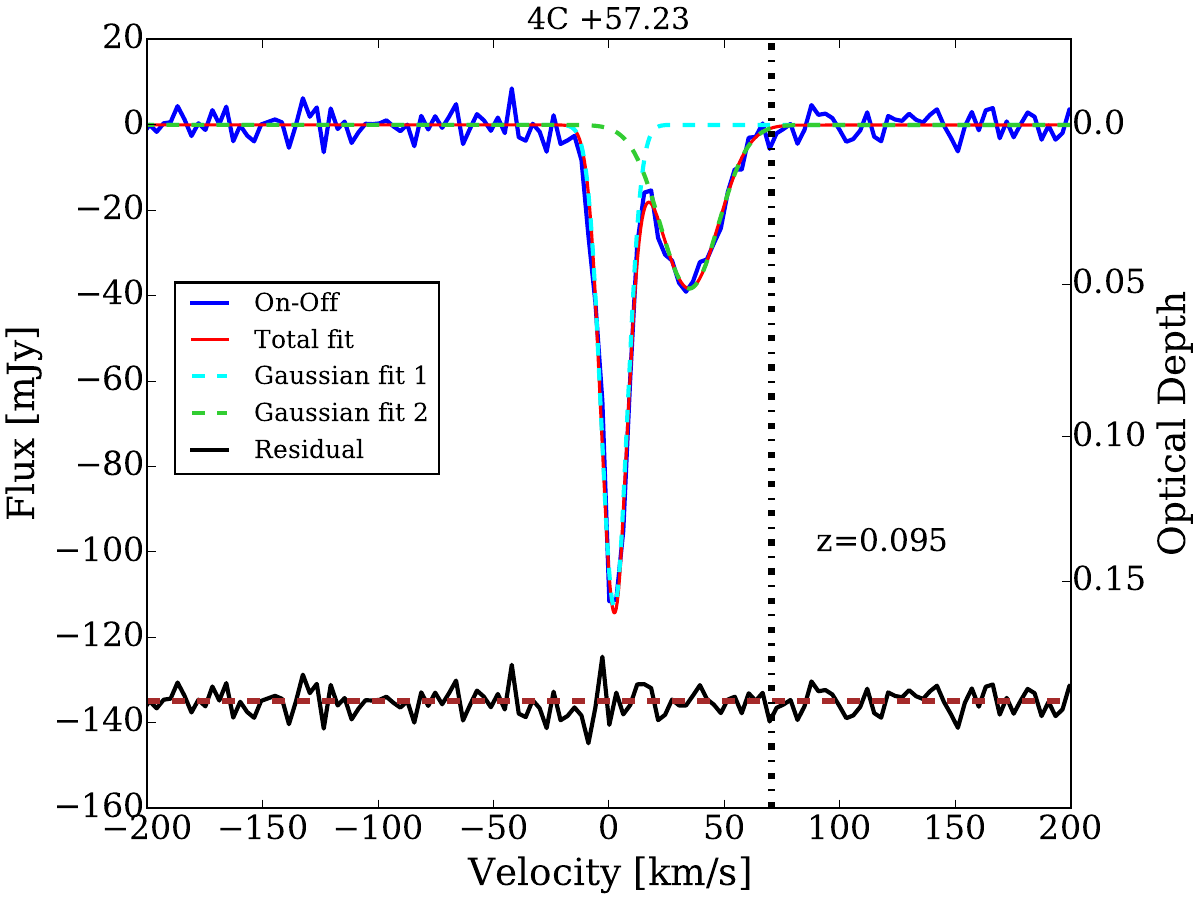}
    \includegraphics[width=0.25\textwidth]{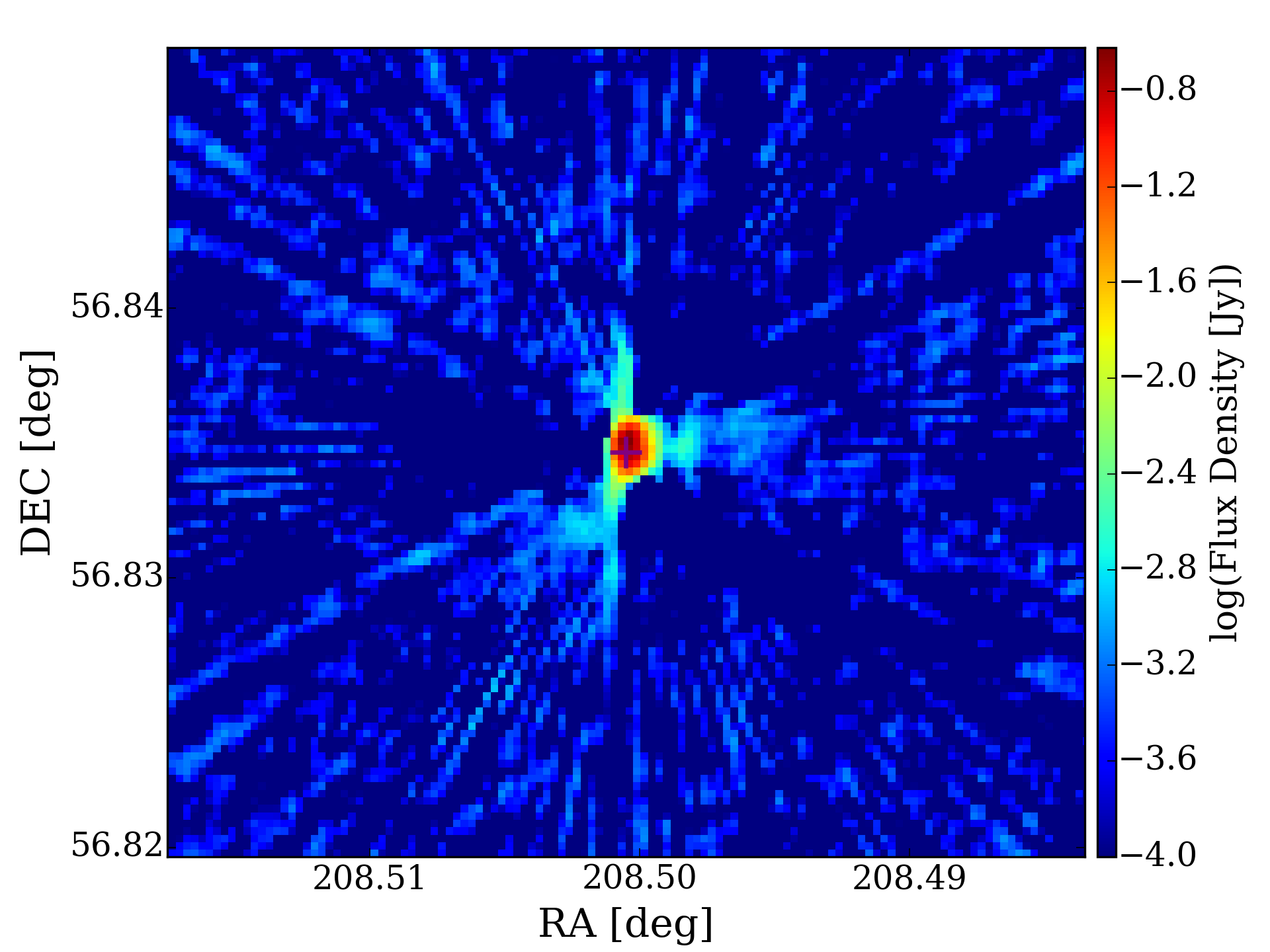}
    \includegraphics[width=0.25\textwidth]{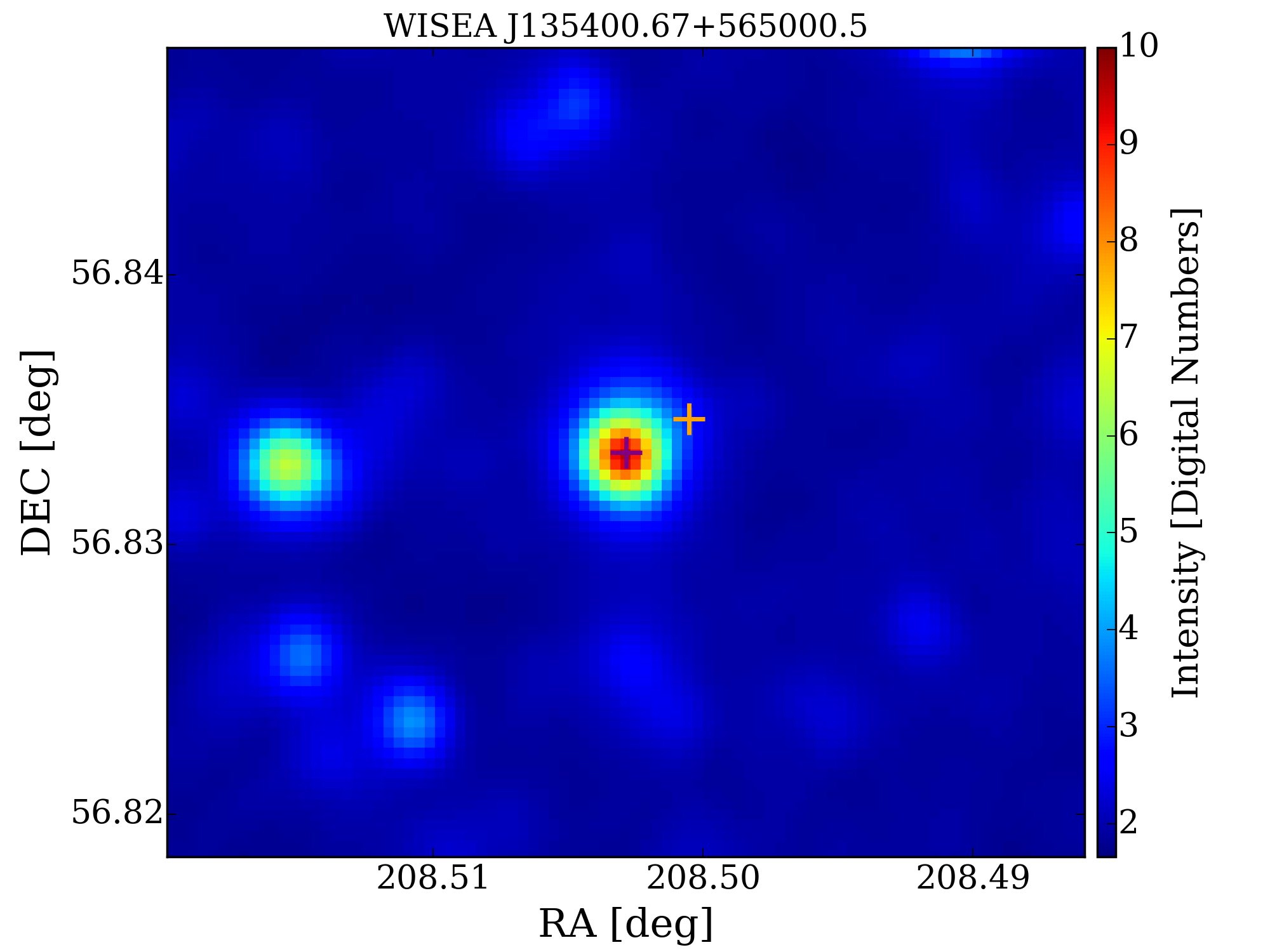}
    \includegraphics[width=0.18\textwidth]{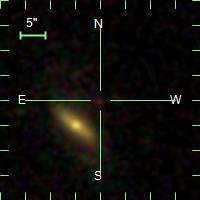}
    \caption{Left: same as Figure~\ref{4C+31.04_fit}, but for 4C\,+57.23. Middle left: radio map from VLASS centered at 4C\,+57.23. Middle right: W1 band infrared map of WISEA J135400.67\allowbreak+565000.5 (shown as a purple cross) and 4C\,+57.23 (shown as an orange cross). Right: SDSS optical map centered at the optical counterpart of 4C\,+57.23, the bright galaxy below is the optical counterpart of WISEA J135400.67\allowbreak+565000.5.}
    \label{4C+57.23_fit}
\end{figure*}

The \hi absorption observed toward 4C +57.23 (Figure~\ref{4C+57.23_fit}) is identified as intervening absorption. The SDSS counterpart linked to the foreground galaxy is SDSS J135400.68\allowbreak+565000.3. Its magnitudes are u=19.783, g=17.964, r=16.976, i=16.546 and z=16.200. Based on the SDSS image, this galaxy appears to be an edge-on spiral galaxy, likely of morphological type Sa. The presence of \hi absorption in SDSS J135400.68\allowbreak+565000.3 toward 4C+57.23 was initially documented by \citet{2015MNRAS.453.1268Z}. In our blind detection, we reconfirmed this absorption and obtained measurements closely aligned with \citet{2015MNRAS.453.1268Z}, who reported a peak optical depth of $\tau_{\rm peak} \sim 0.14$ and $\int\tau dv = 3.03 \kms$.

The WISE counterpart to the foreground SDSS J135400.68\allowbreak+565000.3 is WISEA J135400.67\allowbreak+565000.5 according to NED. WISE W1[\SI{3.4}{\micro\metre}], W2[\SI{4.6}{\micro\metre}], W3[\SI{12.1}{\micro\metre}] and W4[\SI{22.2}{\micro\metre}] magnitudes for WISEA J135400.67\allowbreak+565000.5 are 13.901 $\pm$ 0.026, 13.803 $\pm$ 0.035, 12.464 $\pm$ 0.278 and 9.192, respectively. W1-W2 color of WISEA J135400.67\allowbreak+565000.5 is 0.098, which suggests that the mid-IR emission comes mainly from stars. According to the W2-W3 value of 1.339 mag, WISEA J135400.67\allowbreak+565000.5 is located in the intersection of the Spirals region and the Ellipticals region in the WISE color-color diagram. 

%The WISE counterpart to 4C\,+57.23 is WISEA\,J135400.67+565000.5 as shown in the NASA/IPAC Extragalactic Database. The WISE W1[\SI{3.4}{\micro\metre}], W2[\SI{4.6}{\micro\metre}], W3[\SI{12.1}{\micro\metre}] and W4[\SI{22.2}{\micro\metre}] magnitudes for WISEA\,J135400.67+565000.5 are 13.901 $\pm$ 0.026, 13.803 $\pm$ 0.035, 12.464 $\pm$ 0.278 and 9.192, respectively. The W1-W2 color of WISEA\,J135400.67+565000.5 is 0.098, which means that the mid-IR emission comes mainly from the stars. According to the W2-W3 value of 1.339 mag, WISEA\,J135400.67+565000.5 is located in the intersection of the Spirals region and the Ellipticals region in the WISE color-color diagram. 

\subsubsection{NVSS J141314-031227 (NGC 5506)}

\begin{figure*}[hbt!]
    \centering
    \includegraphics[width=0.25\textwidth]{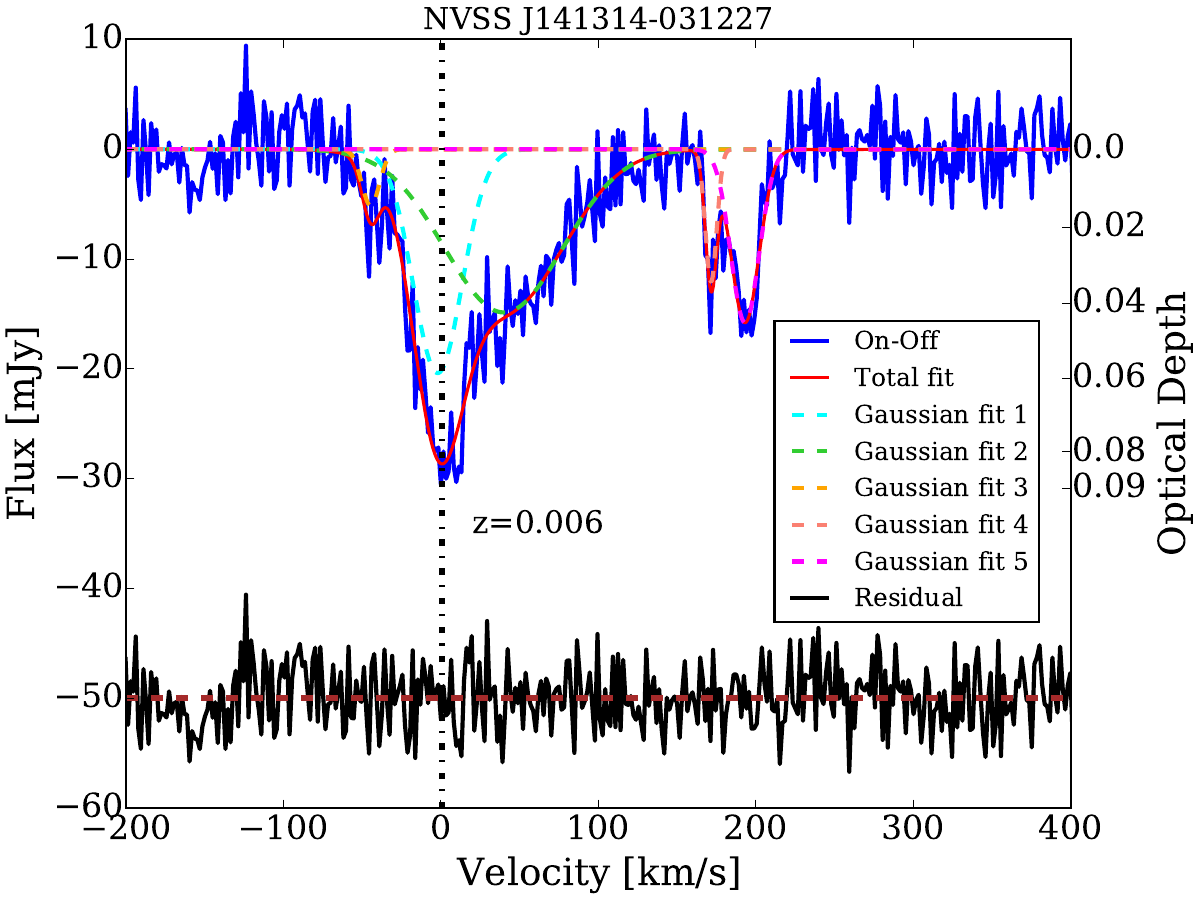}
    \includegraphics[width=0.25\textwidth]{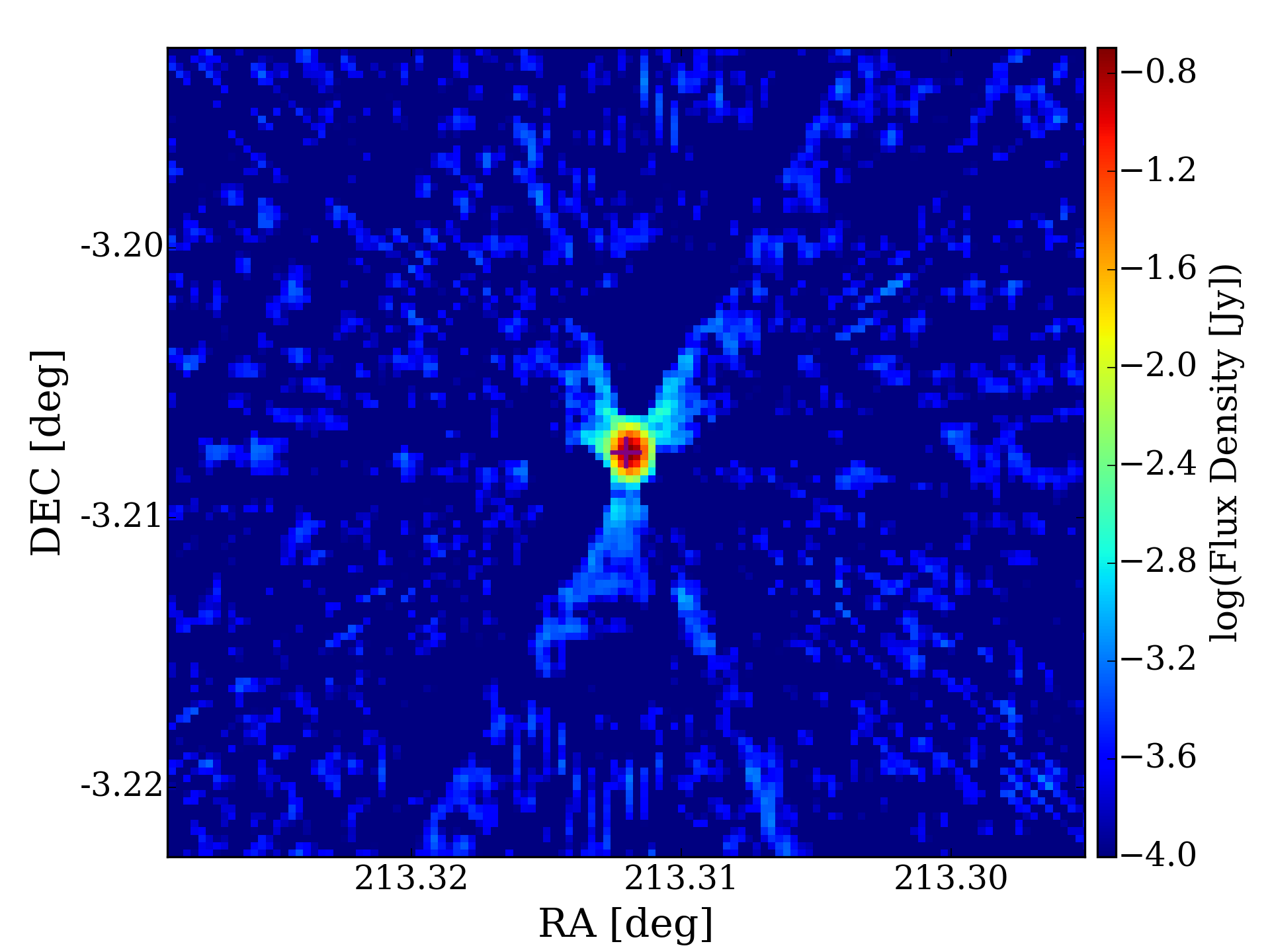}
    \includegraphics[width=0.25\textwidth]{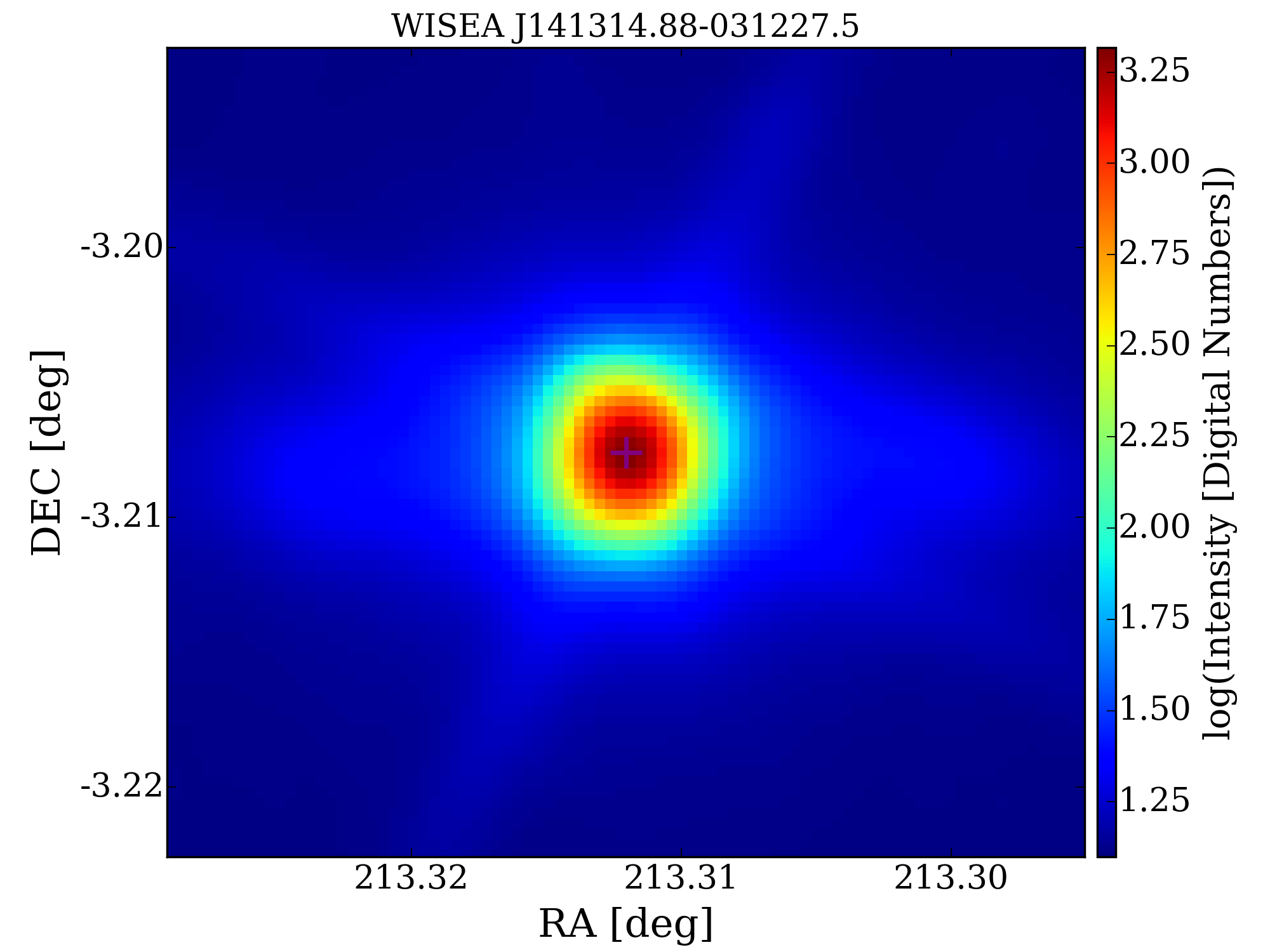}
    \includegraphics[width=0.18\textwidth]{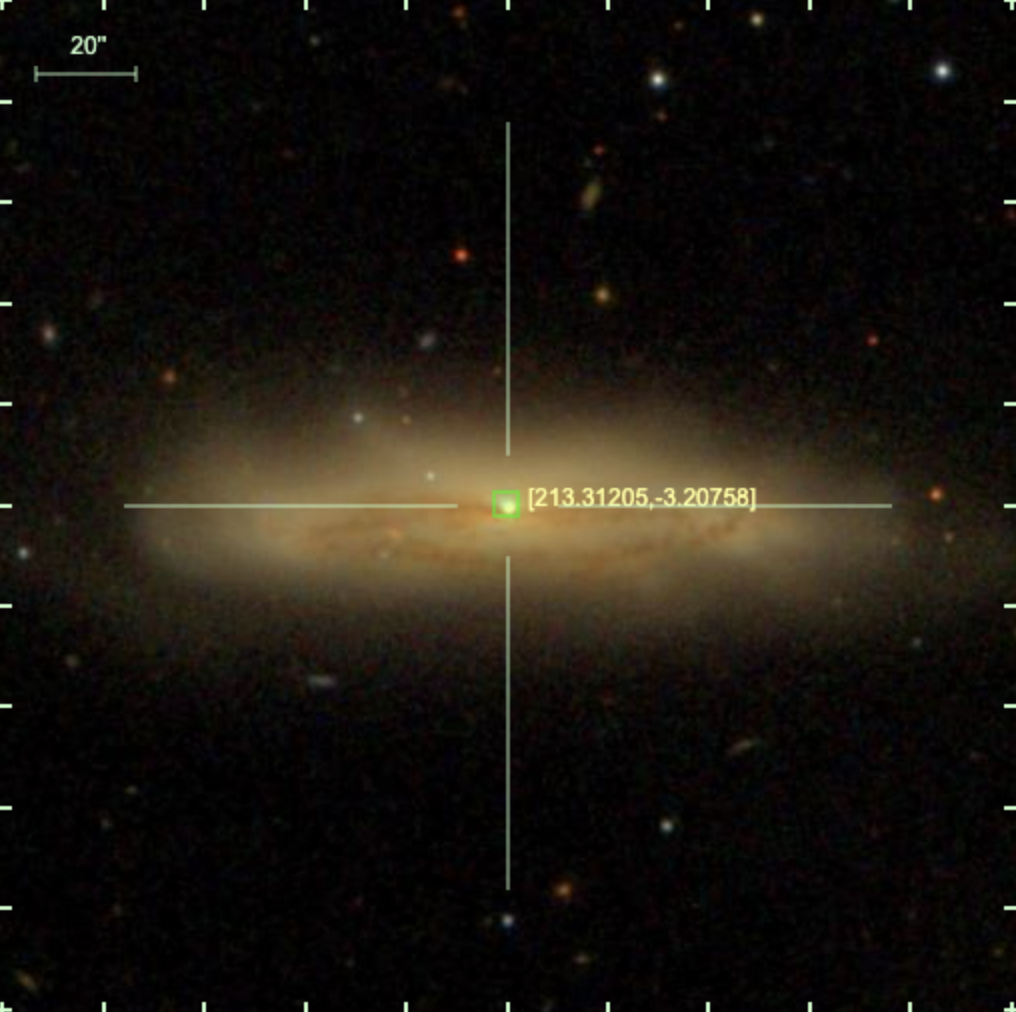}
    \caption{Left: same as Figure~\ref{4C+31.04_fit}, but for NVSS J141314\allowbreak-031227. Middle left: radio map from VLASS centered at NVSS J141314\allowbreak-031227. Middle right: W2 band infrared map of WISEA J141314.88\allowbreak-031227.5 (WISE counterpart to NVSS J141314\allowbreak-031227) from WISE. Right: SDSS optical map of the optical counterpart of WISEA J141314.88\allowbreak-031227.5.}
    \label{NVSS_J141314-031227_fit}
\end{figure*}

NGC\,5506 (Figure~\ref{NVSS_J141314-031227_fit}), identified as an X-ray bright Seyfert II galaxy \citep{1976MNRAS.177..673W,1978ApJ...224L..55R} within the local Virgo supercluster, hosts an AGN classified as a radio-quiet narrow-line Seyfert I \citep{2002A&A...391L..21N}. The galaxy shows a smooth underlying stellar distribution without any apparent HII regions. Initial reports of \hi absorption linked to NGC 5506 were presented by \citet{1982ApJ...252..125T} and subsequently mapped through a VLA imaging survey by \citet{1999ApJ...524..684G}. We blindly re-detected the \hi absorption associated with NGC\,5506, the discerned \hi profile exhibits two components: a deeper one at 1796 \kms and a shallower one at 1989 \kms. We measured the absorption with $\tau_{peak} = 0.08$ and an integrated flux of $\int S_{\hi}dv=2724.20$ mJy$\kms$, which is consistent with the measurements from \citet{1999ApJ...524..684G}, where $\tau_{peak} = 0.109\pm0.003$ and $\int S_{\hi}dv=3.0\pm0.10$ Jy$\kms$.

The optical image in Figure~\ref{NVSS_J141314-031227_fit} reveals NGC 5506's nearly edge-on orientation, characterized by a conspicuous east-west dust lane that bisects the galaxy and a luminous, resolved nucleus, partly concealed by the dust lane. It is plausible that the predominant absorption arises from foreground disk gas, while the redshifted velocities of the line suggest that the shallower component may originate from an in-falling object.

%The WISE counterpart to NVSS\,J141314-031227 is WISEA\,J141314.88-031227.5 as shown in the NASA/IPAC Extragalactic Database. The WISE W1[\SI{3.4}{\micro\metre}], W2[\SI{4.6}{\micro\metre}], W3[\SI{12.1}{\micro\metre}] and W4[\SI{22.2}{\micro\metre}] magnitudes for WISEA\,J141314.88-031227.5 are 7.399 $\pm$ 0.027, 6.244 $\pm$ 0.019, 3.692 $\pm$ 0.015 and 1.068 $\pm$ 0.014, respectively. The W1-W2 color of WISEA\,J141314.88-031227.5 is 1.155, which means that the mid-IR emission comes mainly from the AGN. According to the W2-W3 value of 2.552 mag, WISEA\,J141314.88-031227.5 is located in the intersection of the Seyferts region and the QSO region in the WISE color-color diagram.

\subsubsection{NVSS J141558+132024 (PKS 1413+135)}

\begin{figure*}[hbt!]
    \centering
    \includegraphics[width=0.25\textwidth]{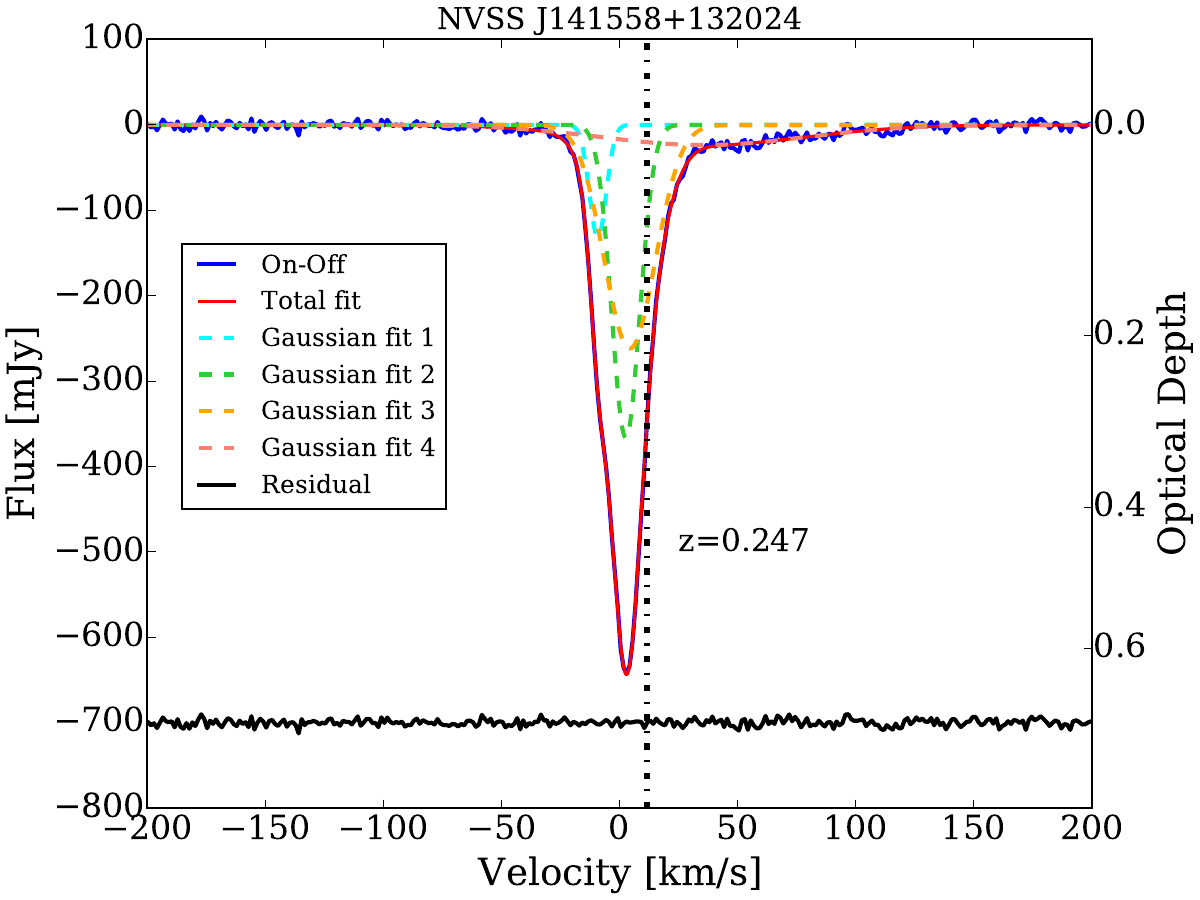}
    \includegraphics[width=0.25\textwidth]{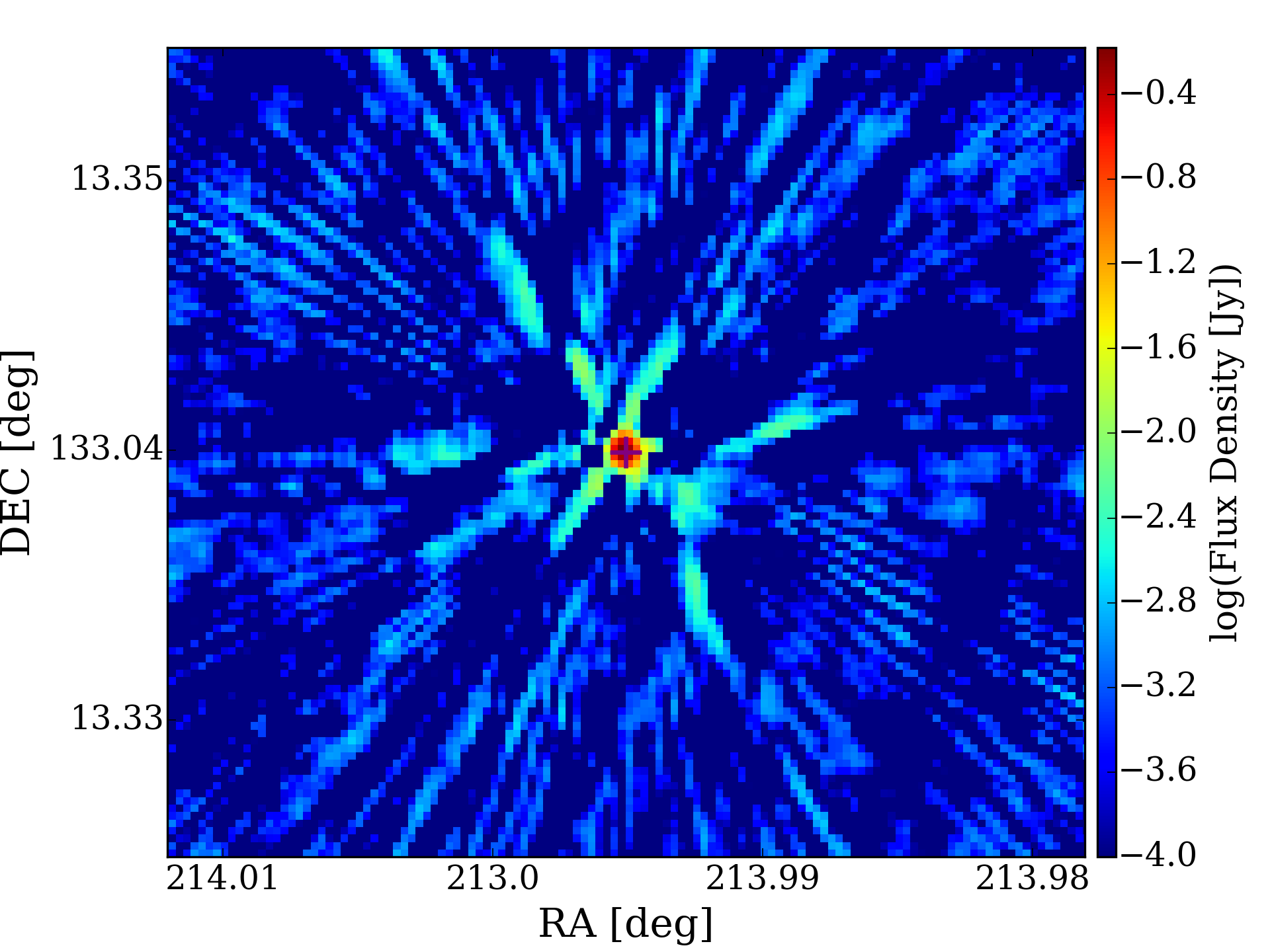}
    \includegraphics[width=0.25\textwidth]{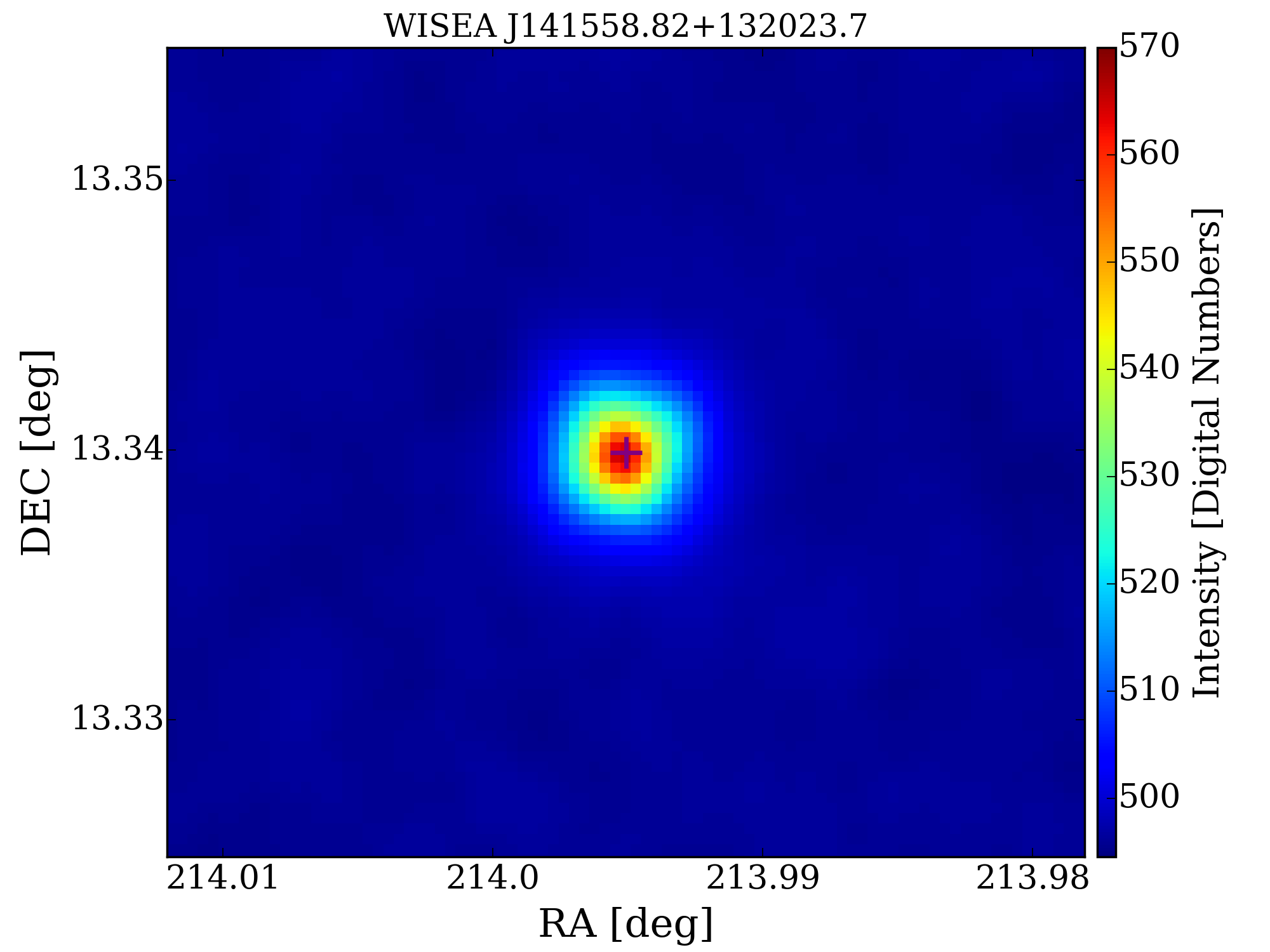}
    \includegraphics[width=0.18\textwidth]{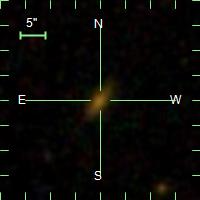}
    \caption{Left: same as Figure~\ref{4C+31.04_fit}, but for NVSS J141558\allowbreak+132024. Middle left: radio map from VLASS centered at NVSS J141558\allowbreak+132024. Middle right: W3 band infrared map of WISEA J141558.82\allowbreak+132023.7 from WISE. Right: SDSS optical map of the optical counterpart of NVSS J141558\allowbreak+132024.}
    \label{NVSS_J141558+132024_fit}
\end{figure*}

The radio source PKS 1413\allowbreak+135 (Figure~\ref{NVSS_J141558+132024_fit}) stands out as one of the most puzzling blazars. Despite being categorized as a BL Lac object \citep{1981Natur.293..711B,1981Natur.293..714B}, PKS 1413\allowbreak+135 appears to be linked to a disk galaxy at a redshift of 0.25 \citep{1991MNRAS.249..742M}. This is intriguing given that the majority of BL Lac objects and radio-loud quasars are typically associated with elliptical galaxies rather than spiral galaxies. Furthermore, while BL Lac objects usually exhibit jet axes aligned closely with the line of sight, PKS 1413\allowbreak+135 has also been identified as a CSO \citep{1996AJ....111.1839P}. These objects have jet axes that are not aligned closely with the line of sight. 

Employing multi-epoch very long baseline interferometry (VLBI), radio monitoring observations, and analyzing the infrared spectrum of PKS 1413\allowbreak+135, \citet{2021ApJ...907...61R} demonstrated that the orientation of the jet axis closely aligns with the line of sight. The jetted AGN is highly likely to be a background source at $z<0.5$, rather than being situated in the spiral galaxy at z=0.247. The intervening spiral galaxy at z=0.247 is identified as a Seyfert II early-type spiral galaxy, observed in an edge-on perspective.

The initial identification of \hi 21 cm absorption in PKS 1413\allowbreak+135 was made by \citet{1992ApJ...400L..13C} and recently revisited with Meerkat by \citet{2023A&A...671A..43C}. We re-detected \hi absorption toward PKS 1413\allowbreak+135 in a blind survey and showcase its highest-resolution spectrum here. Our measurements closely align with those of \citet{2023A&A...671A..43C}, who documented a peak optical depth of $\tau_{\rm peak} \sim 0.463$ and $\int\tau dv = 10.86 \kms$. The \hi spectrum stands out due to its distinct narrow central component and a redshifted wing that extends up to 110 \kms. This redshifted wing is presumably associated with an outer gaseous ring orbiting the galaxy. Additionally, it has been observed that \hi absorption is absent towards the radio core but is predominantly observed from a counterjet knot located 100 pc northeast of the core along the minor axis of the galaxy \citep{2002AJ....124.2401P}. 

From the WISE and optical image, only one source is resolved. According to the color of the WISE counterpart, this source is classified as QSOs, indicating the WISE map is dominated by the background source PKS 1413\allowbreak+135.

%The WISE counterpart to NVSS\,J141558+132024 is WISEA J141558.82+132023.7 as shown in the NASA/IPAC Extragalactic Database. The WISE W1[\SI{3.4}{\micro\metre}], W2[\SI{4.6}{\micro\metre}], W3[\SI{12.1}{\micro\metre}] and W4[\SI{22.2}{\micro\metre}] magnitudes for WISEA J141558.82+132023.7 are 12.160 $\pm$ 0.023, 10.839 $\pm$ 0.020, 8.058 $\pm$ 0.018 and 5.755 $\pm$ 0.032, respectively. The W1-W2 color of WISEA J141558.82+132023.7 is 1.321, which means that the mid-IR emission comes mainly from the AGN. According to the W2-W3 value of 2.781 mag, WISEA J141558.82+132023.7 locates in the QSO region in the WISE color-color diagram. 

\subsubsection{NVSS J160332+171158 (NGC 6034)}

\begin{figure*}[hbt!]
    \centering
    \includegraphics[width=0.25\textwidth]{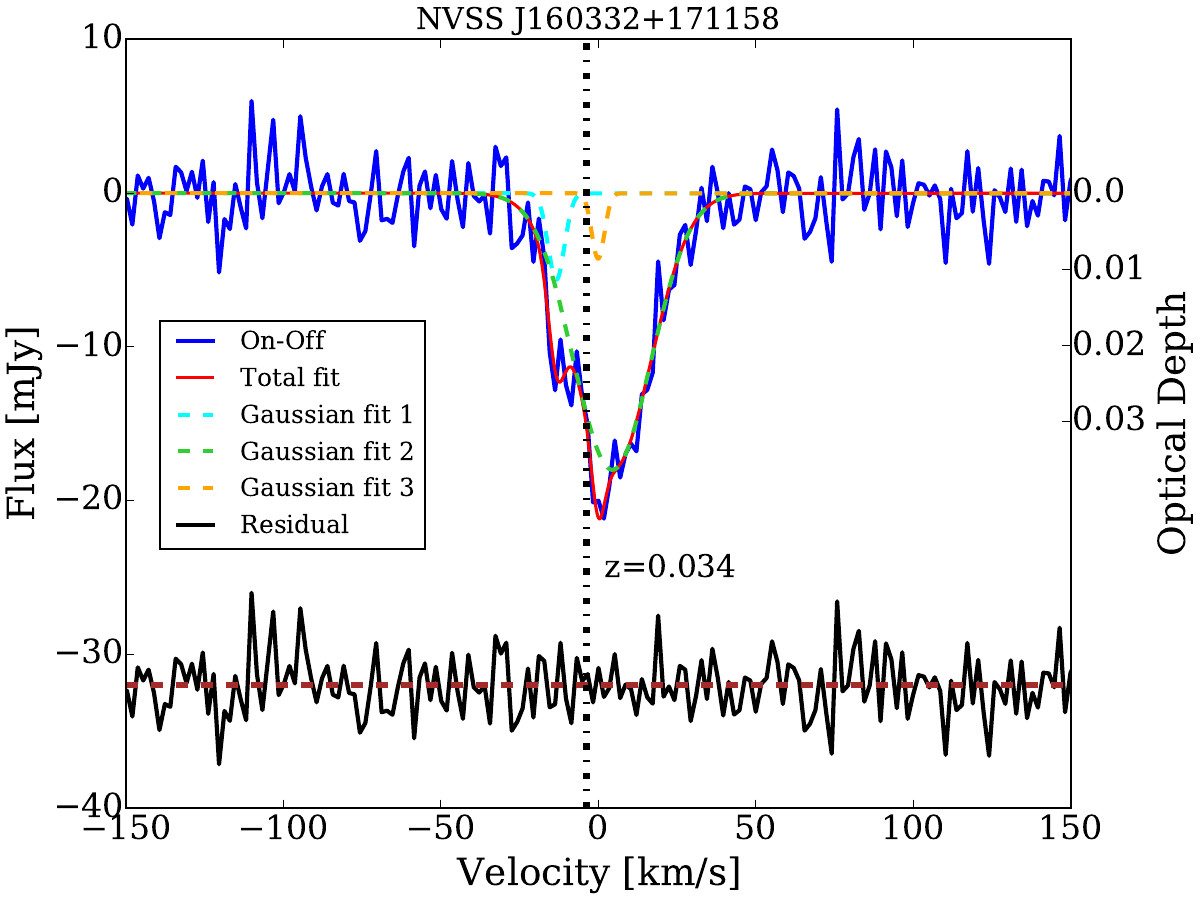}
    \includegraphics[width=0.25\textwidth]{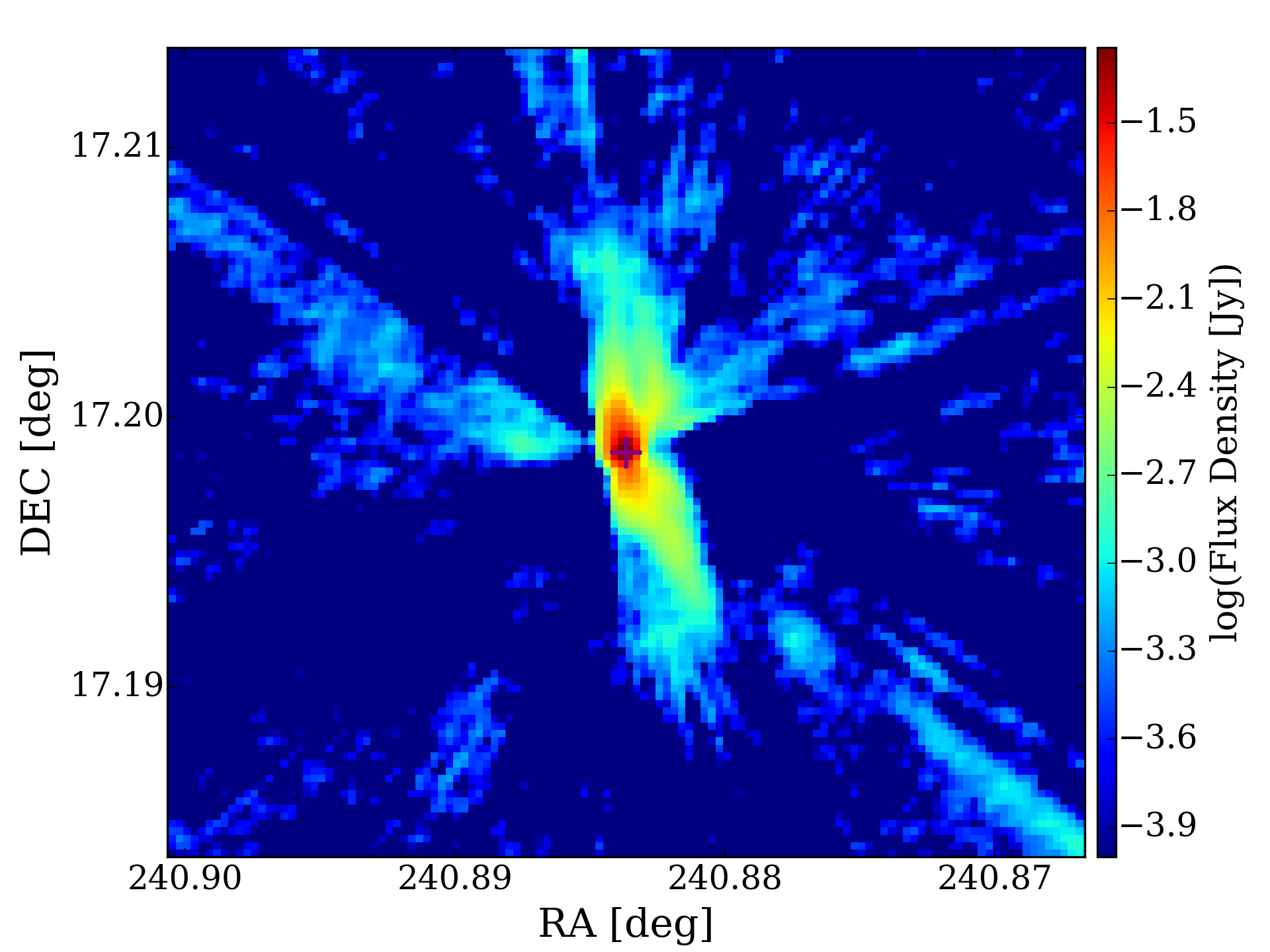}
    \includegraphics[width=0.25\textwidth]{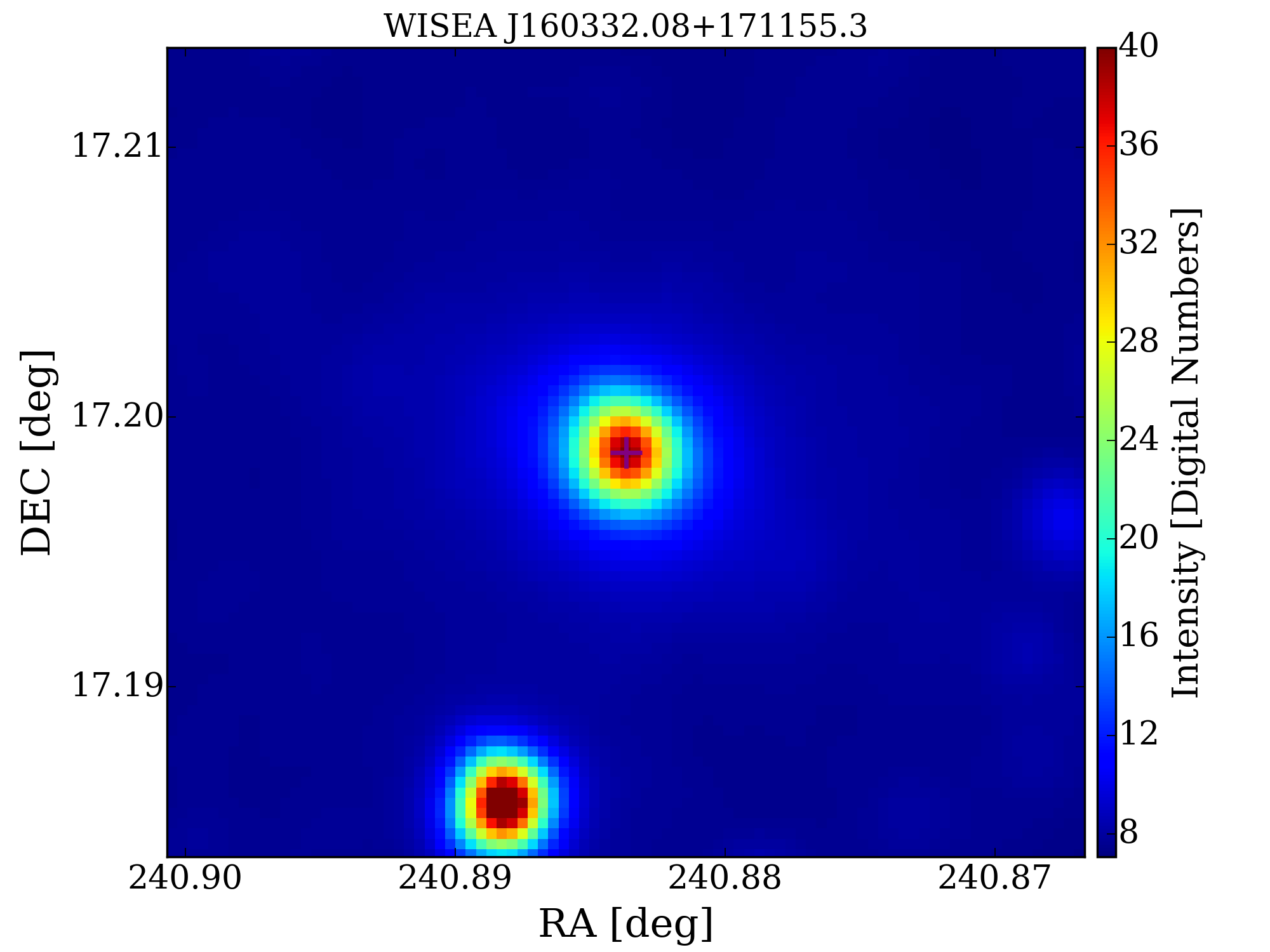}
    \includegraphics[width=0.18\textwidth]{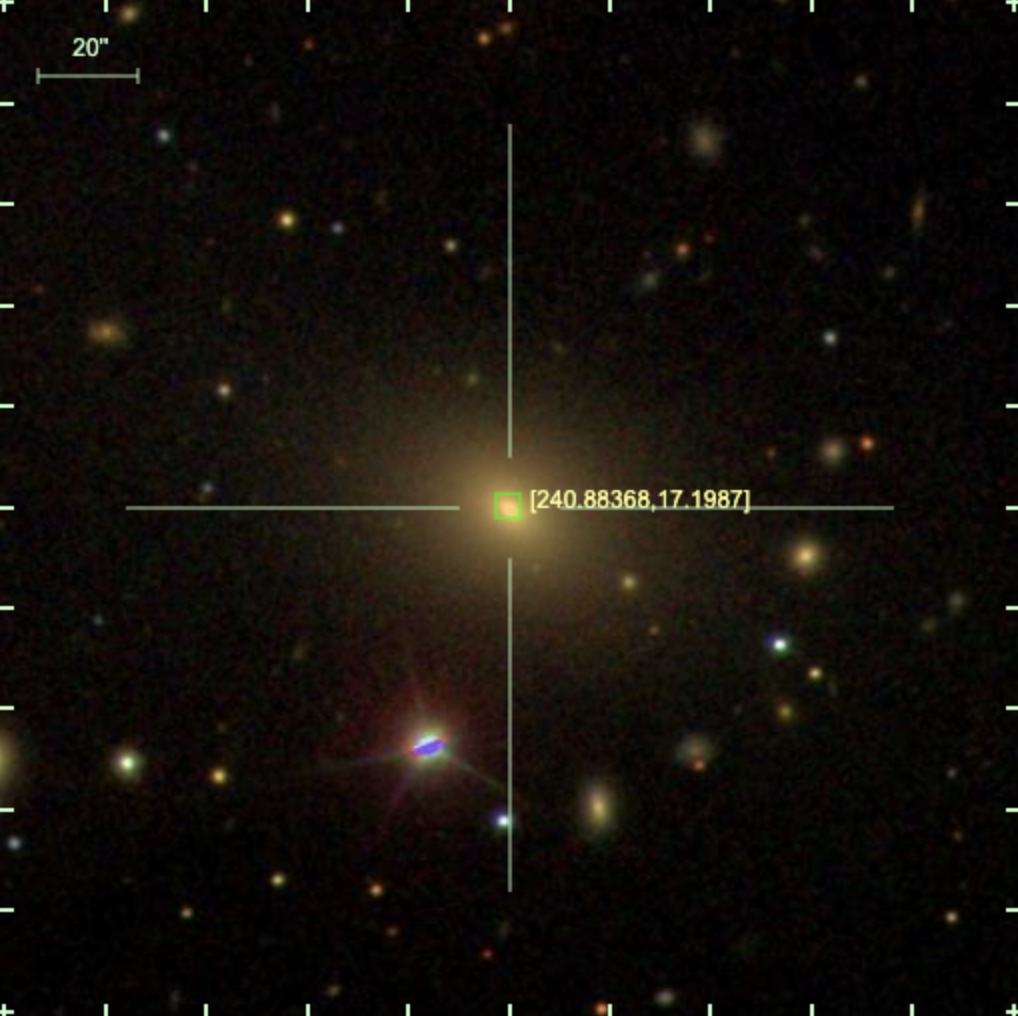}
    \caption{Left: same as Figure~\ref{4C+31.04_fit}, but for NVSS J160332\allowbreak+171158. Middle left: radio map from VLASS centered at NVSS J160332\allowbreak+171158. Middle right: W2 band infrared map of WISEA J160332.08\allowbreak+171155.3 (WISE counterpart to NVSS J160332\allowbreak+171158) from WISE. Right: SDSS optical map of the optical counterpart of WISEA J160332.08\allowbreak+171155.3.}
    \label{NVSS_J160332+171158_fit}
\end{figure*}

NGC 6034 (Figure~\ref{NVSS_J160332+171158_fit}) is identified as a luminous E/S0 radio galaxy within Abell 2151, which is part of the Hercules supercluster. This radio source is extended, featuring two jets emerging in the northern and southern directions. The absorption line observed in this galaxy is narrow and precisely centered at the velocity of the system, suggesting a potential \hi rotating disk within the host galaxy. The presence of neutral hydrogen in NGC 6034 was initially detected through VLA observations conducted by \citet{1997AJ....113.1939D} and subsequently confirmed by the Westerbork Synthesis Radio Telescope (WSRT) observations \citep{2015A&A...575A..44G}. Our independent blind detection of the \hi absorption line yielded a consistent line profile, aligning with previous observations. Notably, the optical depth we obtain differs from that given in \citet{2015A&A...575A..44G}, mainly due to their reliance on the continuum flux measured by WSRT, which is 278 mJy, while we utilize the flux measured in the line-free region near the position of the absorption, amounting to 501.78 mJy. The FAST flux might be overestimated due to its 3-arcminute beam, which could encompass other bright sources.
%We note that the flux difference could be attributed to the different frequencies at which the measurements were taken, as well as the varying resolutions of WSRT(75$\times$11 arcsec) and FAST(3 arcmin). According to VLASS measurements at 3GHz, NGC 6034 has a major axis of 12.32 arcsec and a minor axis of 3.33 arcsec.

%The WISE counterpart to NVSS\,J160332+171158 is WISEA\,J160332.08+171155.3 as shown in the NASA/IPAC Extragalactic Database. The WISE W1[\SI{3.4}{\micro\metre}], W2[\SI{4.6}{\micro\metre}], W3[\SI{12.1}{\micro\metre}] and W4[\SI{22.2}{\micro\metre}] magnitudes for WISEA J160332.08+171155.3 are 10.939 $\pm$ 0.024, 10.966 $\pm$ 0.022, 10.372 $\pm$ 0.063 and 8.476, respectively. The W1-W2 color of WISEA J160332.08+171155.3 is -0.027, which means that the mid-IR emission comes mainly from the stars. According to the W2-W3 value of 0.594 mag, WISEA J160332.08+171155.3 is located in the Ellipticals region in the WISE color-color diagram. 

\subsection{New Absorbers}

\subsubsection{NVSS J004219+570836}

\begin{figure*}[hbt!]
    \centering
    \includegraphics[width=0.25\textwidth]{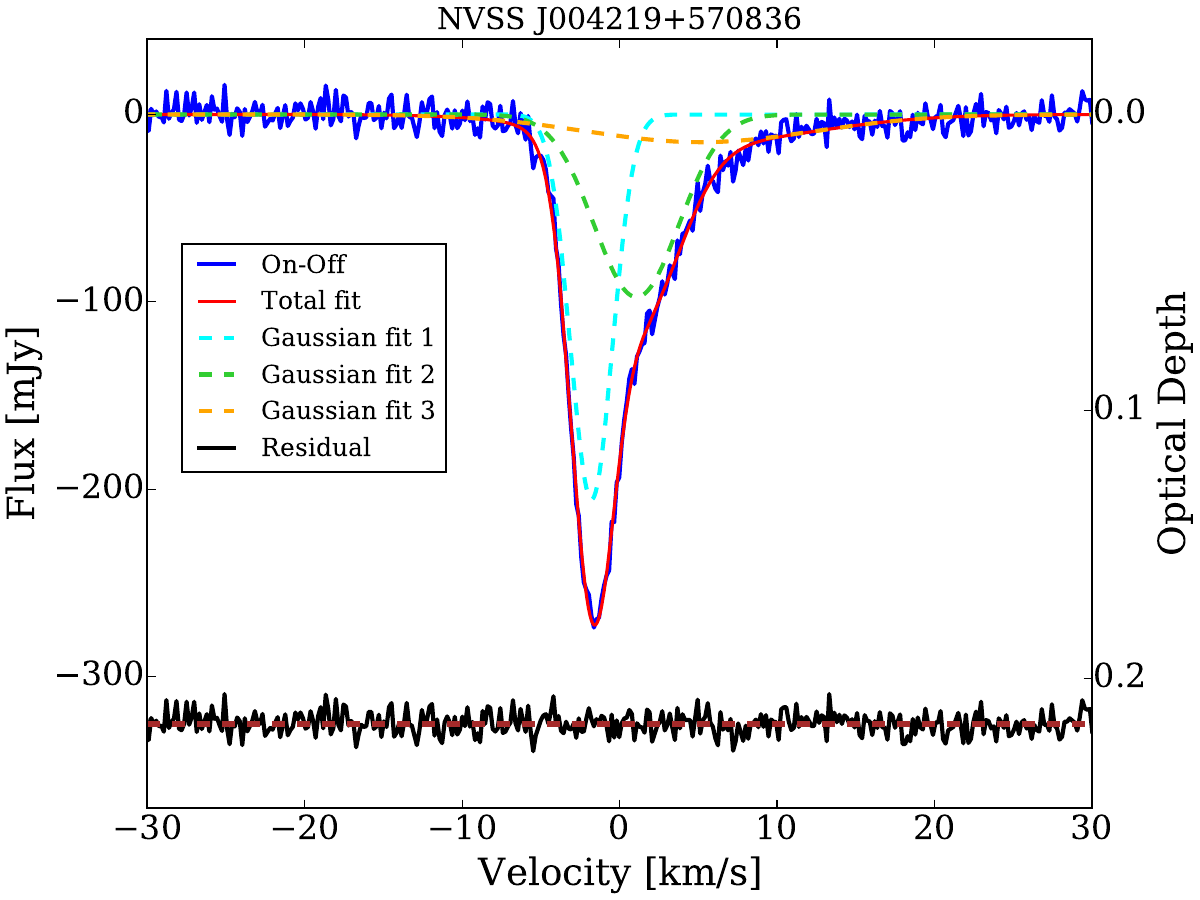}
    \includegraphics[width=0.25\textwidth]{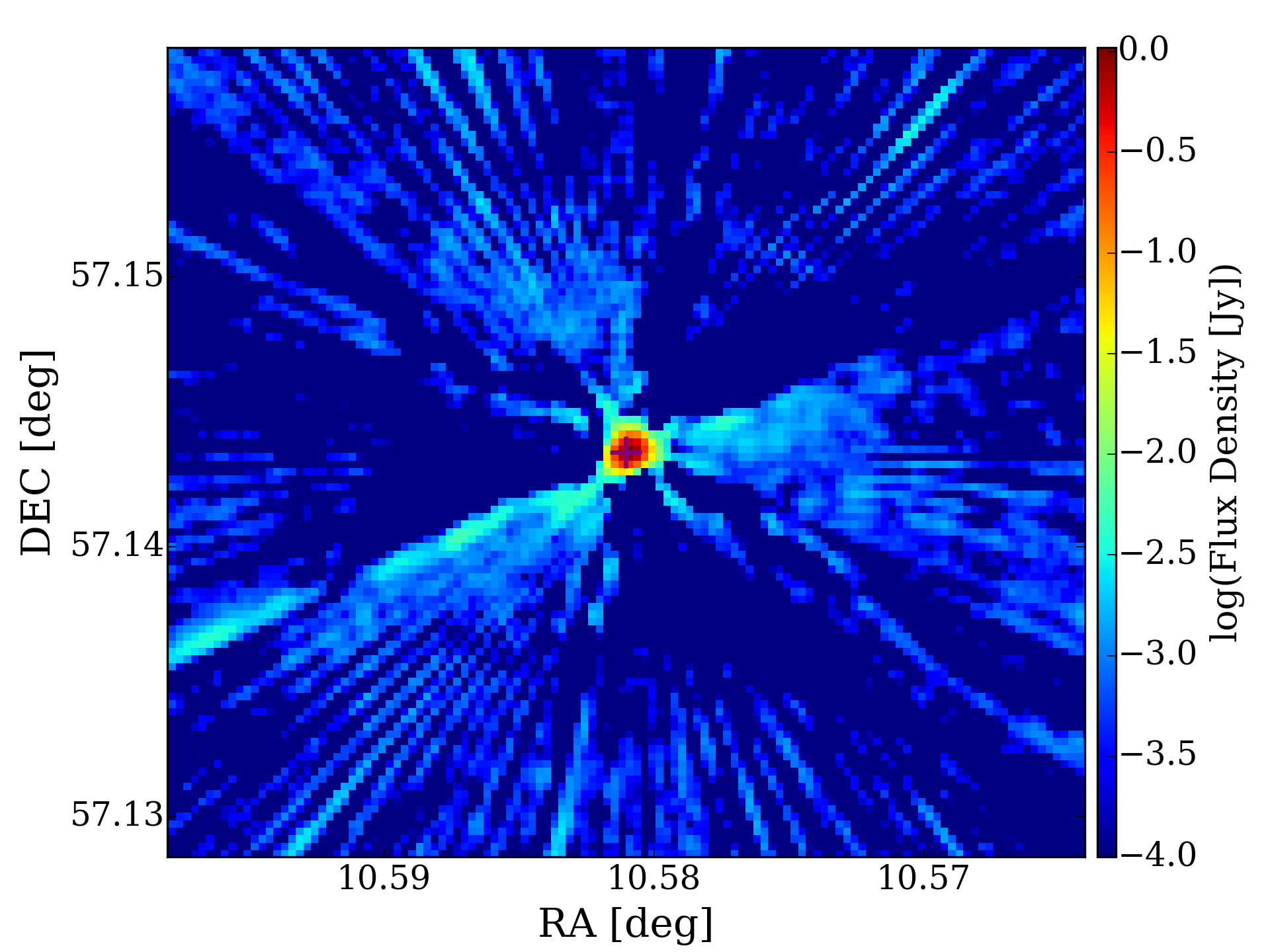}
    \includegraphics[width=0.25\textwidth]{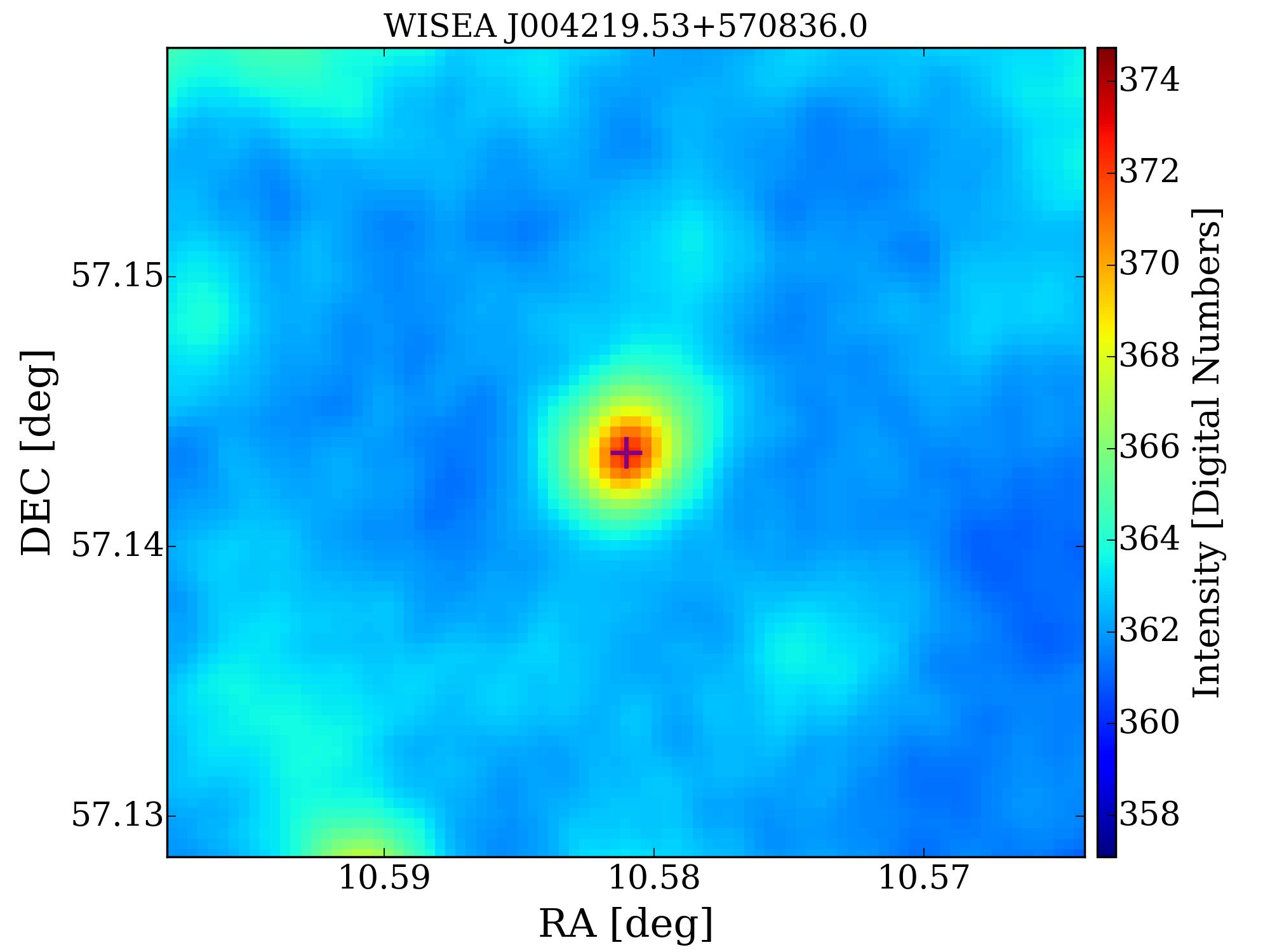}
    \caption{Left: same as Figure~\ref{4C+56.02_fit}, but for NVSS J004219\allowbreak+570836. Middle: radio map from VLASS centered at NVSS J004219\allowbreak+570836. Right: WISE W3 band infrared map centred at NVSS J004219\allowbreak+570836.}
    \label{NVSS_J004219+570836_fit}
\end{figure*}

NVSS J004219\allowbreak+570836 (Figure~\ref{NVSS_J004219+570836_fit}) is a blazar-like source located at z=1.141 and has been measured in K, Q and L band. 
%Additional high-resolution follow-up observations are necessary to establish clarity for the foreground source. 
The \hi absorption profile can be modeled using a three-component Gaussian function. This function comprises a more pronounced, narrower, and symmetric component, possibly originating from the gas disk. Additionally, there are two less pronounced, broader, and redshifted components, indicating the existence of unsettled gas structures and suggesting a potential gas accretion.

\subsubsection{NVSS J011322+251852}

\begin{figure*}[hbt!]
    \centering
    \includegraphics[width=0.25\textwidth]{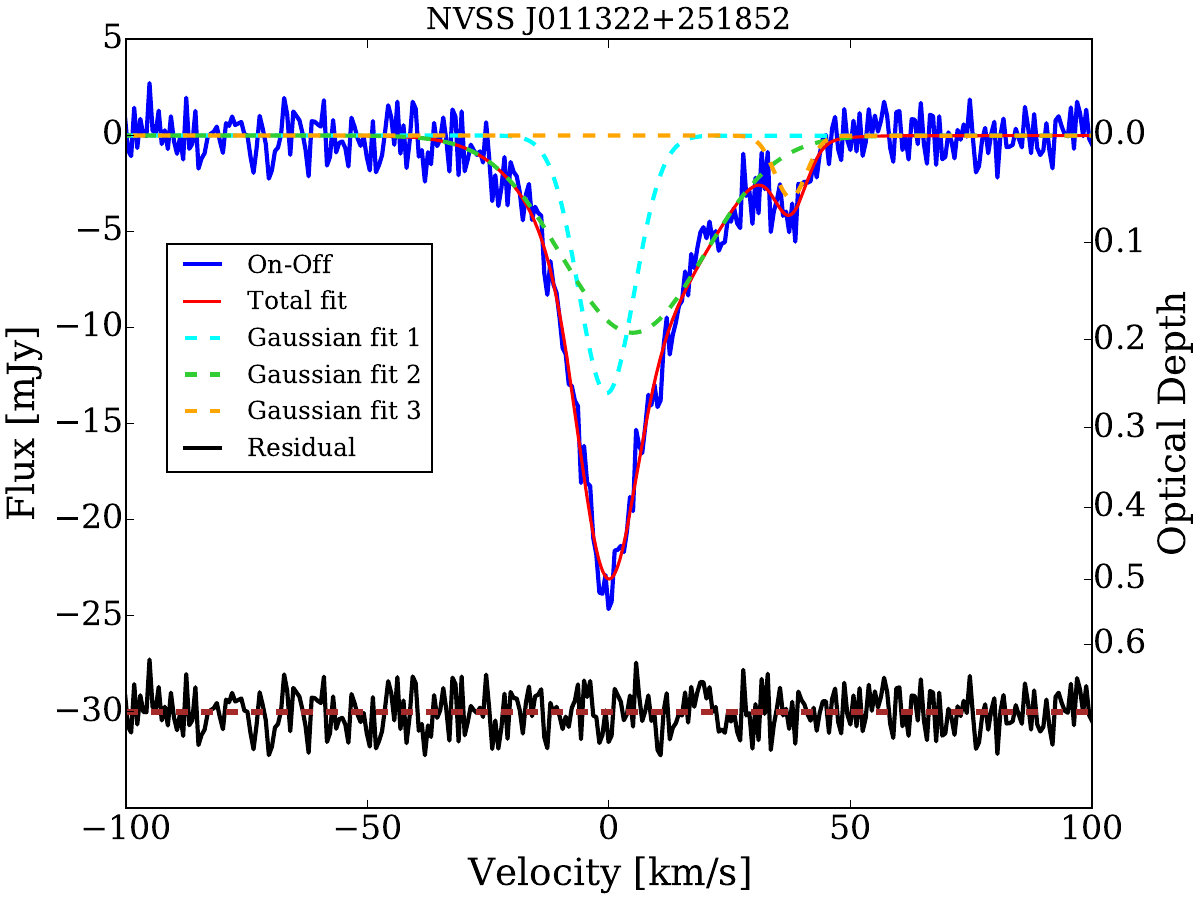}
    \includegraphics[width=0.25\textwidth]{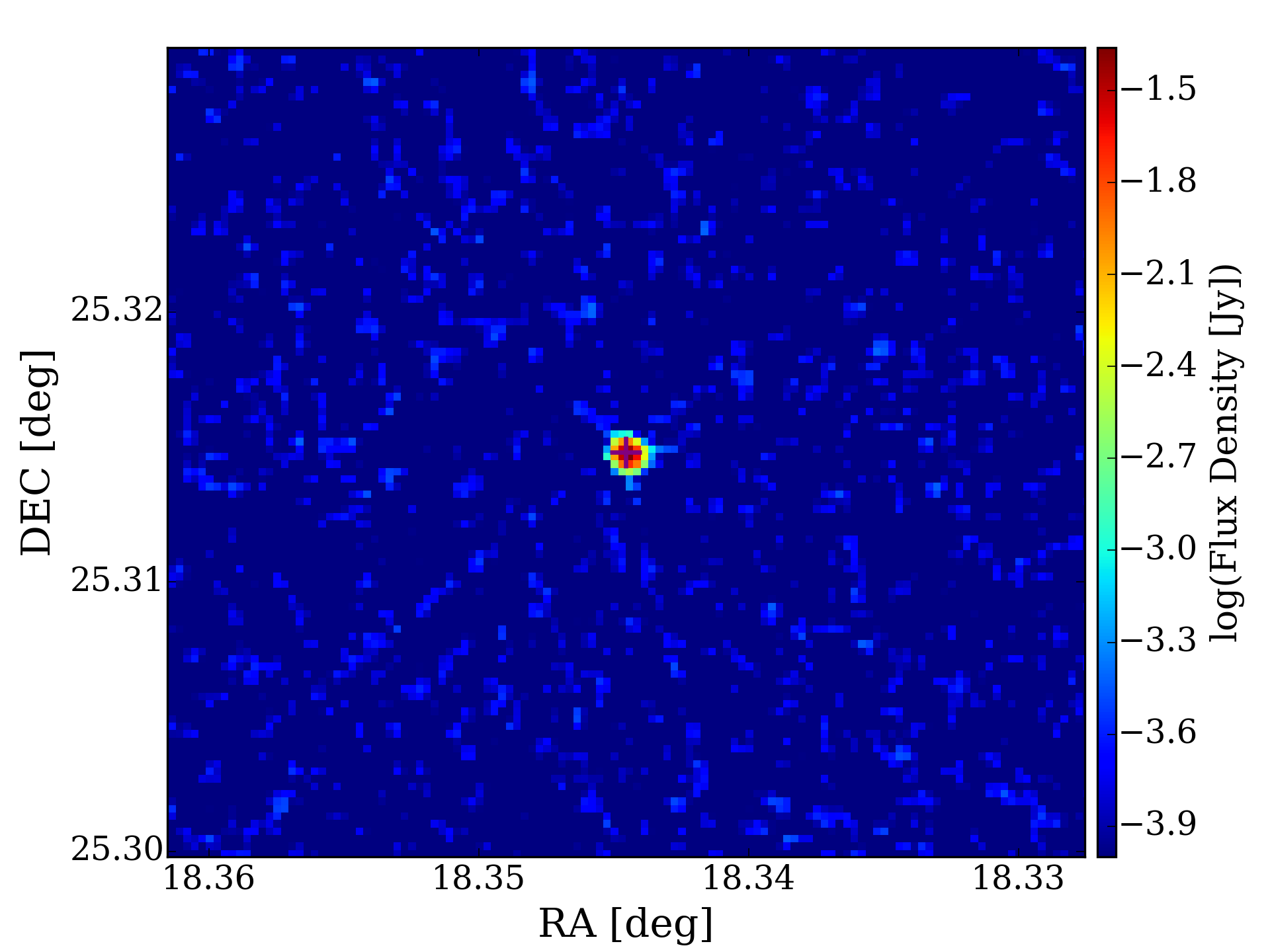}
    \includegraphics[width=0.25\textwidth]{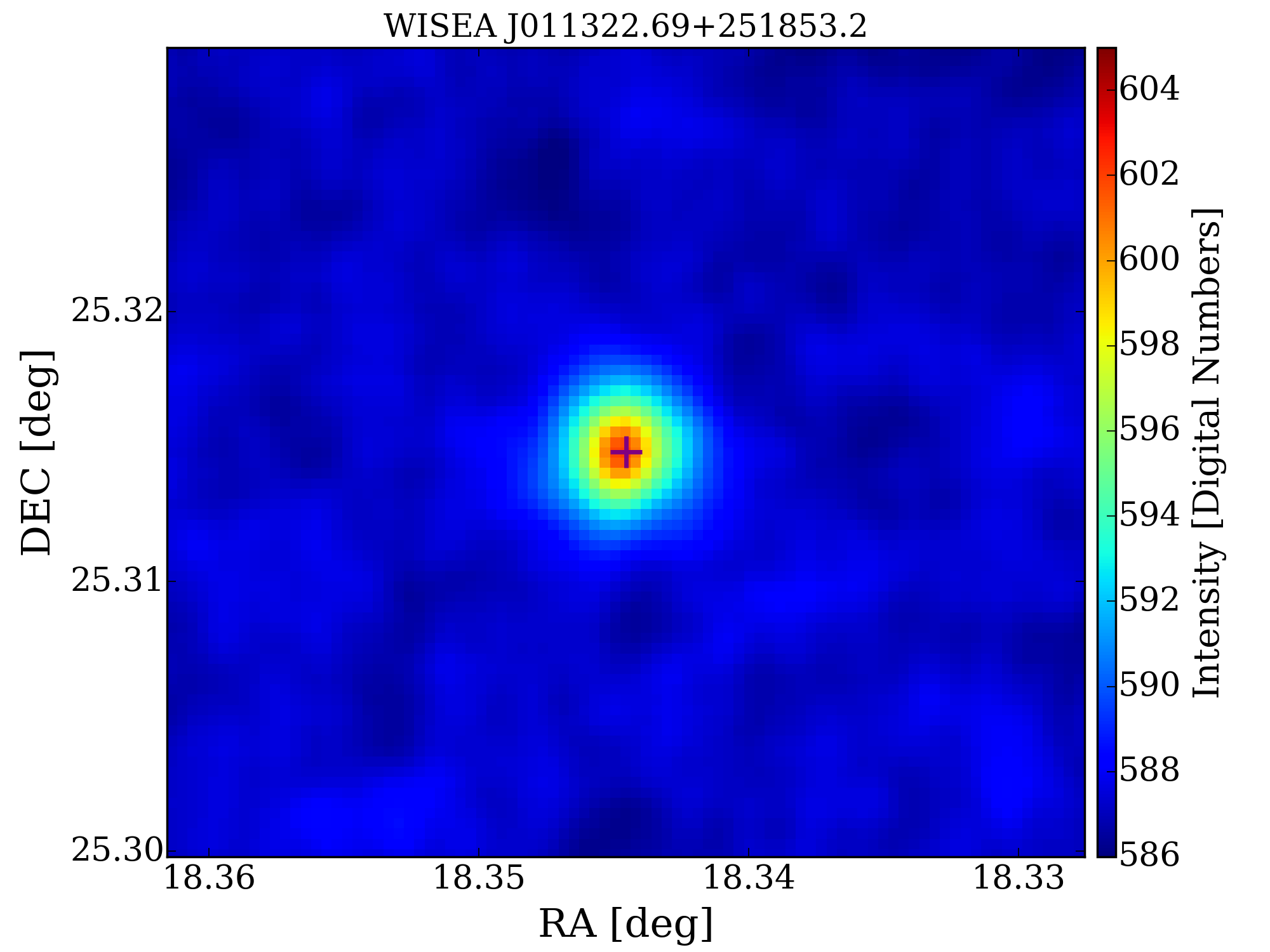}
    \hspace*{0.25cm}
    \includegraphics[width=0.18\textwidth]{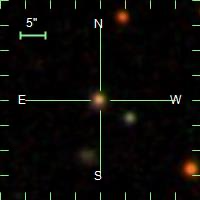}
    \caption{Left: same as Figure~\ref{4C+56.02_fit}, but for NVSS J011322\allowbreak+251852. Middle left: radio map from VLASS centered at NVSS J011322\allowbreak+251852. Middle right: WISE W3 band infrared map centered at NVSS J011322\allowbreak+251852. Right: SDSS optical map centered at NVSS J011322\allowbreak+251852.}
    \label{NVSS_J011322+251852_fit}
\end{figure*}

NVSS J011322\allowbreak+251852 (Figure~\ref{NVSS_J011322+251852_fit}) is a blazar-like source located at z=1.589 and has been measured in X-ray (Burst Alert Telescope, \citealt{2018ApJS..235....4O}), ultraviolet (GALEX, \citealt{2005ApJ...619L...1M}), infrared (WISE and 2MASS) and radio band (NVSS). Optical spectroscopic data is available from the SDSS-IV Extended Baryon Oscillation Spectroscopic Survey (eBOSS \citealt{2015ApJS..221...27M}). 
%Additional high-resolution follow-up observations are necessary to establish clarity for the foreground source. 
The \hi absorption profile displays a more intense and symmetric component, likely stemming from the gas disk. Furthermore, a shallower redshifted wing is present, suggesting the presence of unsettled gas structures and hinting at a potential accretion of gas.

\subsubsection{4C +26.07}

\begin{figure*}[hbt!]
    \centering
    \includegraphics[width=0.25\textwidth]{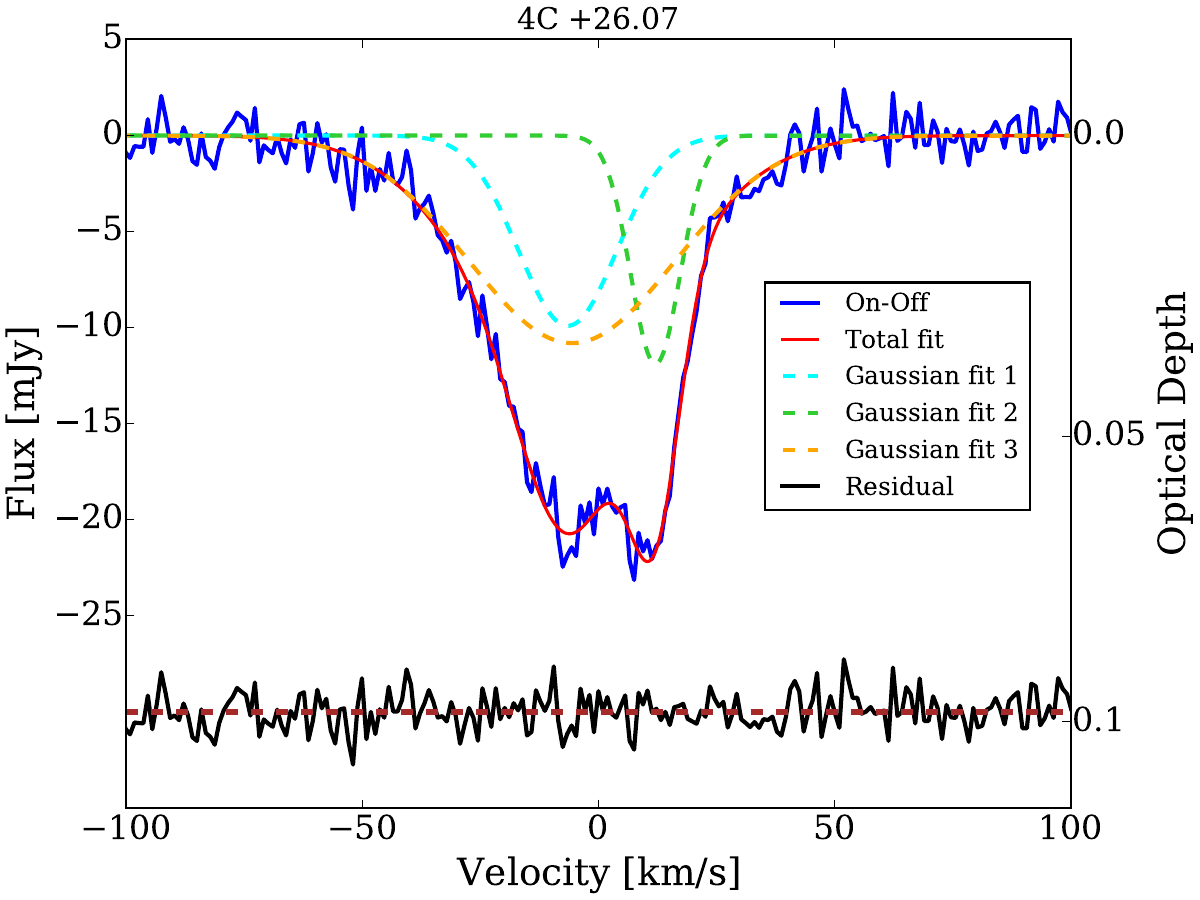}
    \includegraphics[width=0.25\textwidth]{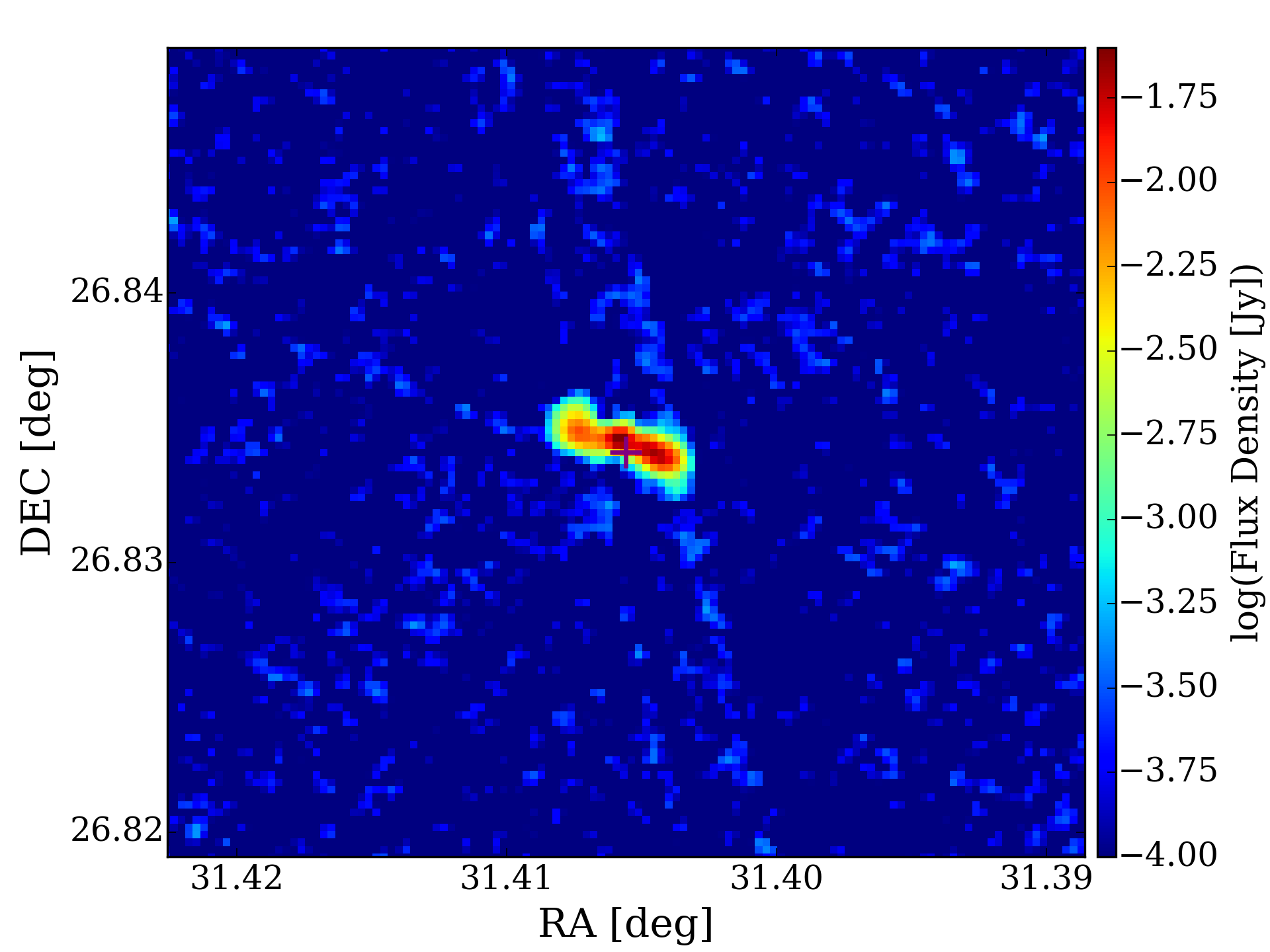}
    \includegraphics[width=0.25\textwidth]{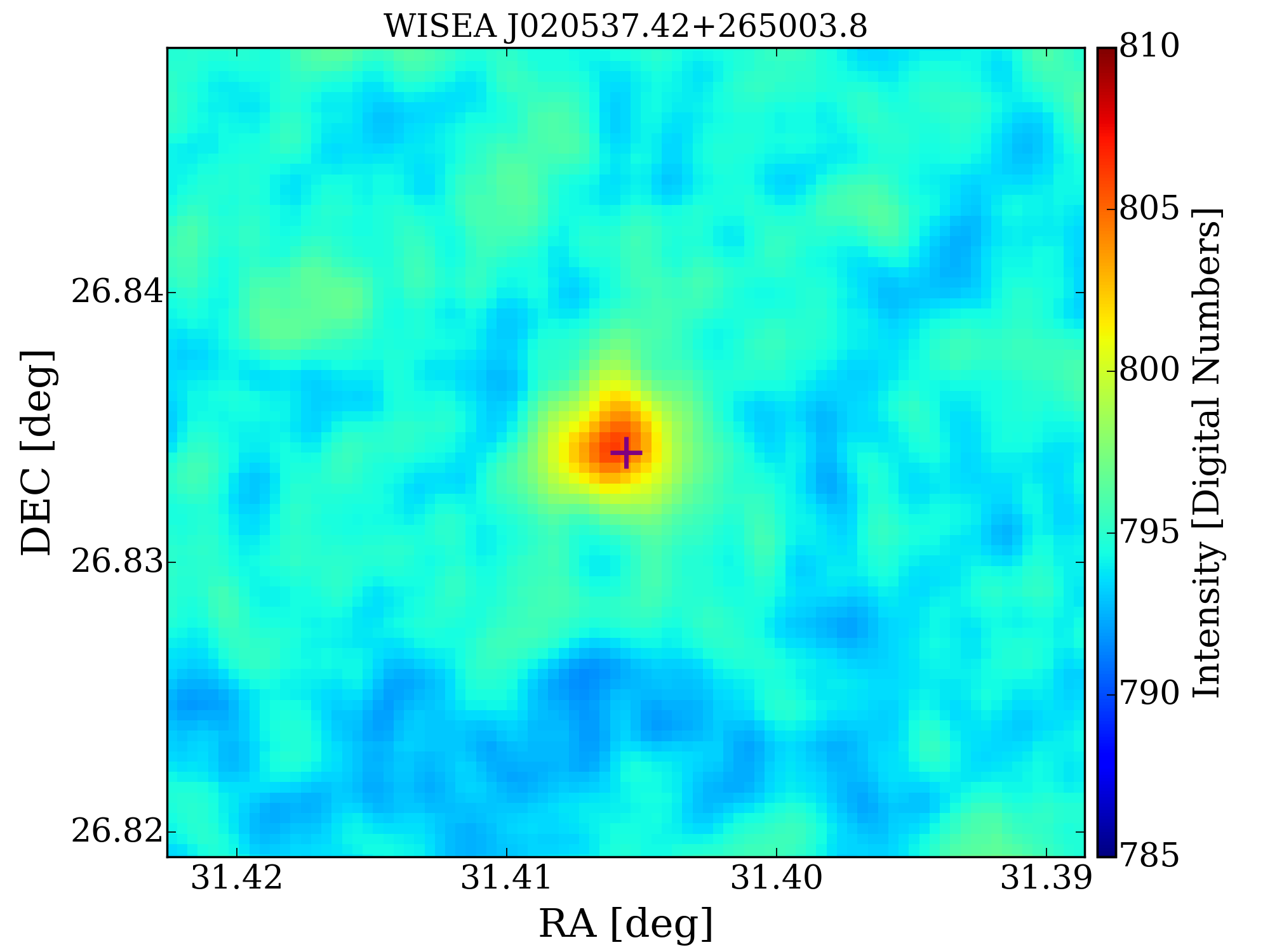}
    \hspace*{0.25cm}
    \includegraphics[width=0.18\textwidth]{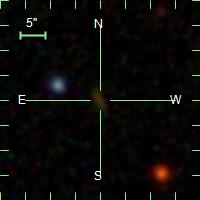}
    \caption{Left: same as Figure~\ref{4C+56.02_fit}, but for 4C\,+26.07. Middle left: radio map from VLASS centered at 4C\,+26.07. Middle right: WISE W3 band infrared map centered at 4C\,+26.07. Right: SDSS optical map of the optical counterpart of 4C\,+26.07.}
    \label{4C+26.07_fit}
\end{figure*}

4C\,+26.07 (Figure~\ref{4C+26.07_fit}) is a radio source which has been detected in the VLA Low-Frequency Sky Survey (VLSS, \citealt{2007AJ....134.1245C}) and NVSS. Precise redshift details for 4C\,+26.07 are currently unavailable, resulting in uncertainty in the identification of the counterpart. A three-component Gaussian function can effectively model the \hi absorption profile. It comprises two narrower components at the center, accompanied by a broad wing signifying a possible gas disk.

%Its optical and mid-infrared photometric flux density has been measured in WISE and SDSS. The NASA/IPAC Extragalactic Database shows that the WISE counterpart to 4C\,+26.07 is WISEA J020537.42+265003.8. The WISE W1[\SI{3.4}{\micro\metre}], W2[\SI{4.6}{\micro\metre}], W3[\SI{12.1}{\micro\metre}] and W4[\SI{22.2}{\micro\metre}] magnitudes for WISEA J020537.42+265003.8 are 14.942 $\pm$ 0.034, 13.343 $\pm$ 0.039, 9.912 $\pm$ 0.094 and 7.476 $\pm$ 0.154, respectively. The W1-W2 colour of WISEA J020537.42+265003.8 is 1.599, implying WISEA J020537.42+265003.8 may host an AGN. Combined with the W2-W3 value of 3.431 mag, WISEA J020537.42+265003.8 lies in the QSOs region in the WISE color-color diagram. 

%We blindly detected its \hi absorption in beam 12 in the CRAFTS drift-scan data and verified it in the follow-up observation in ON-OFF mode. The absorption spectra towards 4C\,+26.07 and its four-components Gaussian fitting are presented in the left panel of Figure~\ref{4C+26.07_fit}. Under the assumption $T_{\rm s} \ll c_{\rm f}T_{\rm c}$, our measurements of the \hi absorption towards 4C\,+26.07 give a flux density depth of $S_{\hi,\rm peak} \sim -20.23 \mJy$, an FWHM of $\sim 41.89 \kms$, $\int\tau dv\sim 3.84 \kms$ and $N_{\hi} \sim 0.070T_{s}$10$^{20}$cm$^{-2}$K$^{-1}$.

%The middle left, middle right panel, and right panel of Figure~\ref{4C+26.07_fit} show the image centered at 4C\,+26.07, constructed using radio data at S-band from VLASS, infrared data at 12.1 microns (W3) from WISE and optical data from SDSS, respectively. The straight cross shows the position of 4C\,+26.07.

\subsubsection{NVSS J033529+195621}

\begin{figure*}[hbt!]
    \centering
    \includegraphics[width=0.25\textwidth]{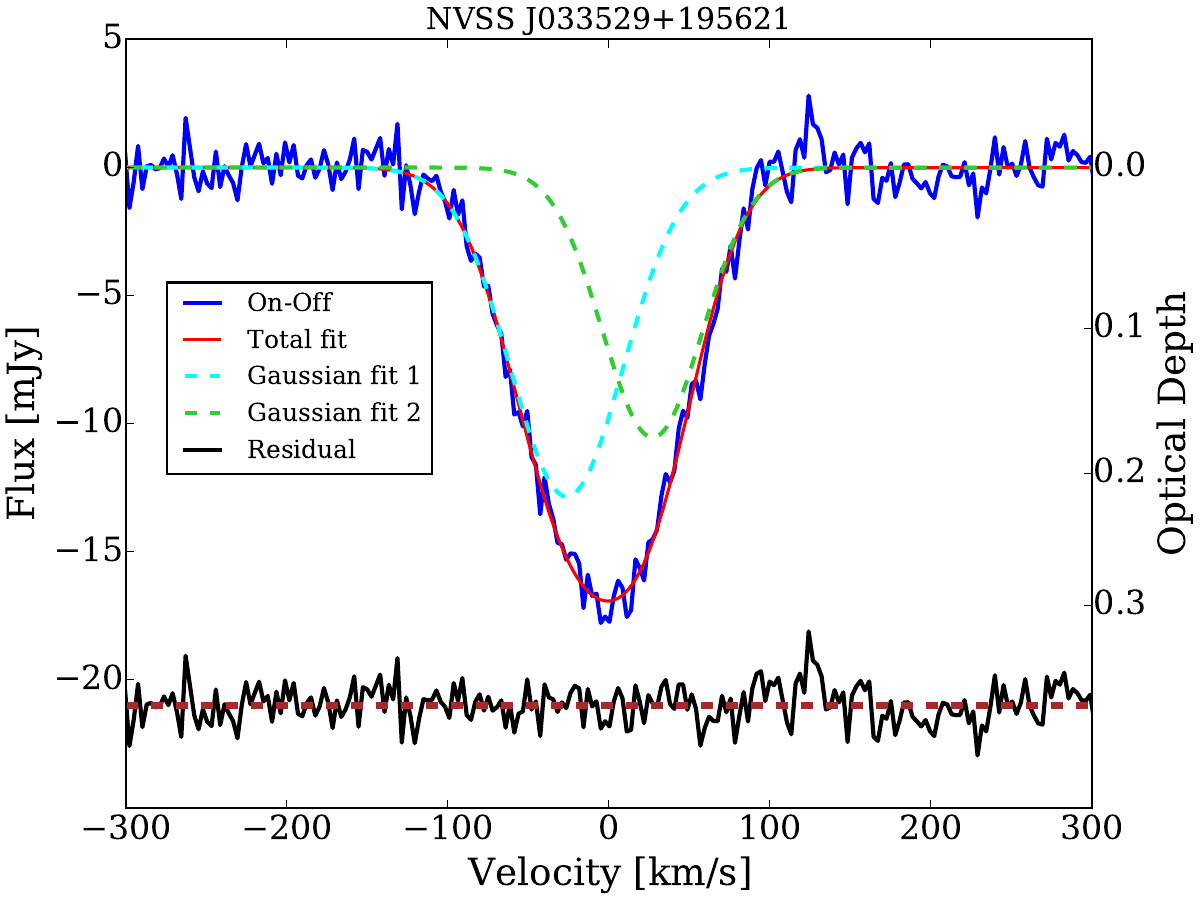}
    \includegraphics[width=0.25\textwidth]{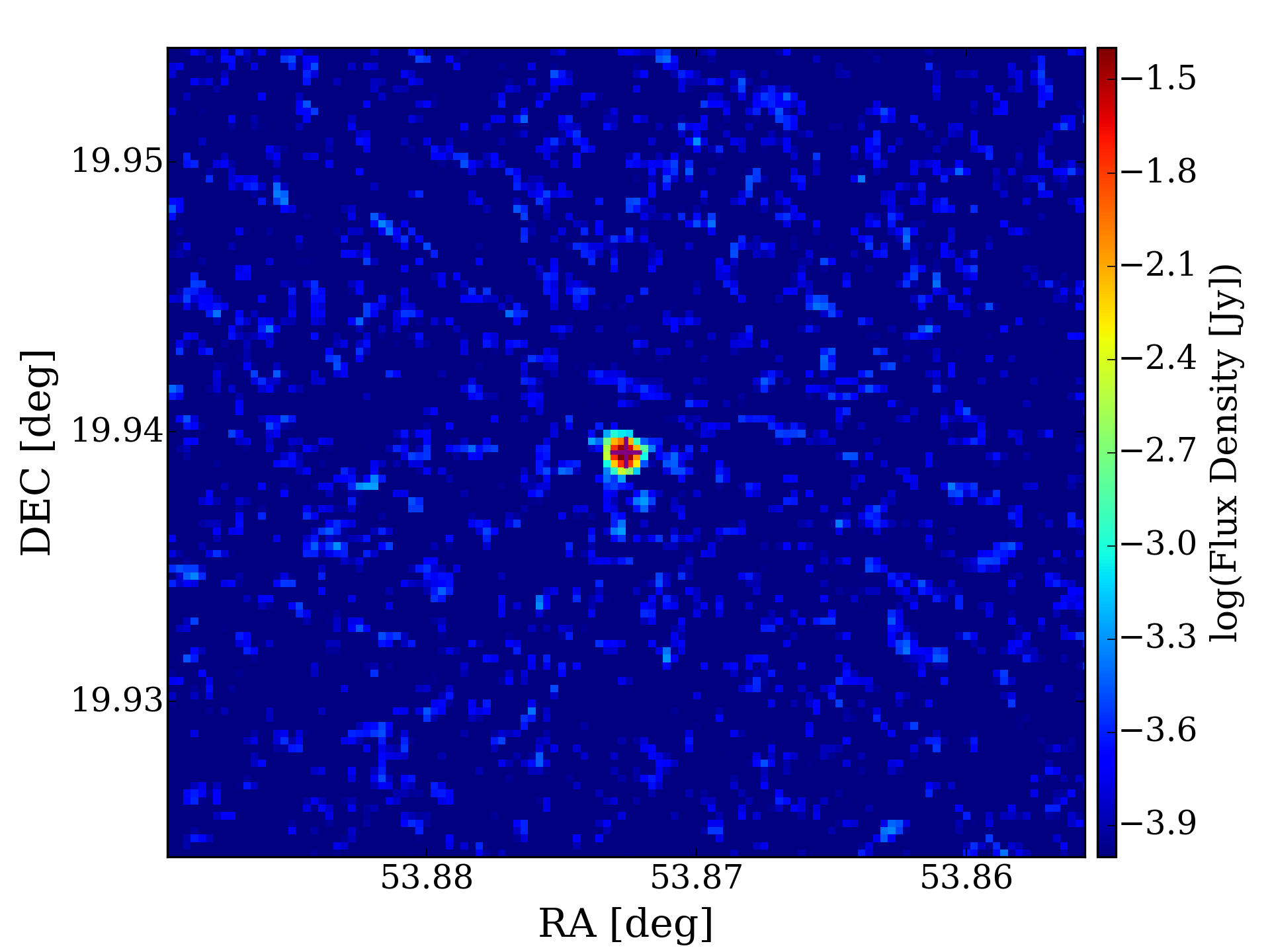}
    \includegraphics[width=0.25\textwidth]{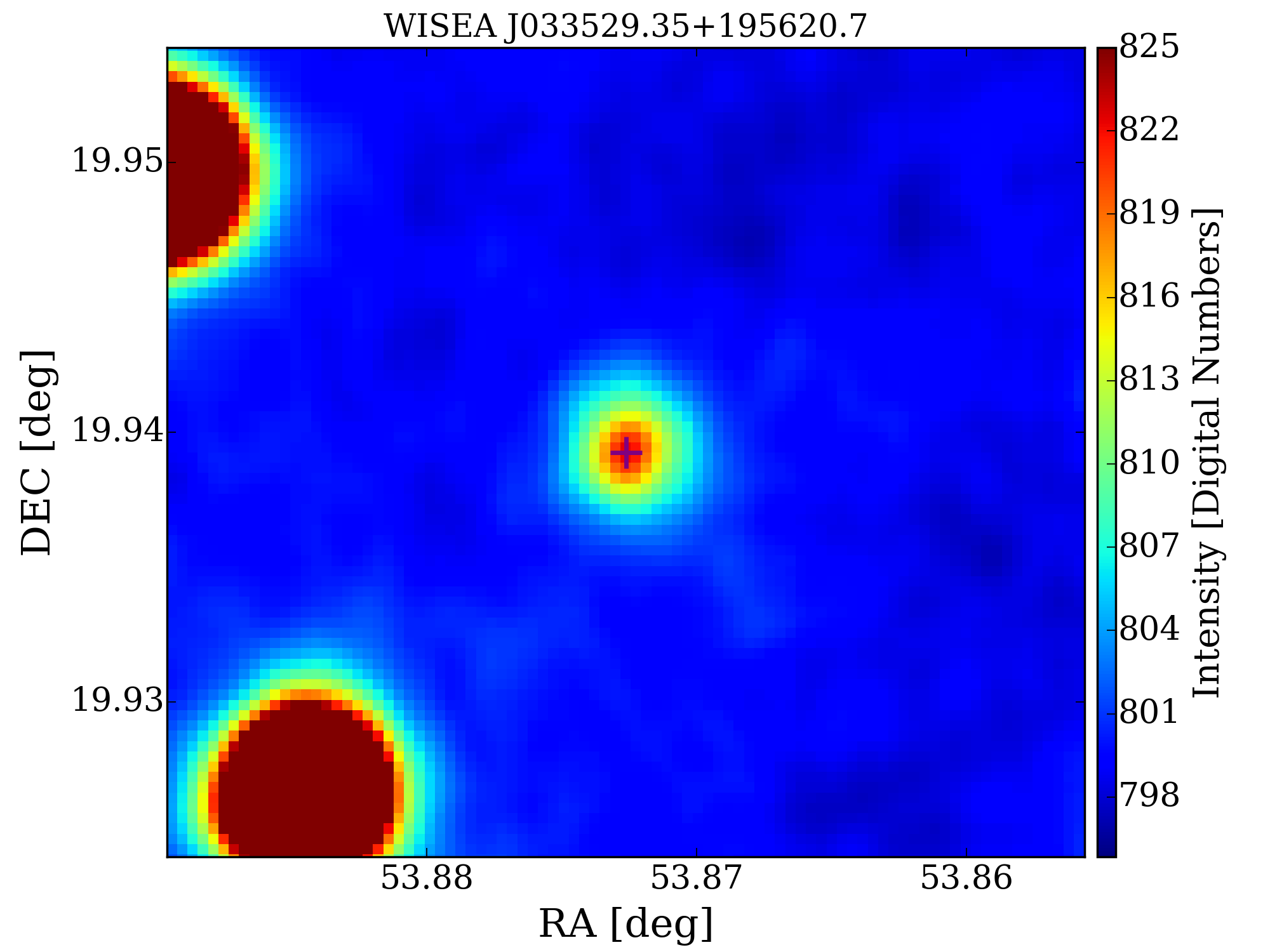}
    \caption{Left: same as Figure~\ref{4C+56.02_fit}, but for NVSS J033529\allowbreak+195621. Middle: radio map from VLASS centered at NVSS J033529\allowbreak+195621. Right: WISE W3 band infrared map centered at NVSS J033529\allowbreak+195621.}
    \label{NVSS_J033529+195621_fit}
\end{figure*}

NVSS J033529\allowbreak+195621 (Figure~\ref{NVSS_J033529+195621_fit}) is a radio source which is relatively little studied. Its infrared and radio signals have been observed by WISE and NVSS. %The lack of precise redshift information for NVSS J033529\allowbreak+195621 gives rise to uncertainty regarding its status as a foreground or background object. To dispel this ambiguity, additional high-resolution follow-up observations are essential. 
The \hi absorption profile, characterized by its broad and symmetric nature, can be accurately modeled using a two-component Gaussian function.

\subsubsection{NVSS J040845+001306}

\begin{figure*}[hbt!]
    \centering
    \includegraphics[width=0.25\textwidth]{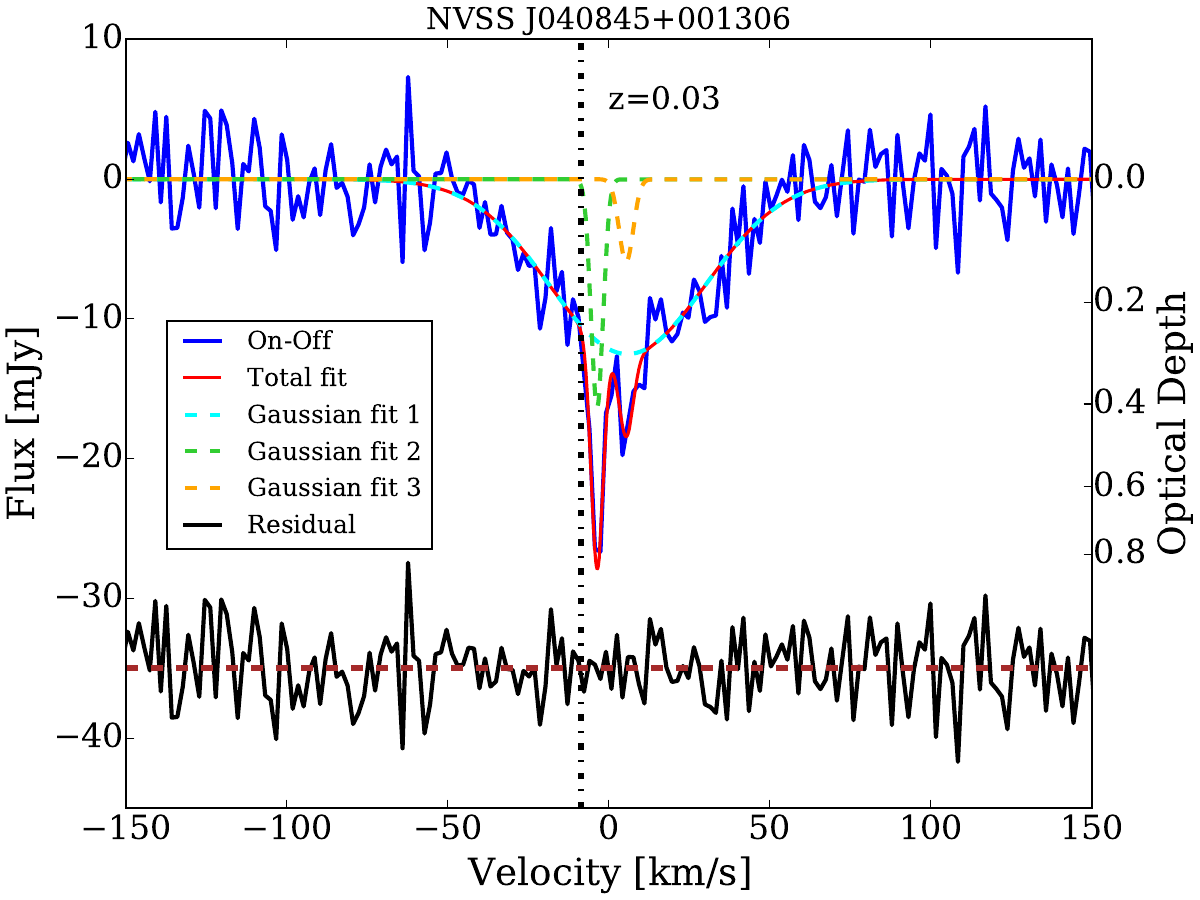}
    \includegraphics[width=0.25\textwidth]{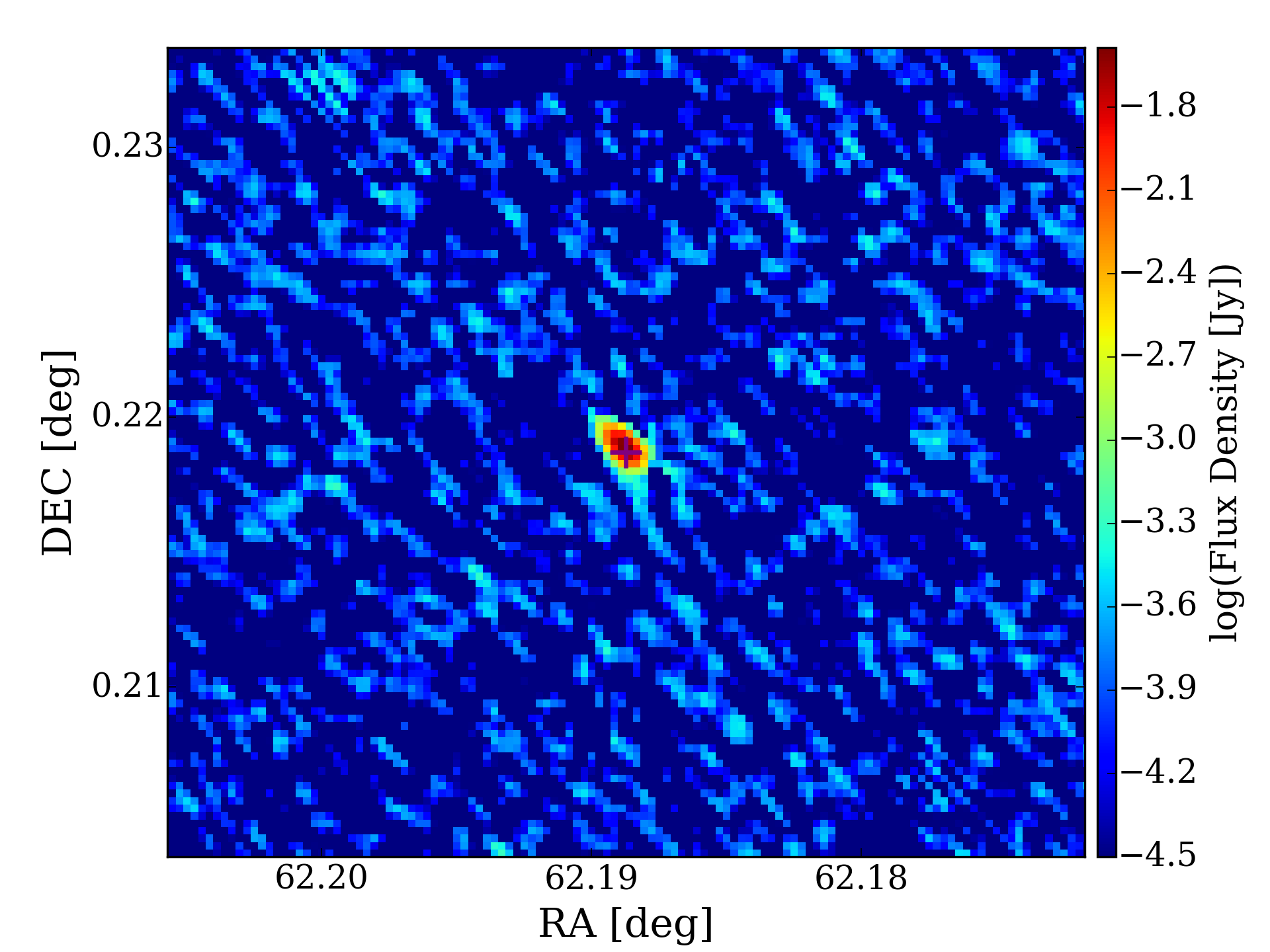}
    \includegraphics[width=0.25\textwidth]{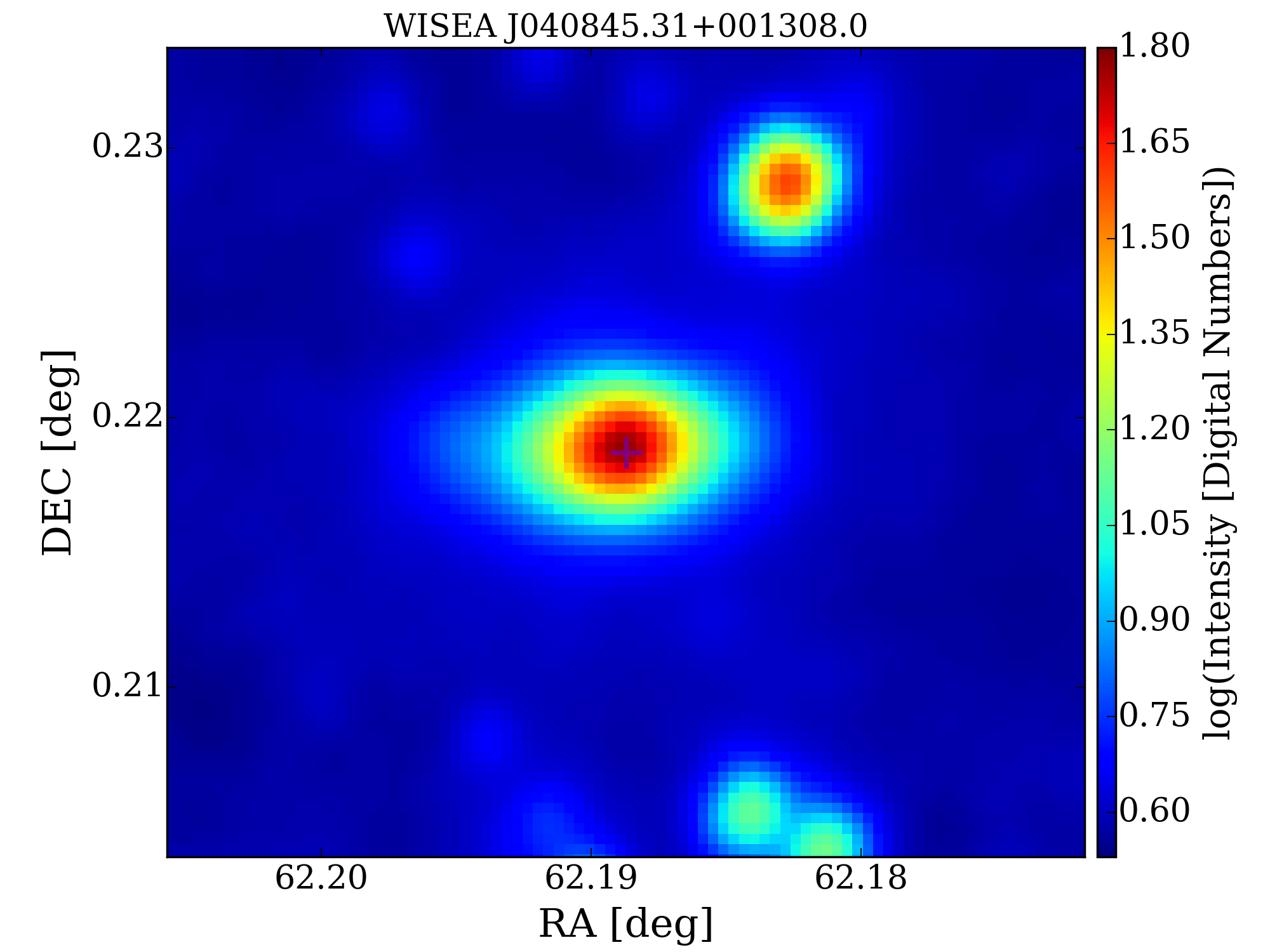}
    \includegraphics[width=0.18\textwidth]{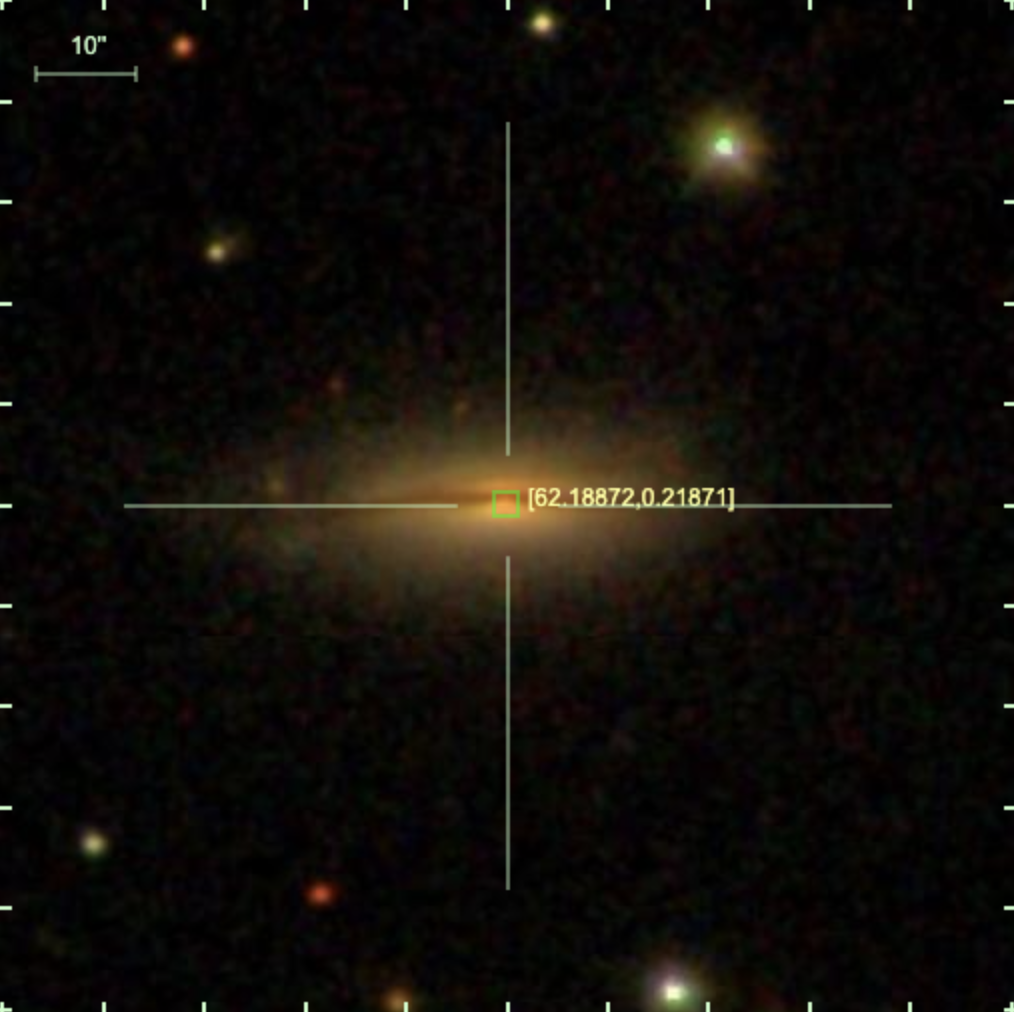}
    \caption{Left: same as Figure~\ref{4C+31.04_fit}, but for NVSS J040845\allowbreak+001306. Middle left: radio map from VLASS centered at NVSS J040845\allowbreak+001306. Middle right: W1 band infrared map of WISEA J040845.31\allowbreak+001308.0. from WISE. Right: SDSS optical map of the optical counterpart of WISEA J040845.31\allowbreak+001308.0.}
    \label{NVSS_J040845+001306_fit}
\end{figure*}

NVSS J040845\allowbreak+001306 (Figure~\ref{NVSS_J040845+001306_fit}) is a galaxy divided by a prominent east-west dust lane. The associated \hi absorption profile comprises a broader Gaussian-shaped component and a distinct double-horn feature. The predominant absorption is likely associated with diffuse gas, while the sharp component may stem from gas in rotational motion within the galaxy.

%The WISE counterpart to NVSS\,J040845+001306 is WISEA\,J040845.31+001308.0 as shown in the NASA/IPAC Extragalactic Database. The WISE W1[\SI{3.4}{\micro\metre}], W2[\SI{4.6}{\micro\metre}], W3[\SI{12.1}{\micro\metre}] and W4[\SI{22.2}{\micro\metre}] magnitudes for WISEA\,J040845.31+001308.0 are 11.511 $\pm$ 0.024, 11.433 $\pm$ 0.022, 8.520 $\pm$ 0.023 and 6.698 $\pm$ 0.065, respectively. The W1-W2 color of WISEA\,J040845.31+001308.0 is 0.078, which means that the mid-IR emission comes mainly from the stars. According to the W2-W3 value of 2.913 mag, WISEA\,J040845.31+001308.0 is located in the Spirals region in the WISE color-color diagram.

\subsubsection{NVSS J055437+271126}

\begin{figure*}[hbt!]
    \centering
    \includegraphics[width=0.25\textwidth]{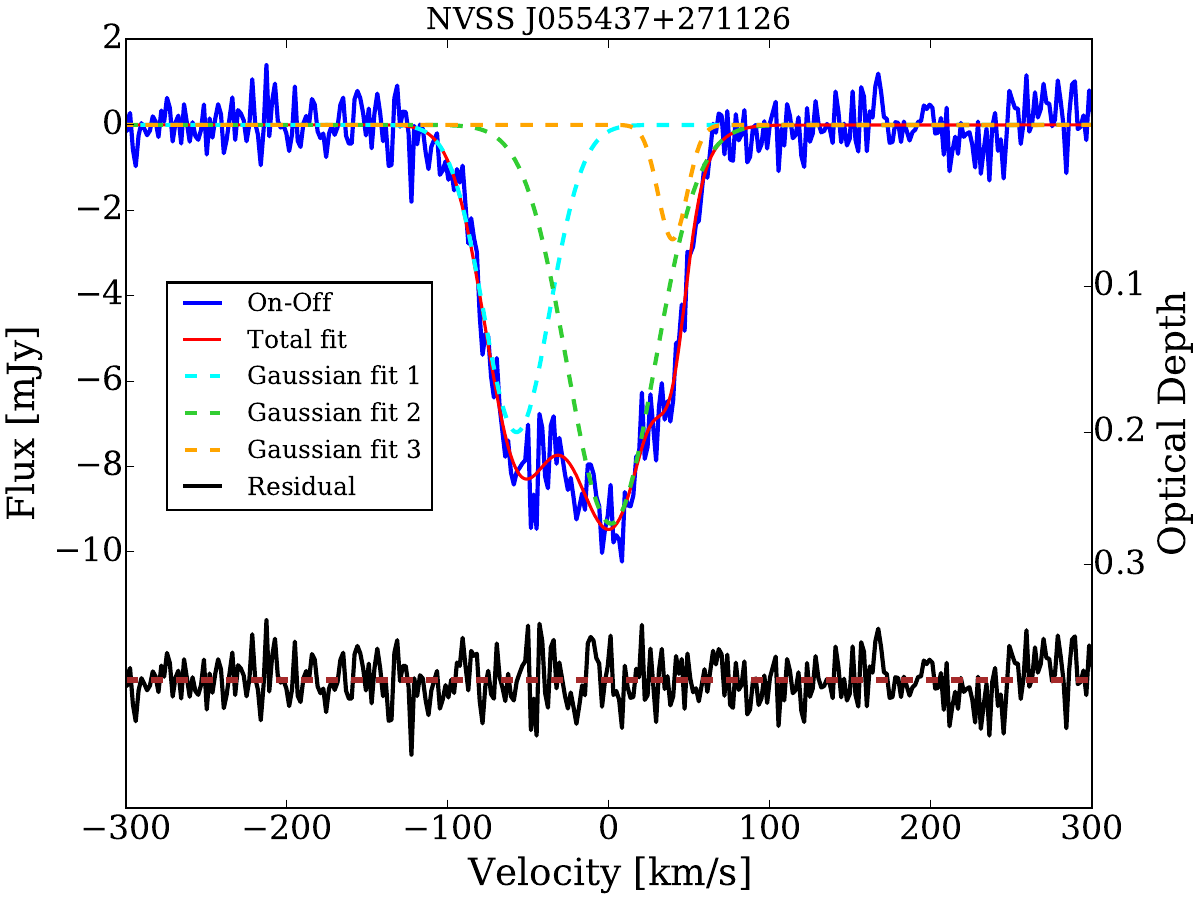}
    \includegraphics[width=0.25\textwidth]{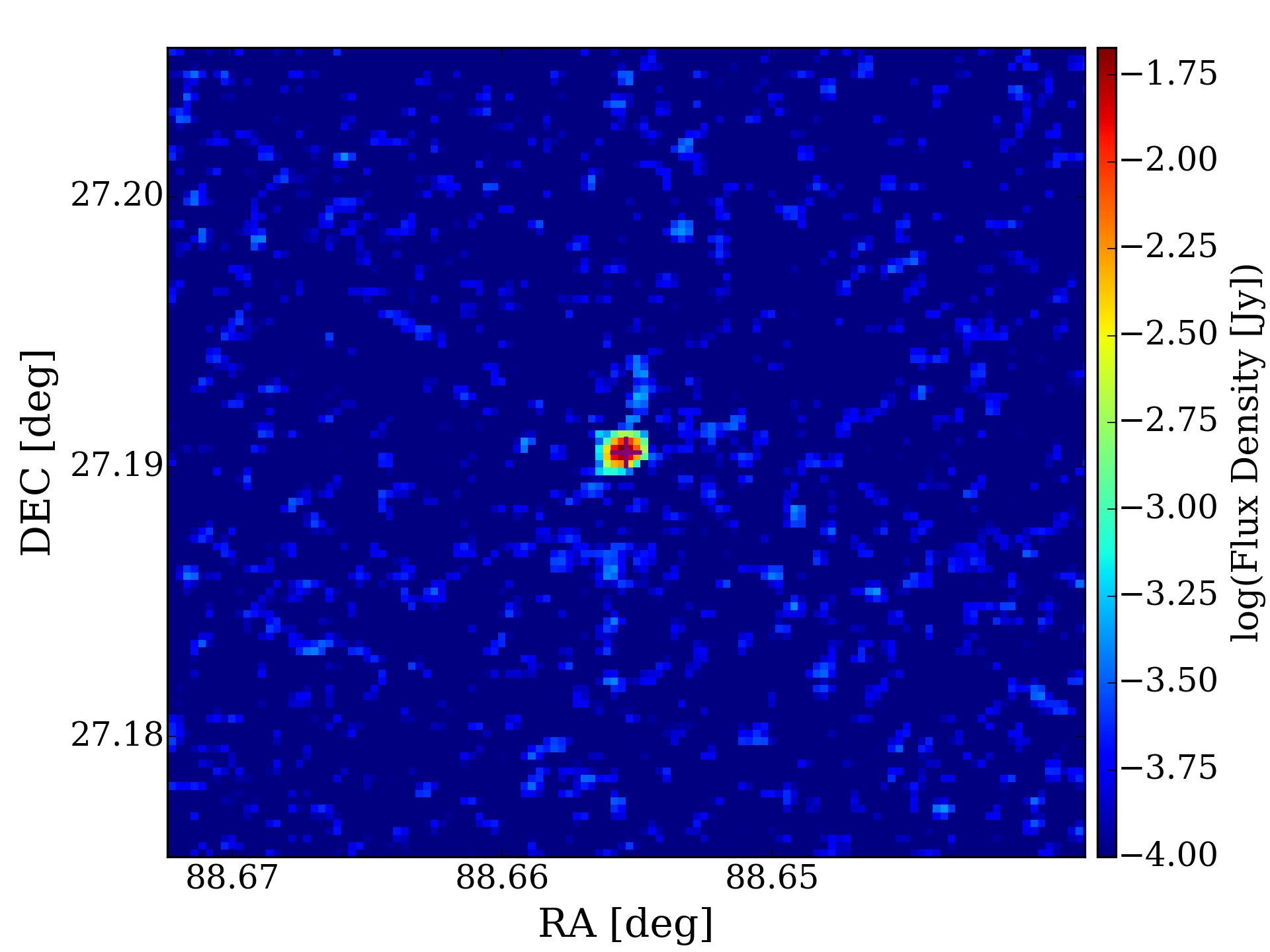}
    \includegraphics[width=0.25\textwidth]{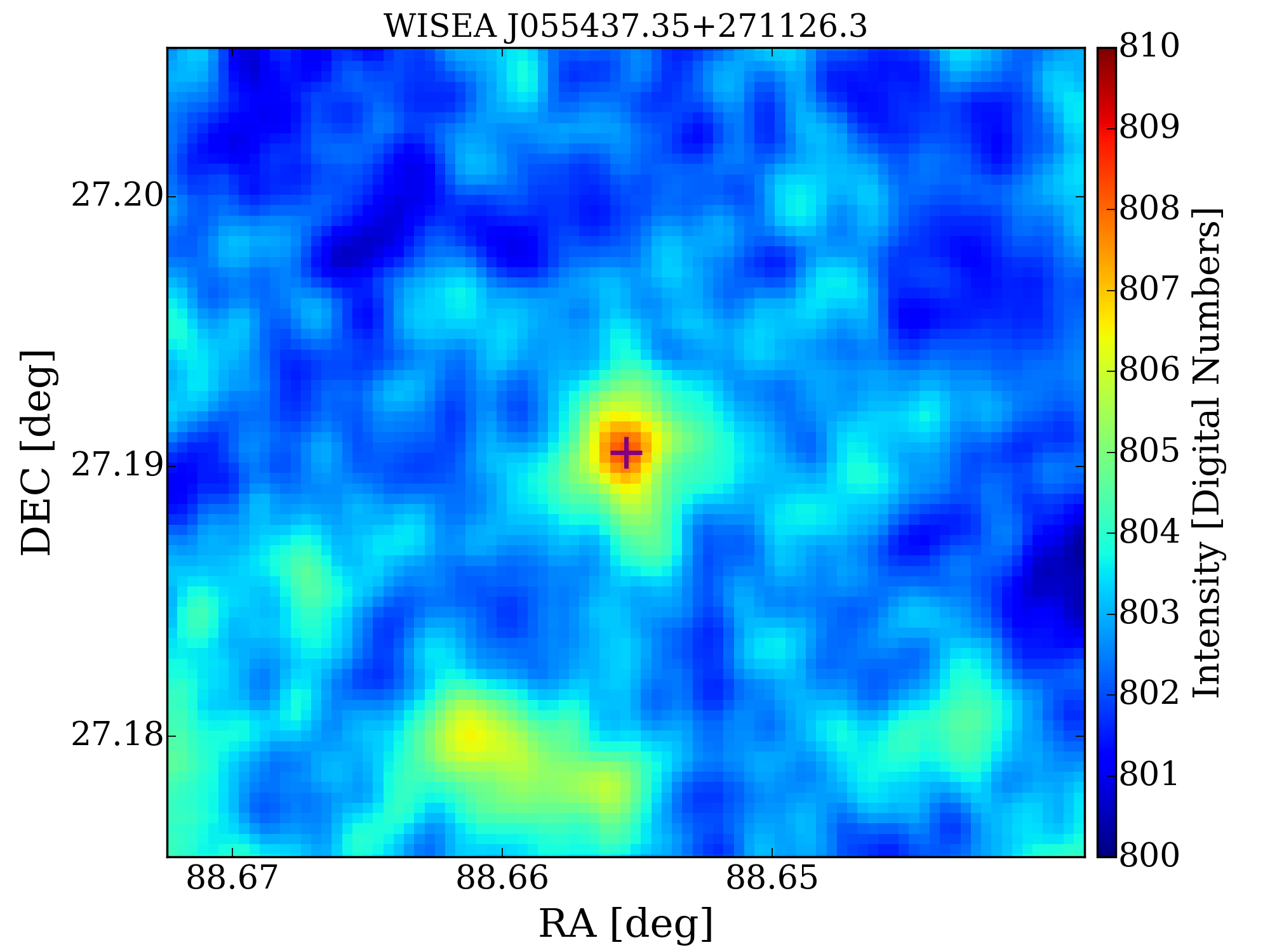}
    \caption{Left: same as Figure~\ref{4C+56.02_fit}, but for NVSS J055437\allowbreak+271126. Middle: radio map from VLASS centered at NVSS J055437\allowbreak+271126. Right: WISE W3 band infrared map centered at NVSS J055437\allowbreak+271126.}
    \label{NVSS_J055437+271126_fit}
\end{figure*}

NVSS J055437\allowbreak+271126 (Figure~\ref{NVSS_J055437+271126_fit}) is a radio source, detected in NRAO VLA Sky Survey and re-detected in mid-infrared band in WISE Survey and Two Micron All Sky Survey (2MASS, \citealt{2006AJ....131.1163S}). Only a photometric redshift with a large uncertainty is available. %To achieve the identification of this system, further high-resolution follow-up observations are required. 
The \hi absorption profile is wide and symmetrical, which can be properly modeled using a three-component Gaussian function.

%The WISE counterpart to NVSS\,J055437+271126 is WISEA J055437.35+271126.3 as shown in the NASA/IPAC Extragalactic Database. The WISE W1[\SI{3.4}{\micro\metre}], W2[\SI{4.6}{\micro\metre}], W3[\SI{12.1}{\micro\metre}] and W4[\SI{22.2}{\micro\metre}] magnitudes for WISEA J055437.35+271126.3 are 12.257 $\pm$ 0.024, 12.113 $\pm$ 0.029, 10.830 $\pm$ 0.152 and 8.318 $\pm$ 0.365, respectively. The W1-W2 color of WISEA J055437.35+271126.3 is 0.144, which means that the mid-IR emission comes mainly from the stars. According to the W2-W3 value of 1.283 mag, WISEA J055437.35+271126.3 is located in the intersection of the center of the Ellipticals region and the blue-end of Spirals region in the WISE color-color diagram. 

%The middle and right panel of Figure~\ref{NVSS_J055437+271126_fit} show the image centered at NVSS\,J055437+271126, constructed using radio data at S-band from VLASS and infrared data at 12.1 microns (W3) from WISE, respectively. The straight cross shows the position of NVSS\,J055437+271126.

\subsubsection{NVSS J073755+264652}

\begin{figure*}[hbt!]
    \centering
    \includegraphics[width=0.25\textwidth]{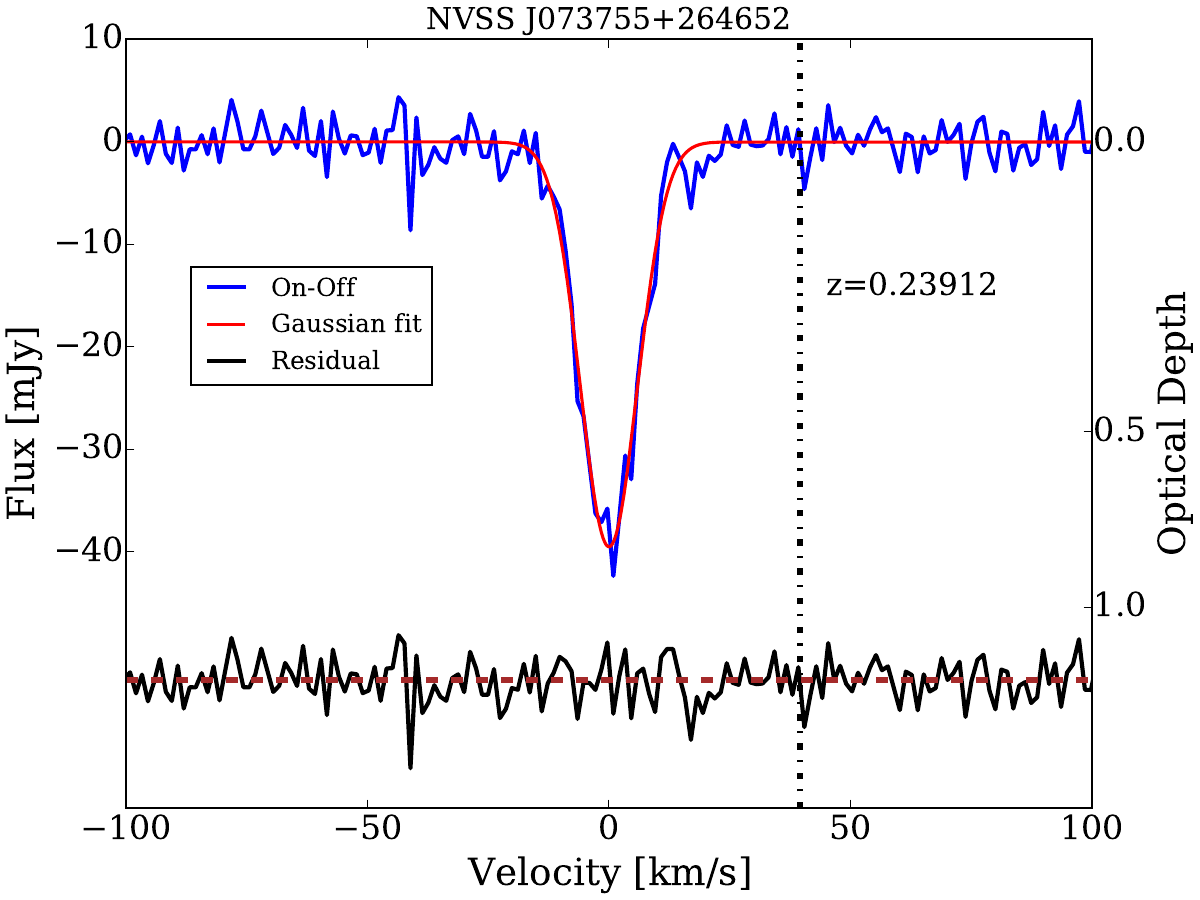}
    \includegraphics[width=0.25\textwidth]{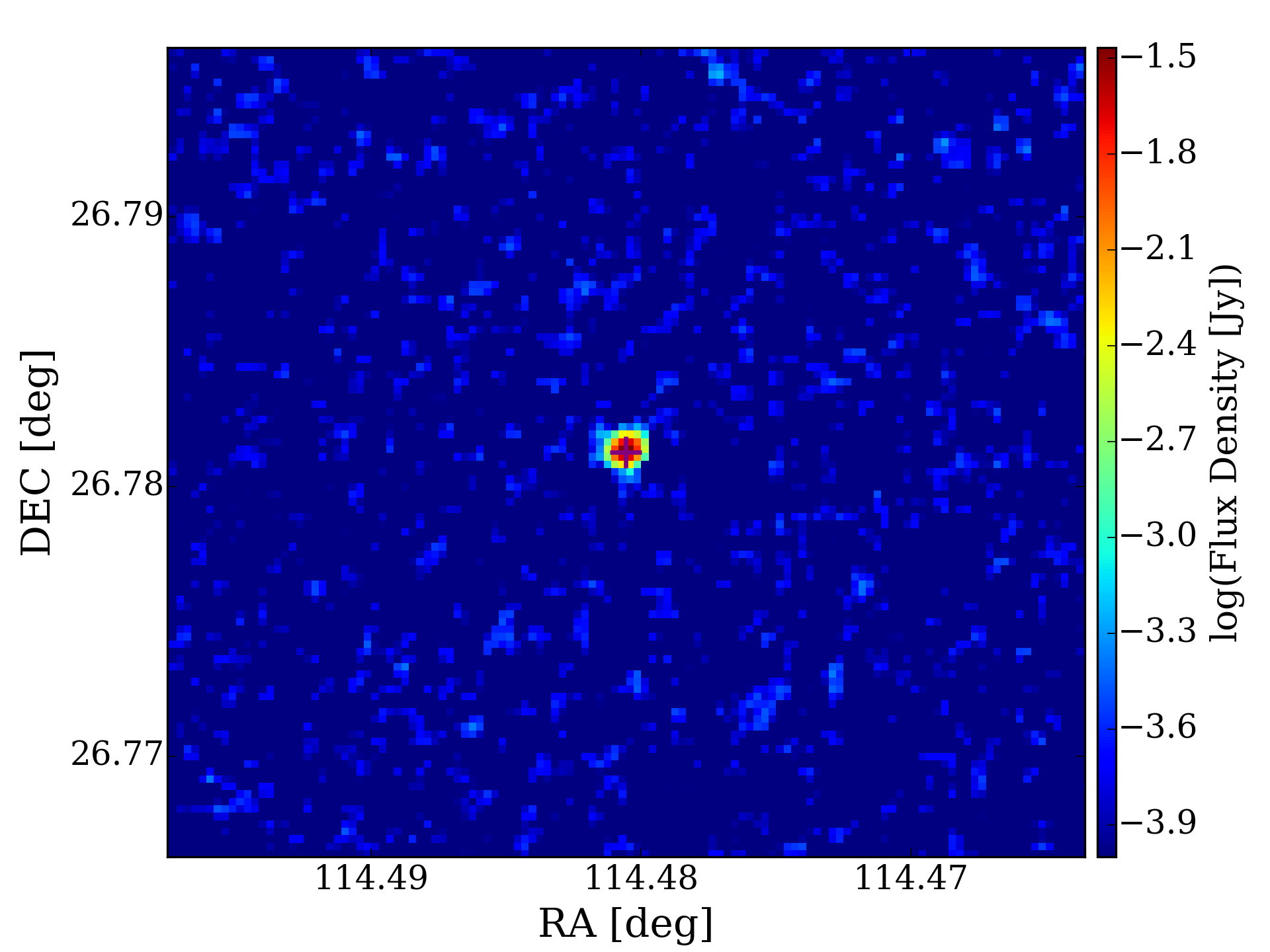}
    \includegraphics[width=0.25\textwidth]{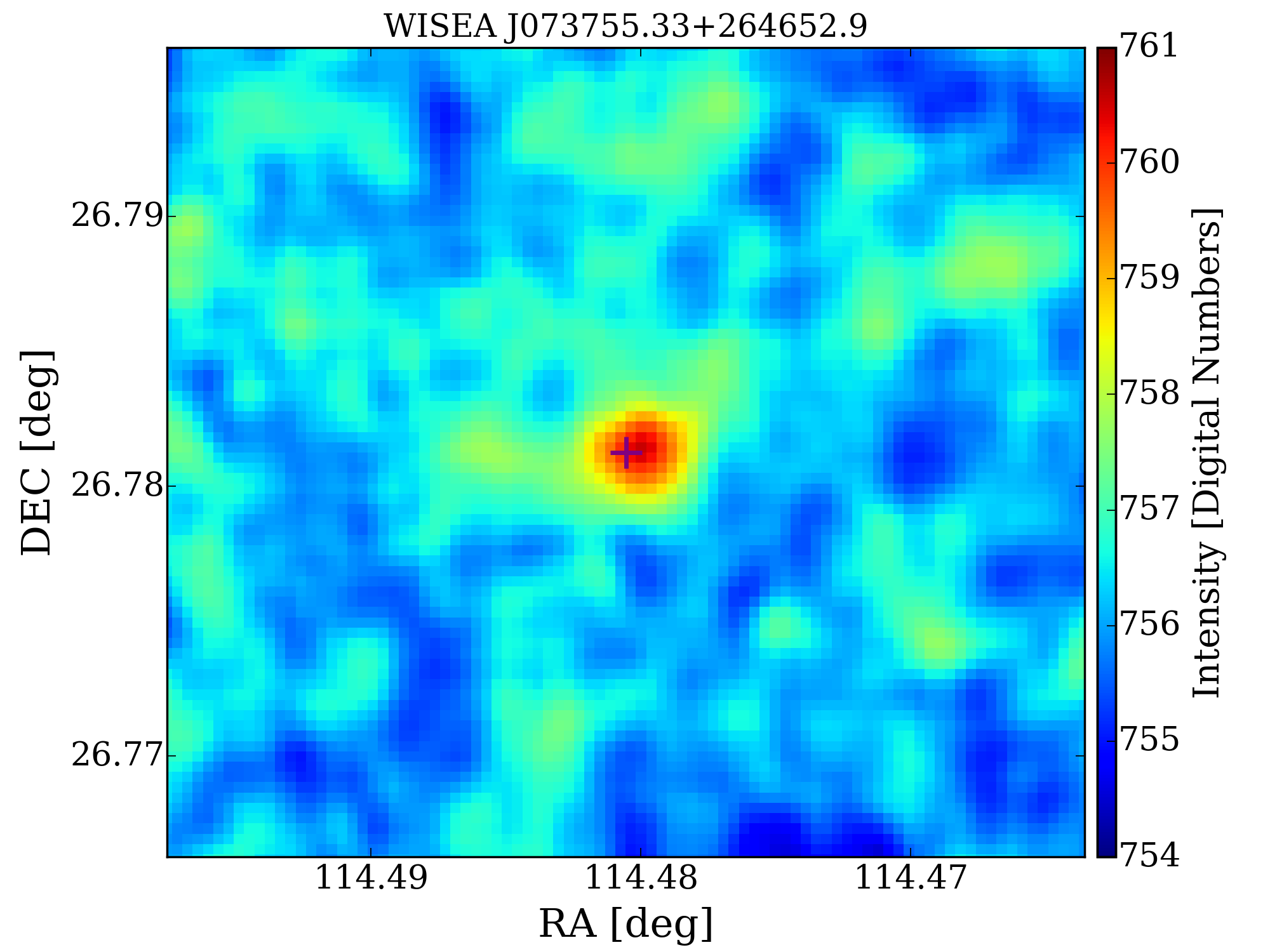}
    \hspace*{0.25cm}
    \includegraphics[width=0.18\textwidth]{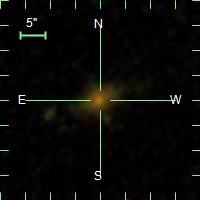}
    \caption{Left: same as Figure~\ref{4C+31.04_fit}, but for NVSS J073755\allowbreak+264652. Middle left: radio map from VLASS centered at NVSS J073755\allowbreak+264652. Middle right: W3 band infrared map of WISEA J073755.33\allowbreak+264652.9 from WISE. Right: SDSS optical map of the optical counterpart of NVSS J073755\allowbreak+264652.}
    \label{NVSS_J073755+264652_fit}
\end{figure*}

NVSS J073755\allowbreak+264652 (Figure~\ref{NVSS_J073755+264652_fit}) is a blue radio galaxy that has been observed in ultraviolet (GALEX), optical (SDSS), infrared (WISE and 2MASS) and radio band (NVSS). The \hi absorption profile demonstrates straightforward symmetry and can be precisely represented by a one-component Gaussian function. This suggests the existence of a settled system with a gas disk.

\subsubsection{NVSS J080101-075121}

\begin{figure*}[hbt!]
    \centering
    \includegraphics[width=0.25\textwidth]{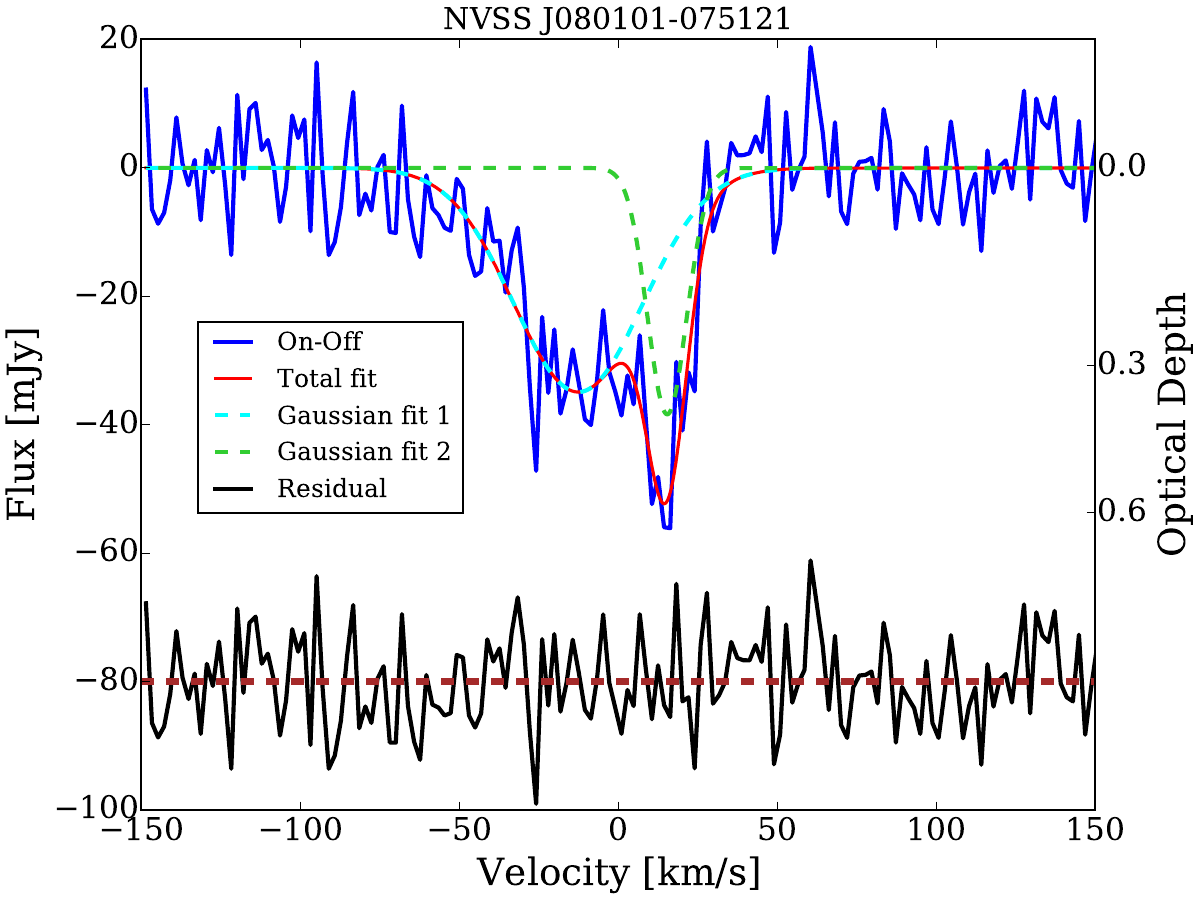}
    \includegraphics[width=0.25\textwidth]{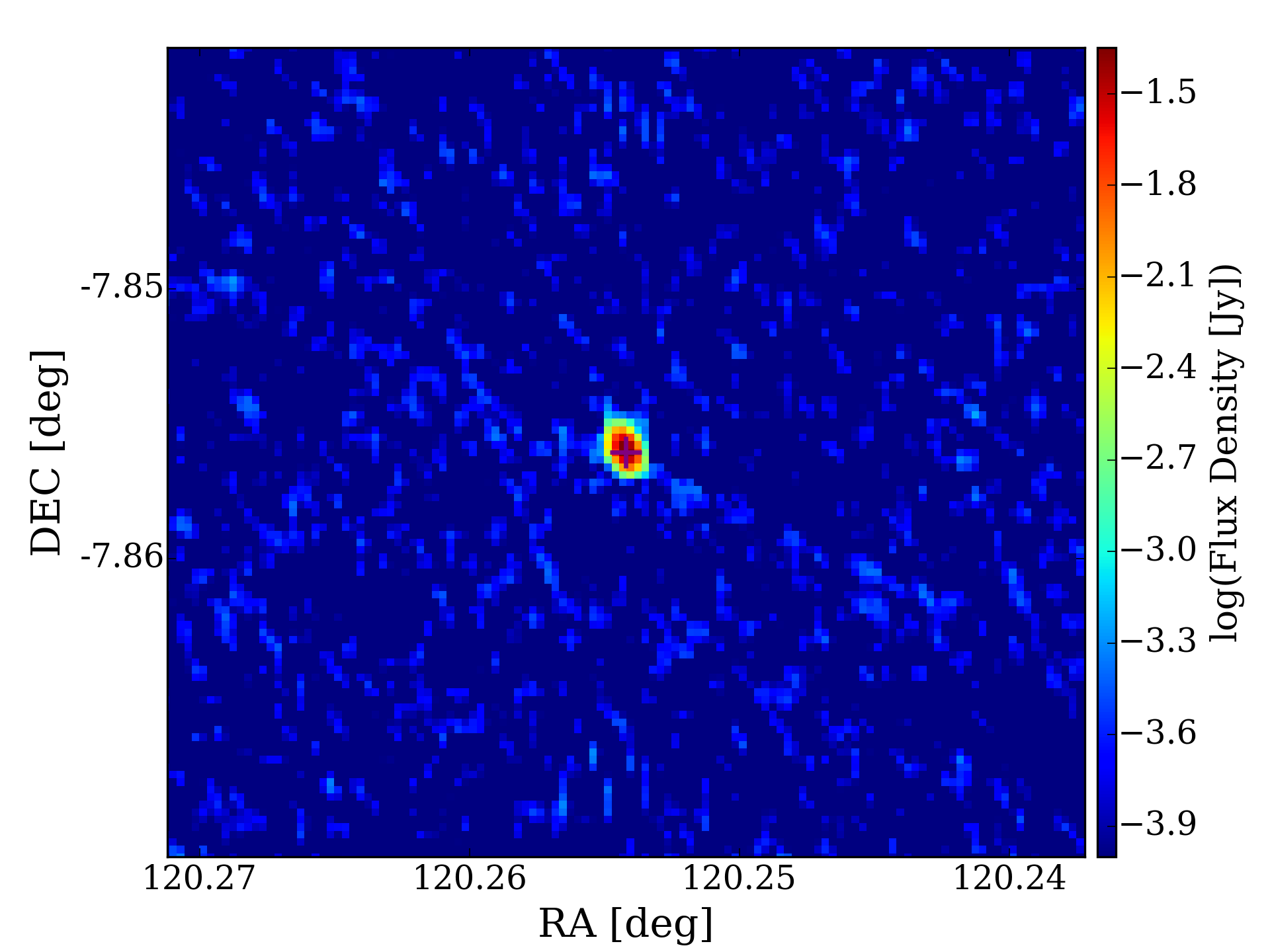}
    \includegraphics[width=0.25\textwidth]{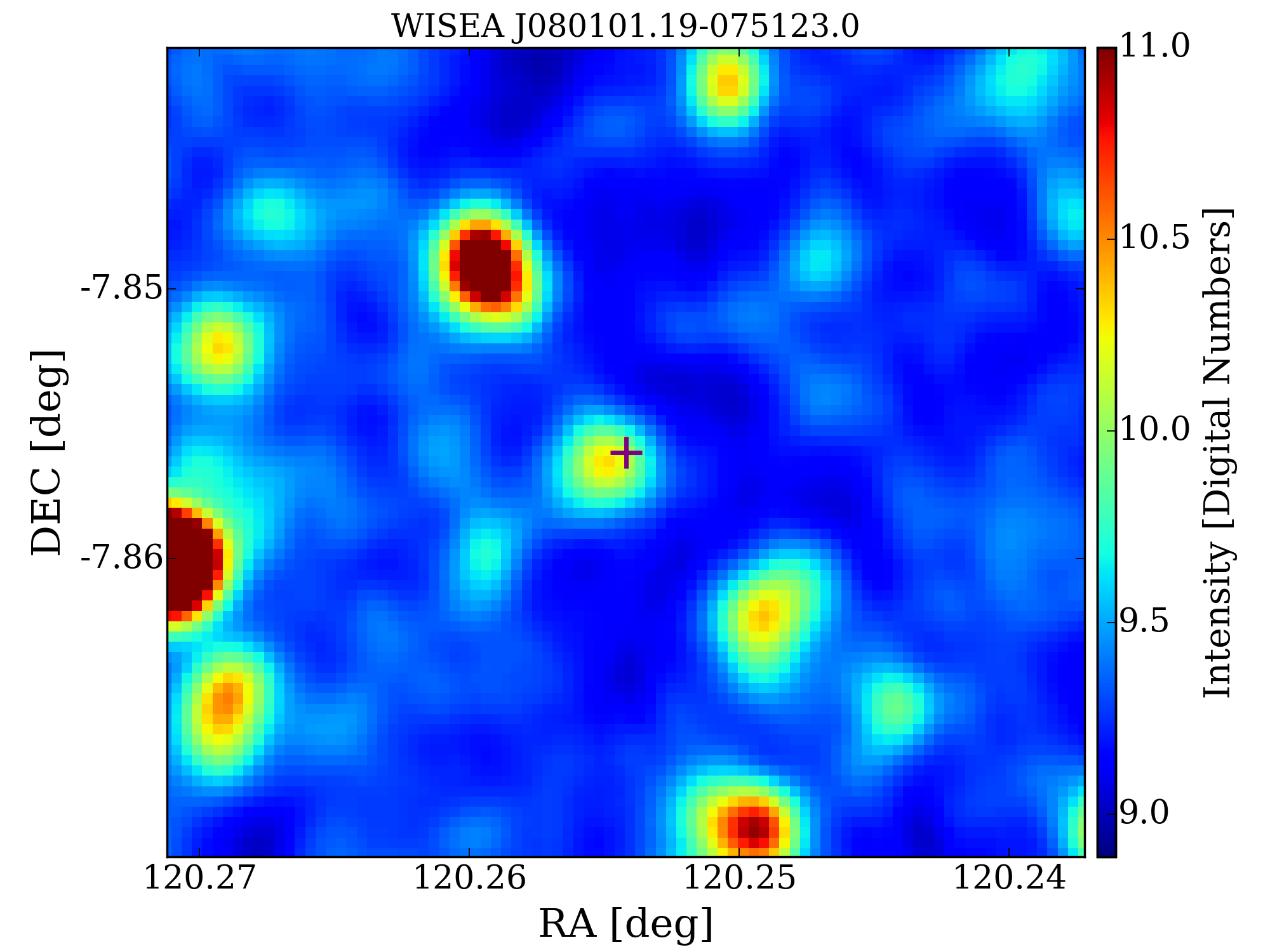}
    \hspace*{0.25cm}
    \includegraphics[width=0.18\textwidth]{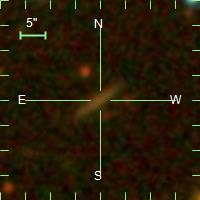}
    \caption{Left: same as Figure~\ref{4C+56.02_fit}, but for NVSS J080101\allowbreak-075121. Middle left: radio map from VLASS centered at NVSS J080101\allowbreak-075121. Middle right: WISE W2 band infrared map centered at NVSS J080101\allowbreak-075121. Right: SDSS optical map centered at NVSS J080101\allowbreak-075121.}
    \label{NVSS_J080101-075121_fit}
\end{figure*}

NVSS J080101\allowbreak-075121 (Figure~\ref{NVSS_J080101-075121_fit}) has been detected at 365 MHz in the Texas Survey of Radio Sources \citep{1996AJ....111.1945D} and at 1.4\,GHz in NVSS, and it is detected in the mid-infrared band by WISE. The absence of redshift for the radio source NVSS J080101\allowbreak-075121 introduces uncertainty in identifying the counterpart in the foreground/background. %To resolve this ambiguity, further high-resolution follow-up observation is required. 
The W1-W2 color of NVSS J080101\allowbreak-075121 is 0.135, which is lower than the value of 0.8 often used to select AGN candidates, and implies that the mid-IR emission comes mainly from stars. Combined with the W2-W3 value of 2.617 mag, NVSS J080101\allowbreak-075121 lies in the region of spiral galaxies in the WISE color-color diagram.

The \hi absorption profile reveals a symmetric, broader component, likely associated with the gas disk, and a narrower, deeper peak, suggesting the presence of distinct, unsettled gas structures.

%The WISE W1[\SI{3.4}{\micro\metre}], W2[\SI{4.6}{\micro\metre}], W3[\SI{12.1}{\micro\metre}] and W4[\SI{22.2}{\micro\metre}] magnitudes for NVSS\,J080101-075121 are 15.097 $\pm$ 0.044, 14.962 $\pm$ 0.096, 12.345 $\pm$ 0.355 and 8.642, respectively. The W1-W2 color of NVSS\,J080101-075121 is 0.135, which is lower than the value of 0.8 often used to select AGN candidates and implies that the mid-IR emission comes mainly from the stars. Combining with the W2-W3 value of 2.617 mag, NVSS\,J080101-075121 plausibly lies in the region of spiral galaxies in the WISE color-color diagram.

\subsubsection{NVSS J092351+281527}

\begin{figure*}[hbt!]
    \centering
    \includegraphics[width=0.25\textwidth]{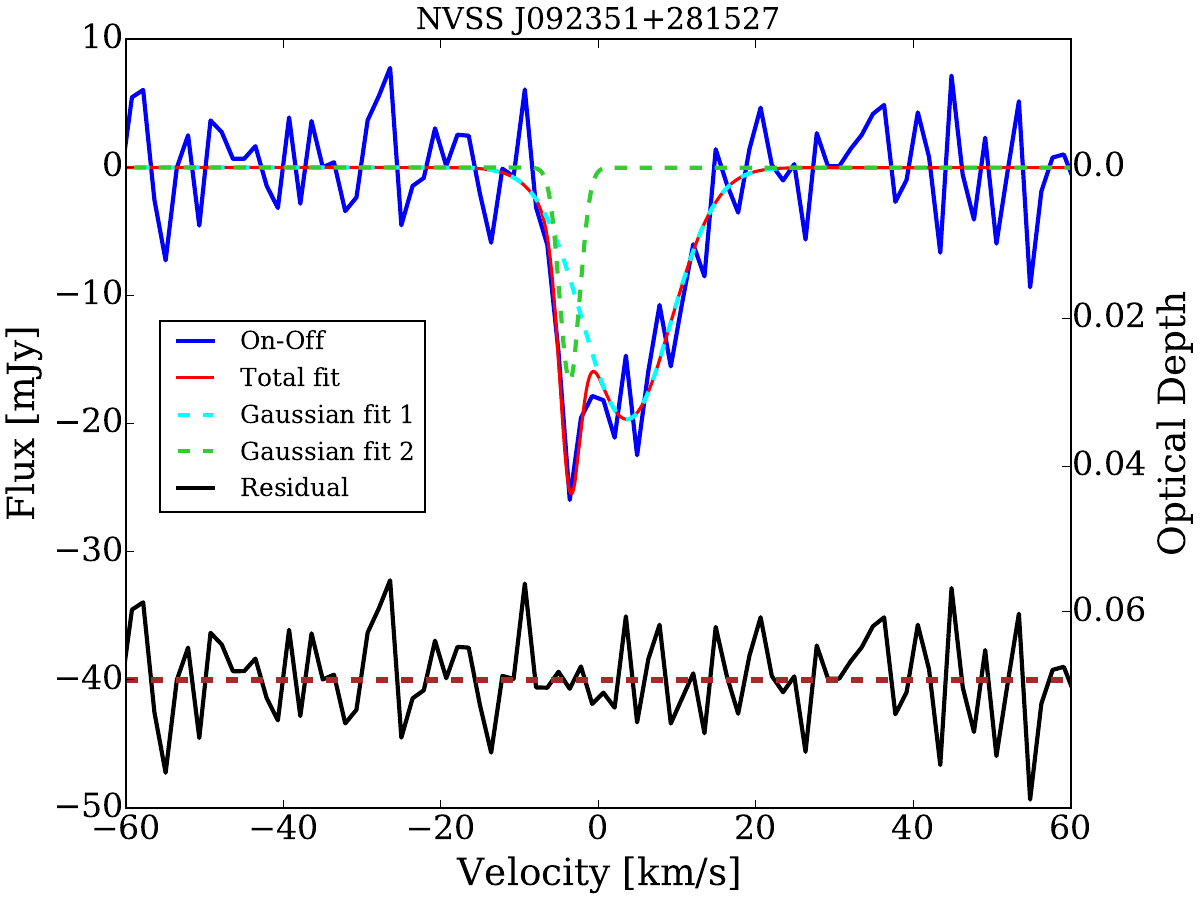}
    \includegraphics[width=0.25\textwidth]{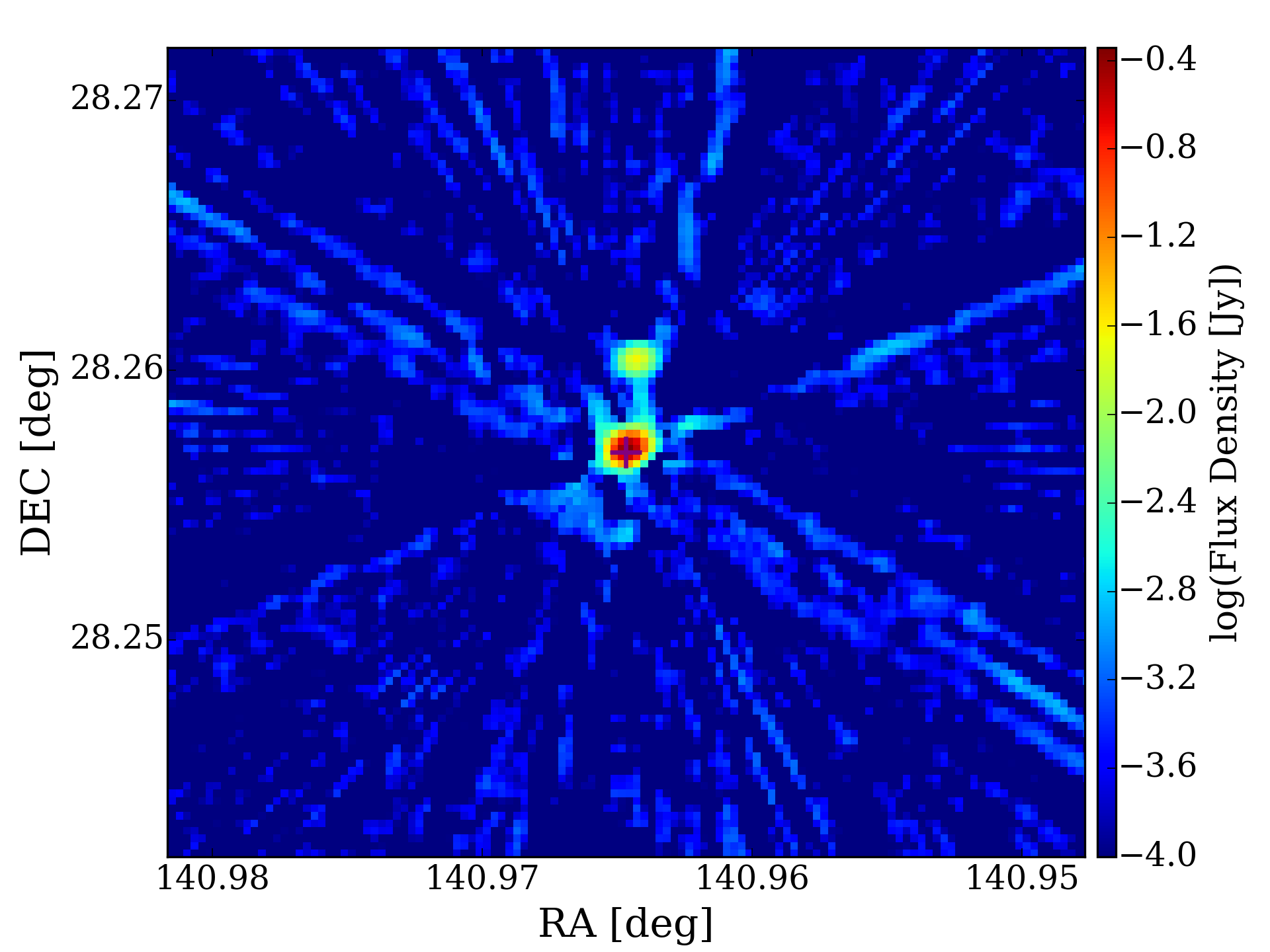}
    \includegraphics[width=0.25\textwidth]{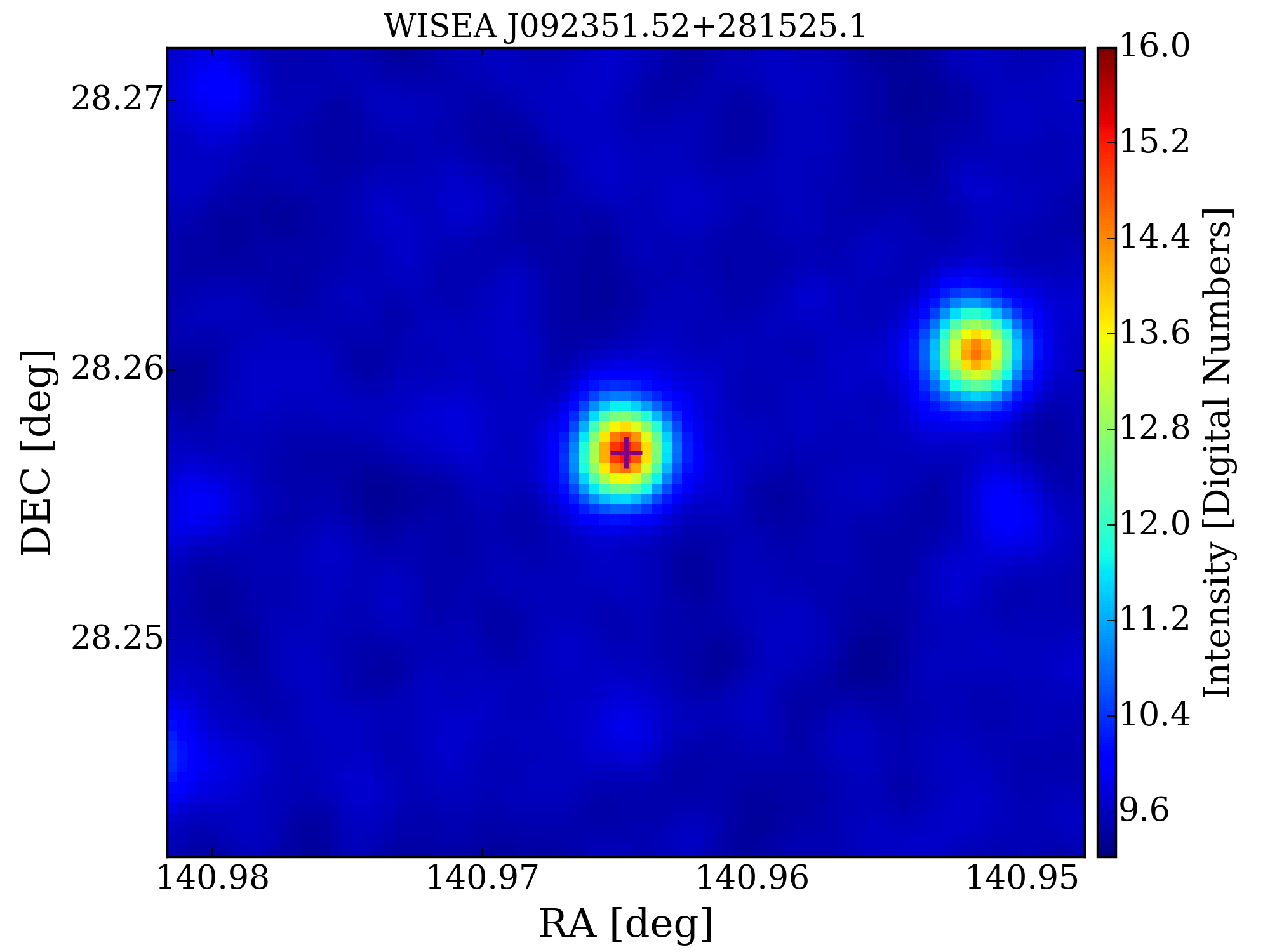}
    \includegraphics[width=0.18\textwidth]{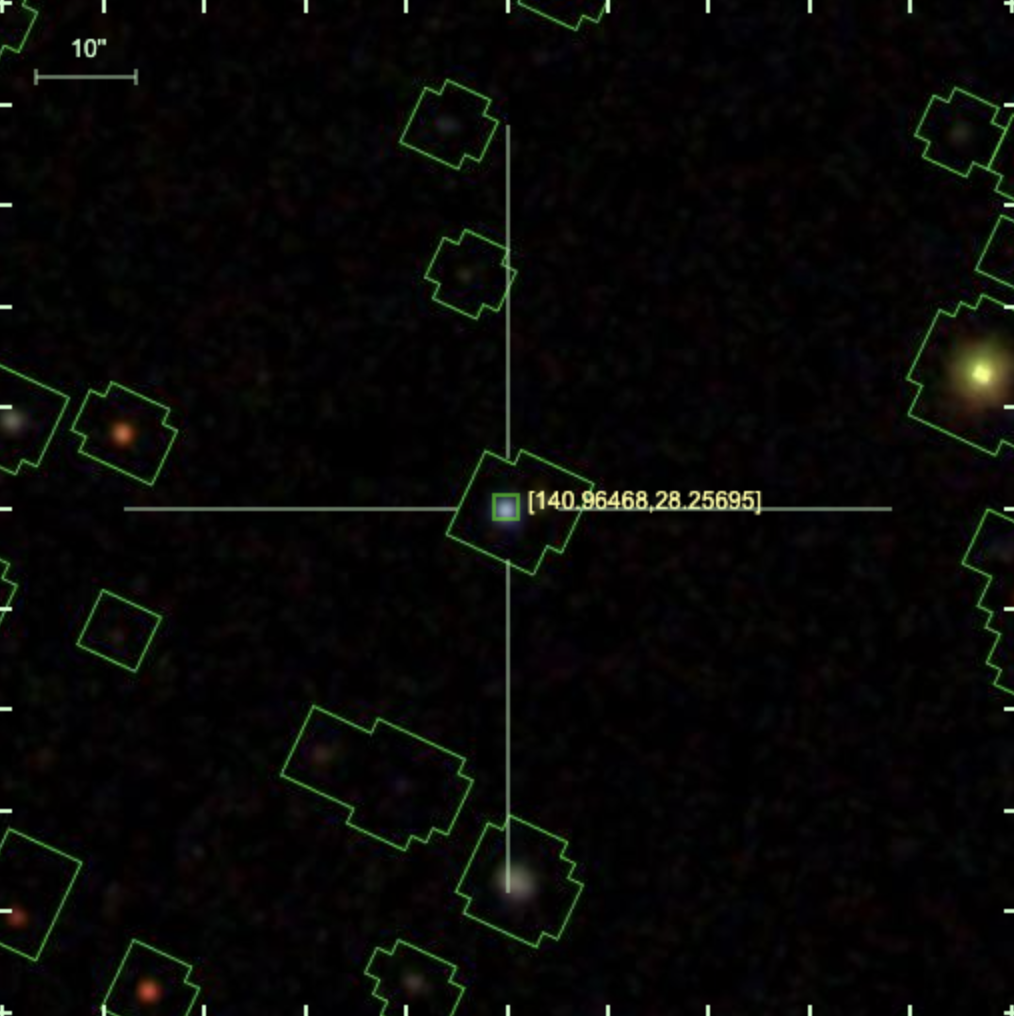}
    \caption{Left: same as Figure~\ref{4C+56.02_fit}, but for NVSS J092351\allowbreak+281527. Middle left: radio map from VLASS centered at NVSS J092351\allowbreak+281527. Middle right: W2 band infrared map centered at NVSS J092351\allowbreak+281527. Right: SDSS optical map centered at NVSS J092351\allowbreak+281527.}
    \label{NVSS_J092351+281527_fit}
\end{figure*}

Positioned at a redshift of 0.745, NVSS J092351\allowbreak+281527 (Figure~\ref{NVSS_J092351+281527_fit}), identified as a quasar, has undergone observations across various spectra, including ultraviolet (GALEX), optical (SDSS), Infrared/Submillimeter (WISE and Planck), and radio bands (Texas, FIRST, and NVSS). Additionally, NVSS J092351\allowbreak+281527 is categorized as a Flat-Spectrum Radio Source in \citet{2007ApJS..171...61H}. The counterpart to the foreground is currently ambiguous, necessitating high-resolution follow-up observations for verification.

%The WISE counterpart to NVSS\,J092351+281527 is WISEA\,J092351.52+281525.1 as given by the SDSS Database. The WISE W1[\SI{3.4}{\micro\metre}], W2[\SI{4.6}{\micro\metre}], W3[\SI{12.1}{\micro\metre}] and W4[\SI{22.2}{\micro\metre}] magnitudes for WISEA\,J092351.52+281525.1 are 14.117 $\pm$ 0.030, 13.045 $\pm$ 0.032, 9.946 $\pm$ 0.049 and 7.721 $\pm$ 0.166, respectively. The W1-W2 color of WISEA\,J092351.52+281525.1 is 1.072, which indicates WISEA\,J092351.52+281525.1 as an AGN candidate. Considering the W2-W3 value of 3.099 mag, WISEA\,J092351.52+281525.1 is located in the QSOs region in the WISE color-color diagram.

\subsubsection{NVSS J093150+254034}

\begin{figure*}[hbt!]
    \centering
    \includegraphics[width=0.25\textwidth]{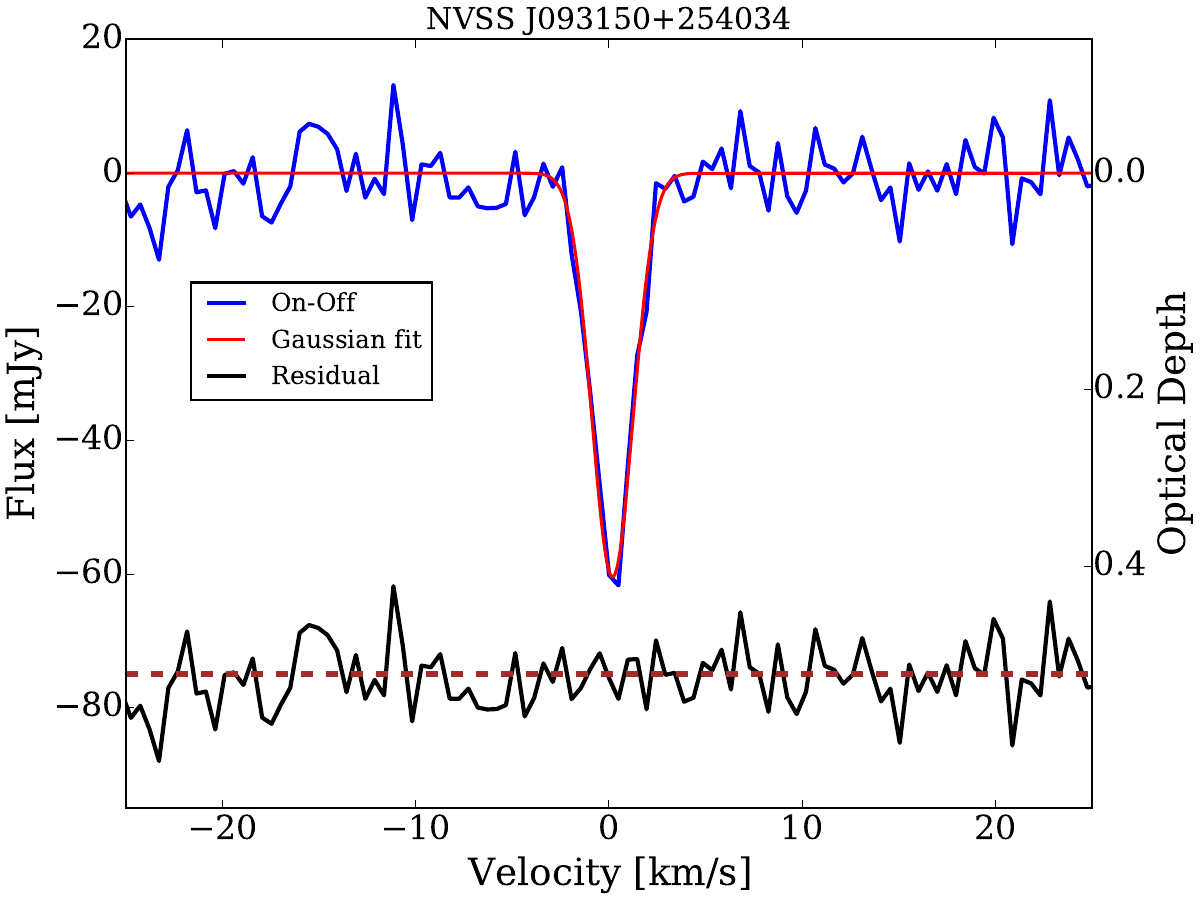}
    \includegraphics[width=0.25\textwidth]{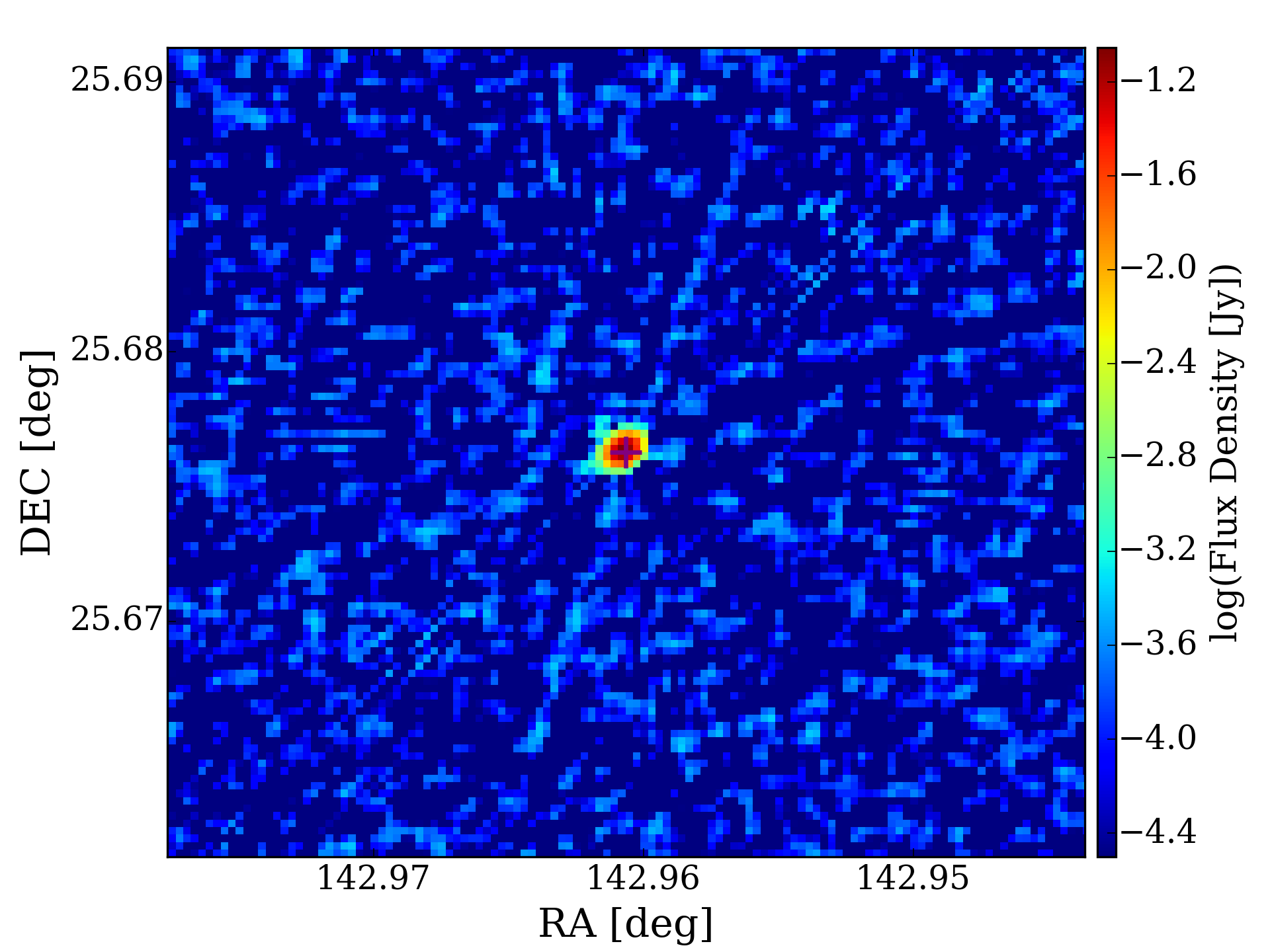}
    \includegraphics[width=0.25\textwidth]{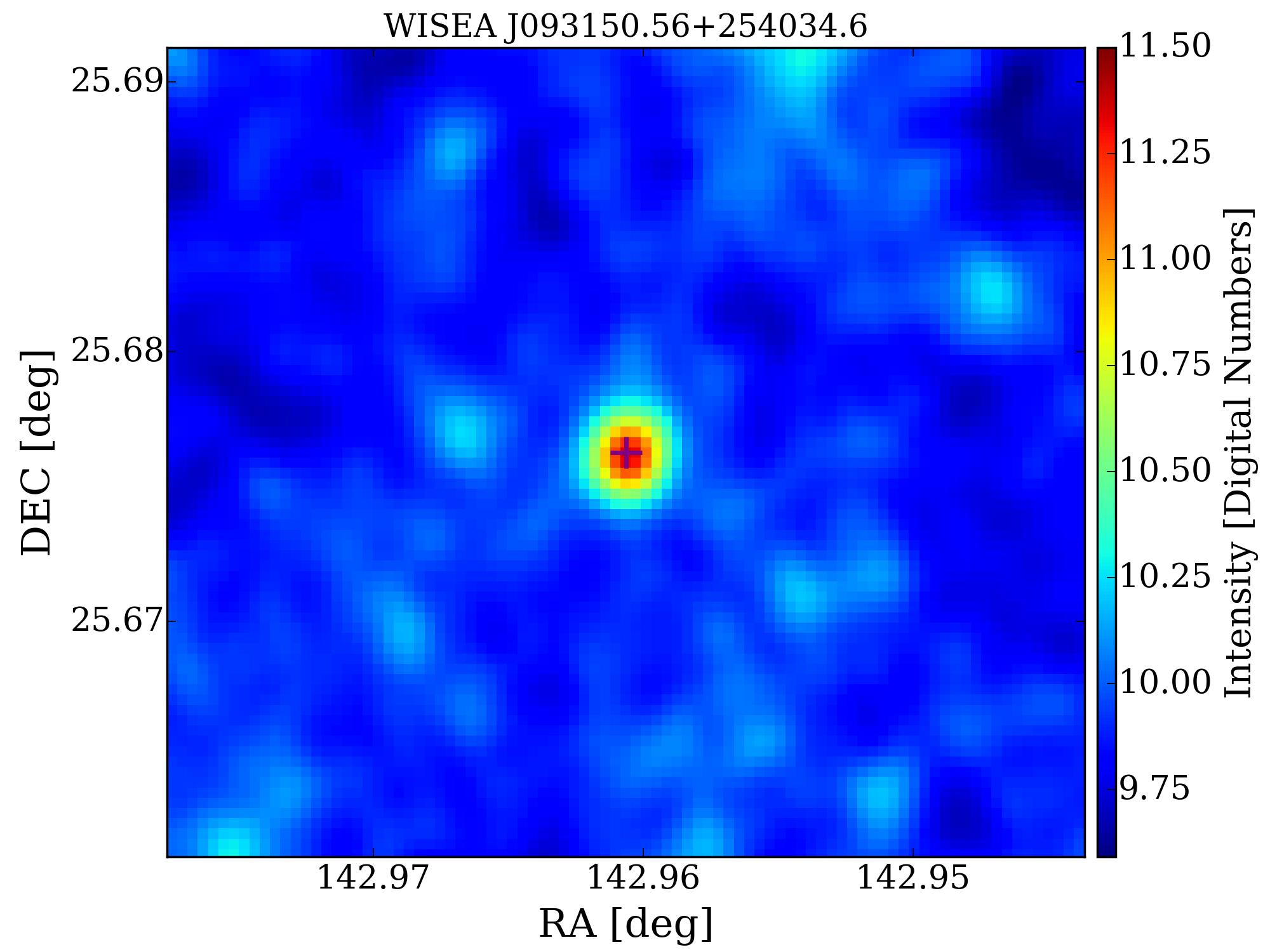}
    \caption{Left: same as Figure~\ref{4C+56.02_fit}, but for NVSS J093150\allowbreak+254034. Middle: radio map from VLASS centered at NVSS J093150\allowbreak+254034. Right: WISE W2 band infrared map centred at NVSS J093150\allowbreak+254034.}
    \label{NVSS_J093150+254034_fit}
\end{figure*}

NVSS J093150\allowbreak+254034 (Figure~\ref{NVSS_J093150+254034_fit}) is a galaxy located at a redshift of 0.812, which has been observed in infrared (WISE) and radio bands (NVSS). The counterpart to the foreground is currently uncertain. %and requires further high-resolution follow-up observations for confirmation. 
The \hi absorption profile from this intervening system exhibits a slender and symmetrical shape, suggesting the presence of a gas disk.
%The WISE counterpart to NVSS\,J093150+254034 is WISEA\,J093150.56+254034.6 as given by the SDSS Database. The WISE W1[\SI{3.4}{\micro\metre}], W2[\SI{4.6}{\micro\metre}], W3[\SI{12.1}{\micro\metre}] and W4[\SI{22.2}{\micro\metre}] magnitudes for WISEA\,J093150.56+254034.6 are 15.339 $\pm$ 0.051, 14.590 $\pm$ 0.080, 11.552 $\pm$ 0.200 and 8.263 $\pm$ 0.268, respectively. The W1-W2 color of WISEA\,J093150.56+254034.6 is 0.749, which indicates WISEA\,J093150.56+254034.6 as an AGN candidate. Considering the W2-W3 value of 3.038 mag, WISEA\,J093150.56+254034.6 is located in the Seyferts region in the WISE color-color diagram.

\subsubsection{NVSS J094208+135152}

\begin{figure*}[hbt!]
    \centering
    \includegraphics[width=0.25\textwidth]{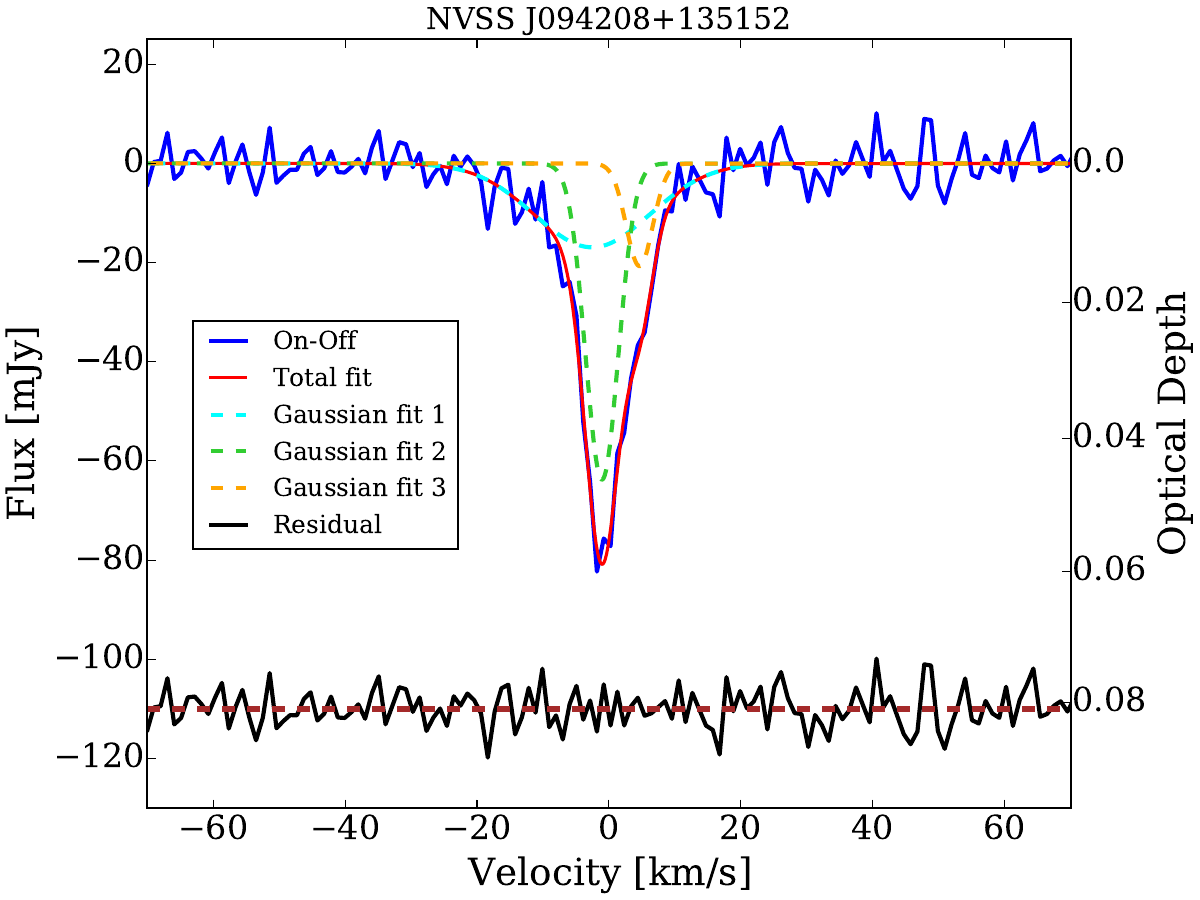}
    \includegraphics[width=0.25\textwidth]{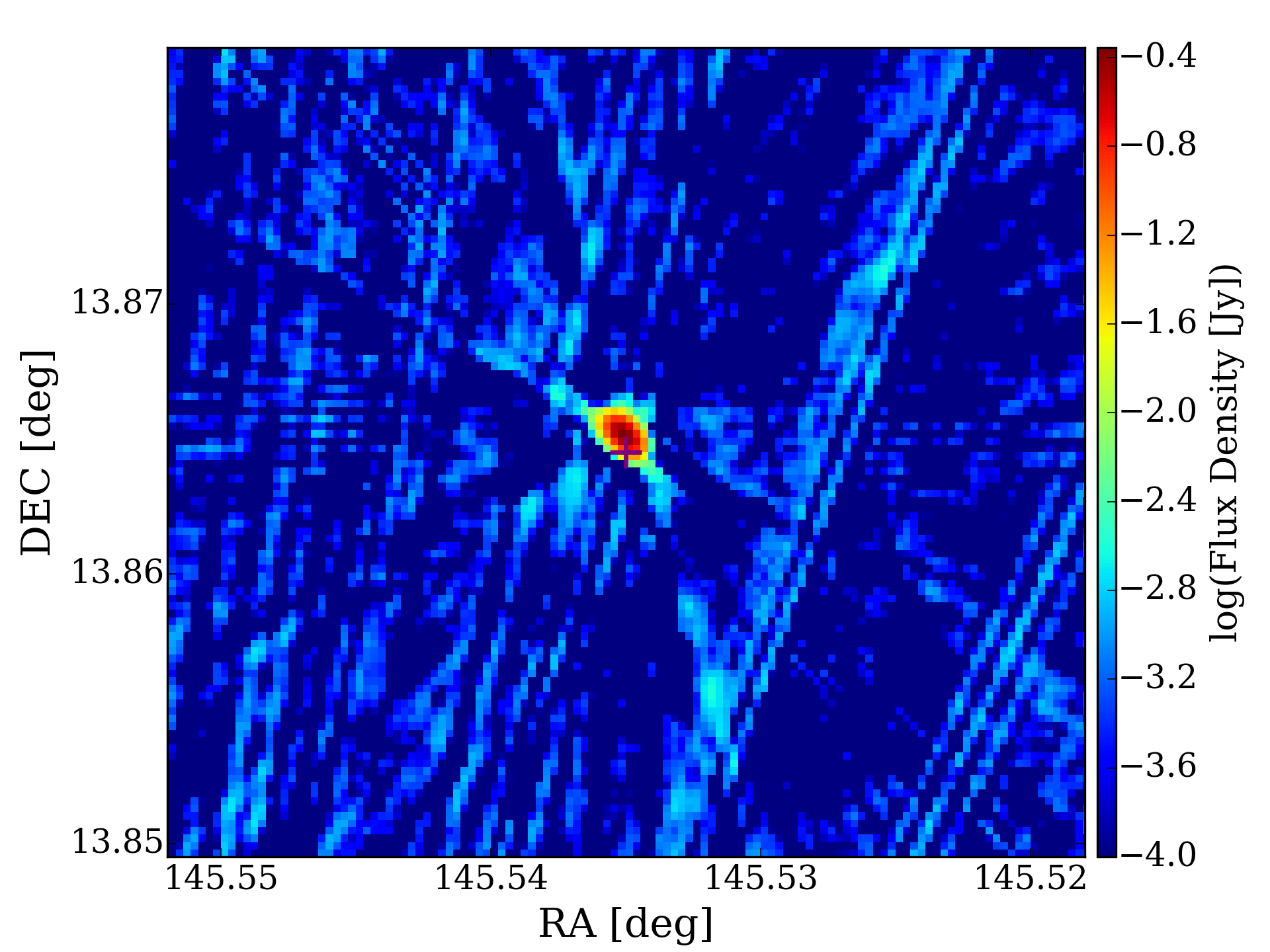}
    \includegraphics[width=0.25\textwidth]{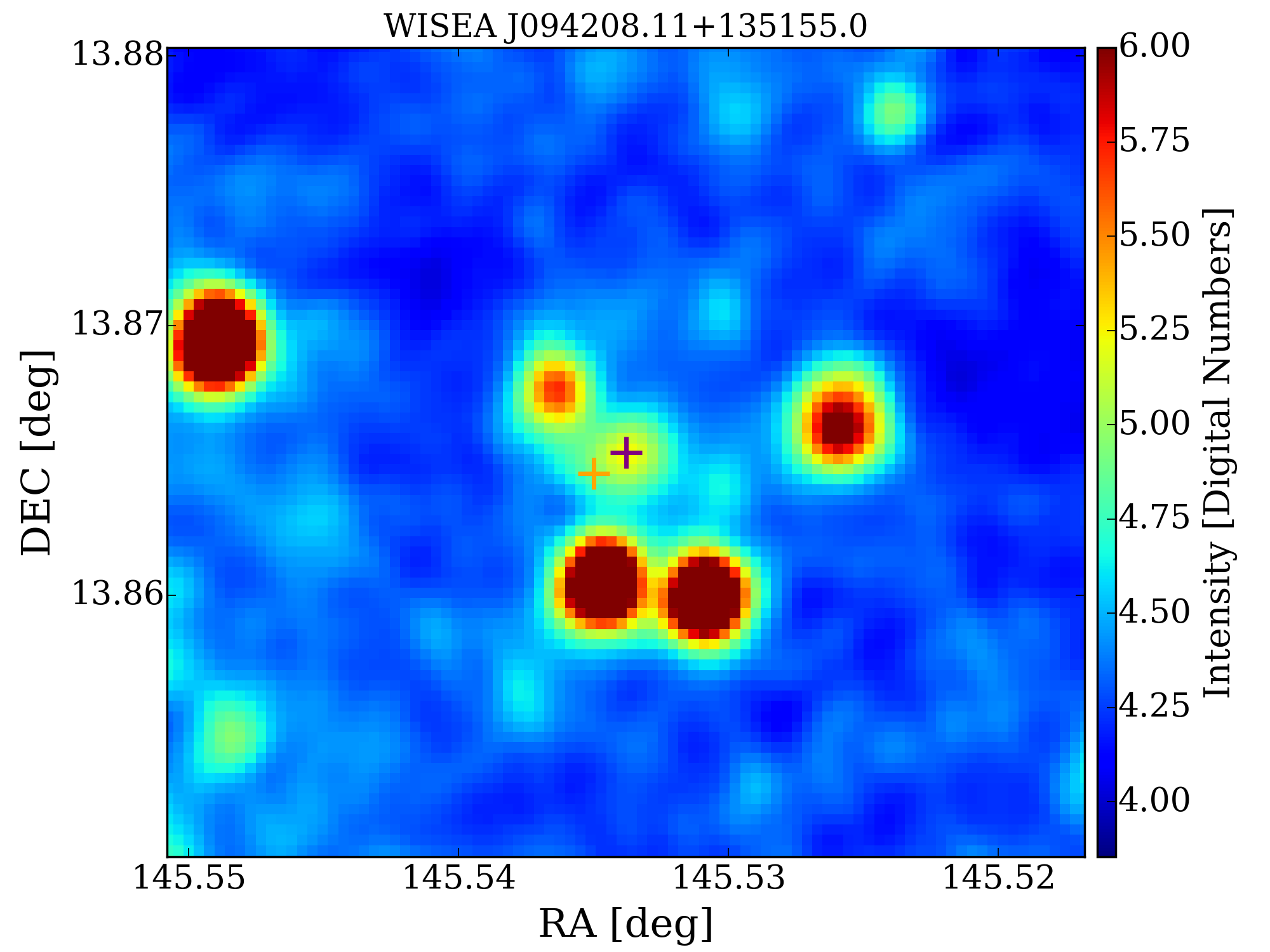}
    \includegraphics[width=0.18\textwidth]{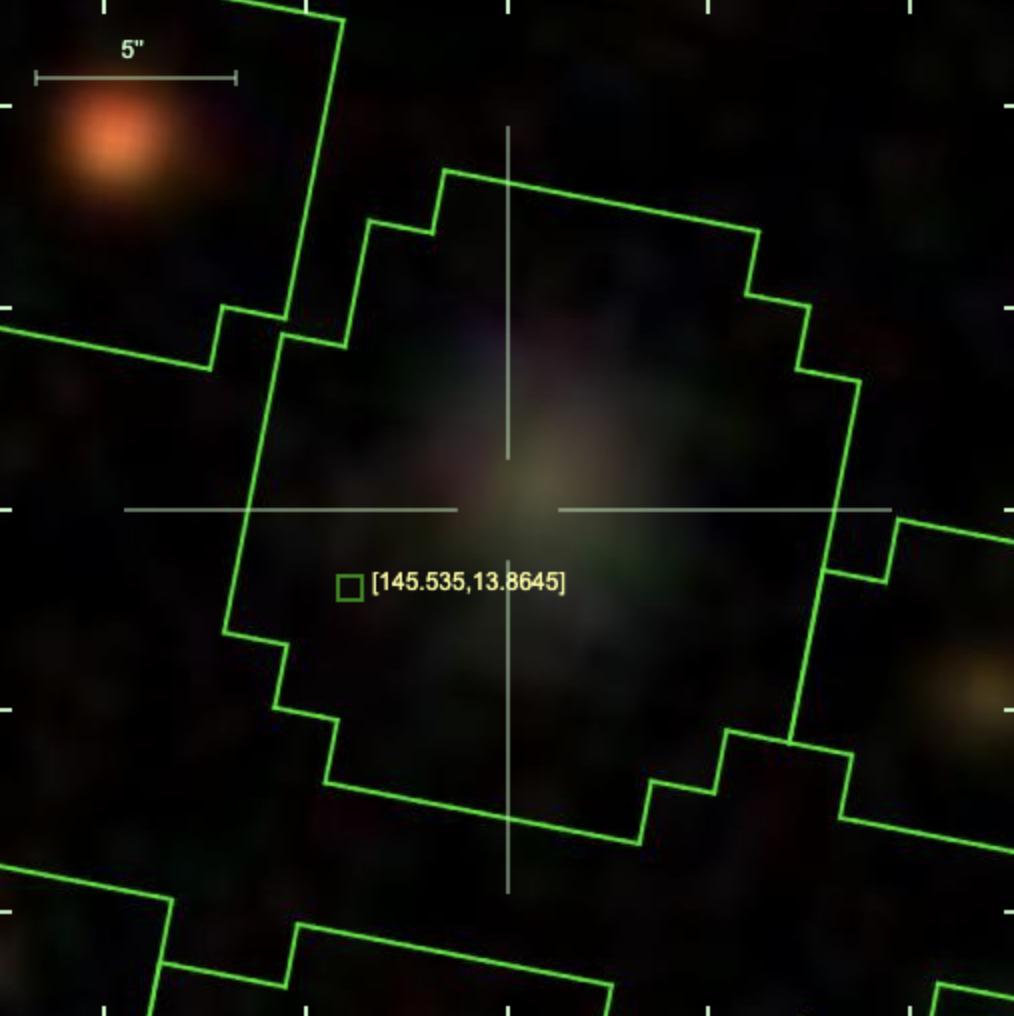}
    \caption{Left: same as Figure~\ref{4C+56.02_fit}, but for NVSS J094208\allowbreak+135152. Middle left: radio map from VLASS centered at NVSS J094208\allowbreak+135152. Middle right: W1 band infrared map centered at WISEA J094208.11\allowbreak+135155.0 (shown as a purple cross). The position of the background is shown as an orange cross). Right: SDSS optical map of the SDSS J094208.05\allowbreak+135154.9. The green square shows the position of the background radio source.}
    \label{NVSS_J094208+135152_fit}
\end{figure*}

NVSS J094208\allowbreak+135152 (Figure~\ref{NVSS_J094208+135152_fit}) is categorized as a radio source, close to 3C\,225A (located at z = 1.565), with a mere 0.039 arcmin separation. Since only one radio source is identified in the same region by VLASS, NVSS J094208\allowbreak+135152 likely corresponds to 3C\,225A. In the HUBBLE SPACE TELESCOPE snapshot survey presented by \citet{1997ApJS..112..415M}, the identification of 3C\,225A is very faint and is confused with a foreground spiral in ground-based images. The SDSS counterpart to the foreground galaxy is SDSS J094208.05\allowbreak+135154.9, its magnitudes are u=20.724, g=19.471, r=18.889, i=18.614 and z=18.536. 

The WISE counterpart to foreground galaxy SDSS J094208.05\allowbreak+135154.9 is WISEA J094208.11\allowbreak+135155.0 as shown in the NED. The WISE W1[\SI{3.4}{\micro\metre}], W2[\SI{4.6}{\micro\metre}], W3[\SI{12.1}{\micro\metre}] and W4[\SI{22.2}{\micro\metre}] magnitudes for WISEA J094208.11\allowbreak+135155.0 are 16.307 $\pm$ 0.101, 15.839 $\pm$ 0.225, 11.331 $\pm$ 0.173 and 8.316, respectively. The W1-W2 color of WISEA J094208.11\allowbreak+135155.0 is 0.468, which means that the mid-IR emission comes mainly from the stars. According to the W2-W3 value of 4.508 mag, WISEA J094208.11\allowbreak+135155.0 is located in the ULIRGs/LINERs region in the WISE color-color diagram. %The WISE counterpart to NVSS\,J094208+135152(3C\,225A) is not found. 

\subsubsection{NVSS J095058+375758}

\begin{figure*}[hbt!]
    \centering
    \includegraphics[width=0.25\textwidth]{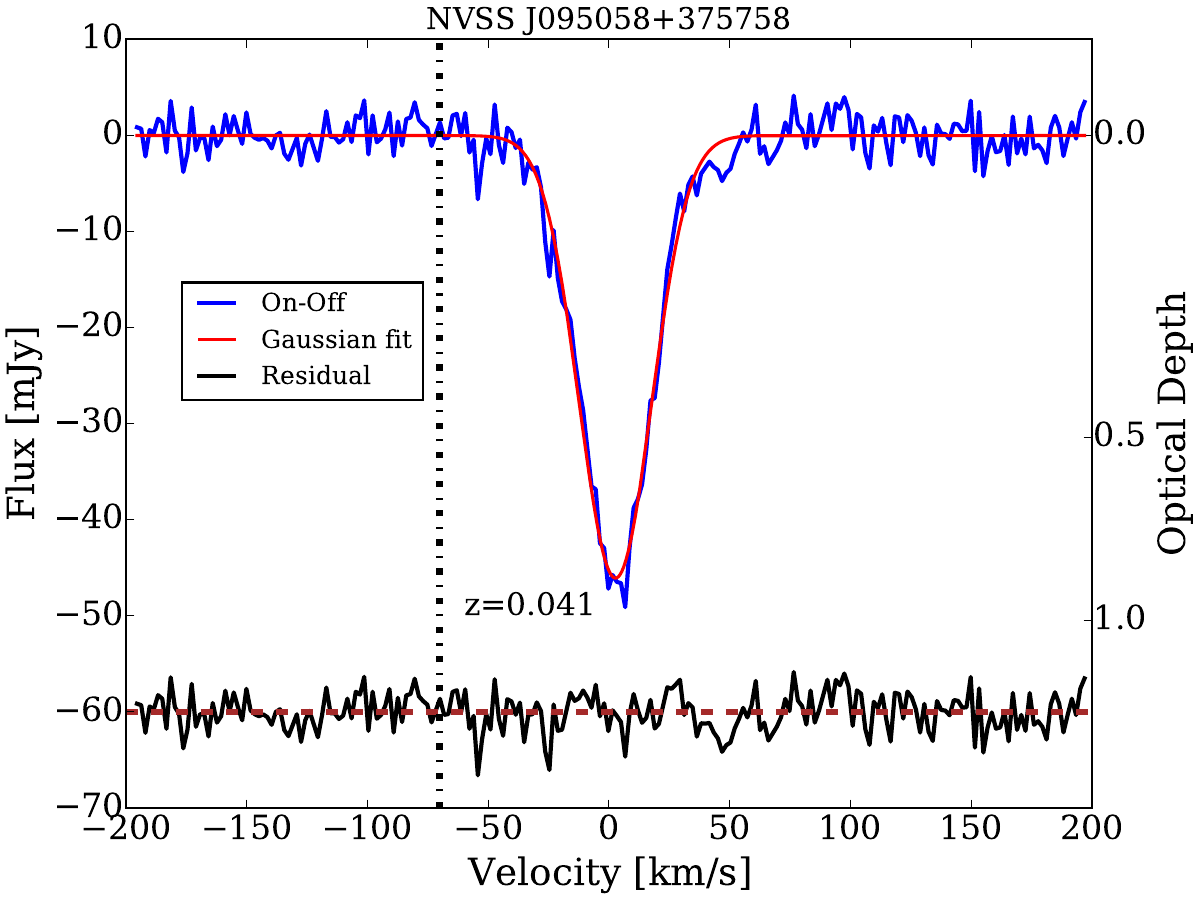}
    \includegraphics[width=0.25\textwidth]{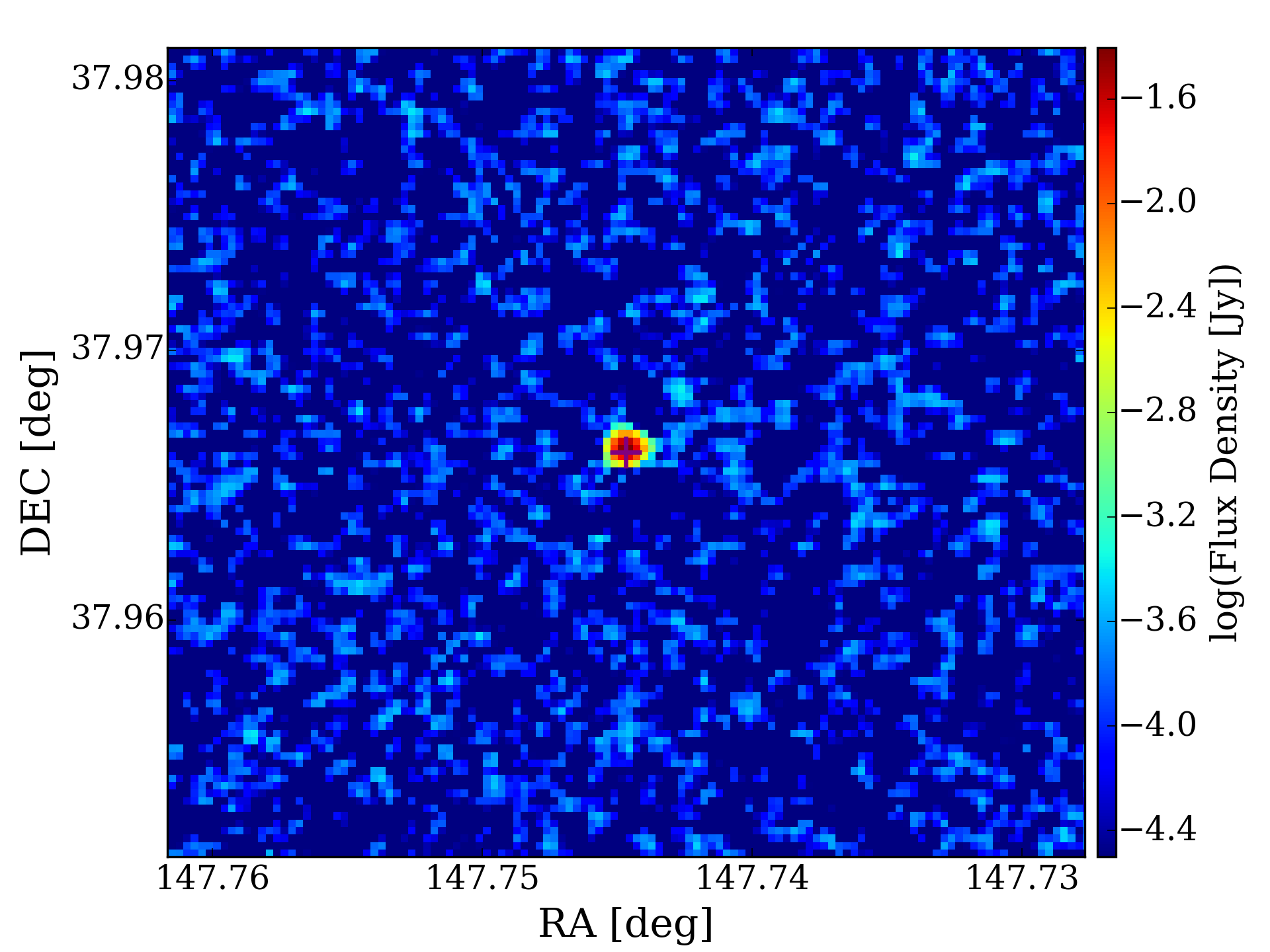}
    \includegraphics[width=0.25\textwidth]{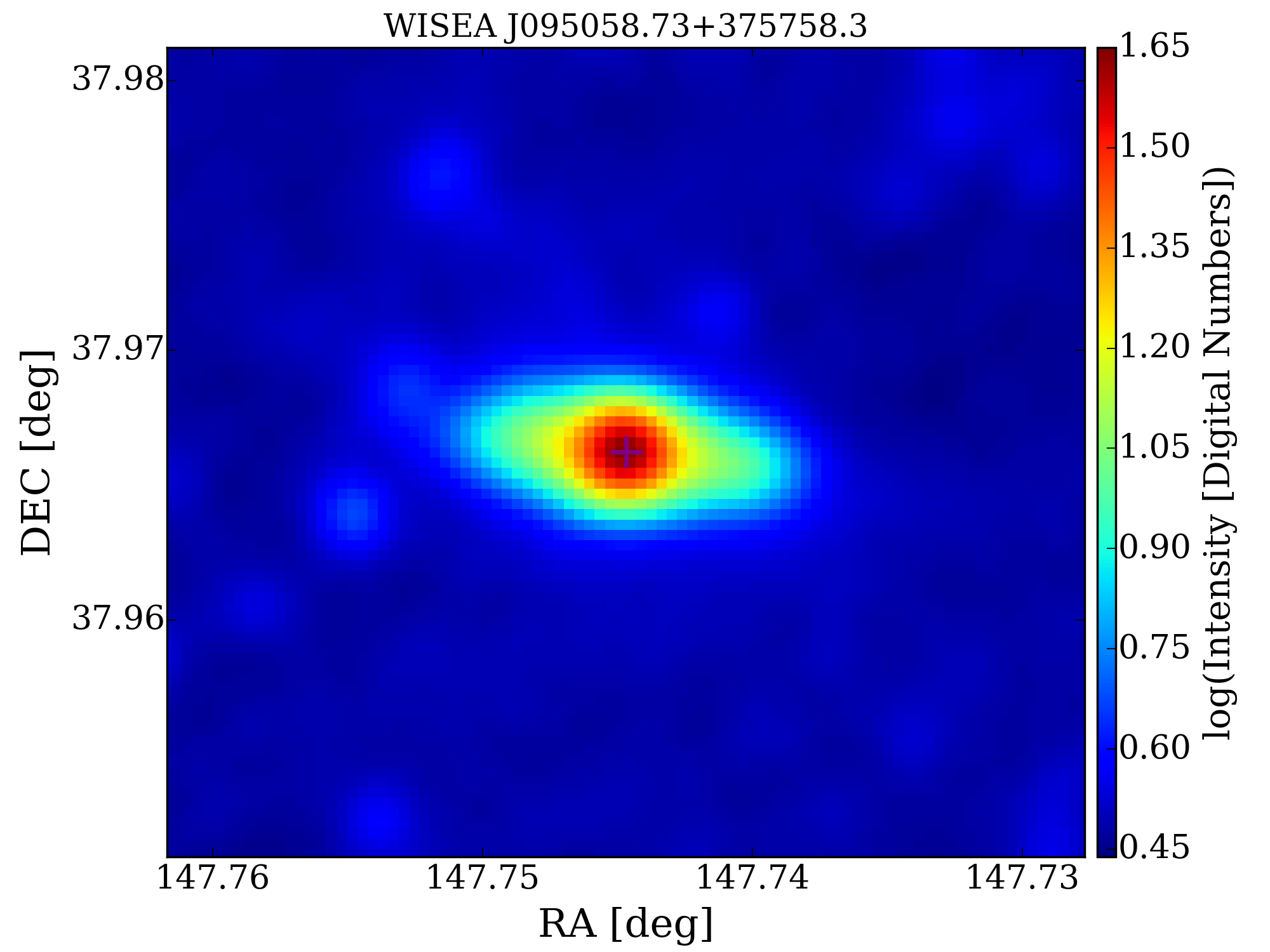}
    \includegraphics[width=0.18\textwidth]{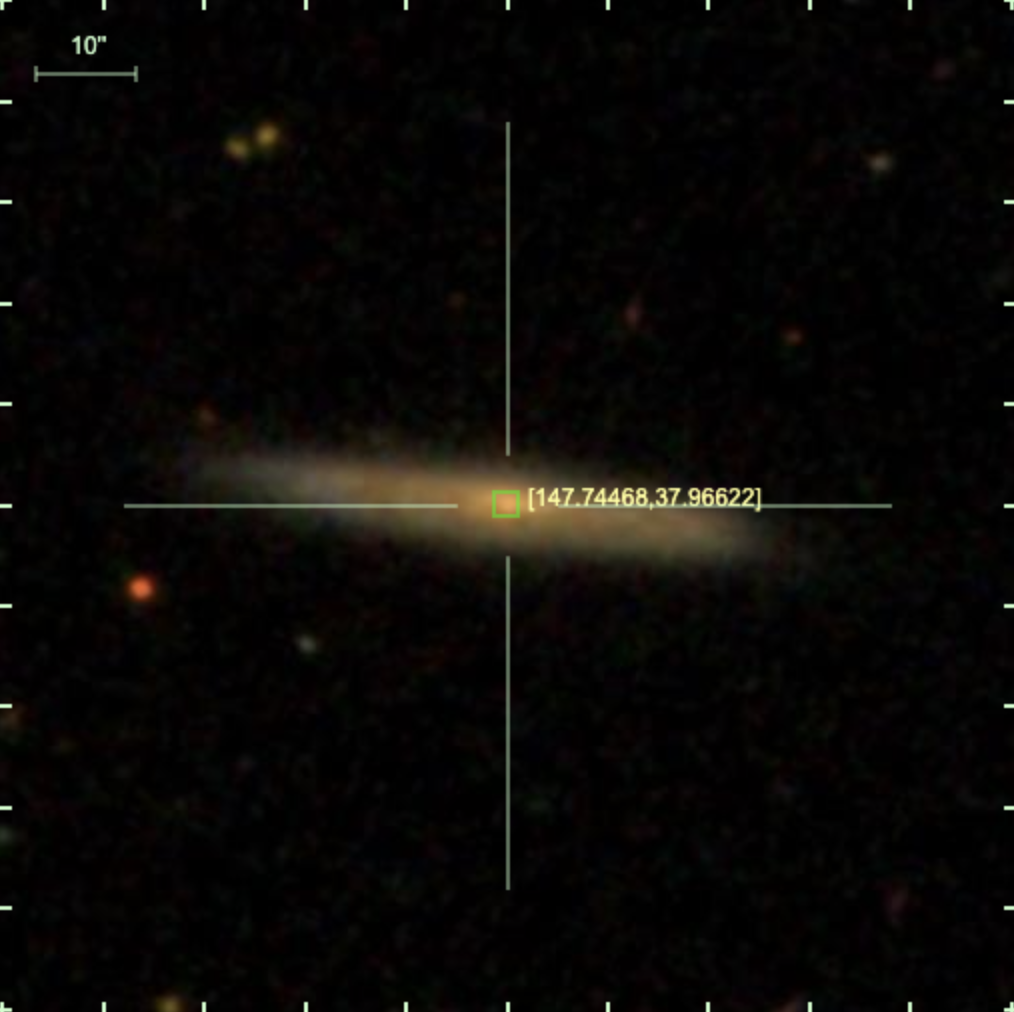}
    \caption{Left: same as Figure~\ref{4C+31.04_fit}, but for NVSS J095058\allowbreak+375758. Middle left: radio map from VLASS centered at NVSS J095058\allowbreak+375758. Middle right: W1 band infrared map of WISEA J095058.73\allowbreak+375758.3. from WISE. Right: SDSS optical map of the optical counterpart of WISEA J095058.73\allowbreak+375758.3.}
    \label{NVSS_J095058+375758_fit}
\end{figure*}

NVSS J095058\allowbreak+375758 (Figure~\ref{NVSS_J095058+375758_fit}) is a nearby galaxy hosting AGN. Morphologically, it is classified as an edge-on FR0 galaxy \citep{2019ApJS..240...34M}. The associated \hi absorption profile shows symmetry, indicating a gas disk, as the absorption line is centered at the systemic velocity of the galaxy.
%The WISE counterpart to NVSS\,J095058+375758 is WISEA\,J095058.73+375758.3 as shown in the NASA/IPAC Extragalactic Database. The WISE W1[\SI{3.4}{\micro\metre}], W2[\SI{4.6}{\micro\metre}], W3[\SI{12.1}{\micro\metre}] and W4[\SI{22.2}{\micro\metre}] magnitudes for WISEA\,J095058.73+375758.3 are 11.933 $\pm$ 0.023, 11.143 $\pm$ 0.021, 8.138 $\pm$ 0.020 and 5.728 $\pm$ 0.047, respectively. The W1-W2 color of WISEA\,J095058.73+375758.3 is 0.790, which means that the mid-IR emission comes mainly from the stars. According to the W2-W3 value of 3.005 mag, WISEA\,J095058.73+375758.3 is located in the Seyferts region in the WISE color-color diagram.

\subsubsection{NVSS J095812+112643}

\begin{figure*}[hbt!]
    \centering
    \includegraphics[width=0.25\textwidth]{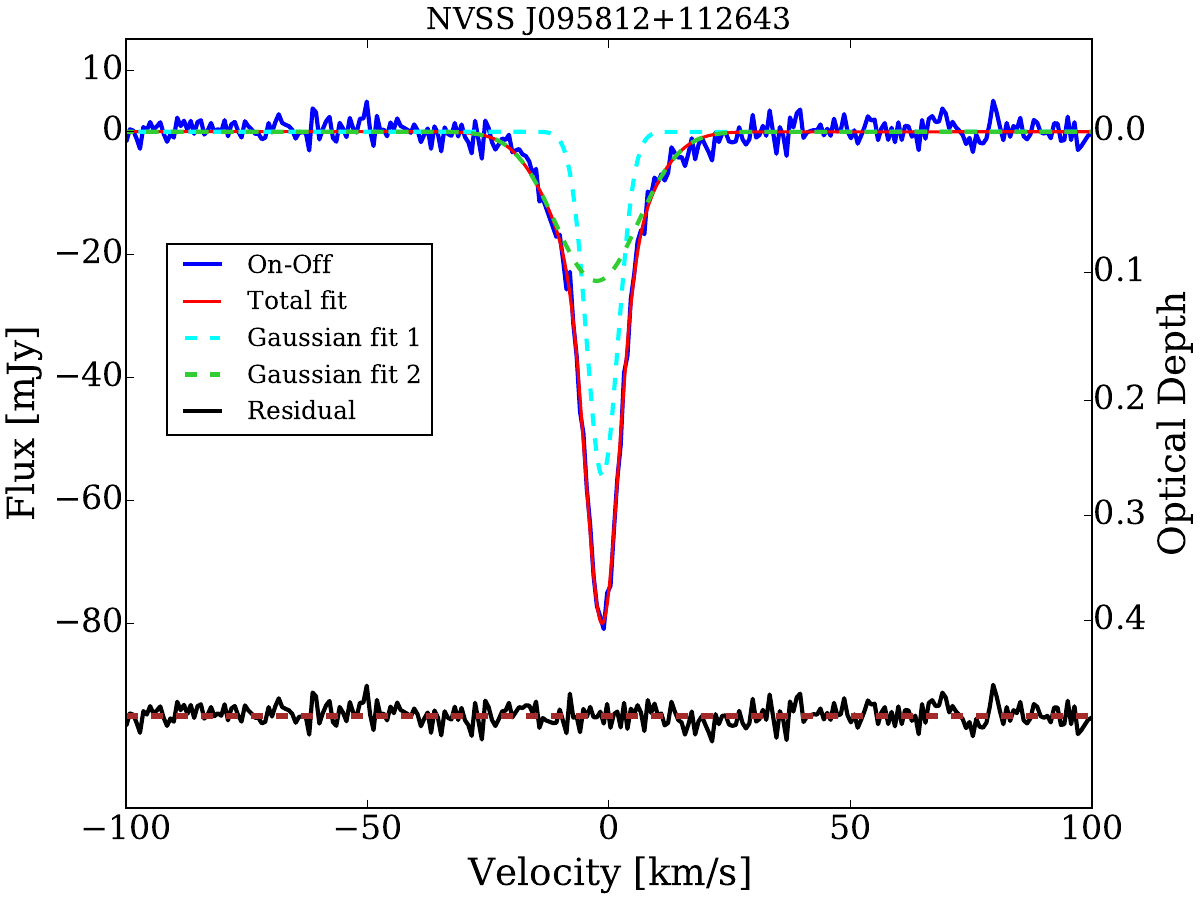}
    \includegraphics[width=0.25\textwidth]{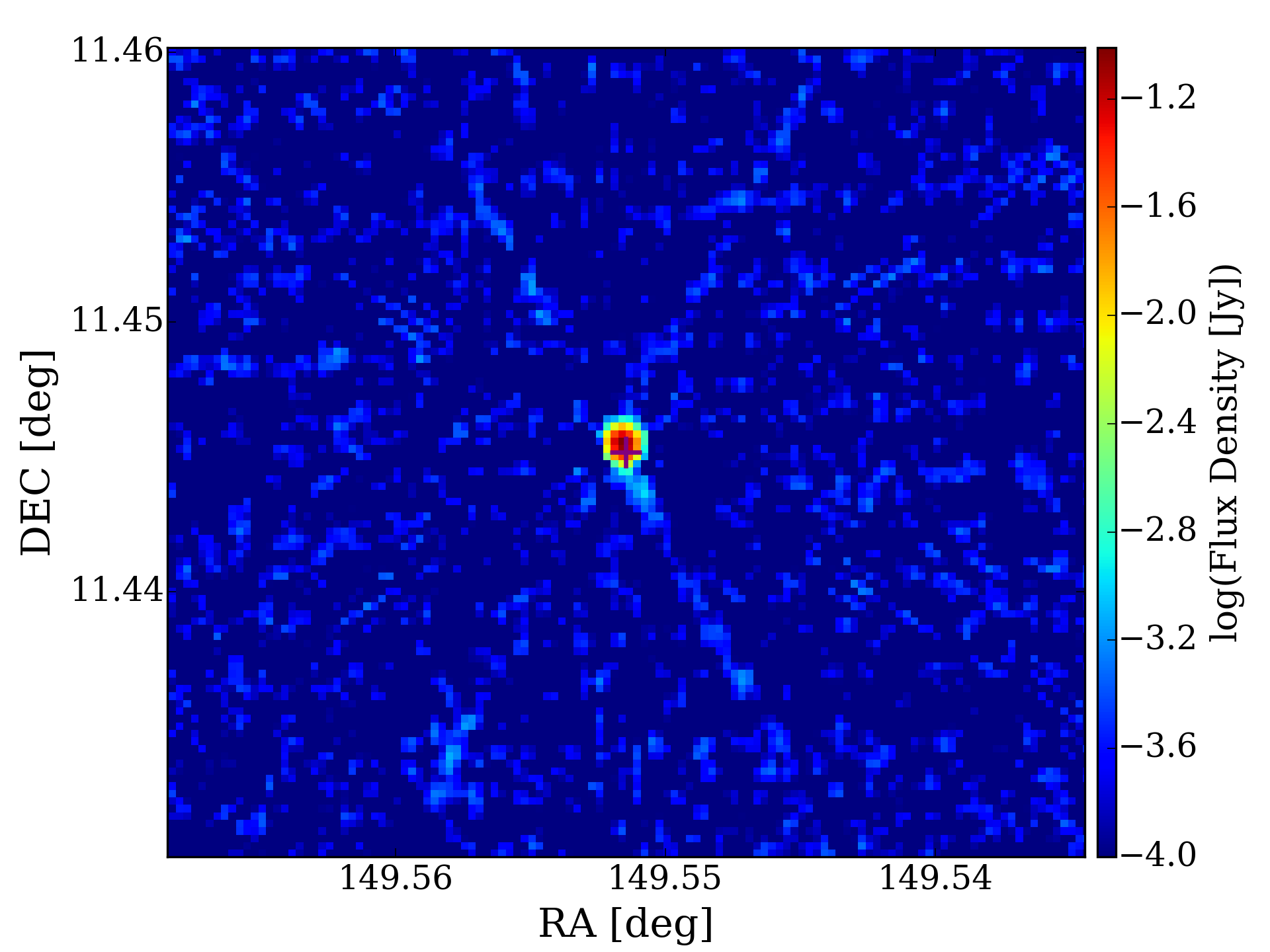}
    \includegraphics[width=0.25\textwidth]{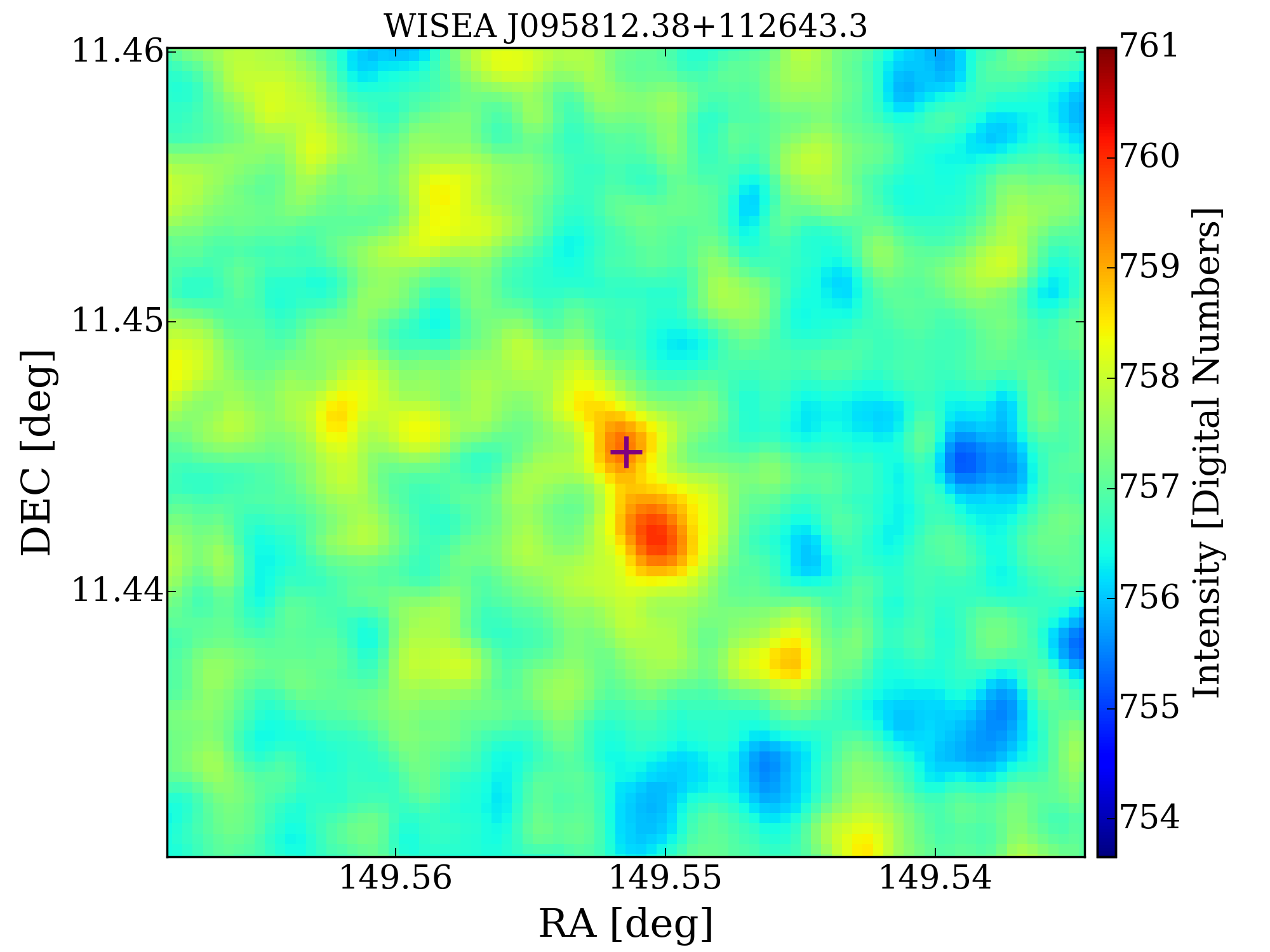}
    \hspace*{0.3cm}
    \includegraphics[width=0.18\textwidth]{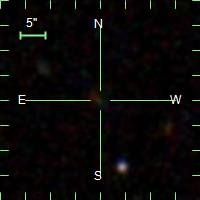}
    \caption{Left: same as Figure~\ref{4C+56.02_fit}, but for NVSS J095812\allowbreak+112643. Middle left: radio map from VLASS centered at NVSS J095812\allowbreak+112643. Middle right: WISE W3 band infrared map centered at NVSS J095812\allowbreak+112643. Right: SDSS optical map centered at NVSS J095812\allowbreak+112643.}
    \label{NVSS_J095812+112643_fit}
\end{figure*}

NVSS J095812\allowbreak+112643 (Figure~\ref{NVSS_J095812+112643_fit}) is a radio galaxy that has been observed in optical (SDSS), infrared (WISE) and radio band (NVSS, VLSS, and Texas \citep{1996AJ....111.1945D}). The lack of precise redshift information for NVSS J095812\allowbreak+112643 gives rise to uncertainty regarding its status in the foreground or background. 
%To dispel this ambiguity, additional high-resolution follow-up observations are imperative. 
The \hi absorption profile observed in this system displays a slim and symmetrical shape, indicating the likely presence of a gas disk.
%The WISE counterpart to NVSS\,J095812+112643 is WISEA J095812.38+112643.3 as given by the NASA/IPAC Extragalactic Database. The WISE W1[\SI{3.4}{\micro\metre}], W2[\SI{4.6}{\micro\metre}], W3[\SI{12.1}{\micro\metre}] and W4[\SI{22.2}{\micro\metre}] magnitudes for WISEA J095812.38+112643.3 are 16.130 $\pm$ 0.074, 15.775 $\pm$ 0.172, 11.327 $\pm$ 0.149 and 8.551 $\pm$ 0.328, respectively. The W1-W2 color of WISEA J095812.38+112643.3 is 0.355, which implies that the mid-IR emission comes mostly from stars. Considering the W2-W3 value of 4.448 mag, WISEA J095812.38+112643.3 locates in the ULIRGs/LINERs and Starburst region in the WISE color–color diagram.

%The middle left, middle right panel and right panel of Figure~\ref{NVSS_J095812+112643_fit} show the image centered at NVSS\,J095812+112643, constructed using radio data at S-band from VLASS, infrared data at 12.1 microns (W3) from WISE and optical data from SDSS, respectively. The straight cross shows the position of NVSS\,J095812+112643.

\subsubsection{NVSS J100755+405519}

\begin{figure*}[hbt!]
    \centering
    \includegraphics[width=0.25\textwidth]{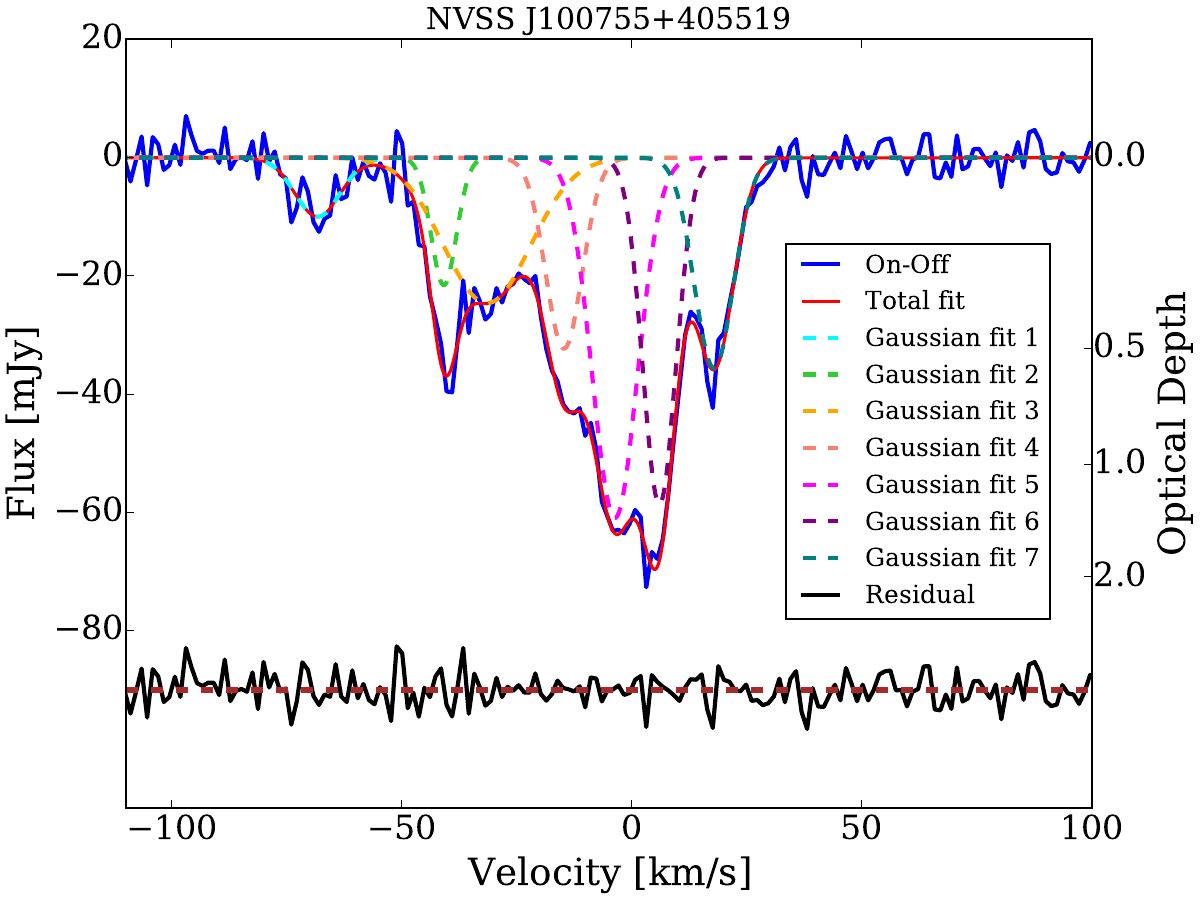}
    \includegraphics[width=0.25\textwidth]{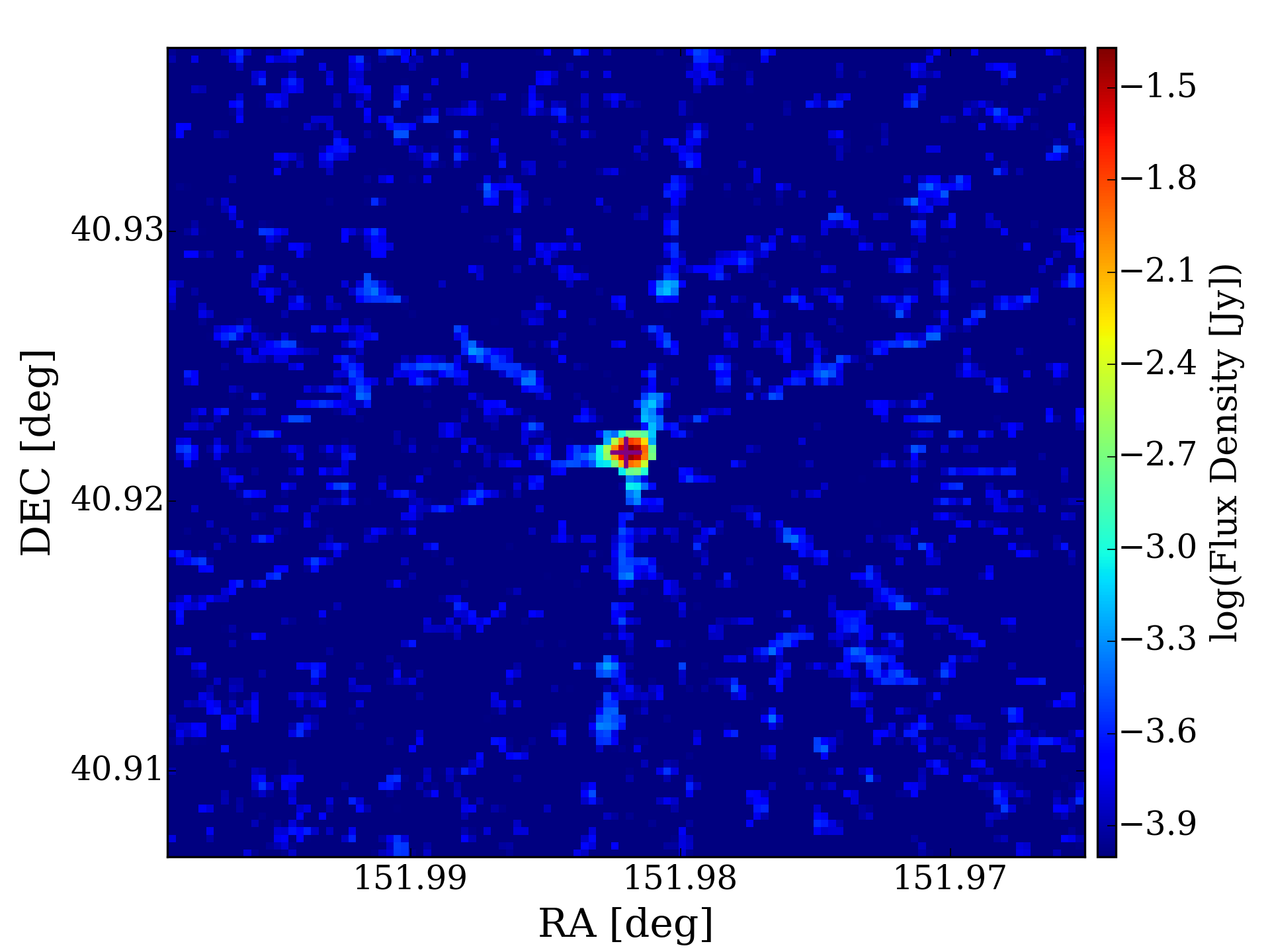}
    \includegraphics[width=0.25\textwidth]{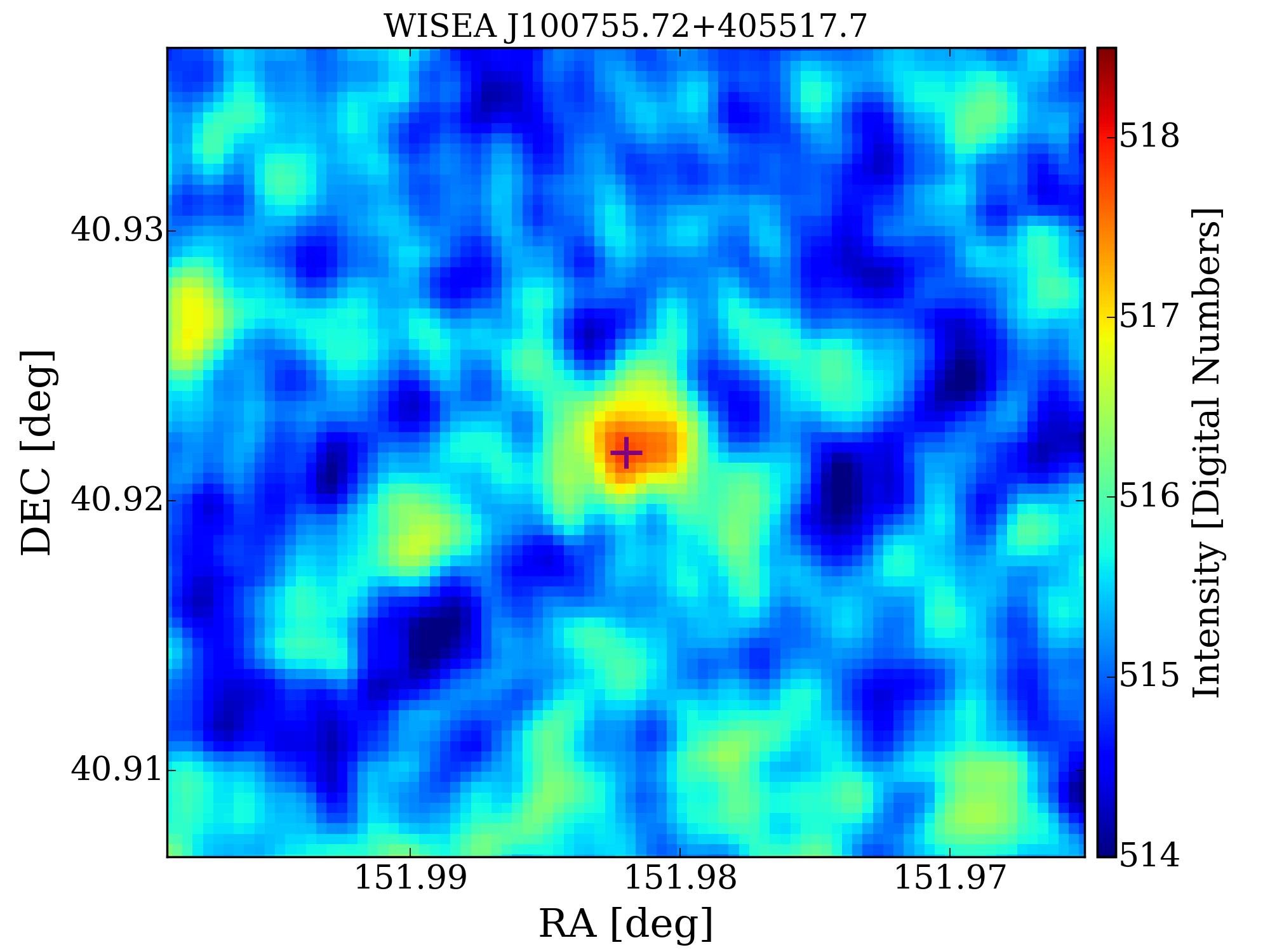}
    \hspace*{0.25cm}
    \includegraphics[width=0.18\textwidth]{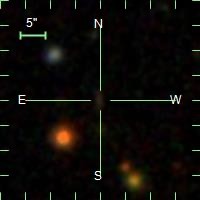}
    \caption{Left: same as Figure~\ref{4C+56.02_fit}, but for NVSS J100755\allowbreak+405519. Middle left: radio map from VLASS centered at NVSS J100755\allowbreak+405519. Middle right: WISE W3 band infrared map centered at NVSS J100755\allowbreak+405519. Right: SDSS optical map of the optical counterpart of NVSS J100755\allowbreak+405519.}
    \label{NVSS_J100755+405519_fit}
\end{figure*}

NVSS J100755\allowbreak+405519 (Figure~\ref{NVSS_J100755+405519_fit}) is a radio source, detected in the Faint Images of the Radio Sky at Twenty Centimeters Survey (FIRST, \citealt{1995ApJ...450..559B}) and NVSS. WISE and SDSS measured its flux density in the mid-infrared band and optical band, respectively. Precise redshift information for NVSS J100755\allowbreak+405519 is currently unavailable, making it difficult to determine whether it is an associated or intervening system.

%To bring clarity to this situation, further high-resolution follow-up observations are required. 
The WISE counterpart to NVSS J100755\allowbreak+405519 is WISEA J100755.72\allowbreak+405517.7 according to the NED. The W1-W2 color of WISEA J100755.72+405517.7 is 1.005, indicating WISEA J100755.72\allowbreak+405517.7 is an AGN candidate. Combined with the W2-W3 value of 3.074 mag, WISEA J100755.72\allowbreak+405517.7 is located in the region of QSOs in the WISE color-color diagram. 

The intricate absorption profile suggests an unsettled system. To characterize this profile, a seven-component Gaussian function is employed. Apart from dominant components at the center, three components exhibit blueshifts with velocity separations of -73 km/s, -46 km/s, and -36 km/s, suggesting a potential gas outflow or the presence of a satellite galaxy. Additionally, a redshifted component implies the possibility of gas accretion.

\subsubsection{NVSS J104941+133255}

\begin{figure*}[hbt!]
    \centering
    \includegraphics[width=0.25\textwidth]{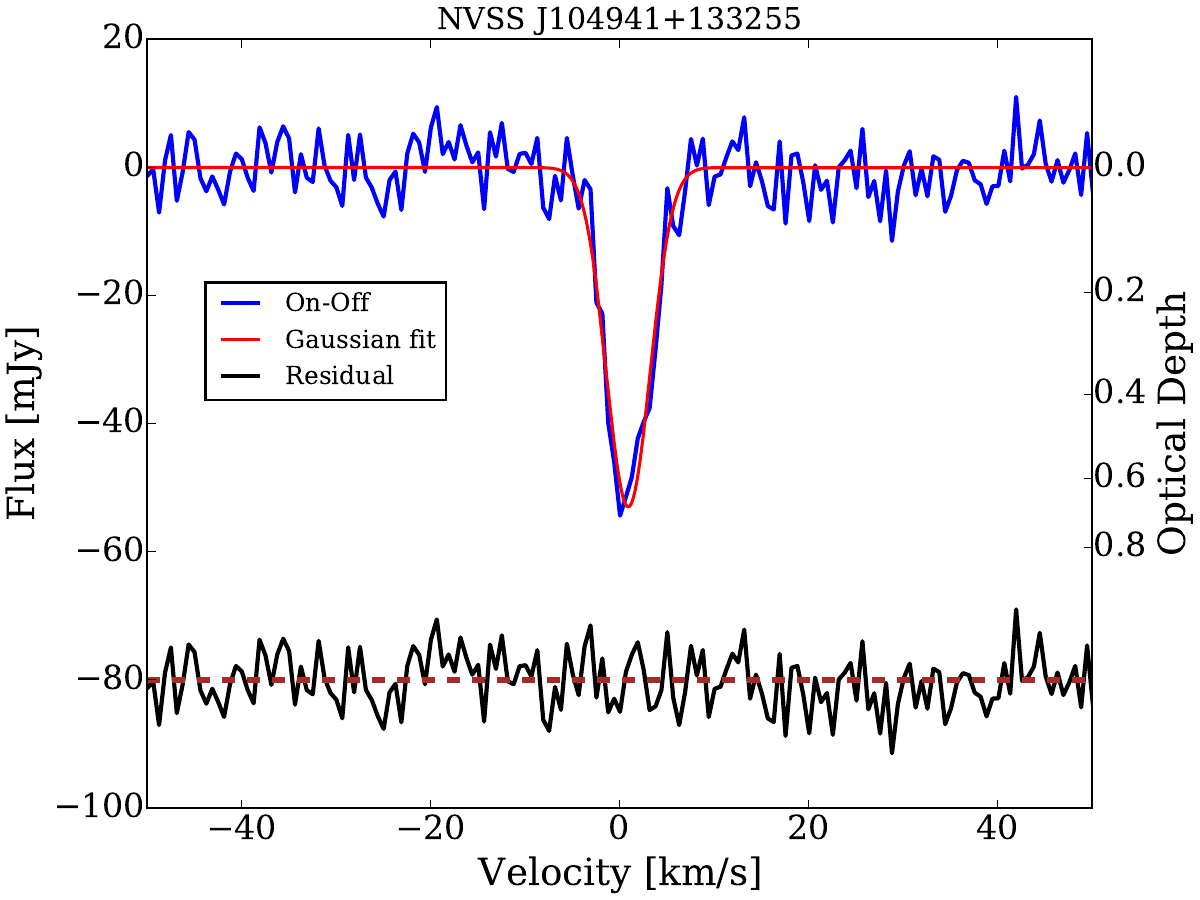}
    \includegraphics[width=0.25\textwidth]{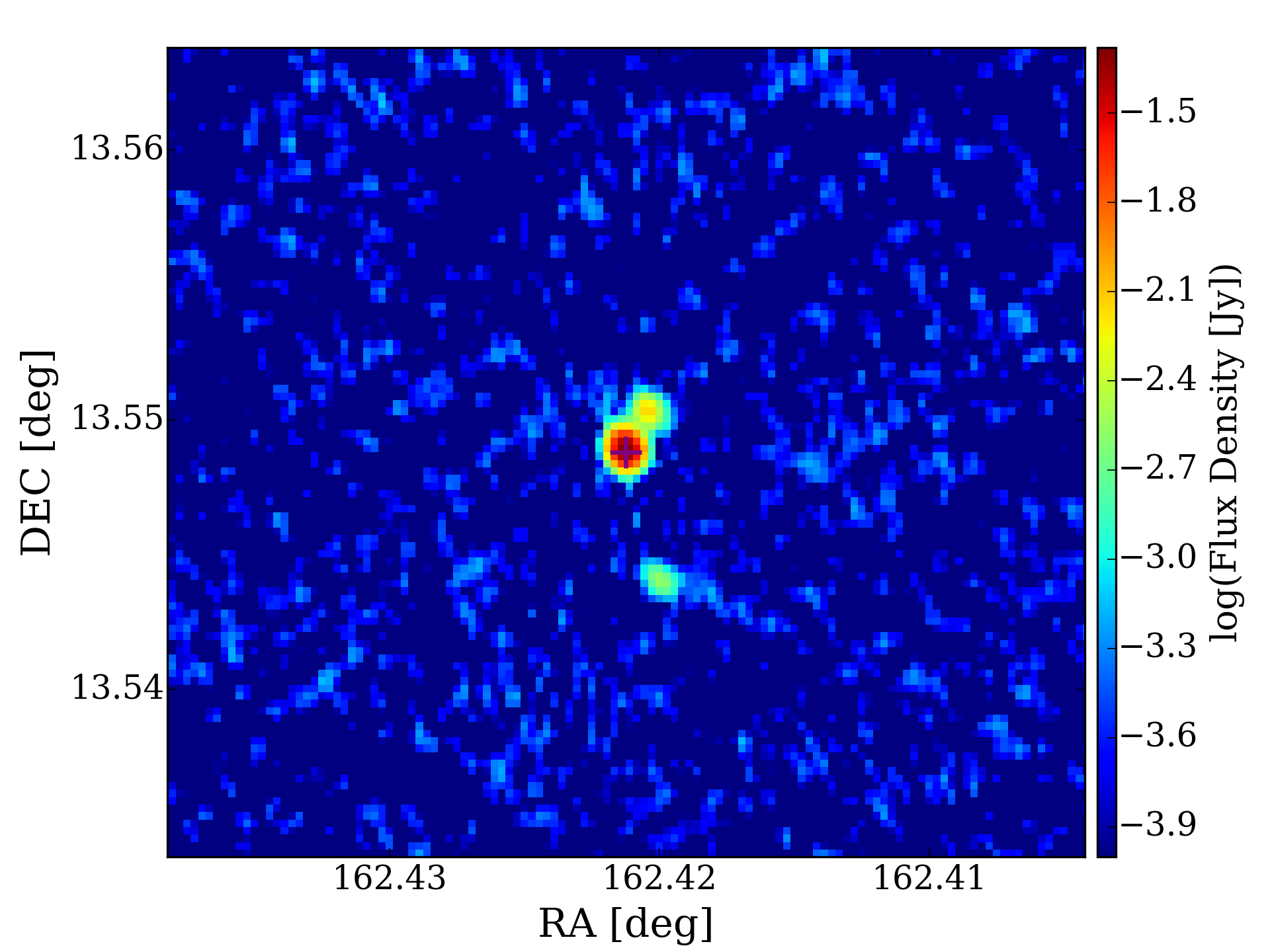}
    \includegraphics[width=0.25\textwidth]{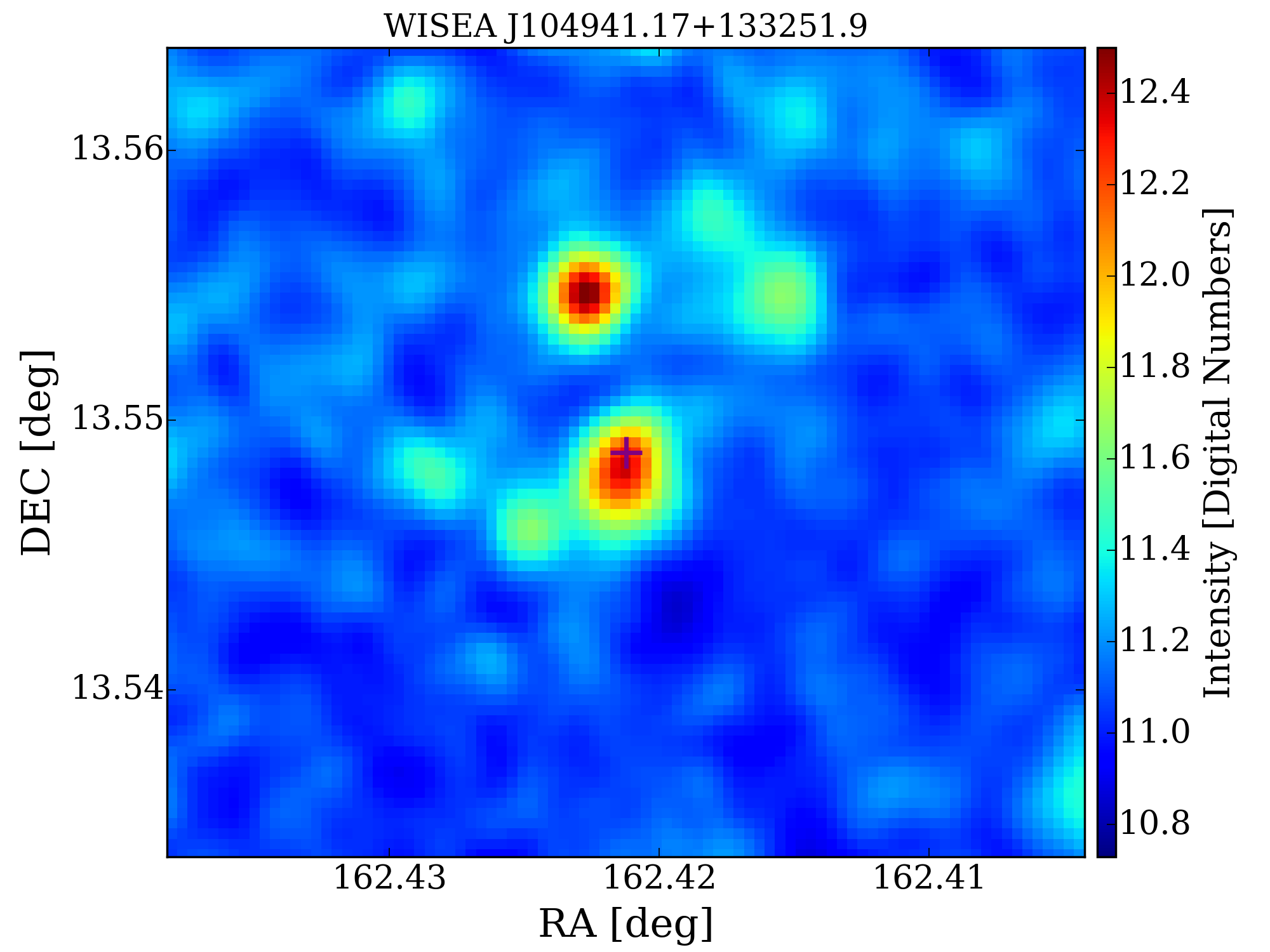}
    \hspace*{0.3cm}
    \includegraphics[width=0.18\textwidth]{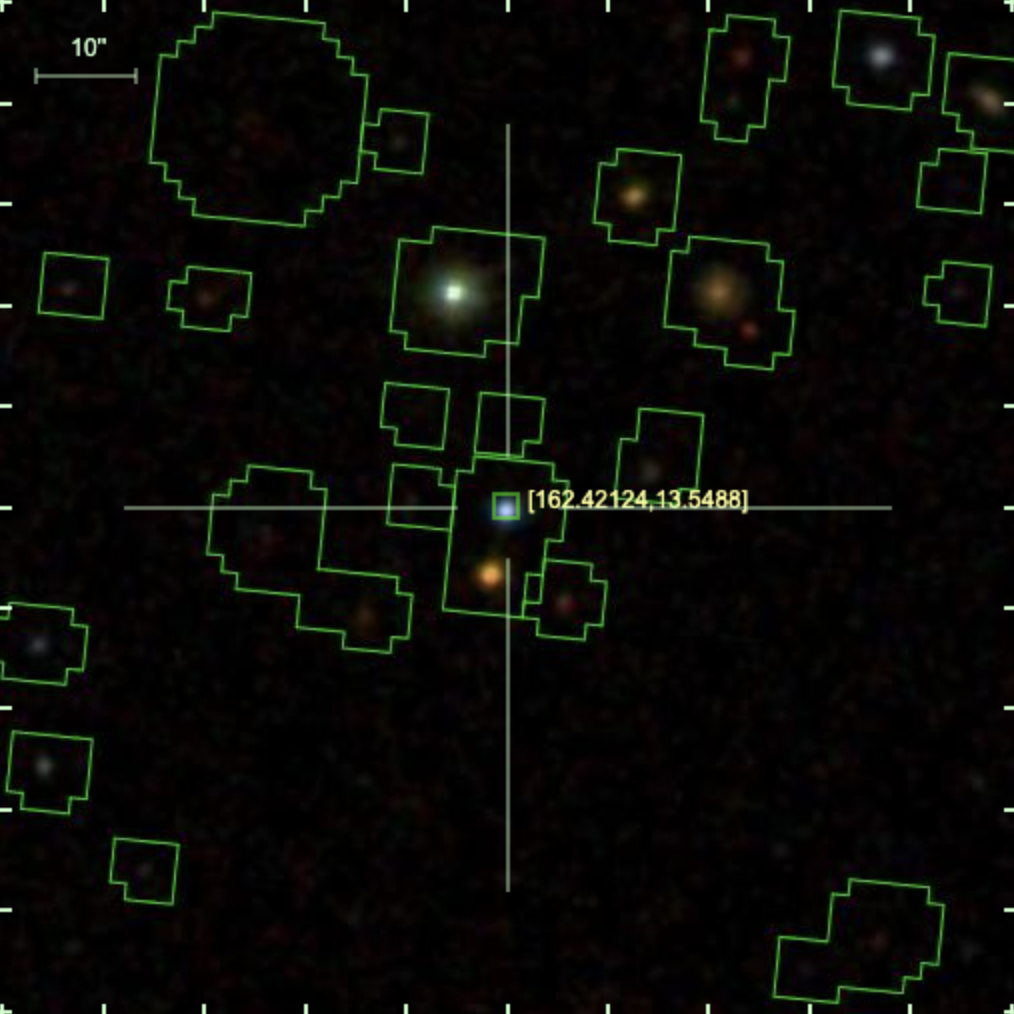}
    \caption{Left: same as Figure~\ref{4C+56.02_fit}, but for NVSS J104941\allowbreak+133255. Middle left: radio map from VLASS centered at NVSS J104941\allowbreak+133255 Middle right: WISE W2 band infrared map centered at NVSS J104941\allowbreak+133255. Right: SDSS optical map centered at NVSS J104941\allowbreak+133255. The bright orange point source below NVSS J104941\allowbreak+133255 is a star.}
    \label{NVSS_J104941+133255_fit}
\end{figure*}

NVSS J104941\allowbreak+133255 (Figure~\ref{NVSS_J104941+133255_fit}) is a quasar located at a redshift of 2.764, it has been observed in optical and Ultraviolet (SDSS), and radio bands (NVSS and Texas Survey). The foreground object remains ambiguous and requires a thorough high-resolution follow-up observation for confirmation. The \hi absorption profile exhibits symmetry and can be accurately modeled using a one-component Gaussian function. 

%The WISE counterpart to NVSS\,J104941+133255 is WISEA\,J104941.17+133251.9 as given by the SDSS Database. The WISE W1[\SI{3.4}{\micro\metre}], W2[\SI{4.6}{\micro\metre}], W3[\SI{12.1}{\micro\metre}] and W4[\SI{22.2}{\micro\metre}] magnitudes for WISEA\,J104941.17+133251.9 are 15.281 $\pm$ 0.047, 14.834 $\pm$ 0.091, 12.309 and 8.586, respectively. The W1-W2 color of WISEA\,J104941.17+133251.9 is 0.447, which indicates WISEA\,J104941.17+133251.9 as an AGN candidate. Considering the W2-W3 value of 2.525 mag, WISEA\,J104941.17+133251.9 is located in the Spirals region in the WISE color-color diagram.

\subsubsection{NVSS J115948+582020}

\begin{figure*}[hbt!]
    \centering
    \includegraphics[width=0.25\textwidth]{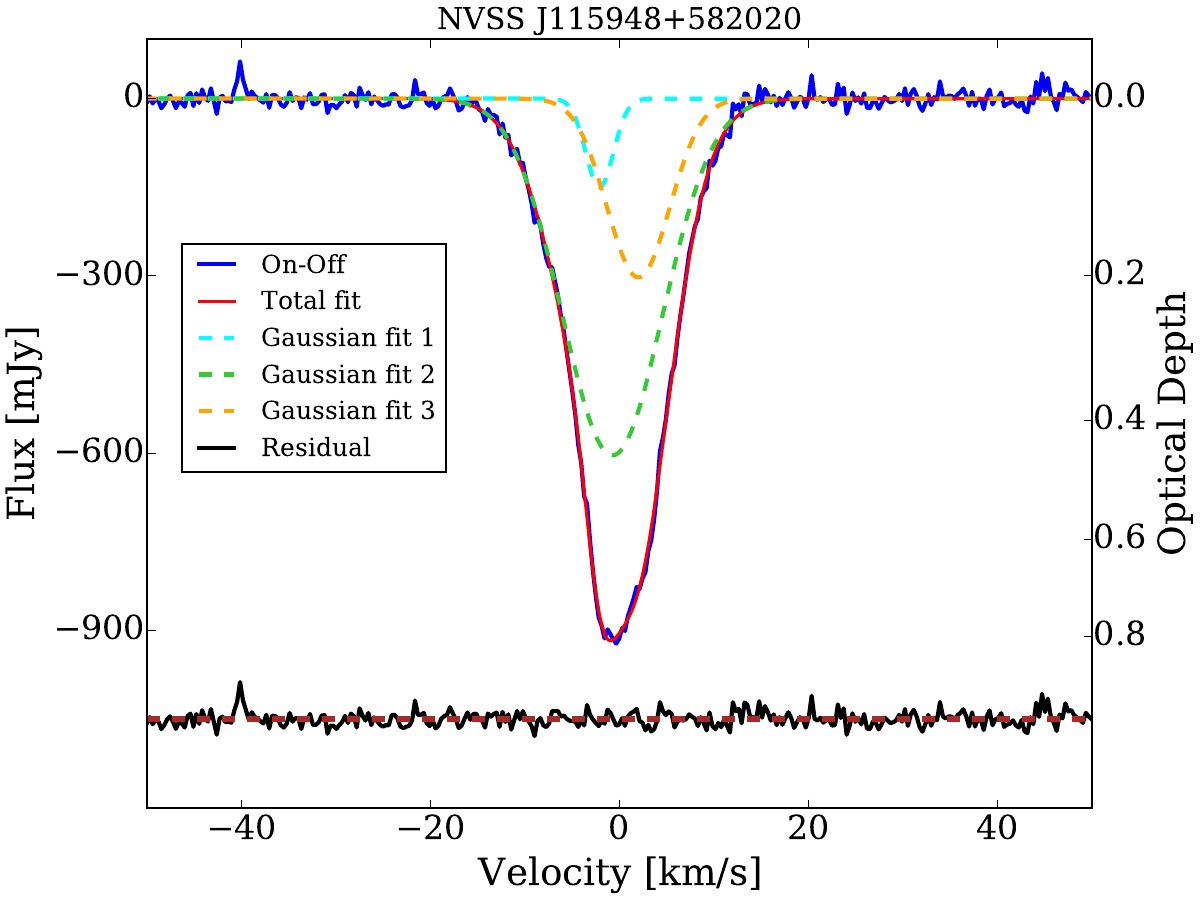}
    \includegraphics[width=0.25\textwidth]{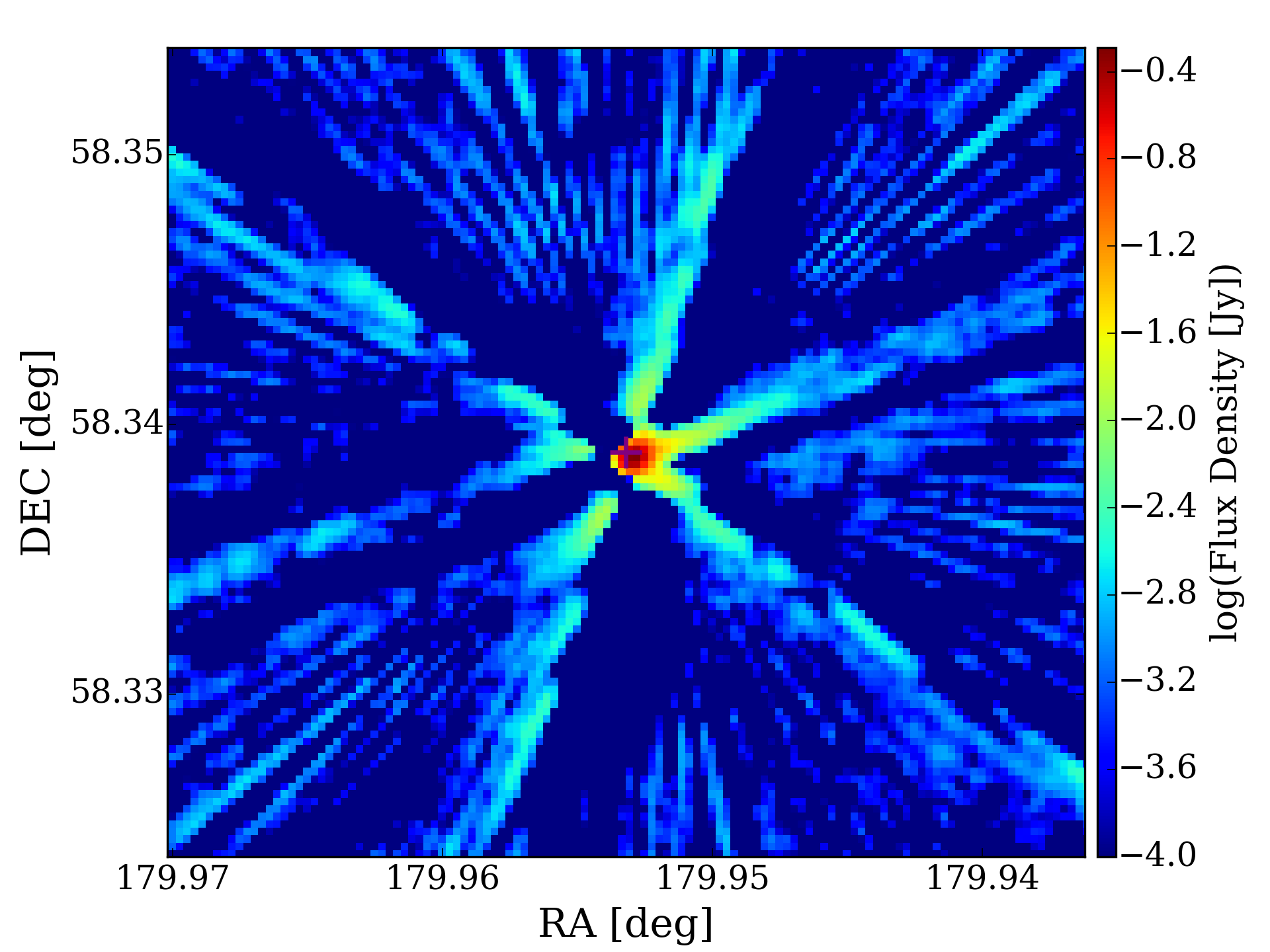}
    \includegraphics[width=0.25\textwidth]{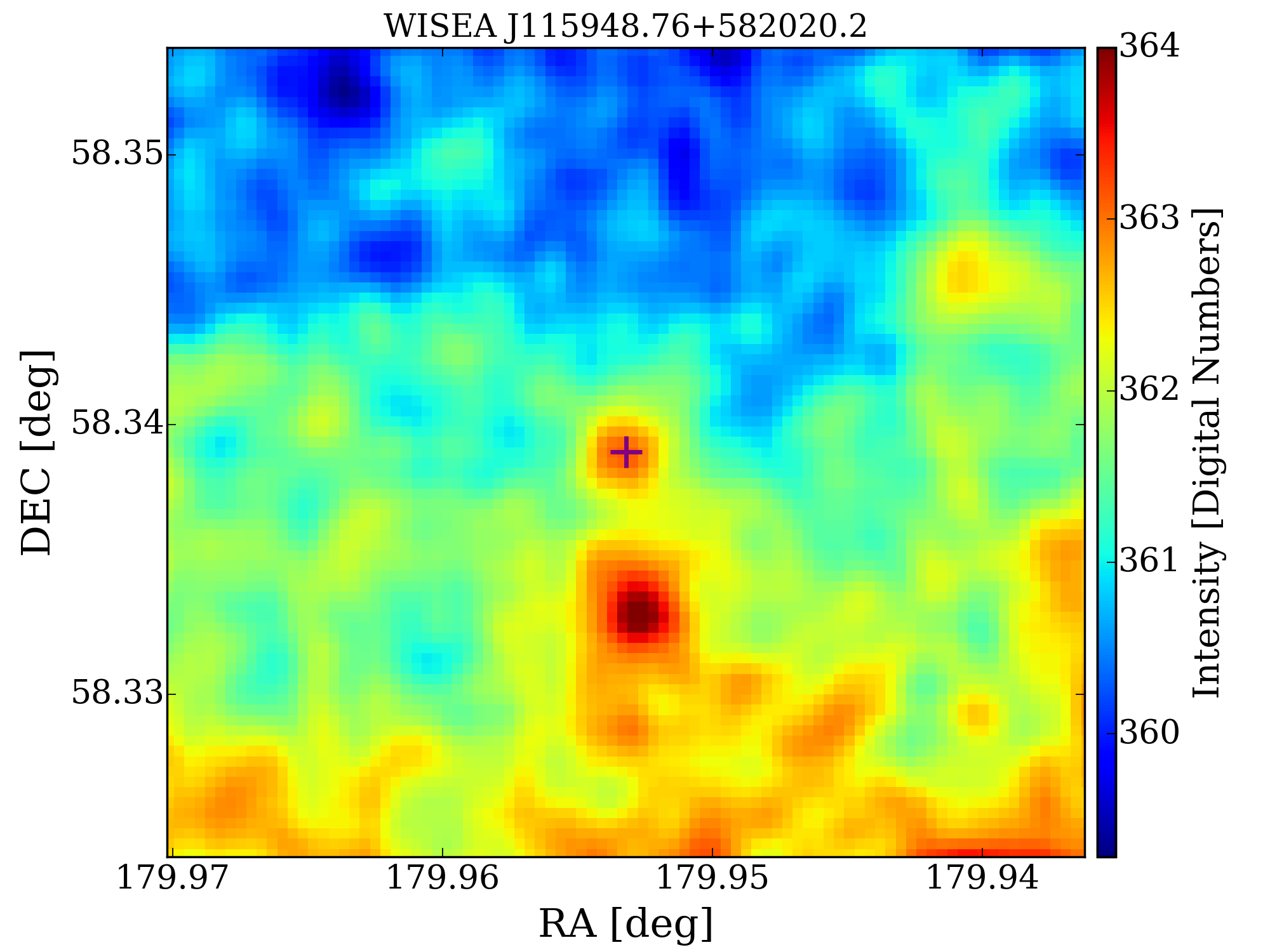}
    \hspace*{0.3cm}
    \includegraphics[width=0.18\textwidth]{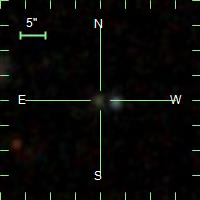}
    \caption{Left: same as Figure~\ref{4C+56.02_fit}, but for NVSS J115948\allowbreak+582020. Middle left: radio map from VLASS centered at NVSS J115948\allowbreak+582020. Middle right: WISE W3 band infrared map centered at NVSS J115948\allowbreak+582020. Right: SDSS optical map centered at NVSS J115948\allowbreak+582020.}
    \label{NVSS_J115948+582020_fit}
\end{figure*}

NVSS J115948\allowbreak+582020 (Figure~\ref{NVSS_J115948+582020_fit}) is a quasar located at z=0.629 and has been observed in optical (eBOSS), infrared (WISE), and radio bands (NVSS, VLBA and Texas Survey). 
%Additional high-resolution follow-up observations are necessary to establish clarity for the foreground source. 
The \hi absorption signal in this intervening system is robust, displaying an almost symmetrical shape that can be effectively modeled using a three-component Gaussian function.

\subsubsection{NVSS J162549+402921 (NGC 6150)}

\begin{figure*}[hbt!]
    \centering
    \includegraphics[width=0.25\textwidth]{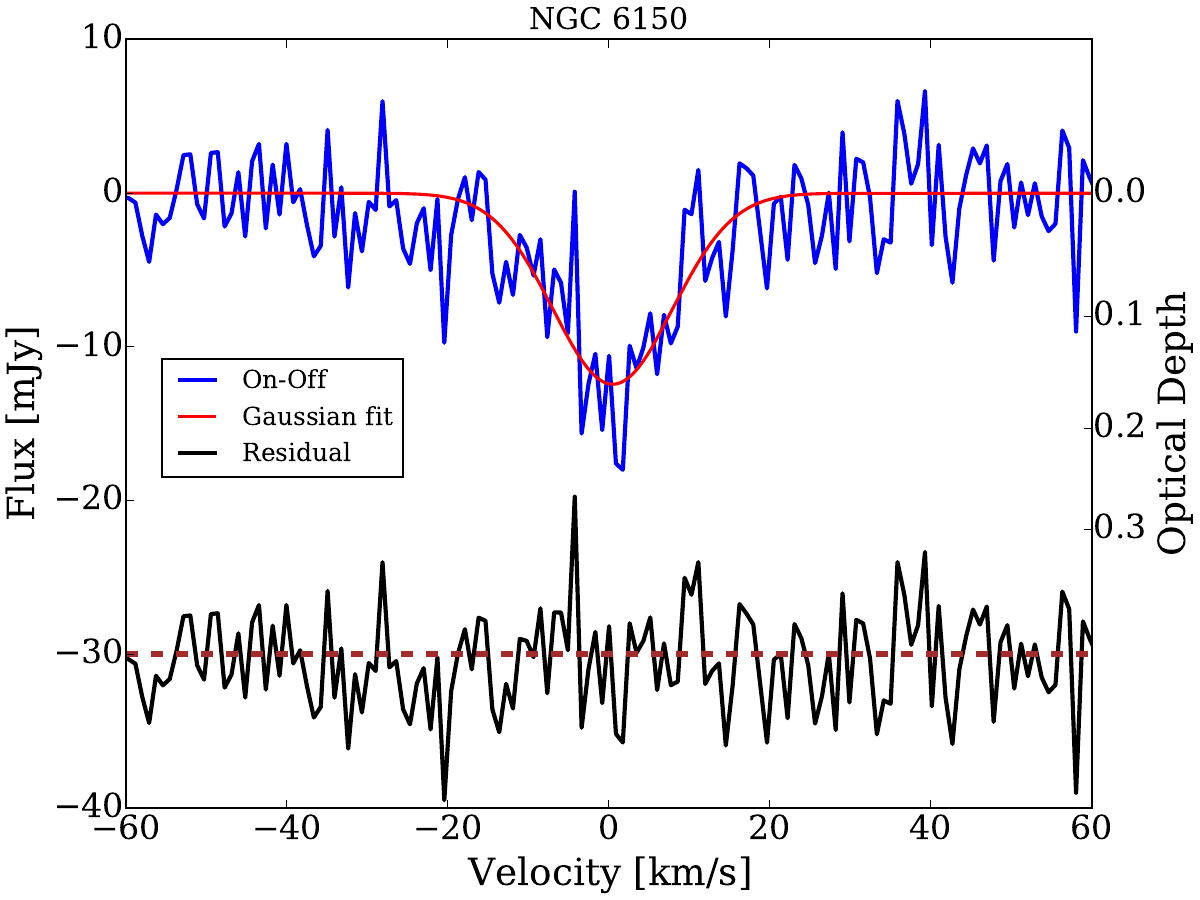}
    \includegraphics[width=0.25\textwidth]{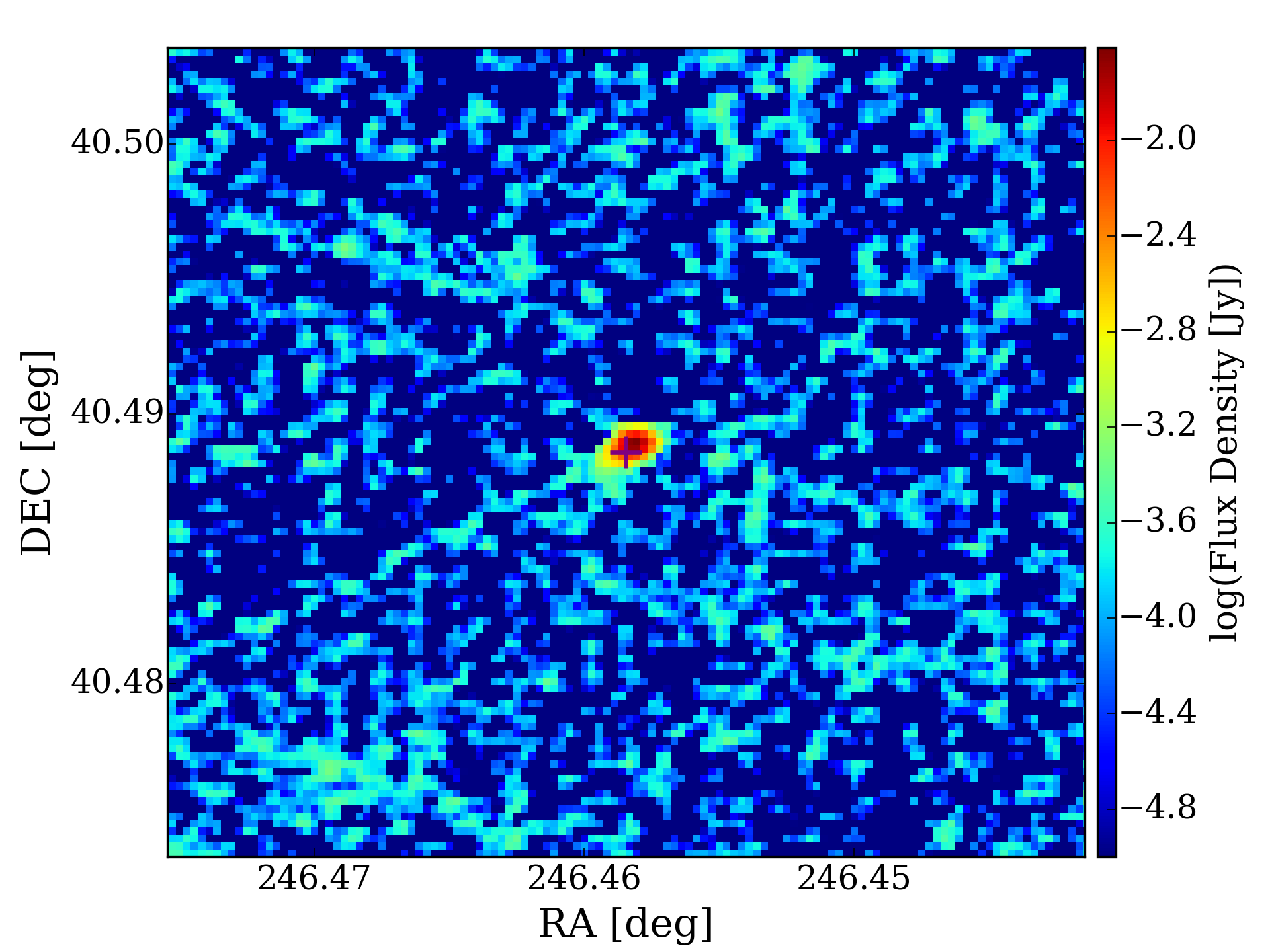}
    \includegraphics[width=0.25\textwidth]{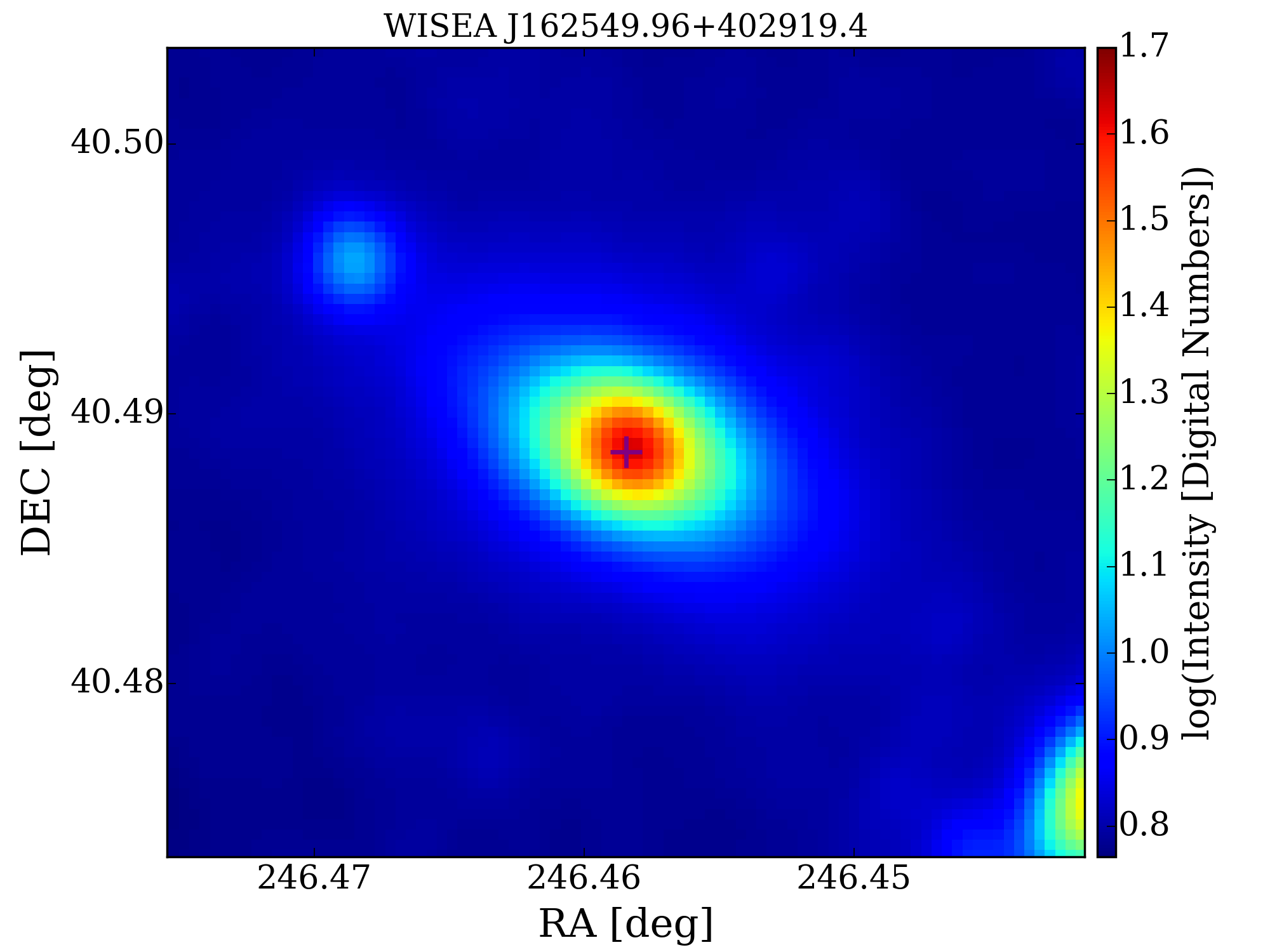}
    \includegraphics[width=0.18\textwidth]{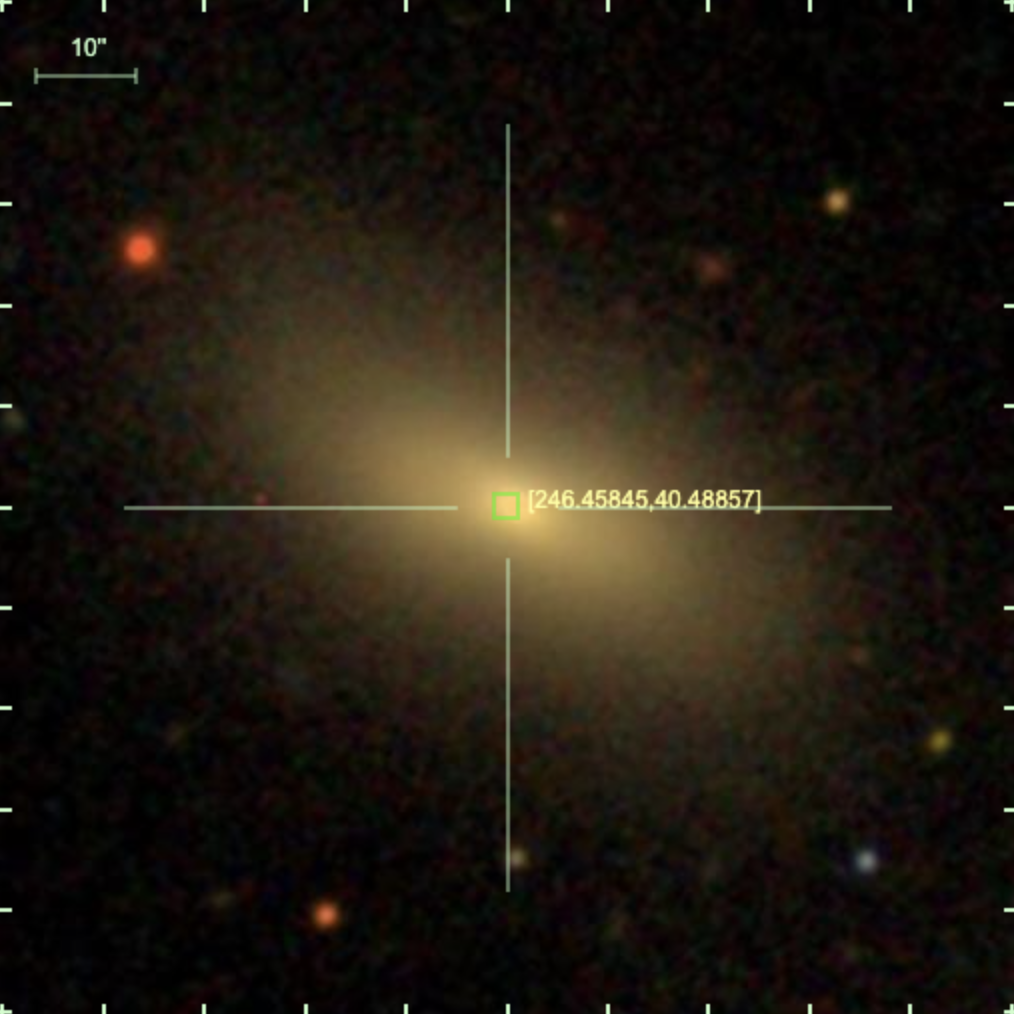}
    \caption{Left: same as Figure~\ref{4C+56.02_fit}, but for NGC\,6150. Middle left: radio map from VLASS centered at NGC\,6150. Middle right: W2 band infrared map centered at NGC\,6150. Right: SDSS optical map of the SDSS J162549.96\allowbreak+402919.3.}
    \label{NGC_6150_fit}
\end{figure*}

NGC\,6150 (Figure~\ref{NGC_6150_fit}) is situated within the Abell\,2197 cluster and is classified as an elliptical galaxy\citep{2010ApJS..186..427N}. As documented by \citet{1987ApJ...316..113O}, NGC\,6150 is identified as a narrow-angle tail (NAT) radio source.

%The WISE counterpart to NGC\,6150 is WISEA\,J162549.96+402919.4 as shown in the NASA/IPAC Extragalactic Database. The WISE W1[\SI{3.4}{\micro\metre}], W2[\SI{4.6}{\micro\metre}], W3[\SI{12.1}{\micro\metre}] and W4[\SI{22.2}{\micro\metre}] magnitudes for WISEA\,J162549.96+402919.4 are 10.764 $\pm$ 0.022, 10.794 $\pm$ 0.020, 10.049 $\pm$ 0.045 and 8.646 $\pm$ 0.259, respectively. The W1-W2 color of WISEA\,J162549.96+402919.4 is -0.03, which means that the mid-IR emission comes mainly from the stars. According to the W2-W3 value of 0.745 mag, WISEA\,J162549.96+402919.4 is located in the Ellipticals region in the WISE color-color diagram.

\subsubsection{NVSS J225900+274356}

\begin{figure*}[hbt!]
    \centering
    \includegraphics[width=0.25\textwidth]{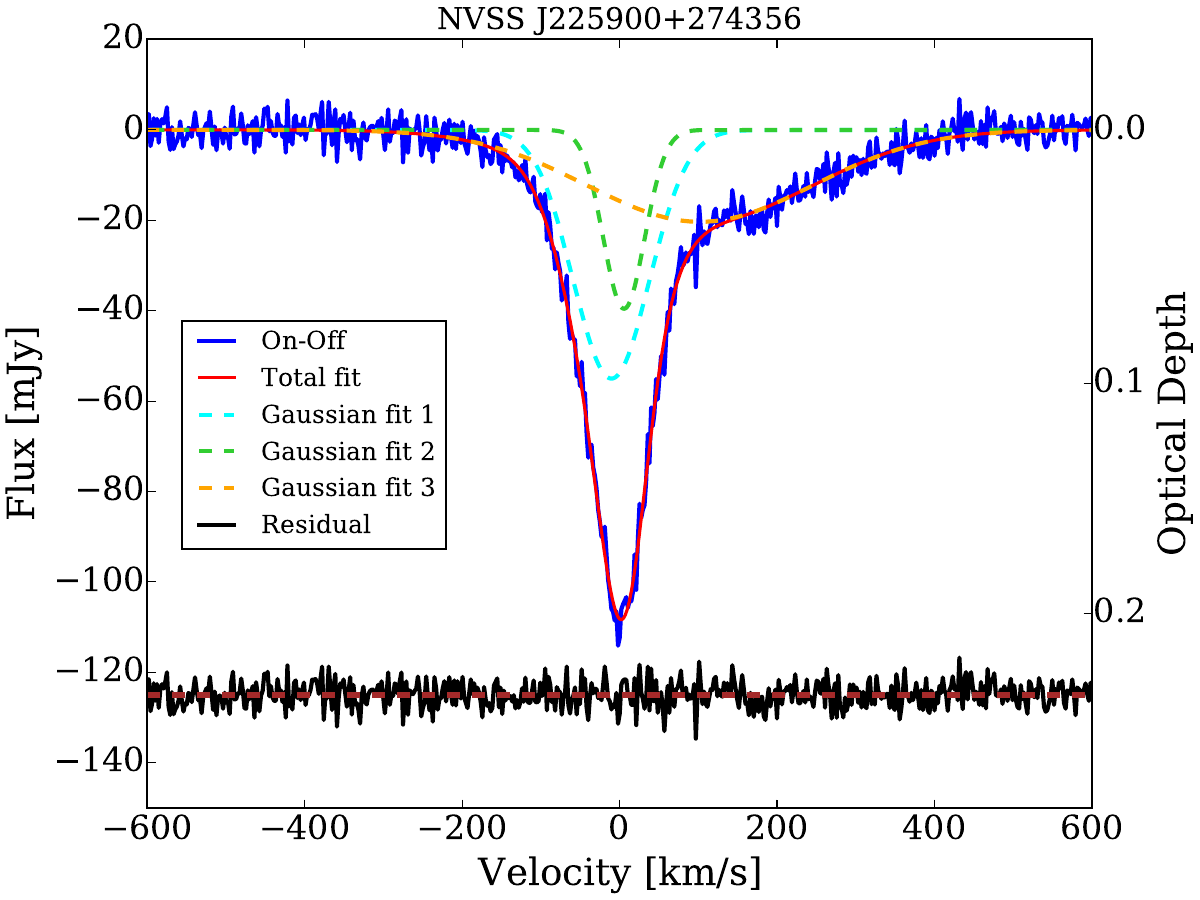}
    \includegraphics[width=0.25\textwidth]{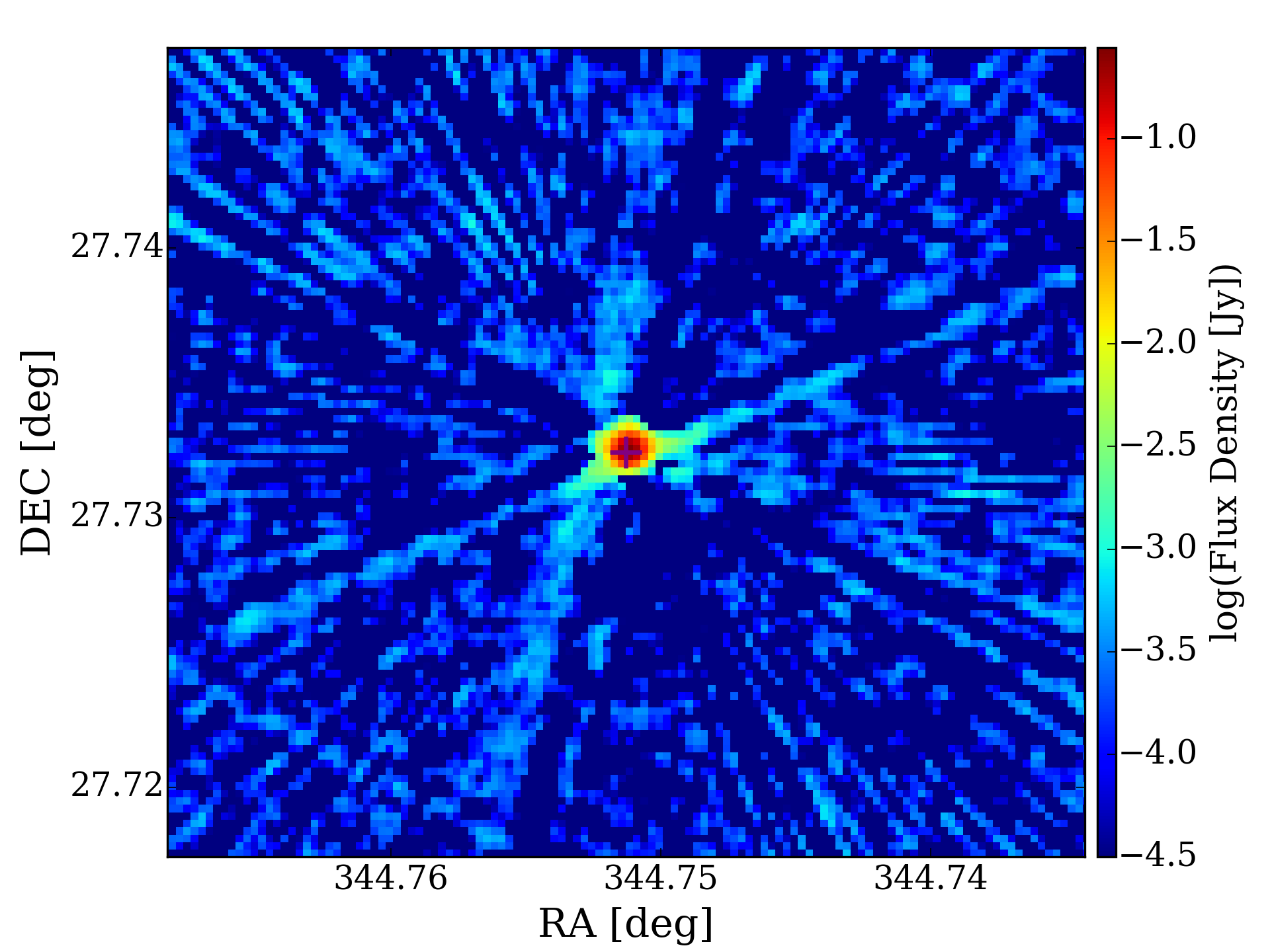}
    \includegraphics[width=0.25\textwidth]{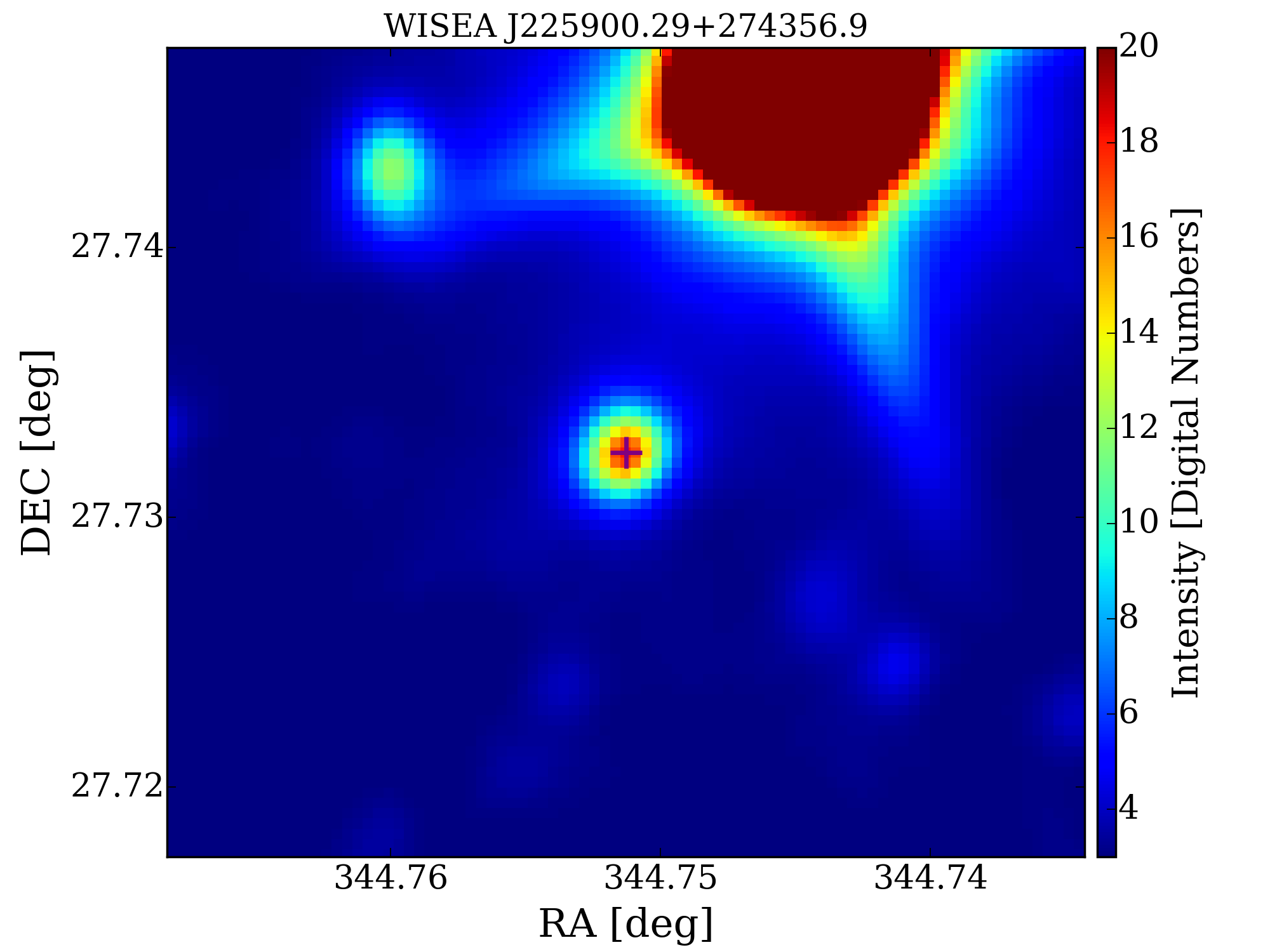}
    \includegraphics[width=0.18\textwidth]{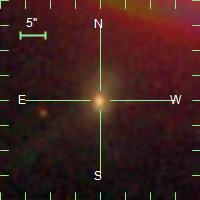}
    \caption{Left: same as Figure~\ref{4C+31.04_fit}, but for NVSS J225900\allowbreak+274356. Middle left: radio map from VLASS centered at NVSS J225900\allowbreak+274356. Middle right: W1 band infrared map of WISEA J225900.29\allowbreak+274356.9. from WISE. Right: SDSS optical map of the optical counterpart of WISEA J225900.29\allowbreak+274356.9.}
    \label{NVSS_J225900+274356_fit}
\end{figure*}

NVSS J225900\allowbreak+274356 (Figure~\ref{NVSS_J225900+274356_fit}) is an AGN-host radio galaxy with a photoz=0.110 $\pm$0.0236, which has been observed in the Infrared/Submillimeter (WISE and 2MASS) and radio bands (Texas and NVSS). The photometric redshift is lower than the redshift of the \hi absorption, suggesting that the \hi gas might originate from an associated infalling object.

The \hi absorption profile exhibits a symmetric, narrower, and more intense component, likely originating from a gas disk. Additionally, there is a broader, shallower redshifted wing, indicating the existence of unsettled gas structures and a potential accretion of gas onto the SMBH.

%The WISE counterpart to NVSS\,J225900+274356 is WISEA\,J225900.29+274356.9. The WISE W1[\SI{3.4}{\micro\metre}], W2[\SI{4.6}{\micro\metre}], W3[\SI{12.1}{\micro\metre}] and W4[\SI{22.2}{\micro\metre}] magnitudes for WISEA\,J225900.29+274356.9 are 13.148 $\pm$ 0.026, 12.429 $\pm$ 0.024, 9.887 $\pm$ 0.043 and 7.551 $\pm$ 0.113, respectively. The W1-W2 color of WISEA\,J225900.29+274356.9 is 0.719, which means that the mid-IR emission comes mainly from the stars. According to the W2-W3 value of 2.542 mag, WISEA\,J225900.29+274356.9 is located in the intersection of the QSO region and the Seyferts region in the WISE color-color diagram. 

\section{\hi Absorption Properties}
\label{sec:Absorption_Properties}

\subsection{\hi Absorption Statistics}
\label{sec:Absorption_Statistics}

Physical properties of associated and intervening absorption are investigated in this section, to understand the nature of \hi absorption systems. We also place our results within the broader context of previous and ongoing \hi 21-cm absorption surveys, including the detections of FLASH\citep{2020MNRAS.494.3627A,2020MNRAS.499.4293S,2022MNRAS.509.1690M,2022MNRAS.512.3638W,2022MNRAS.516.2947S,2024MNRAS.527.8511A}, MALS\citep{2022MNRAS.516.1339S,2022MNRAS.516.2050M,2024A&A...687A..50D} and of the surveys using WSRT\citep{2015A&A...575A..44G,2017A&A...604A..43M}, the Giant Metrewave Radio Telescope (GMRT, \citealt{2009MNRAS.396..385K,2009MNRAS.398..201G,2012A&A...544A..21G,2018MNRAS.473...59A,2018MNRAS.481.1578A,2019MNRAS.482.5597A,2020MNRAS.494.5161C,2021MNRAS.500..998A}) and FAST\citep{2024ApJ...973...48C}.

\subsubsection{\hi absorption surveys}
Using WSRT, \citet{2015A&A...575A..44G,2017A&A...604A..43M} detected 66 associated \hi absorption in 248 radio sources at 0.02 $< z <$ 0.25. These sources were selected by cross-correlating the SDSS DR7 with the FIRST catalog, applying a flux density threshold of S$_{1.4GHz} >$ 30 mJy. \citet{2019MNRAS.482.5597A} observed a sample of 11 radio-bright galaxies, also selected by cross-correlating the SDSS DR7 with the FIRST catalog, using the Band-4 of upgraded GMRT (covering $0.7<z<1.0$), resulting in the detection of four associated \hi 21-cm absorption systems. \citet{2018MNRAS.481.1578A} presented 16 associated \hi absorption systems in 92 AGNs at $z < 3.6$, from the Caltech–Jodrell Bank flat-spectrum (CJF) sample. Additionally, the results of \hi 21-cm absorption searches in 30 GPS sources, including 14 associated detections, are presented in \citet{2018MNRAS.473...59A}. \citet{2020MNRAS.494.5161C} presented seven associated detections from \hi absorption experiment conducted using the GMRT towards 27 low- and intermediate-luminosity AGNs, classified as either low excitation radio galaxies (LERGs) or high excitation radio galaxies (HERGs), with WISE color W2[\SI{4.6}{\micro\metre}]-W3[\SI{12}{\micro\metre}]$>$2. \citet{2024ApJ...973...48C} reported eight associated \hi 21-cm absorption detections from a \hi and OH absorption survey conducted with FAST. This survey targeted 40 radio sources with low-to-intermediate radio luminosity, red mid-infrared colors (W2[\SI{4.6}{\micro\metre}] – W3[\SI{12}{\micro\metre}]$>$2.5 mag), and redshifts up to 0.35. Moreover, several high-redshift intervening and associated \hi 21-cm absorptions were recently detected in the FLASH and MALS surveys. Moreover, for the survey of intervening 21-cm absorption, \citet{2009MNRAS.396..385K} and \citet{2009MNRAS.398..201G,2012A&A...544A..21G} utilized the GMRT to search for \hi 21-cm absorption in strong Mg II absorbers within the redshift range $0.5<z<1.5$, yielding 18 detections of intervening systems.

A total of 116 associated and 24 intervening \hi 21-cm absorption systems from previous and ongoing \hi 21-cm absorption surveys are combined with our catalog to analyze the physical properties of \hi absorption. We retrieve 1.4 GHz flux density, morphology data, WISE magnitudes, and SDSS magnitudes from the NVSS, VLASS, WISE, and SDSS(or from Pan-STARRS \citep{2016arXiv161205560C} if the source is not covered by SDSS) surveys, respectively. In some cases, \hi absorption may lack counterparts in these surveys due to incomplete sky coverage or insufficient redshift information. In each of the following sections, we exclude \hi absorption systems where the necessary data are unavailable. 

\subsubsection{Distribution in Redshift}
\label{sec:distribution_in_redshift}
\begin{figure}[hbt!]
    \centering
    \includegraphics[width=8.5cm]{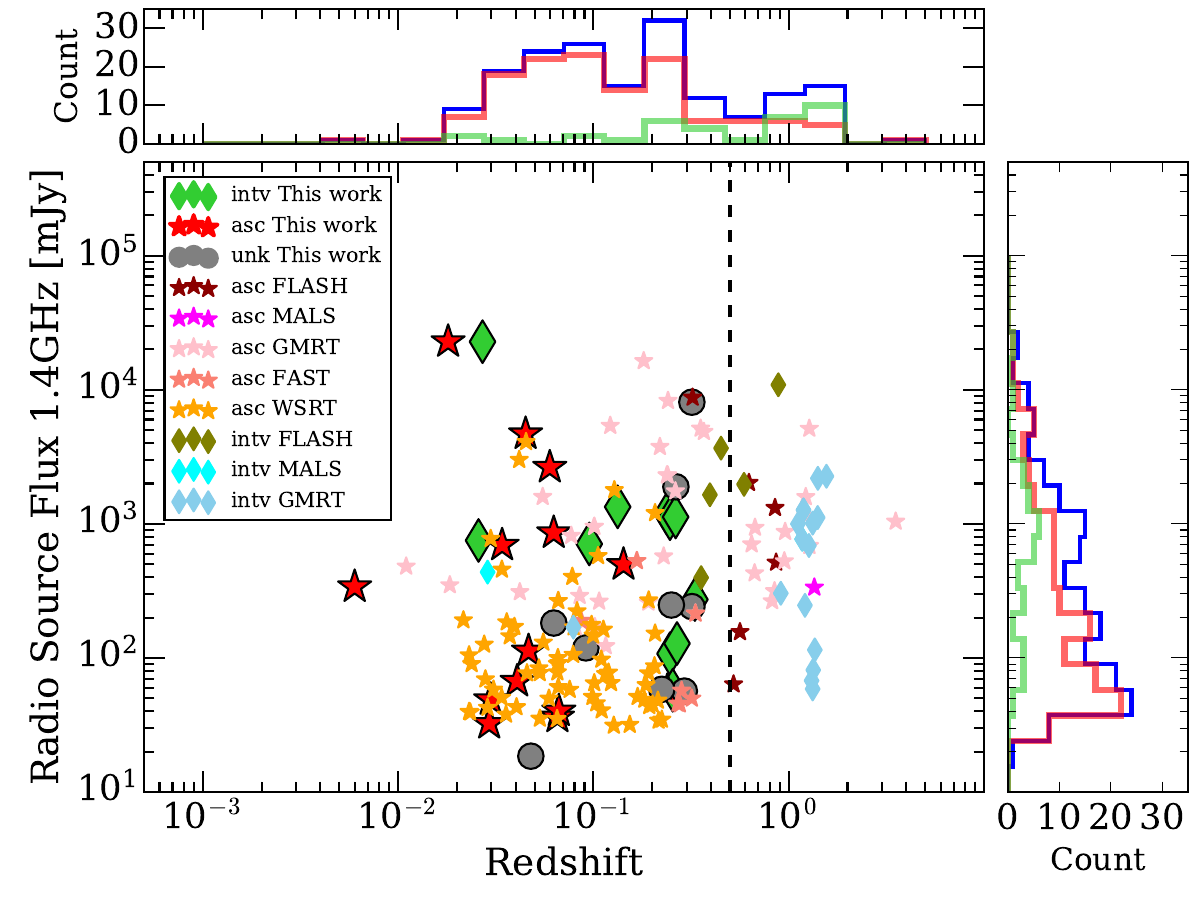}
    \includegraphics[width=8.5cm]{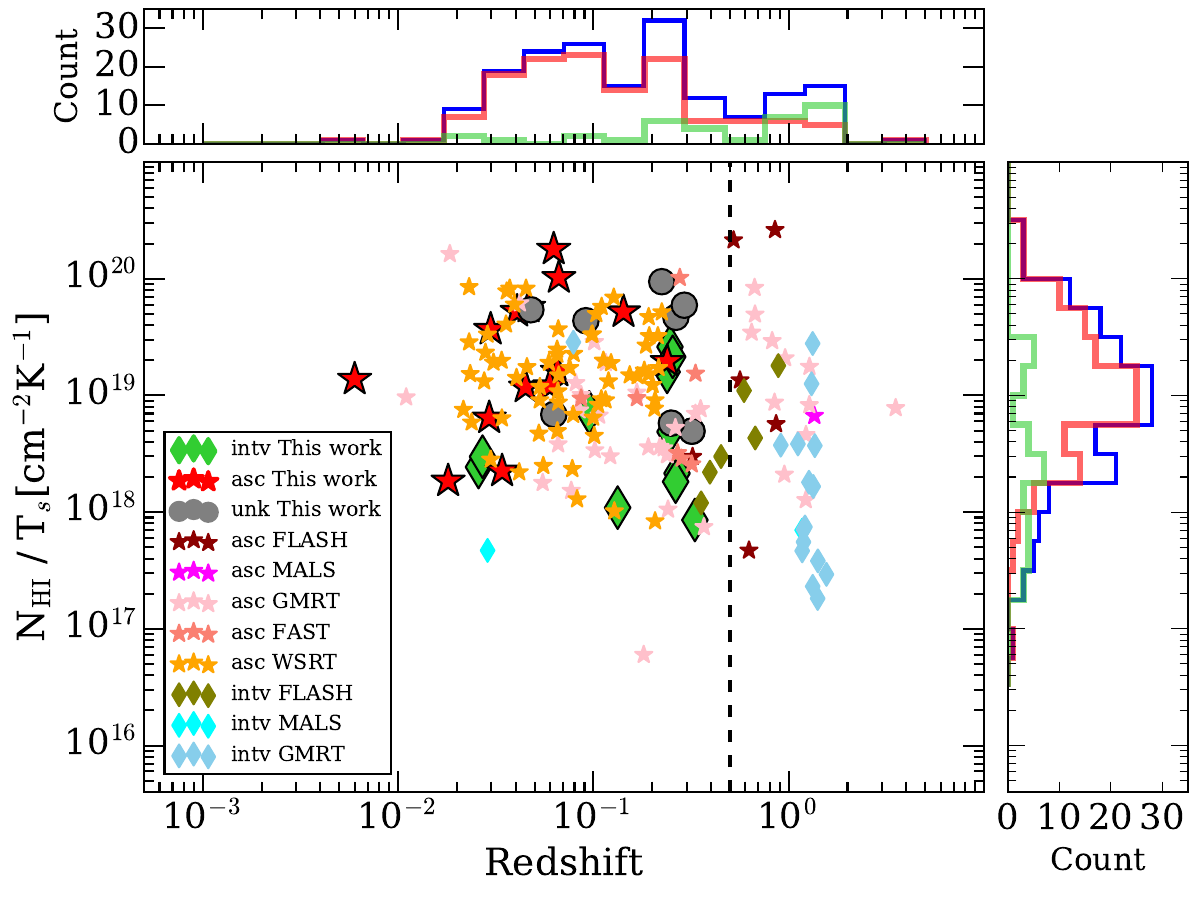}
    \caption{Redshift versus 1.4-GHz flux density of radio sources for the \hi absorption systems (upper) and \hi column density (lower). The black dashed line represents the redshift threshold of 0.5, which we used to distinguish between the higher- and lower-redshift samples discussed in the following sections.}
    \label{redshift_source_flux_N_HI}
\end{figure}

In Figure~\ref{redshift_source_flux_N_HI} we plot the redshift versus 1.4-GHz flux density of the radio sources for \hi absorption systems (upper panel) and \hi column density (lower panel), along with their corresponding histograms for \hi absorption systems across all considered samples. For the \hi absorption detected in this work, the scarcity of \hi absorption systems found between redshift 0.136 and 0.235 is attributed to interference caused by RFIs in the frequency range from 1150 MHz to 1250 MHz. For both intervening and associated \hi absorption systems, there are no evident correlations between redshift and flux densities of radio sources for \hi absorption systems, nor between redshift and N$_{\hi}$. An intriguing result from the distribution of \hi absorption across redshifts reveals a significant dichotomy: the majority of detected associated \hi absorption systems are found at lower redshifts ($z<$0.20), whereas detected intervening \hi absorption systems predominantly occur at higher redshifts ($z>$0.2). This highlights a clear separation in redshift distribution between associated and intervening systems. This phenomenon can be explained by the lower probability of encountering an intervening absorber at a low redshift and the increased difficulty in identifying an absorber as associated at a higher redshift. At low redshifts, the lower density of intervening systems results in a lower probability of encountering absorbers, while at high redshifts, the difficulty in identifying associated absorbers stems from observational challenges and the complex nature of high-redshift environments.

To investigate the properties of \hi absorption at various redshifts, we set a redshift threshold of 0.5 (black-dashed line in Figure~\ref{redshift_source_flux_N_HI}) in the following sections, dividing the absorption systems into two categories: high-redshift and low-redshift samples.

%This phenomenon can be attributed to the fact that at lower redshifts, the majority of \hi gas is contained within galaxies, leading to a higher detection rate of associated systems at these redshifts. Conversely, at higher redshifts, a greater amount of \hi gas is found in Intergalactic Medium (IGM), resulting in a tendency for intervening systems to be more prevalent at these higher redshifts.

\subsubsection{integrated column density versus flux density of radio sources}
\begin{figure}[hbt!]
    \centering
    \includegraphics[width=8.5cm]{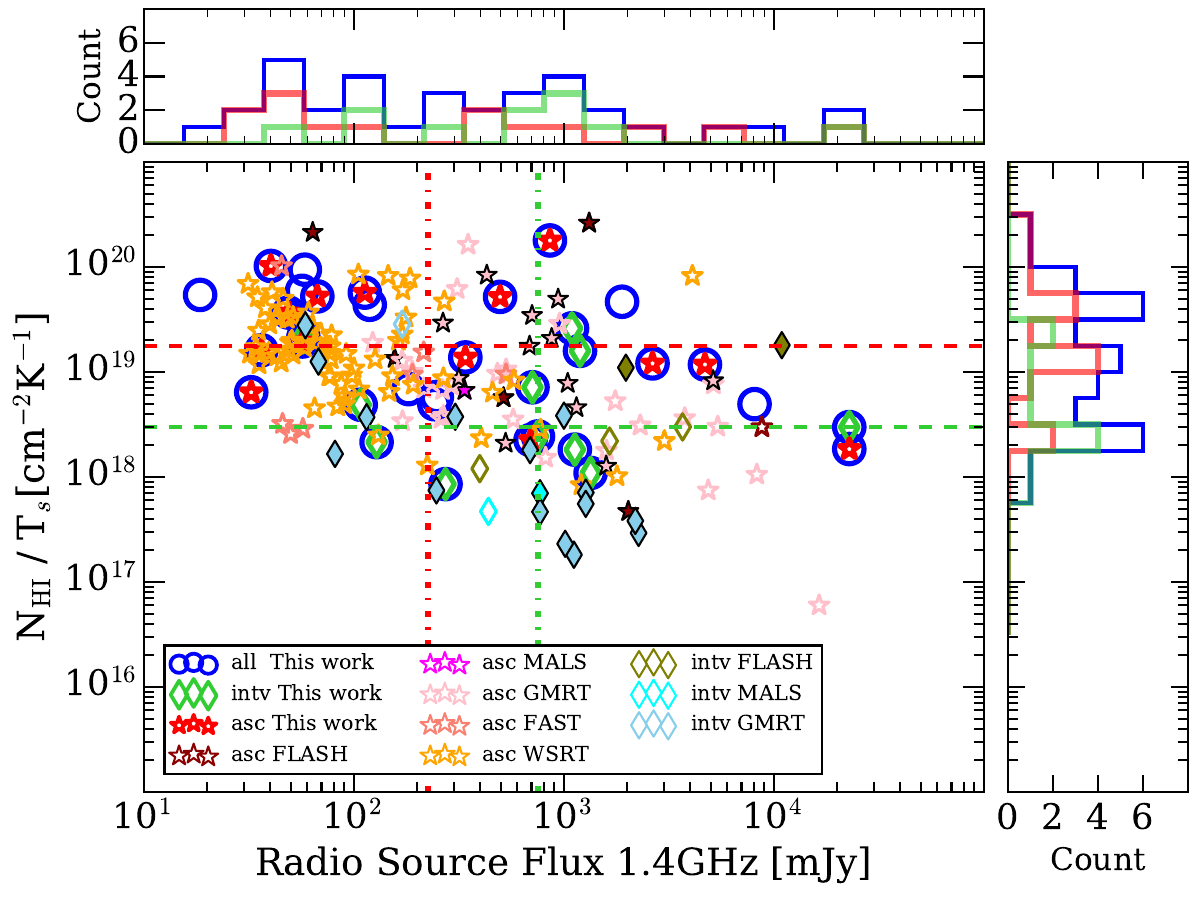}
    \caption{\hi column density versus 1.4-GHz flux density of radio sources for both the \hi absorption systems we detected and those reported in the literature. Higher-($z>0.5$) and lower-redshift ($z\leqslant0.5$) samples are represented by filled points and open points, respectively. Dashed and dot-dashed lines represent median values for intervening (green diamond) and associated (red star) systems detected in this work, respectively.}
    \label{source_flux_N_HI}
\end{figure}

The relationship between integrated \hi column density and 1.4-GHz flux density of radio sources for \hi absorption systems we detected, along with their corresponding histograms and those for \hi absorption systems reported in the literature, is shown in Figure~\ref{source_flux_N_HI}. The flux densities of radio sources for \hi absorption systems we detected span a large range, from $\sim$ 20 mJy to $\sim$ 23 Jy. When comparing intervening and associated systems in our sample, the flux density of radio sources within associated systems exhibits a wider range in both very low (flux $\leqslant$ 60 mJy) and high (flux $\geqslant$ 2 Jy) flux density regions. In terms of \hi column density, associated systems exhibit higher values compared to intervening systems. These patterns are also observed in the relationship between integrated \hi column density and 1.4-GHz flux density of radio sources in \hi absorption systems of all considered surveys outlined at the beginning of Section~\ref{sec:Absorption_Statistics}. Given that the associated systems are typically found at the center of AGNs while the intervening systems are usually associated with spiral galaxies, our measurements suggest that intervening systems may only trace part of the gas, whereas associated systems may capture the entirety of gas components.

\subsubsection{Column Density versus FWHM}
\begin{figure}[hbt!]
    \centering
    \includegraphics[width=8.5cm]{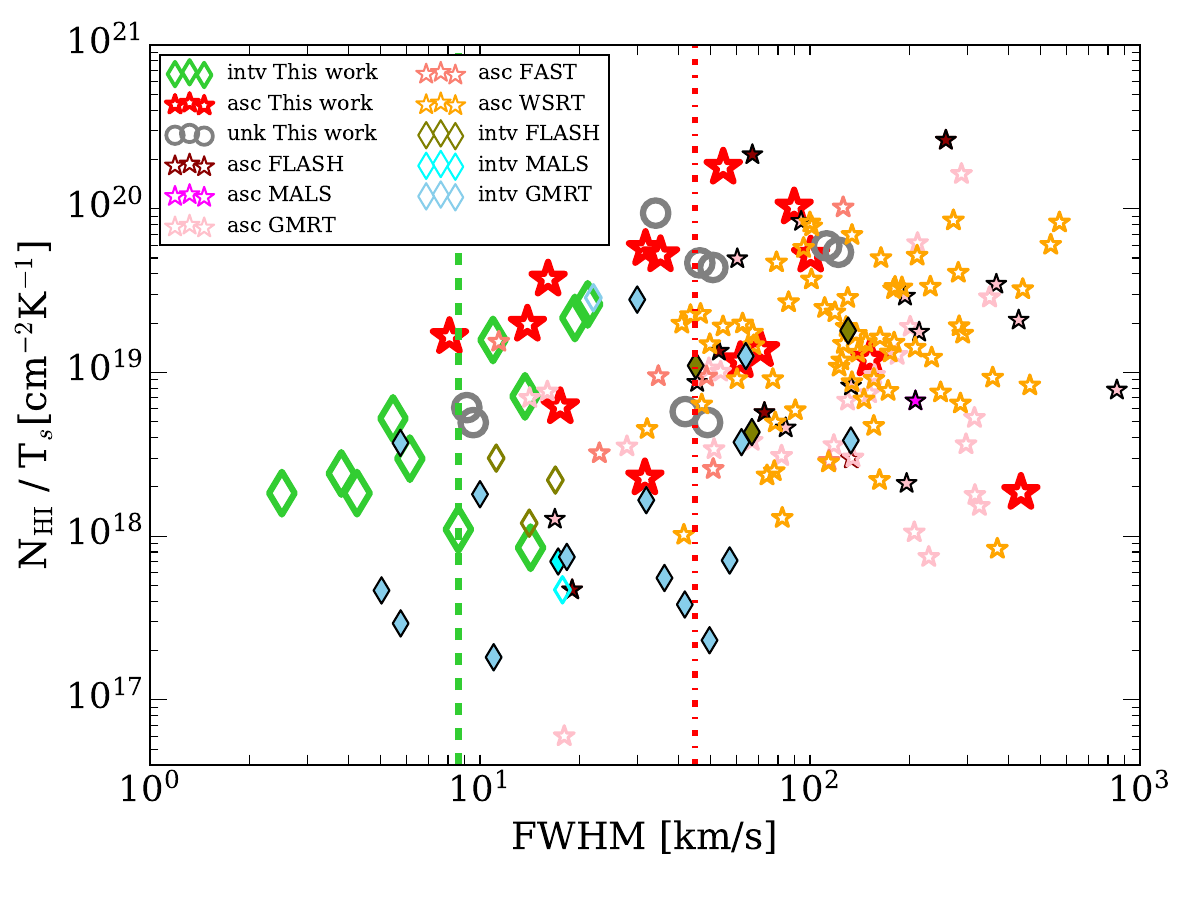}
    \caption{The \hi column density versus FWHM for both our sample and the sample from the literature. Higher-($z>0.5$) and lower-redshift ($z\leqslant0.5$) samples are represented by filled points and open points, respectively. The green and red dashed lines indicate the median values for the intervening and associated systems detected in this study.}
    \label{width_N_HI}
\end{figure}

Figure~\ref{width_N_HI} illustrates the correlation between \hi column density and FWHM of \hi absorption systems. The FWHM values for intervening systems are significantly lower compared to those of associated systems, this is because an intervening galaxy in \hi absorption may be traced only partially and produce narrower lines. For the systems detected in this work, the median FWHM for intervening systems is 9.00 $\pm$ 6.11 $\kms$, whereas for associated systems, it is 44.90 $\pm$ 106.05 $\kms$. Furthermore, there is a noticeable trend indicating that \hi column density increases with FWHM, with this pattern being more pronounced for associated systems compared to intervening systems.

\subsubsection{Column Density versus Morphology}

Enhancing our comprehension of the gas properties associated with different radio source types, such as compact and extended, is vital for understanding the observed features of AGNs. Furthermore, it is proposed that AGN of different types are encased in different gaseous environments and exhibit divergent \hi detection rates. \citet{2014A&A...569A..35G} analyzed a sample of 93 AGNs, selected based on a flux criterion (S$_{\rm 1.4 GHz} >$ 50 mJy) and having SDSS redshifts in the range 0.02 $< z <$ 0.23, categorizing them into compact and extended radio sources. They discovered that the compact sources (with g $-$ r $>$ 0.7) exhibit an \hi absorption detection rate of approximately 42$\%$, in contrast to just about 16$\%$ for extended sources. Through a stacking analysis, it was revealed that the \hi absorption associated with the compact radio sources typically presents higher optical depth, FWHM, and \hi column densities compared to those linked with extended sources \citep{2014A&A...569A..35G}. Furthermore, \citet{2015A&A...575A..44G} reported a tendency for \hi absorption features in compact radio sources to display blueshifted and broad/asymmetric line profiles more frequently. Additionally, \citet{2017A&A...604A..43M} reported that compact sources exhibit broad \hi lines, indicating unsettled kinematics. Both \citet{2017A&A...604A..43M} and \citet{2017MNRAS.465..997C} further noted a higher detection rate of \hi absorption in compact sources compared to extended sources. More recently, \citet{2024MNRAS.527.8511A} presented an ASKAP pilot search for associated \hi 21-cm absorption in a sample of 62 Molonglo Reference Catalog 1-Jy radio galaxies and quasars covering 0.42$<z<$1.00, identifying three new detections of associated \hi 21-cm absorption, all of which were from peaked-spectrum or compact steep-spectrum radio sources.

\begin{figure*}[hbt!]
    \centering
    \includegraphics[width=0.32\textwidth]{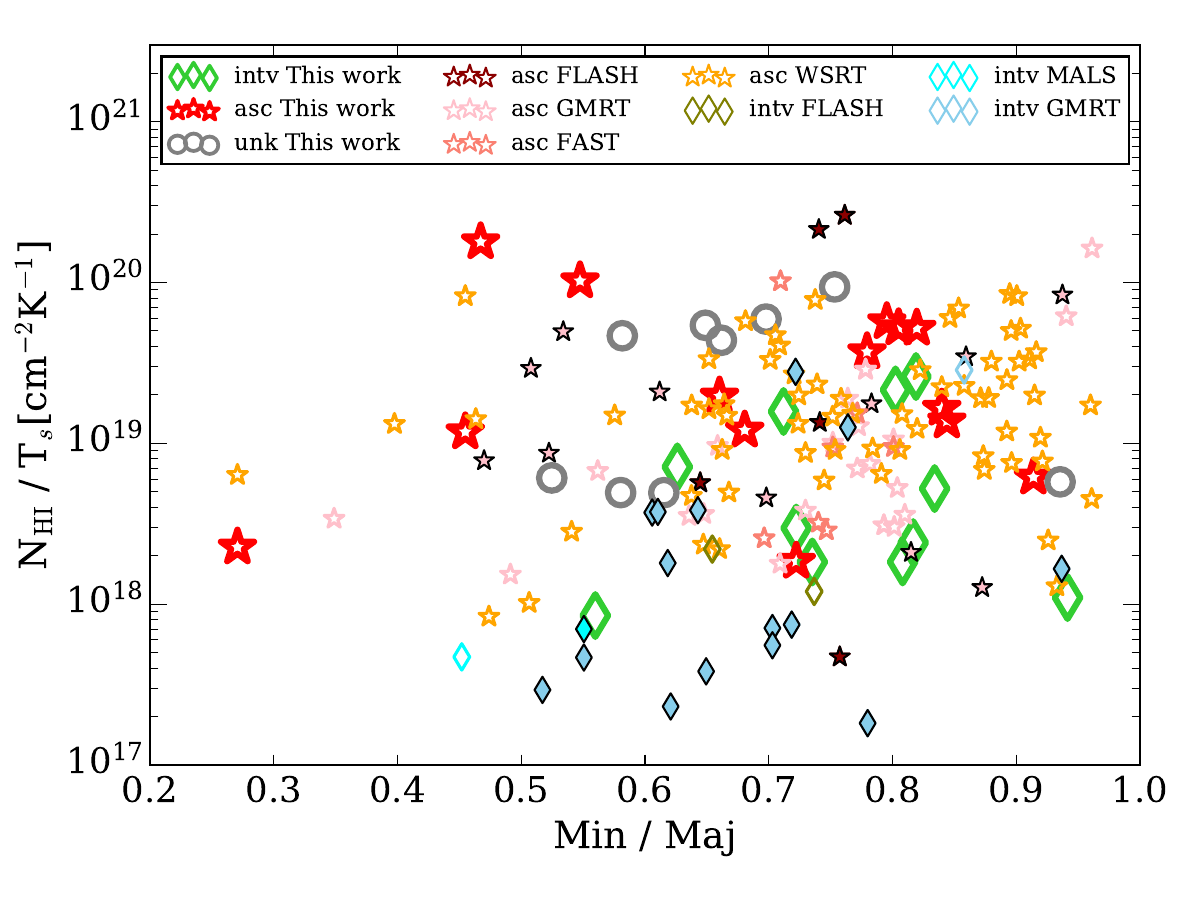}
    \includegraphics[width=0.32\textwidth]{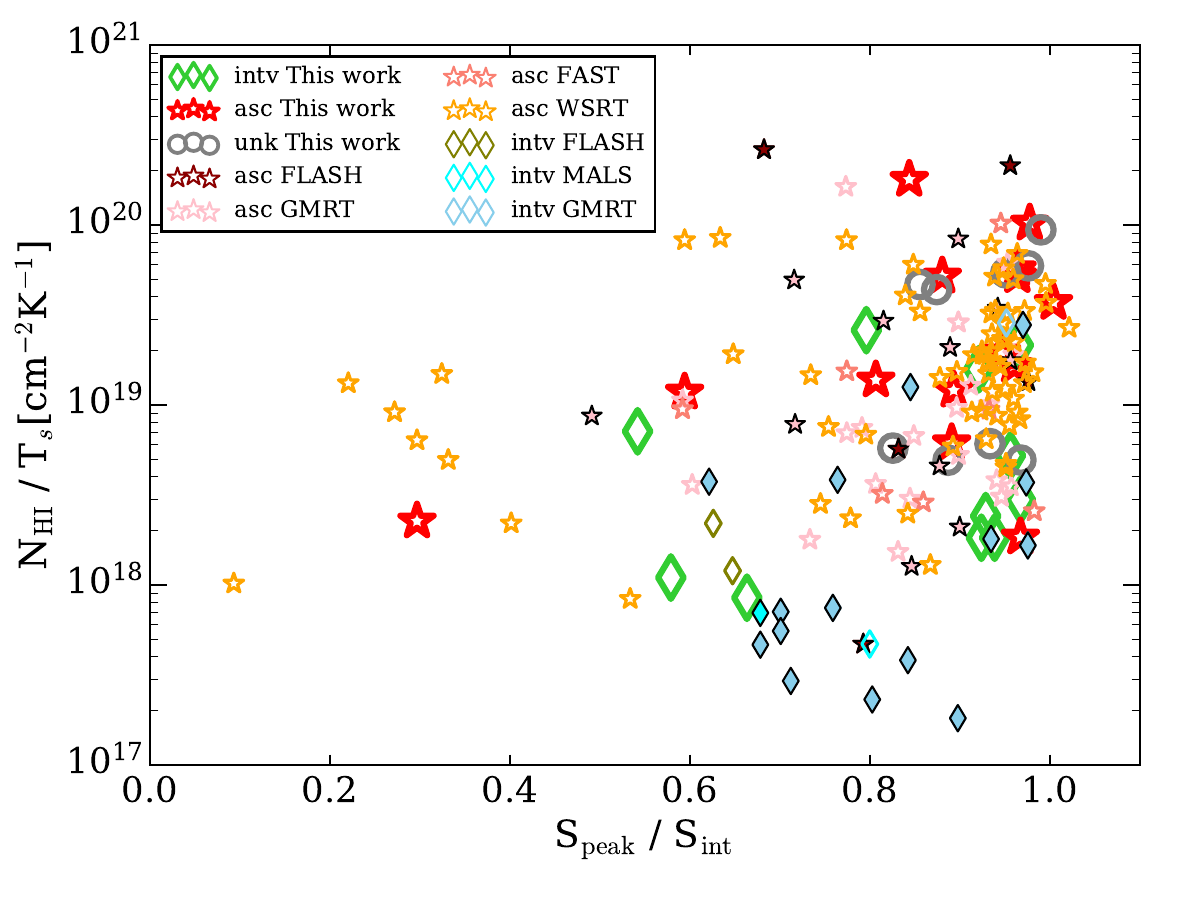}
    \includegraphics[width=0.32\textwidth]{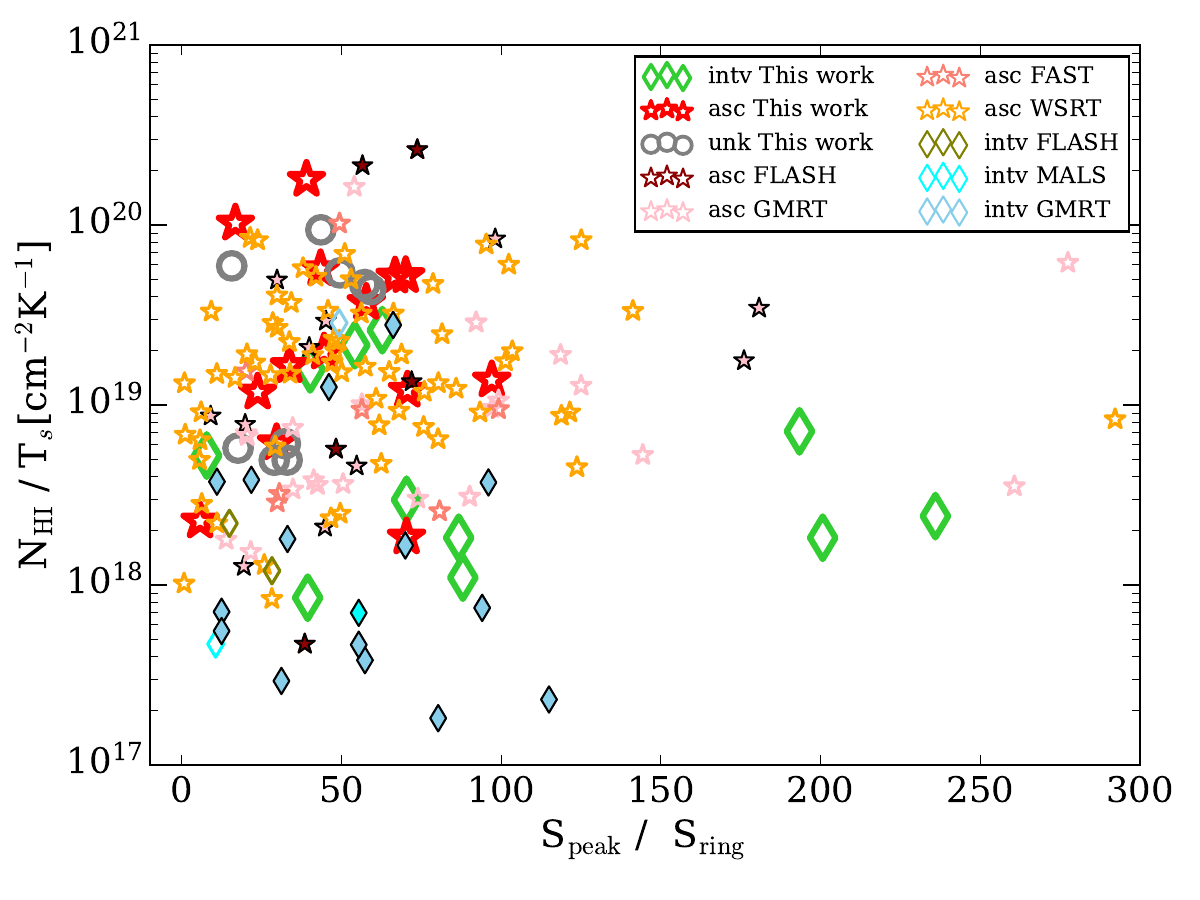}
    \caption{Same as Figure~\ref{width_N_HI}, but for the minor-to-major axis ratio (left), peak-to-integrated flux ratio (center), and peak-to-ring flux ratio (right) versus \hi column density.}
    \label{Morphology_N_HI}
\end{figure*}

We explore N$_{\hi}$ levels in \hi absorption associated with radio sources of varying morphologies in our sample. The morphology is characterized by metrics such as the minor-to-major axis ratio, the peak-to-integrated flux ratio, and the peak-to-ring flux ratio (defined as the ratio of maximum brightness within 5 arcsecs of the component to the maximum brightness in an annulus centered on the component with inner and outer radii of 5 arcsecs and 10 arcsecs). These parameters were obtained from the VLASS catalog. Radio sources (4C -06.18 and 3C 84) located outside the VLASS coverage area were omitted from the morphology-based analysis. The findings are depicted in Figure~\ref{Morphology_N_HI}. For intervening systems, there appears to be no discernible connection between the morphology of background radio sources and N$_{\hi}$. Similarly, for associated systems, no evident relationship is observed between the minor-to-major axis ratio and N$_{\hi}$. However, there are noticeable increasing trends of N$_{\hi}$ with both the peak-to-integrated flux ratio and the peak-to-ring flux ratio, suggesting that associated \hi absorption linked to compact radio sources exhibits higher \hi column densities in comparison to those associated with extended radio sources.

\subsubsection{Column Density versus Color}
\begin{figure}[hbt!]
    \centering
    \includegraphics[width=8.5cm]{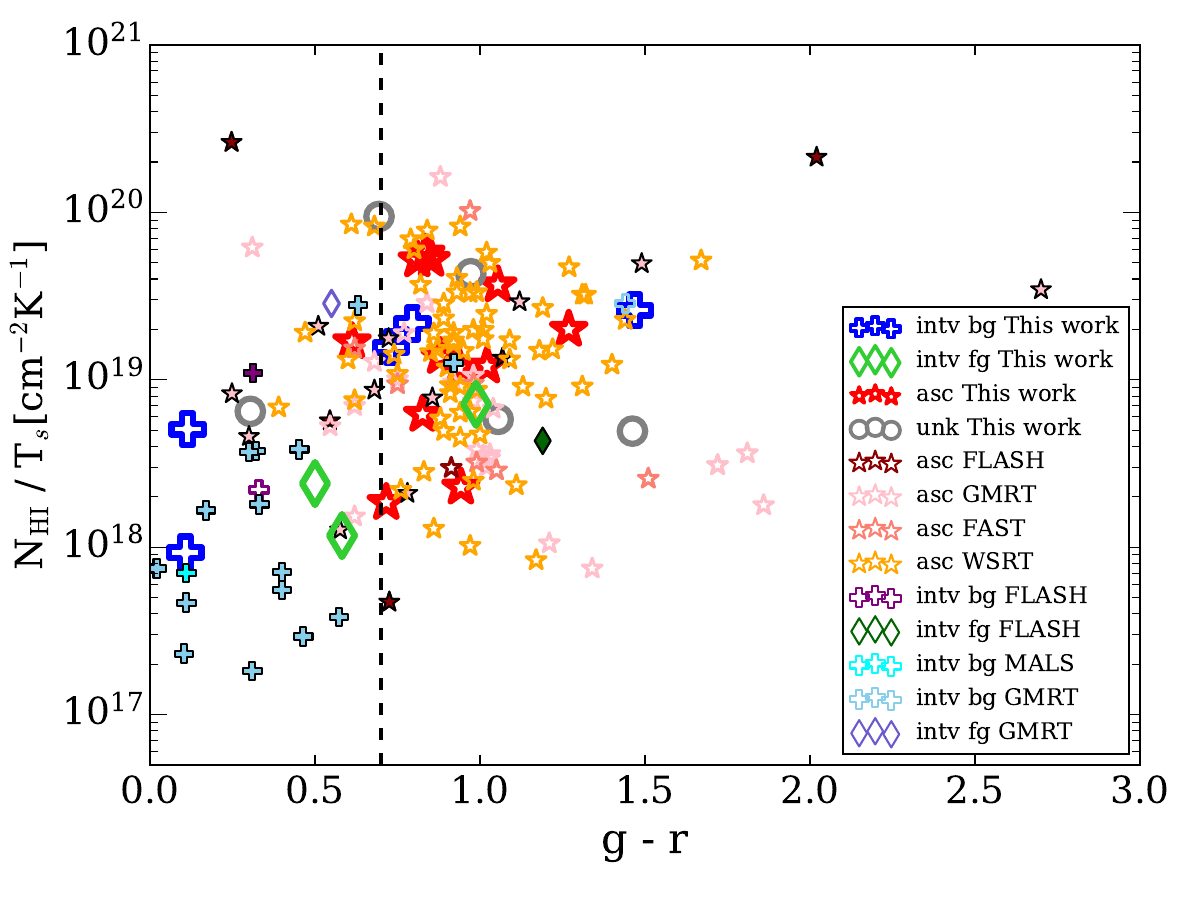}
    \caption{Same as Figure~\ref{width_N_HI}, but for the \hi column density versus the SDSS g-r color. The black-dashed line represents a g-r value of 0.7.}
    \label{g_r_N_HI}
\end{figure}

The N$_{\hi}$ versus the SDSS $g-r$ color of our \hi 21-cm absorption sample, along with those of \hi absorption systems reported in the literature, is shown in Figure~\ref{g_r_N_HI}. For both the associated \hi absorption in our sample and the literature, most of the hosting galaxies are red ($g-r>0.7$) objects. The background sources in the intervening systems within our sample span a wide range of g-r color values, whereas most background sources reported in the literature fall within the blue object region ($g-r<0.7$). Among the three intervening systems in our sample with identified foreground sources, two are classified as blue objects. Out of the five identified foreground sources from our sample and literature, three are also blue objects. A preliminary division in color distribution for foregrounds of intervening and associated systems has been identified, necessitating further validation with a larger sample size, especially concerning intervening systems. This division aligns with the redshift distribution of associated and intervening \hi absorption detailed in Section~\ref{sec:distribution_in_redshift}, reflecting the prevalence of bluer galaxies at higher redshifts due to the process of galaxy evolution. The small sample size of the identified foreground objects in the intervening system in the literature limits the ability to draw a high-confidence conclusion. For the associated systems, those at higher redshift ($z>0.5$) show less concentration along the color axis and exhibit a broader range of g-r colors.

Figure~\ref{g_r_N_HI} illustrates that only a small fraction of intervening absorption systems have identified foreground sources. Identifying these foreground objects is challenging because their signal is often obscured by the intense signal of the bright background. A blind radio survey can assist in detecting intervening absorption systems with backgrounds that are less bright in the optical band.

\subsection{WISE color-color Classification}
\label{sec:WISE_color_color}
The WISE color-color diagram is often used to illustrate the locations of interesting classes of objects and has also been employed to classify \hi 21-cm absorption. \citet{2017MNRAS.465..997C} identified a clear distinction in the W2-W3 color distributions between sources with and without \hi detections, with detections having larger W2-W3 values than non-detections. They also observed a high detection rate of \hi absorption in galaxies with WISE infrared colors W2-W3$>$2, which are typically gas-rich systems and possess compact radio structures. \citet{2017A&A...604A..43M} used WISE colors to differentiate between dust-poor sources and mid-infrared (MIR) bright sources. They found that dust-poor galaxies predominantly exhibit narrow, deep \hi absorption lines, usually centered at the systemic velocity, indicating that most \hi in these galaxies is settled in a rotating disk. In contrast, MIR bright sources showed more frequent \hi detections compared to dust-poor sources.

\begin{figure}[hbt!]
    \centering
    \includegraphics[width=9.5cm]{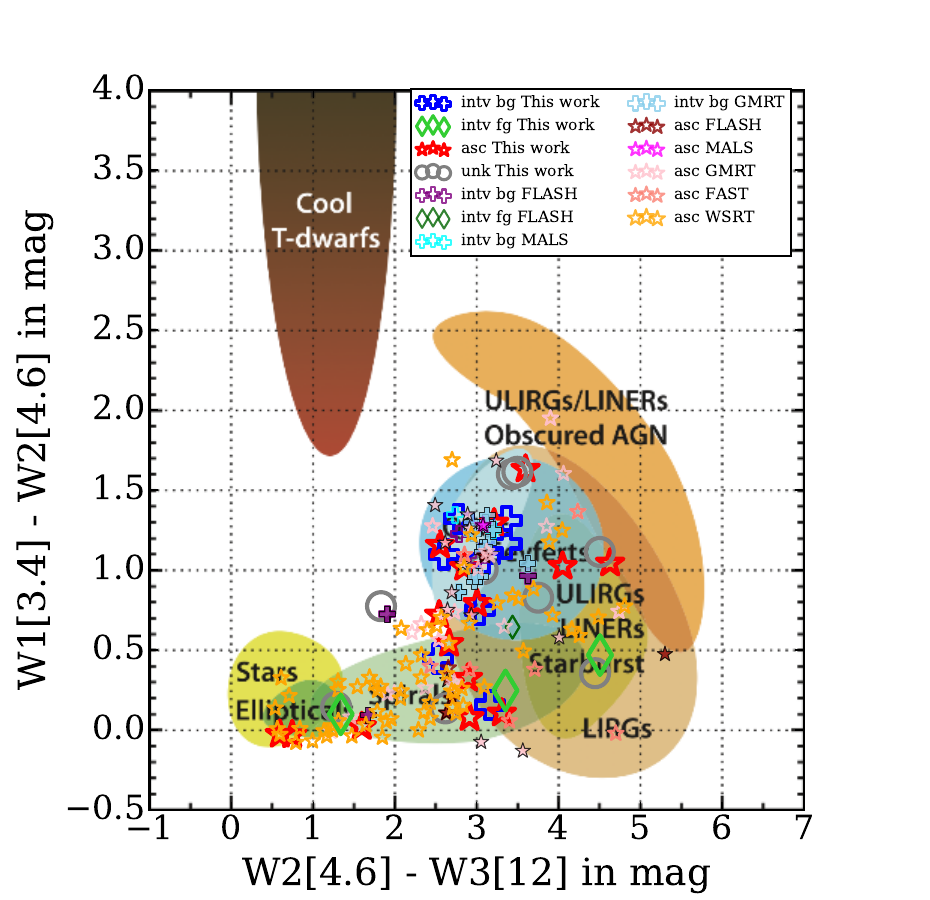}
    \caption{The WISE color-color diagram showing the locations of interesting classes of objects, over-plotted with samples from our work and the literature. Higher-($z>0.5$) and lower-redshift ($z\leqslant0.5$) samples are represented by filled points and open points, respectively.}
    \label{wise_color_color}
\end{figure}

We use the WISE color to infer possible classes of foreground and background sources, the positions of our systems in the color-color diagram are shown in Figure~\ref{wise_color_color}. Among the intervening systems we detected, eight background radio sources have corresponding WISE counterparts. Of these, five are located in the region of QSOs, while two are found in areas corresponding to either spiral or elliptical galaxies. Almost all the background radio sources in the intervening systems in the quoted GMRT survey position in the QSOs region. These findings are further supported by the discoveries from GMRT, FLASH, and MALS. Additionally, four (including the one from FLASH) foreground radio sources in the intervening systems were identified in the WISE survey, each in distinct regions: three in a spiral galaxy, and the fourth in the ULIRGs/LINERs starburst region. For the associated systems, the radio sources spread a broader range of locations, encompassing the QSO region, Seyferts region, spiral galaxy region, elliptical galaxy region, and the ULIRGs/LINERs starburst region, with the spiral galaxy region representing the largest portion. This distribution pattern is consistent in both our sample and those reported in the literature. Literature detections indicate that the distribution of associated absorption varies significantly with redshift: higher-redshift systems are primarily found in the QSO region, whereas lower-redshift systems are more concentrated in the spiral galaxy region.

\begin{figure}[hbt!]
    \centering
    \includegraphics[width=8.5cm]{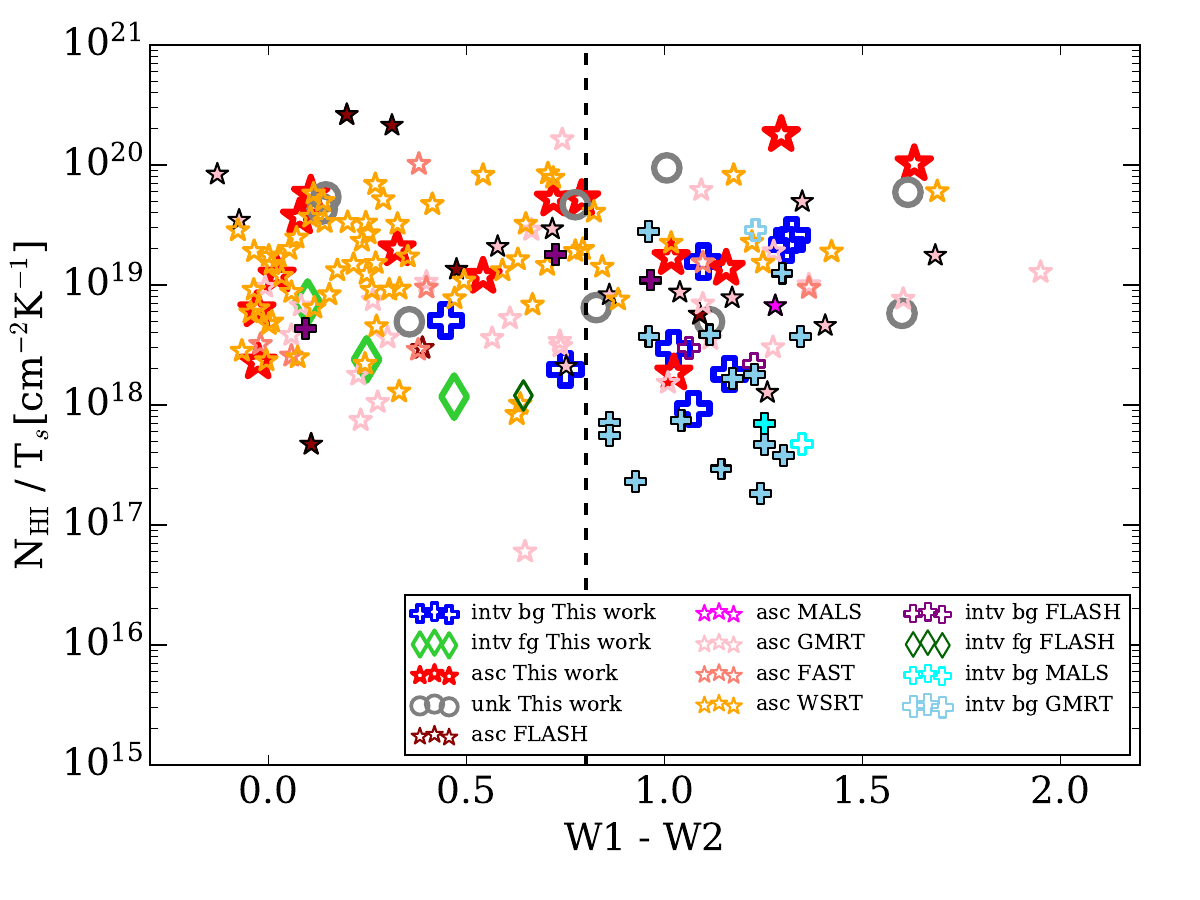}
    \caption{Same as Figure~\ref{width_N_HI}, but for WISE magnitudes difference between W1 and W2 bands versus the \hi column density. The black-dashed line represents a W1-W2 value of 0.8.}
    \label{W1_W2_N_HI}
\end{figure}

The color difference between W1[\SI{3.4}{\micro\metre}] and W2[\SI{4.6}{\micro\metre}], with a value of 0.8, is commonly utilized as a basic mid-infrared color criterion to identify AGN candidates \citep{2012ApJ...753...30S}. Figure~\ref{W1_W2_N_HI} illustrates the difference in WISE magnitudes between the W1 and W2 bands plotted against the \hi column density. It reveals no apparent correlation between the W1-W2 magnitudes and \hi column density in associated systems. However, for the foreground sources in the intervening systems, a potential trend suggests that \hi column density diminishes as W1-W2 magnitudes increase. Notably, all identified foreground sources exhibit W1-W2 magnitudes less than 0.8, suggesting that these sources are likely galaxies with minimal or no AGN activity. Figure~\ref{W1_W2_N_HI} illustrates that most lower-redshift associated absorption is found at W1-W2$<0.8$, implying that at low redshifts, most associated \hi absorption likely originates from galaxies with little AGN activity. In contrast, the associated absorption at higher redshifts shows a broader distribution in W1-W2 values, ranging from -0.1 to 1.7, with nearly half of the systems positioned at W1-W2$>0.8$, potentially reflecting galaxy evolution. Additionally, Figure~\ref{W1_W2_N_HI} highlights that backgrounds of intervening systems are more likely to have W1-W2$>0.8$, suggesting they are AGN candidates.

\subsection{Spectral Stacking Results}

Spectral stacking is a technique that combines multiple spectra to enhance weak signals and effectively replicates the overall spectral characteristics \citep{2019MNRAS.489.1619H,2020MNRAS.493.1587H,2021MNRAS.507.5580H}. In this section, we examine \hi properties of both associated and intervening \hi absorption systems. This is achieved by stacking \hi absorption spectra from, for the first time, a blind catalog of \hi absorbers that was compiled without bias. 

The stacking technique used in this paper is similar to that described in \citet{2019MNRAS.489.1619H}. Flux density spectra were converted to optical depth spectra using Eq.~(\ref{tau}). We introduce a weight factor that equals the inverse of completeness. The weight of ith \hi absorption system is expressed as: 
\begin{eqnarray}
w_{i} = 1/C(W_{i}, \mathrm{SN}_{i}, \nu_{i}), 
\label{weight}
\end{eqnarray}
where $C(W_{i}, \mathrm{SN}_{i}, \nu_{i})$ refers to the completeness for a \hi absorbers with FWHM of $W_{i}$, S/N of $\mathrm{SN}_{i}$ and at the frequency of $\nu_{i}$. The completeness was estimated by adding mock absorption lines to real CRAFTS and FASHI spectra and calculating the fraction detected using our search method and selection threshold. 

The averaged final stacked spectrum is obtained from:
\begin{eqnarray}
    \langle \tau_{\rm HI}(\nu)\rangle = \frac{\sum_{i=1}^{n}w_{i}\tau_{\rm HI,i}}{\sum_{i=1}^{n}w_{i}}.
    \label{weighted_spectrum}
\end{eqnarray}

The integrated $\tau$ of a stack, or $\langle \tau\rangle$, is then defined as integral along the velocity axis over $\tau$ spectrum:
\begin{eqnarray}
    \tau_{\hi} = \int_{-\Delta v}^{\Delta v}\langle \tau_{\hi}(v)\rangle dv,
    \label{integrated_mass}
\end{eqnarray}
where $\Delta v$ is large enough to capture all signals from the stack (we will later use $\Delta v = 200\kms$). The integrated N$_{\hi}$ of a stack ($\langle$ N$_{\hi}\rangle$)is then calculated by 1.82 $\times$ 10$^{18}$ T$_{\rm s}\langle \tau\rangle$. The errors for measurements were estimated through jackknife resampling.

\begin{figure}[hbt!]
    \centering
    \includegraphics[width=8.5cm]{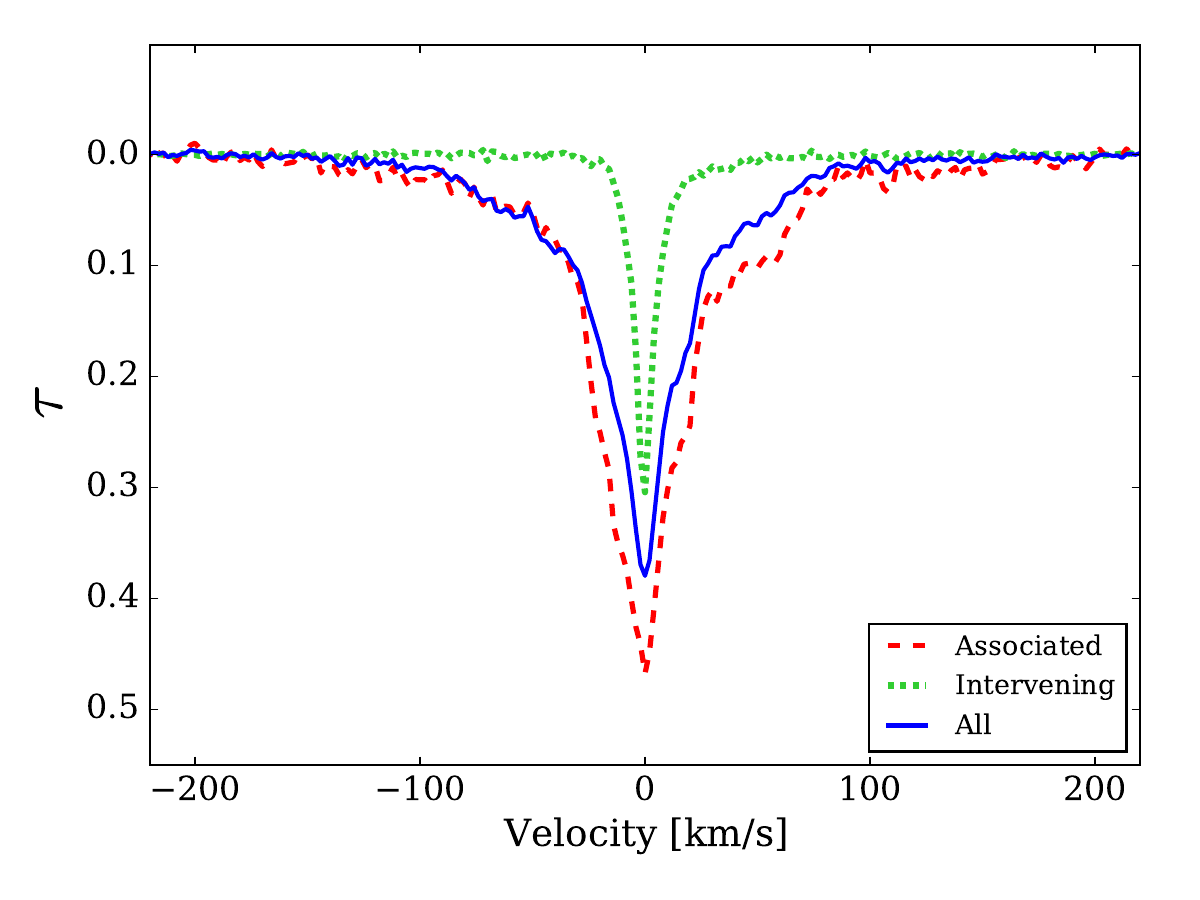}
    \caption{Stack of \hi absorption spectra of the associated, intervening, and all \hi absorption samples detected in this work.}
    \label{stack_spectrum}
\end{figure}

The \hi absorption used for stacking consists of 33 systems except for NVSS J112832\allowbreak+583346 (emission and absorption are indistinguishable), including 14 associated systems, 10 intervening systems, and 9 unknown-type systems. Stacking optical depth spectra from our sample results in strong detections. The stacked spectra are shown in Figure~\ref{stack_spectrum}. The mean peak optical path, mean velocity-integrated optical path $\langle \tau\rangle$, mean FWHM and mean \hi column density $\langle$ N$_{\hi}\rangle$ are measured to be 0.47 $\pm$ 0.10, 0.30 $\pm$ 0.07 and 0.38 $\pm$ 0.06; 27.19 $\pm$ 8.71 $\kms$, 4.36 $\pm$ 1.46 $\kms$ and 19.89 $\pm$ 4.83 $\kms$; 42.61 $\pm$ 15.62 $\kms$, 9.33 $\pm$ 2.08 $\kms$ and 34.58 $\pm$ 12.34 $\kms$; 0.49 $\pm$ 0.16, 0.08 $\pm$ 0.03 and 0.36 $\pm$ 0.08 T$_{s} \times$ 10$^{20}$cm$^{-2}$K$^{-1}$, for associated, intervening and all \hi absorption samples.

\section{Discussions}
\label{sec:Discussion}

\subsection{Comoving Absorption Path}

The absorption path length defines the moving interval of the survey that is sensitive to intervening absorbers. We estimate total comoving absorption path length ($\Delta X$) for the data (CRAFTS and FASHI) used here (1050-1150 MHz and 1250-1450 MHz), adopting the method described in \citet{2015MNRAS.453.1249A,2020MNRAS.494.3627A,2021MNRAS.503..985A} and Paper I.

\begin{figure}[hbt!]
    \centering
    \includegraphics[width=8.5cm]{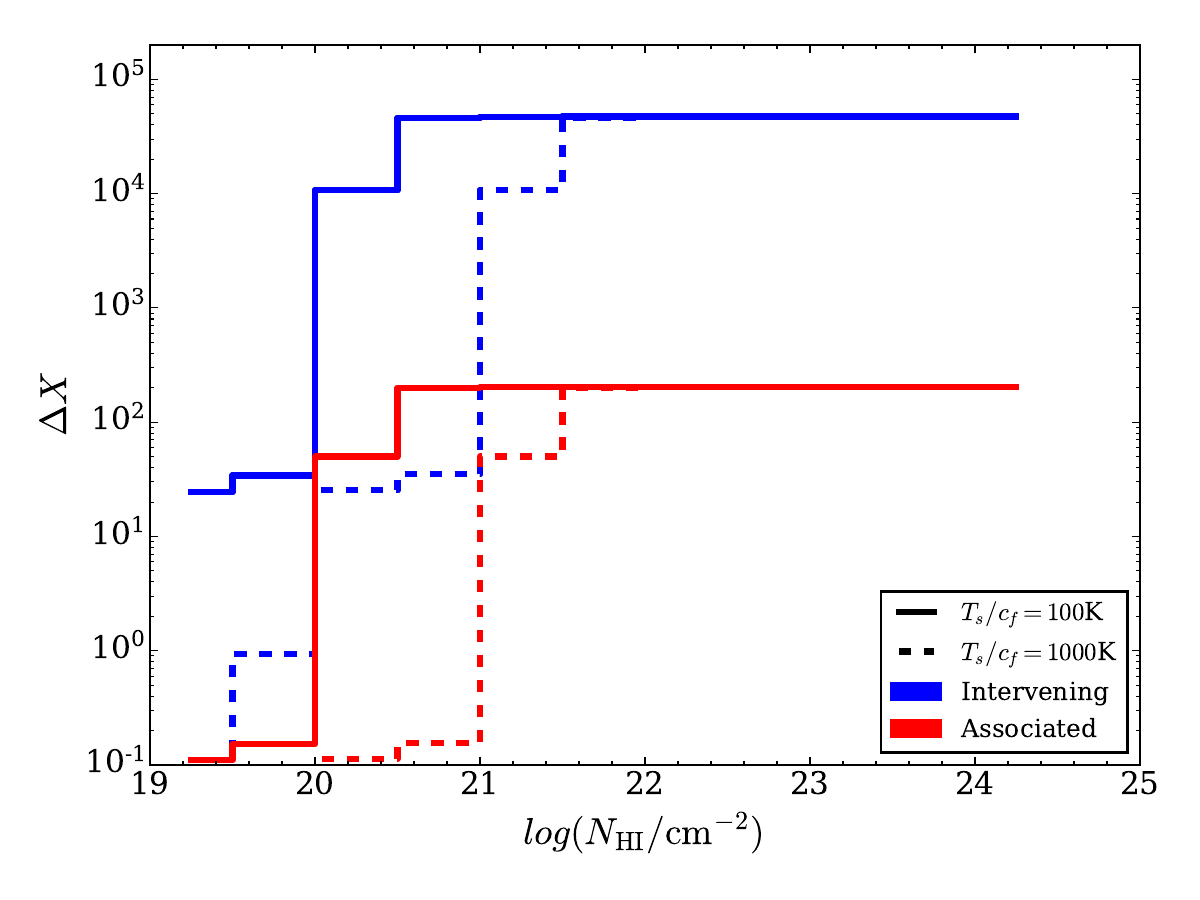}
    \caption{The comoving absorption path length ($\Delta X$) spanned by our data as a function of \hi column density sensitivity. The results for spin temperature to source covering fraction ratios of 100 K and 1000 K are shown as blue and red lines, respectively.}
    \label{comoving_absorption_path}
\end{figure}

Assuming the covering factor $c_{\rm f} = 1$, velocity FWHM $\Delta v_{50} = 30$ \kms (the mean velocity width for all intervening 21-cm absorbers detected \citep{2016MNRAS.462.4197C,2016MNRAS.462.1341A}), and a detection threshold of $5.5 \sigma$, we show the comoving absorption path length for our data as a function of HI column density in Figure~\ref{comoving_absorption_path}. We present results for $T_{s}/c_{f}$ = 100 and 1000 K, which are the typical spin temperature of the cold (CNM) and the warm neutral medium (WNM) respectively \citep{2018ApJS..238...14M}. Under the assumption of $T_{s}/c_{f}$ = 100 K, the total comoving absorption path length spanned by our data is $\Delta X^{\mathrm{inv}}$ = 4.72$\times10^{4}$ ($\Delta z^{\mathrm{inv}} = 3.74\times10^{4}$) and $\Delta X^{\mathrm{asc}}$ = 2.04$\times10^{2}$ ($\Delta z^{\mathrm{asc}} = 1.48\times10^{2}$). The comoving absorption paths sensitive to the DLAs ($N_{\hi} \geqslant 2\times10^{20} \cm^{-2}$) are $\Delta X^{\mathrm{inv}}$ = 4.07$\times10^{4}$ ($\Delta z^{\mathrm{inv}} = 3.19\times10^{4}$) for the intervening absorption and $\Delta X^{\mathrm{asc}}$ = 1.81$\times10^{2}$ ($\Delta z^{\mathrm{asc}} = 1.31\times10^{2}$) for the associated absorption, respectively.

\subsection{Detection Rate}

Our survey is an unbiased blind survey, and it is informative to compare the detection fractions in our sample with those reported in other studies from the literature. The integrated optical depth of the weakest associated and intervening \hi absorptions detected in our sample is approximately 1\kms and 0.5\kms, respectively. %For our estimates, we only consider associated and intervening detections with an integrated optical depth greater than 1\kms and 0.5\kms, respectively. 
To ensure a fair comparison with previous surveys conducted at different redshifts, we base the detection rate estimates on observations where the 3$\sigma$ upper limits for the velocity-integrated optical depth are $\leqslant$1\kms and $\leqslant$0.5\kms for associated and intervening \hi absorption, respectively, in both our survey and other surveys. The 3$\sigma$ upper limits to the velocity integrated optical depth is estimated as $\int \tau dv = -\mathrm{ln} (1-3\sigma_{\mathrm{rms}} /S_{\mathrm{cont}}) \times \Delta v$, where $S_{\mathrm{cont}}$ represents the continuum flux density of the targeted radio sources, $\sigma_{\mathrm{rms}}$ is the root mean square(RMS) per noise velocity channel $\Delta$v, $\Delta$v is the velocity channel resolution, which is selected to be 100 \kms for associated \hi absorption. For cases where the literature reports RMS noise values for different velocity resolutions, the RMS noise is adjusted by dividing by a factor of $\sqrt{100/\Delta v_{\mathrm{rms}}}$, where $\Delta v_{\mathrm{rms}}$ is the velocity channel width corresponding to the reported RMS noise. In the case of intervening systems, we adopt a velocity resolution $\Delta v_{\mathrm{intv}}$ of 30\kms and apply a 3$\sigma$ upper limit for the velocity-integrated optical depth of $\leqslant$0.5\kms. To ensure completeness in our search, we correct the total number of detections by a completeness factor: N$^{\mathrm{asc}}_{\mathrm{total}}= \sum_{i}^{N^{\mathrm{asc}}} 1/$C${i}$ and N$^{\mathrm{intv}}_{\mathrm{total}}= \sum_{i}^{N^{\mathrm{intv}}} 1/$C${i}$, where $N^{\mathrm{asc}}$ and $N^{\mathrm{intv}}$ represent the total number of detected \hi 21-cm associated and intervening absorptions, respectively.

\begin{figure}
    \centering
    \includegraphics[width=8.5cm]{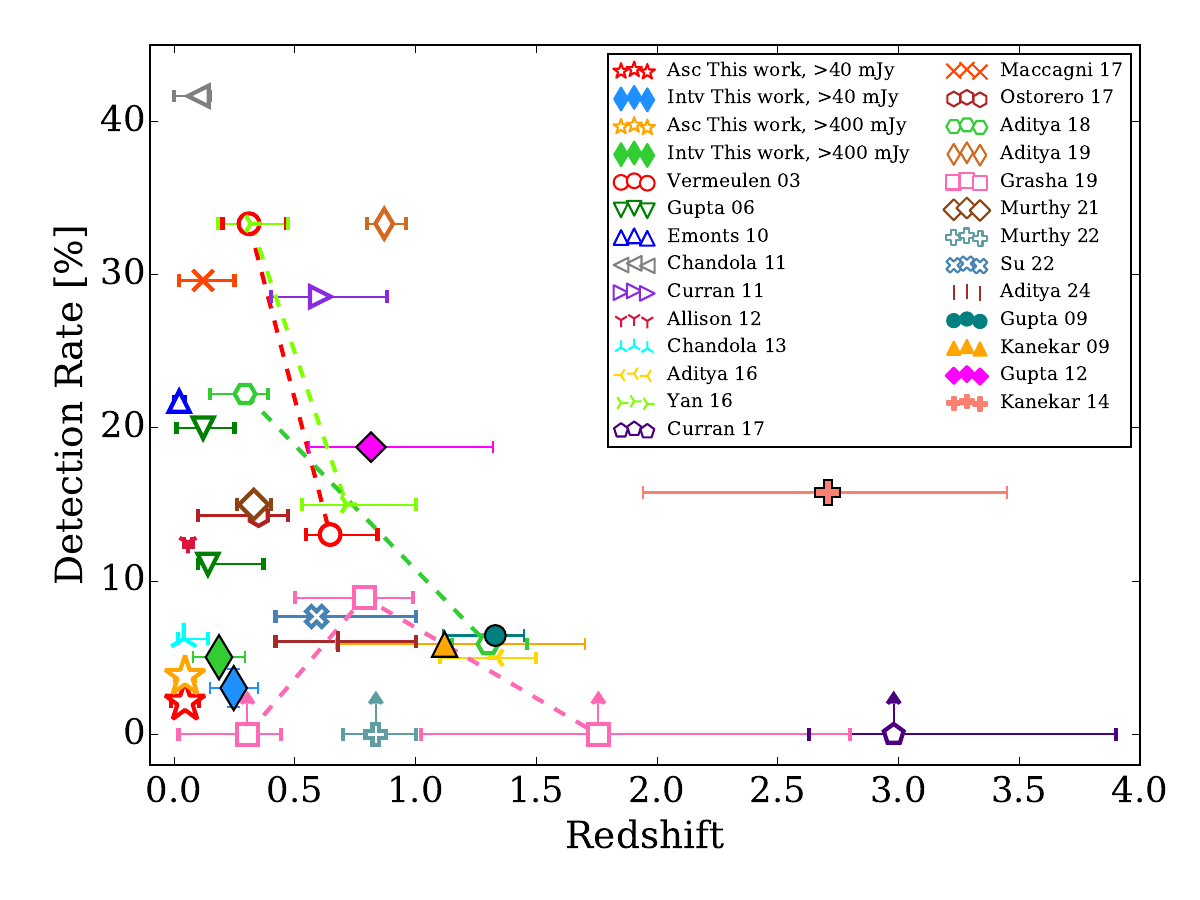}
    \caption{The detection rate of \hi 21-cm absorbers from this study and previous literature across different redshift intervals. Only the intervening and associated \hi 21-cm absorptions with integrated optical depth limits exceeding 0.5\kms and 1\kms, respectively, are considered here. Detection rates from the same survey at various redshifts are linked with dashed lines. The detection rates for associated and intervening systems are indicated by open and filled points, respectively. The references for the literature are:\citet{2003A&A...404..861V,2006MNRAS.373..972G,2010MNRAS.406..987E,2011MNRAS.418.1787C,2013MNRAS.429.2380C,2011MNRAS.413.1165C,2017MNRAS.470.4600C,2012MNRAS.423.2601A,2016MNRAS.455.4000A,2024MNRAS.527.8511A,2016AJ....151...74Y,2017A&A...604A..43M,2017ApJ...849...34O,2018MNRAS.481.1578A,2019MNRAS.482.5597A,2019ApJS..245....3G,2021A&A...654A..94M,2022A&A...659A.185M,2022MNRAS.516.2947S,2009MNRAS.396..385K,2014MNRAS.438.2131K,2009MNRAS.398..201G,2012A&A...544A..21G}}
    \label{detection_rate}
\end{figure}

The detection rates from our survey, along with those from the literature for associated(open points) and intervening(filled points) \hi absorption, are shown in Figure~\ref{detection_rate}. If we focus on background sources with a flux density $\geqslant$40 mJy at the \hi absorption frequency, the detection rate for associated \hi absorption in our study is 2.11$\pm$0.55$\%$ at $z\sim$0.045, and for intervening \hi absorption, it is 3.02$\pm$1.23$\%$ at $z\sim$0.247. When we limit the analysis to sources with flux density $\geqslant$400 mJy at the \hi absorption frequency, the detection rates rise to 3.79$\pm$0.46$\%$ for associated \hi absorption at $z\sim$0.045 and 5.04$\pm$0.85$\%$ for intervening \hi absorption at $z\sim$0.186. While the detection rate for intervening \hi absorption is similar to that of associated \hi absorption, fewer surveys have targeted intervening systems. This disparity exists because associated \hi absorption at low redshift is generally easier to detect, as discussed in Section~\ref{sec:distribution_in_redshift}. However, with the advent of more powerful telescopes like the SKA, its precursors and pathfinders(MALS and FLASH), and the FAST core array\citep{2024AstTI...1...84J}, we anticipate the detection of more intervening systems in the near future.

Compared to other surveys, our survey shows a lower detection rate for both associated and intervening \hi 21-cm absorption. This is primarily because ours is a blind survey, whereas the others specifically target sources with additional information (such as precise redshift data) or particular types of sources (like compact steep spectrum sources or MgII absorbers). %Moreover, we reconfirm the decreasing trend in detection rate with increasing redshift, which aligns with findings from FLASH \citep{2022MNRAS.516.2947S,2024MNRAS.527.8511A}.

\subsection{Statistic Completeness}

In this study, we examine characteristics of associated and intervening \hi 21-cm absorption using extensive statistical analysis. It is important to acknowledge that the outcomes are significantly influenced by sample completeness. To assess our sample's completeness, we introduced mock absorption into actual data and conducted the search process again. We identified that the most significant factor affecting completeness was the bandpass fluctuations, primarily due to standing waves. In the FAST data, these standing waves exhibit a frequency width of approximately 1 MHz (equivalent to about 200 $\kms$ at the redshift of 0), posing a challenge to detect \hi absorption features with an FWHM of around 200 $\kms$. Nonetheless, the average velocity width for the intervening \hi 21-cm absorption is recorded at 30 $\kms$, and the FWHM for most associated \hi 21-cm absorption remains below 200 $\kms$ \citep{2015A&A...575A..44G}. The \hi absorption features with an FWHM of around 200 $\kms$ should be faint absorption systems without peak narrow components. Typical absorption profiles, even with high-speed outflows will keep the FWHM lower than 200 $\kms$ because these features are shallow, and more visible at the W20 (the width of the spectral line measured at 20$\%$ of the peak intensity) level. Given that our analysis has been adjusted for completeness, the findings of this work can still faithfully reflect the characteristics of the \hi 21-cm absorption with FWHM less than 200 $\kms$.

\citet{2023ApJ...954..139L} introduced a data processing pipeline for the FAST \hi intensity mapping drift-scan survey, employing a low-pass filter to mitigate noise and diminish the presence of standing waves. Their calibration approach effectively corrects both the temporal and bandpass gain variations over the 4-hour drift-scan sessions. More recently, \citet{2024SCPMA..6759514J} unveiled HiFAST, a comprehensive calibration and imaging pipeline specifically designed for FAST \hi data, which includes a method for standing wave elimination based on FFT. We plan to incorporate the standing wave removal technique in our future data processing efforts, which is expected to expand our catalog and enhance statistical analyses across a broader parameter space.

\begin{table*}
    \fontsize{8}{10}\selectfont
	\centering
	\caption{Some basic physical parameters of the associated/background radio sources. The positional values are obtained from NED, while the flux density is measured at the redshifted \hi 21-cm frequency in our calibrated spectra.}
	\label{radiosource_table}
        \begin{threeparttable}
	\begin{tabular}{|c|c|c|c|c|c|}
	\hline
        Radio Source & Source Type & Ra(J2000) & Dec(J2000) & cz (\kms) & $S_{\nu}$ (mJy)\\
	  \hline
        NVSS\,J004219+570836\tnote{$\diamond$} &  QSO & 00h 42m 19.45s & +57d 08m 36.59s & 342063.24 & 1660.31 $\pm$ 27.94 \\
        \hline
        4C\,+56.02\tnote{$\otimes$} &  Radio Source & 01h 10m 57.55s & +56d 32m 16.98s & - & 2104.72 $\pm$ 34.93 \\
        \hline
        NVSS\,J011322+251852\tnote{$\diamond$} &  Radio Galaxy & 01h 13m 22.69s & +25d 18m 53.26s & 208355.79 & 58.69 $\pm$ 1.42 \\
        \hline
        4C\,+31.04\tnote{$\star$} & Double Radio System & 01h 19m 35.00s & +32d 10m 50.05s & 18047.51 $\pm$ 59.96 & 2898.42 $\pm$ 47.85 \\
        \hline
        4C\,+26.07\tnote{$\otimes$} &  Radio Source & 02h 05m 37.34s & +26d 50m 02.70s & 107625.49 $\pm$ 43409.95 & 320.72 $\pm$ 5.37 \\
        \hline
        3C\,84\tnote{$\star$} & Radio Galaxy & 03h 19m 48.16s  & +41d 30m 42.11s & 5264 $\pm$ 11 km/s & 18736.21 $\pm$ 310.17 \\
	  \hline
        NVSS\,J033529+195621\tnote{$\otimes$} &  Radio Source & 03h 35m 29.43s & +19d 56m 21.3s & - & 65.89 $\pm$ 1.25 \\
        \hline
        NVSS\,J040845+001306\tnote{$\star$} & Radio Source & 04h 08m 45.29s & +00d 13m 07.35s & 8945 $\pm$ 35  & 48.79 $\pm$ 2.42 \\
        \hline
        NVSS\,J055437+271126\tnote{$\otimes$} & Radio Galaxy & 05h 54m 37.30s & +27d 11m 25.9s & 27855.52 & 39.77 $\pm$ 0.82\\
        \hline
        NVSS\,J073755+264652\tnote{$\star$} &  Radio Galaxy & 07h 37m 55.34s & +26d 46m 52.57s & 71685.18 $\pm$ 17.69 & 71.91 $\pm$ 2.34 \\
        \hline
        4C\,-06.18\tnote{$\otimes$} &  Radio Source & 07h 44m 21.66s & -06d 29m 35.93s & - & 12326.00 $\pm$ 203.46 \\
        \hline
        NVSS\,J080101-075121\tnote{$\otimes$} & Radio Source & 08h 01m 01.01s & -07d 51m 21.80s & 41371.36 $\pm$ 12111.62 & 118.8 $\pm$ 3.6 \\
        \hline
        NVSS\,J085521+575143\tnote{$\diamond$} & Radio Source & 08h 55m 21.36s & +57d 51m 44.09s & 132598.20 & 751.55 $\pm$ 119.75 \\
        \hline
        NVSS\,J092351+281527\tnote{$\diamond$} & QSO & 09h 23m 51.52s & +28d 15m 25.02s & 223196 $\pm$ 35  & 596.19 $\pm$ 9.88 \\
        \hline
        NVSS\,J093150+254034\tnote{$\diamond$} & Radio Galaxy & 09h 31m 50.56s & +25d 40m 34.61s & 243552 $\pm$ 35 & 178.59 $\pm$ 5.08 \\
        \hline
        NVSS\,J094208+135152\tnote{$\diamond$} & Radio Source & 09h 42m 08.40s & +13d 51m 52.2s & - & 1414.06 $\pm$ 22.40 \\
        \hline
        NVSS\,J095058+375758\tnote{$\star$} & Radio Galaxy & 09h 50m 58.72s & +37d 57m 58.40s & 12149 $\pm$ 4 & 79.78 $\pm$ 2.31 \\
        \hline
        NVSS\,J095812+112643\tnote{$\otimes$} &  Radio Galaxy & 09h 58m 12.35s & +11d 26m 42.62s & 94134.83 $\pm$ 23024.06 & 423.66 $\pm$ 6.82 \\
        \hline
        NVSS\,J100755+405519\tnote{$\otimes$} & Radio Source & 10h 07m 55.69s & +40d 55m 18.50s & 57260.36 $\pm$ 44369.28 & 81.99$\pm$3.44 \\
        \hline
        NVSS\,J104941+133255\tnote{$\diamond$} & QSO & 10h 49m 41.10s & +13d 32m 55.68s & 828592 $\pm$ 75 & 145.90 $\pm$ 5.16 \\
        \hline
        NVSS\,J112832+583346\tnote{$\star$} & Radio Galaxy & 11h 28m 32.30s & +58d 33m 43.0s & 3121 $\pm$ 3 & 675.48$\pm$11.53 \\
        \hline
        NVSS\,J115948+582020\tnote{$\diamond$} &  Radio Galaxy & 11h 59m 48.77s & +58d 20m 20.28s & 188419.59 & 1651.43 $\pm$ 28.61 \\
        \hline
        NVSS\,J134035+444817\tnote{$\star$} & Radio Galaxy & 13h 40m 35.20s & +44d 48m 17.40s & 19619 $\pm$ 2 & 41.04$\pm$2.69 \\
        \hline 
        4C\,+57.23\tnote{$\diamond$} & Radio Source & 13h 54m 00.12s & +56d 50m 04.74s & 62956 & 767.88 $\pm$ 13.04 \\
        \hline
        NVSS\,J141314-031227\tnote{$\star$} & Radio Galaxy & 14h 13m 14.89s & -03d 12m 27.28s & 1824 $\pm$ 3 & 353.07 $\pm$ 6.52 \\
        \hline
        NVSS\,J141558+132024\tnote{$\diamond$} & QSO & 14h 15m 58.82s & +13d 20m 23.71s & 73962 $\pm$ 3 & 1359.21 $\pm$ 22.75 \\
        \hline
        NVSS\,J160332+171158\tnote{$\star$} & Radio Galaxy & 16h 03m 32.08s & +17d 11m 55.31s & 10198 $\pm$ 2 & 501.78 $\pm$ 8.52 \\
        \hline
        NVSS\,J162549+402921\tnote{$\star$} & Radio Source & 16h 25m 50.03s & +40d 29m 18.84s & 8747 $\pm$ 3  & 84.42 $\pm$ 3.50 \\
        \hline
        NVSS\,J225900+274356\tnote{$\star$} & Radio Source & 22h 59m 00.31s & +27d 43m 56.70s & 32977 $\pm$ 6895 & 589.74 $\pm$ 9.62 \\
        \hline
        \end{tabular}

        \begin{tablenotes}
           \item[$\star$] Associated \hi absorption type.
           \item[$\diamond$] Intervening \hi absorption type.
           \item[$\otimes$] Unknown \hi absorption type.
        \end{tablenotes}
        \end{threeparttable}
\end{table*}

\begin{table*}
    %\fontsize{8}{10}\selectfont
	\centering
	\caption{WISE magnitudes for the WISE counterparts of background radio sources (WISE magnitudes for foreground sources of intervening systems, if available, is presented in the context). The identifications of counterparts for NVSS\,J094208+135152, and 4C\,+57.23 are unclear due to the proximity between foreground and background objects.}
	\label{WISE_counterpart_table}
	\begin{tabular}{|c|c|c|c|c|c|c|}
	\hline
        Radio Source & WISE Counterpart & W1 & W2 & W3 & W4 & WISE color-color classification \\
	  \hline
        NVSS\,J004219+570836 & WISEA\,J004219.53+570836.0 & 14.577 & 13.413 & 10.047 & 7.579 & QSOs $\&$ Seyferts \\
        \hline
        4C\,+56.02 & WISEA\,J011057.55+563216.8 & 14.795 & 14.021 & 12.190 & 8.956 & - \\
        \hline
        NVSS\,J011322+251852 & WISEA\,J011322.69+251853.2 & 14.355 & 13.047 & 9.679 & 7.269 & QSOs \\
        \hline
        4C\,+31.04 & WISEA\,J011935.00+321050.2 & 11.641 & 11.620 & 10.024 & 7.578 & Spirals \\
        \hline
        4C\,+26.07 & WISEA J020537.42+265003.8 & 14.942 & 13.343 & 9.912 & 7.476 & QSOs \\
        \hline
        3C\,84 & WISEA\,J031948.16+413042.3 & 9.136 & 8.113 & 4.062 & 1.150 & Seyferts $\&$ Starburst \\
        \hline
        NVSS\,J033529+195621 & WISEA\,J033529.35+195620.7 & 14.360 & 12.746 & 9.248 & 6.276 & QSOs \\
        \hline
        NVSS\,J040845+001306 & WISEA\,J040845.31+001308.0 & 11.511 & 11.433 & 8.520 & 6.698 & Spirals \\
        \hline
        NVSS\,J055437+271126 & WISEA\,J055437.35+271126.3 & 12.257 & 12.113 & 10.830 & 8.318 & Ellipticals/Spirals \\
        \hline
        NVSS\,J073755+264652 & WISEA\,J073755.33+264652.9 & 14.356 & 14.032 & 11.138 & 8.279 & Spirals \\
        \hline
        4C\,-06.18 & WISEA\,J074421.66-062935.7 & 15.440 & 14.327 & 9.821 & 6.232 & LIRG \\
        \hline
        NVSS\,J080101-075121 & WISEA\,J080101.19-075123.0 & 15.097 & 14.962 & 12.345 & 8.642 & Spirals \\
        \hline
        NVSS\,J085521+575143 & WISEA\,J085521.34+575144.5 & 15.126 & 14.971 & 11.808 & 8.772 & Spirals/LIRGs \\
        \hline
        NVSS\,J092351+281527 & WISEA\,J092351.52+281525.1 & 14.117 & 13.045 & 9.946 & 7.721 & QSOs \\
        \hline
        NVSS\,J093150+254034 & WISEA\,J093150.56+254034.6 & 15.339 & 14.590 & 11.552 & 8.263 & Seyferts \\
        \hline
        NVSS\,J094208+135152 & - & - & - & - & - & - \\
        \hline
        NVSS\,J095058+375758 & WISEA\,J095058.73+375758.3 & 11.933 & 11.143 & 8.138 & 5.728 & Seyferts \\
        \hline
        NVSS\,J095812+112643 & WISEA\,J095812.38+112643.3 & 16.130 & 15.775 & 11.327 & 8.551 & ULIRGs/LINERs/Starburst \\
        \hline
        NVSS\,J100755+405519 & WISEA J100755.72+405517.7 & 15.927 & 14.922 & 11.848 & 8.360 & QSOs \\
        \hline 
        NVSS\,J104941+133255 & WISEA\,J104941.17+133251.9 & 15.281 & 14.834 & 12.309 & 8.586 & Spirals \\
        \hline
        NVSS\,J112832+583346 & WISEA\,J112833.59+583346.5 & 9.342 & 8.299 & 3.667 & 0.726 & ULIRGs/LINERs \\
        \hline
        NVSS\,J115948+582020 & WISEA\,J115948.76+582020.2 & 16.207 & 15.110 & 12.511 & 9.314 & Seyferts \\
        \hline
        NVSS\,J134035+444817 & WISEA\,J134035.20+444817.3 & 11.066 & 10.050 & 7.202 & 4.728 & QSOs \\
        \hline
        4C\,+57.23 & - & - & - & - & - & - \\
        \hline
        NVSS\,J141314-031227 & WISEA\,J141314.88-031227.5 & 7.399 & 6.244 & 3.692 & 1.068 & Seyferts/QSOs \\
        \hline
        NVSS\,J141558+132024 & WISEA\,J141558.82+132023.7 & 12.160 & 10.839 & 8.058 & 5.755 & QSOs \\
        \hline
        NVSS\,J160332+171158 & WISEA\,J160332.08+171155.3 & 10.939 & 10.966 & 10.372 & 8.476 & Ellipticals \\
        \hline
        NVSS\,J162549+402921 & WISEA\,J162549.96+402919.4 & 10.764 & 10.794 & 10.049 & 8.646 & Ellipticals \\
        \hline
        NVSS\,J225900+274356 & WISEA\,J225900.29+274356.9 & 13.148 & 12.429 & 9.887 & 7.551 & Seyferts/QSOs \\
        \hline
        \end{tabular}
\end{table*}

\begin{table*}
    \fontsize{8}{10}\selectfont
	%\centering
	\caption{SDSS magnitudes for the SDSS counterparts of background sources (SDSS magnitudes for foreground sources of intervening systems, if available, is presented in the context). Several background sources lack SDSS counterparts because they fall outside the coverage area of SDSS.}
	\label{SDSS_counterpart_table}
	\begin{tabular}{|c|c|c|c|c|c|c|}
	\hline
        Radio Source & SDSS Counterpart & u & g & r & i & z\\
	  \hline
        NVSS\,J004219+570836 & - & - & - & - & - & -\\
        \hline
        4C\,+56.02 & - & - & - & - & - & -\\
        \hline
        NVSS\,J011322+251852 & SDSS J011322.69+251853.2 & 20.650 & 19.921 & 19.126 & 18.586 & 18.242 \\
        \hline
        4C\,+31.04 & SDSS J011934.99+321050.0 & 17.266 & 15.077 & 14.056 & 13.552 & 13.191 \\
        \hline
        4C\,+26.07 & SDSS J020537.32+265001.5 & 22.497 & 21.744 & 20.689 & 20.066 & 20.411 \\
        \hline
        3C\,84 & SDSS J031948.15+413042.1 & 13.492 & 12.030 & 11.314 & 10.981 & 10.720 \\
        \hline
        NVSS\,J033529+195621 & - & - & - & - & - & -\\
        \hline
        NVSS\,J040845+001306 & SDSS J040845.31+001307.2 & 17.278 & 15.288 & 14.234 & 13.644 & 13.133 \\
        \hline
        NVSS\,J055437+271126 & - & - & - & - & - & -\\
        \hline
        NVSS\,J073755+264652 & SDSS J073755.33+264652.6 & 21.214 & 19.746 & 18.477 & 17.902 & 17.394 \\
        \hline
        4C\,-06.18 & - & - & - & - & - & -\\
        \hline
        NVSS\,J080101-075121 & SDSS\,J080101.20-075123.3 & 24.094 & 20.079 & 19.107 & 18.560 & 18.200 \\
        \hline
        NVSS\,J085521+575143 & SDSS\,J085521.37+575144.1 & 22.514 & 21.667 & 20.140 & 19.547 & 19.198 \\
        \hline
        NVSS\,J092351+281527 & SDSS J092351.52+281525.1 & 19.698 & 19.198 & 19.091 & 19.021 & 18.808 \\
        \hline
        NVSS\,J093150+254034 & - & - & - & - & - & -\\
        \hline
        NVSS\,J094208+135152 & SDSS J094208.05+135154.9 & 20.724 & 19.471 & 18.889 & 18.614 & 18.536 \\
        \hline
        NVSS\,J095058+375758 & SDSS J095058.69+375758.8 & 17.847 & 16.194 & 15.342 & 14.852 & 14.450 \\
        \hline
        NVSS\,J095812+112643 & SDSS J095812.32+112642.5 & 24.484 & 22.480 & 21.018 & 20.518 & 19.505 \\
        \hline
        NVSS\,J100755+405519 & SDSS J100755.68+405518.5 & 24.644 & 22.095 & 21.401 & 20.895 & 21.325 \\
        \hline
        NVSS\,J104941+133255 & SDSS J104941.09+133255.6 & 19.841 & 18.827 & 18.713 & 18.741 & 18.569 \\
        \hline
        NVSS\,J112832+583346 & SDSS J112833.41+583346.2 & 14.333 & 13.368 & 12.559 & 12.326 & 11.979 \\
        \hline
        NVSS\,J115948+582020 & SDSS J115948.76+582020.0 & 22.127 & 21.688 & 20.957 & 20.674 & 20.773 \\
        \hline
        NVSS\,J134035+444817 & SDSS J134035.20+444817.3 & 17.999 & 16.776 & 16.164 & 15.739 & 15.457 \\
        \hline
        4C\,+57.23 & SDSS J135400.09+565004.9 & 24.108 & 23.173 & 22.603 & 21.732 & 22.675 \\
        \hline
        NVSS\,J141314-031227 & SDSS J141314.87-031227.3 & 14.644 & 12.898 & 12.019 & 11.385 & 11.268 \\
        \hline
        NVSS\,J141558+132024 & SDSS J141558.81+132023.7 & 22.064 & 20.518 & 19.049 & 18.496 & 18.056 \\
        \hline
        NVSS\,J160332+171158 & SDSS J160332.08+171155.3 & 16.276 & 14.374 & 13.432 & 12.981 & 12.716 \\
        \hline
        NVSS\,J162549+402921 & SDSS J162549.96+402919.3 & 15.697 & 13.800 & 12.974 & 12.536 & 12.204 \\
        \hline
        NVSS\,J225900+274356 & SDSS J225900.30+274356.7 & 19.479 & 17.964 & 17.154 & 16.642 & 16.390 \\
        \hline
        \end{tabular}
\end{table*}

\section{Summary}
\label{sec:Summary}
In this paper, we present a purely blind search for extragalactic \hi 21-cm absorption lines in drift-scan data from FAST. We have examined 1325.6\, hours of data from the CRAFTS and FASHI surveys, which covered 6072.0\,deg$^{2}$ of the sky, with a total of 84533 radio sources with a flux density greater than 12 mJy were searched for \hi absorber. The data of both linear polarization of all 19 beams were processed. All data in the frequency range of 1.05 GHz to 1.45 GHz were searched. Assuming an \hi spin temperature to source covering fraction ratio of $T_{s}/c_{f}$ = 100 K, the total comoving absorption path length spanned by our data and sensitive to the Damped Lyman $\alpha$ Absorbers (DLAs; $N_{\hi} \geqslant 2\times10^{20} \cm^{-2}$) is $\Delta X^{\mathrm{inv}}$ = 4.07$\times10^4$
($\Delta z^{\mathrm{inv}} = 3.19\times10^{4}$) for intervening absorption. For associated absorption, the corresponding value is $\Delta X^{\mathrm{asc}}$ = 1.81$\times10^{2}$ ($\Delta z^{\mathrm{asc}} = 1.31\times10^{2}$). A matched-filtering approach was used to detect \hi absorption profiles. 

We successfully detected 14 known \hi absorbers and 20 new \hi 21-cm absorbers, comprising 15 associated systems, 10 intervening systems, and 9 systems with undetermined classifications. We fit the \hi profiles with multi-components Gaussian functions, and calculate the redshift, width, flux density, optical depth, and \hi column densities for each source. In our study, the detection rate for associated \hi absorption with integrated optical depth limits above 1\kms is 2.11$\pm$0.55$\%$ at $z\sim$0.045, and for intervening \hi absorption with integrated optical depth limits above 0.5\kms, it is 3.02$\pm$1.23$\%$ at $z\sim$0.247. Statistical analysis reveals the following results:
\begin{itemize}[leftmargin=10pt]
\setlength\itemsep{0.em}
\item Most associated \hi absorption systems are observed at lower redshifts ($z<$0.2) while intervening systems appear more often at higher redshifts ($z>$0.2).
\item In our sample the associated systems tend to be hosted by red ($g-r>$0.7) galaxies at lower redshifts, whereas the galaxies hosting intervening \hi absorption are typically found at higher redshifts and are of a bluer ($g-r\leqslant$0.7) type.
\item It has been demonstrated that associated \hi 21-cm absorption connected to compact radio sources display higher N$_{\hi}$ compared to those tied to extended radio sources.
\item Only a small fraction of intervening absorption systems currently have identified foreground sources. With the upcoming capabilities of the SKA and the FAST core array, we expect to detect more of these intervening systems.
\item Most (three out of four WISE-identified foregrounds) foreground absorption systems appear in the spiral galaxy region of the WISE color-color diagram, while the backgrounds cluster around the QSO region. Host galaxies of associated absorptions spread widely, with most in the spiral region, though high-redshift hosts ($z>$0.5) tend to occupy the QSO region.
\item For foreground sources within the intervening systems in our sample, there appears to be a potential trend indicating that \hi column density decreases as W1-W2 magnitudes increase. All identified foreground sources display W1-W2 magnitudes below 0.8, implying these sources are likely galaxies with little or no AGN activity. In contrast, backgrounds of intervening systems are more likely to have W1-W2 magnitudes above 0.8, suggesting they are AGN candidates. For associated absorption, most low-redshift systems show W1-W2 values below 0.8, while associated absorption at higher redshifts ($z>$0.5) exhibits a wider range of W1-W2 values.
\item Through spectral stacking, mean peak optical path, mean velocity-integrated optical path $\langle \tau\rangle$, mean FWHM and mean \hi column density $\langle$ N$_{\hi}\rangle$ are measured to be 0.47 and 0.30; 27.19 $\kms$ and 4.36 $\kms$; 42.61 $\kms$ and 9.33 $\kms$; 0.49 and 0.08 T$_{s} \times$ 10$^{20}$cm$^{-2}$K$^{-1}$, for associated and intervening samples, respectively.
\end{itemize}

\begin{table*}
    \fontsize{6.8}{8}\selectfont
	\centering
	\caption{Some basic physical parameters for known absorption are shown in this paper. The associated, intervening, and unknown types are labeled with star, diamond, and cross, respectively.}
	\label{known_absorption_table_1}
	\begin{tabular}{|c|c|c|c|c|c|c|c|c|}
	\hline
	Radio Source & Comp & cz$_{\rm peak}$ & FWHM & $S_{\hi, \rm peak}$ & $\int S_{\hi}dv$ & $\tau_{\rm peak}\times10^{2}$ & $\int\tau dv$ & $N_{\hi}/T_{s}$\\
	& & (\kms) & (\kms) & (mJy) & (mJy\kms) &  & (\kms) & (10$^{18}$cm$^{-2}$K$^{-1}$)\\
        \hline

        & 1 & 79415.66$\pm$0.18 & 11.28$\pm$0.45 & -40.13$\pm$1.47 & -481.66$\pm$26.40 & 1.92$\pm$0.06 & 0.23$\pm$0.01 & 0.42$\pm$0.02\\
        & 2 & 79375.09$\pm$0.47 & 38.01$\pm$0.88 & -98.48$\pm$2.05 & -3984.32$\pm$128.49 & 4.79$\pm$0.06 & 1.93$\pm$0.05 & 3.50$\pm$0.10\\
        & 3 & 79314.52$\pm$0.35 & 29.45$\pm$2.19 & -581.73$\pm$31.25 & -18235.41$\pm$1680.70 & 32.35$\pm$1.95 & 9.67$\pm$0.91 & 17.59$\pm$1.66\\
        4C\,+56.02$^{\otimes}$ & 4 & 79297.70$\pm$0.08 & 15.90$\pm$0.32 & -443.39$\pm$31.77 & -7505.57$\pm$562.49 & 23.66$\pm$1.86 & 3.87$\pm$0.31 & 7.04$\pm$0.56\\
        & 5 & 79311.73$\pm$0.78 & 58.67$\pm$1.49 & -184.12$\pm$24.28 & -11497.78$\pm$1547.26 & 9.15$\pm$1.25 & 5.64$\pm$0.78 & 10.27$\pm$1.42\\
        & 6 & 79332.25$\pm$0.14 & 15.79$\pm$0.47 & -237.52$\pm$24.62 & -3992.35$\pm$431.67 & 11.97$\pm$1.30 & 1.98$\pm$0.22 & 3.60$\pm$0.40\\
        & Total & 79301.40$\pm$0.41 & 46.32$\pm$1.43 & -885.85$\pm$29.18 & -45697.09$\pm$845.04 & 54.63$\pm$2.08 & 25.82$\pm$0.21 & 46.99$\pm$0.39\\

        \hline
        
        & 1 & 17823.24$\pm$1.81 & 49.68$\pm$4.55 & -16.78$\pm$2.42 & -887.36$\pm$151.90 & 0.58$\pm$0.08 & 0.31$\pm$0.05 & 0.56$\pm$0.10\\
        & 2 & 17904.56$\pm$1.50 & 54.78$\pm$7.01 & -20.08$\pm$5.03 & -1170.81$\pm$329.73 & 0.70$\pm$0.17 & 0.40$\pm$0.11 & 0.74$\pm$0.21\\
        & 3 & 17946.14$\pm$2.95 & 142.00$\pm$2.76 & -101.32$\pm$2.77 & -15314.23$\pm$549.60 & 3.56$\pm$0.08 & 5.35$\pm$0.17 & 9.73$\pm$0.32\\
        4C\,+31.04$^{\star}$ & 4 & 18020.67$\pm$0.90 & 27.36$\pm$2.79 & -9.17$\pm$0.79 & -267.24$\pm$35.87 & 0.32$\pm$0.03 & 0.09$\pm$0.01 & 0.17$\pm$0.02\\
        & 5 & 18131.48$\pm$0.47 & 6.73$\pm$1.12 & -8.57$\pm$1.16 & -61.40$\pm$13.22 & 0.30$\pm$0.04 & 0.02$\pm$0.00 & 0.04$\pm$0.01\\
        & 6 & 18150.80$\pm$0.26 & 14.79$\pm$ 0.43 & -48.14$\pm$3.05 & -758.07$\pm$53.77 & 1.67$\pm$0.10 & 0.26$\pm$0.02 & 0.48$\pm$0.03\\
        & 7 & 18154.00$\pm$0.11 & 6.04$\pm$0.37 & -44.90$\pm$3.16 & -288.57$\pm$27.17 & 1.56$\pm$0.11 & 0.10$\pm$0.01 & 0.18$\pm$0.02\\
        & Total & 17926.39$\pm$2.61 & 151.94$\pm$2.97 & -108.96$\pm$3.00 & -18747.67$\pm$377.37 & 3.83$\pm$0.09 & 6.55$\pm$0.08 & 11.93$\pm$0.15\\
        
        \hline

        & 1 & 8118.94$\pm$0.55 & 23.52$\pm$0.78 & -381.80$\pm$26.41 & -9560.18$\pm$739.80 & 2.06$\pm$0.14 & 0.51$\pm$0.04 & 0.94$\pm$0.07\\
        & 2 & 8120.26$\pm$0.08 & 3.82$\pm$0.24 & -467.78$\pm$27.02 & -1901.94$\pm$164.65 & 2.53$\pm$0.14 & 0.10$\pm$0.01 & 0.19$\pm$0.02\\
        3C\,84 HVS$^{\diamond}$ & 3 & 8113.26$\pm$0.02 & 5.56$\pm$0.06 & -2604.56$\pm$54.06 & -15424.63$\pm$395.06 & 14.97$\pm$0.20 & 0.87$\pm$0.02 & 1.58$\pm$0.03\\
        & 4 & 8105.77$\pm$0.09 & 4.25$\pm$0.25 & -468.40$\pm$24.89 & -2118.82$\pm$170.27 & 2.53$\pm$0.13 & 0.11$\pm$0.01 & 0.21$\pm$0.02\\
        & Total & 8113.30$\pm$0.20 & 6.13$\pm$0.32 & -2929.92$\pm$56.54 & -29005.56$\pm$562.80 & 17.01$\pm$0.19 & 1.62$\pm$0.02 & 2.94$\pm$0.03\\

        \hline
        
        &  1 & 5159.24$\pm$29.14 & 271.67$\pm$56.13 & -32.67$\pm$3.03 & -9445.91$\pm$2154.60 & 0.17$\pm$0.02 & 0.50$\pm$0.11 & 0.91$\pm$0.21\\
        3C\,84$^{\star}$ & 2 & 5417.57$\pm$15.13 & 208.67$\pm$26.59 & -49.58$\pm$4.91 & -11012.97$\pm$1801.26 & 0.26$\pm$0.03 & 0.59$\pm$0.10 & 1.07$\pm$0.17\\
        & Total & 5408.06$\pm$21.60 & 435.66$\pm$40.23 & -52.49$\pm$4.08 & -20458.88$\pm$624.17 & 0.28$\pm$0.02 & 1.09$\pm$0.03 & 1.98$\pm$0.05\\
        
        \hline
        
        & 1 & 95443.11$\pm$1.45 & 50.67$\pm$3.90 & -39.16$\pm$2.60 & -2112.01$\pm$214.61 & 0.32$\pm$0.02 & 0.17$\pm$0.02 & 0.31$\pm$0.03\\
        & 2 & 95566.33$\pm$0.45 & 24.10$\pm$1.44 & -77.85$\pm$4.01 & -1997.22$\pm$157.71 & 0.63$\pm$0.03 & 0.16$\pm$0.01 & 0.30$\pm$0.02\\
        & 3 & 95587.42$\pm$4.97 & 154.69$\pm$14.36 & -76.15$\pm$2.81 & -12539.51$\pm$1252.49 & 0.62$\pm$0.02 & 1.02$\pm$0.10 & 1.86$\pm$0.18\\
        4C\,-06.18$^{\otimes}$ & 4 & 95684.00$\pm$2.69 & 29.69$\pm$11.03 & -22.76$\pm$10.35 & -719.28$\pm$422.43 & 0.18$\pm$0.08 & 0.06$\pm$0.03 & 0.11$\pm$0.06\\
        & 5 & 95727.47$\pm$0.24 & 16.85$\pm$0.79 & -143.78$\pm$6.50 & -2578.97$\pm$168.04 & 1.17$\pm$0.05 & 0.21$\pm$0.01 & 0.38$\pm$0.02\\
        & 6 & 95721.51$\pm$3.08 & 85.96$\pm$3.63 & -151.93$\pm$10.83 & -13902.37$\pm$1152.62 & 1.24$\pm$0.09 & 1.13$\pm$0.09 & 2.06$\pm$0.17\\
        & Total & 95727.15$\pm$3.30 & 48.93$\pm$7.43 & -301.76$\pm$7.91 & -33849.36$\pm$530.66 & 24.79$\pm$0.05 & 2.76$\pm$0.01 & 5.02$\pm$0.01\\

        \hline
        
        & 1 & 7737.48$\pm$0.04 & 3.40$\pm$0.15 & -152.67$\pm$9.75 & -553.36$\pm$44.37 & 25.34$\pm$1.79 & 0.88$\pm$0.07 & 1.61$\pm$0.13\\
	NVSS\,J085521+575143$^{\diamond}$ & 2 & 7739.17$\pm$0.50 & 6.84$\pm$0.55 & -36.67$\pm$8.82 & -266.84$\pm$67.92 & 5.53$\pm$1.36 & 0.40$\pm$0.10 & 0.73$\pm$0.19 \\
	& Total & 7737.56$\pm$0.19 & 3.80$\pm$0.28 & -183.91$\pm$9.68 & -820.21$\pm$22.36 & 31.42$\pm$1.86 & 1.33$\pm$0.03 
        & 2.41$\pm$0.06\\
        
        \hline

        & 1 & 3093.46$\pm$4.05 & 65.26$\pm$9.31 & -24.26$\pm$2.14 & -1685.23$\pm$285.07 & - & - & -\\
        NVSS\,J112832+583346$^{\star}$ & 2 & 3185.46$\pm$3.76 & 74.43$\pm$9.08 & -27.79$\pm$2.04 & -2201.76$\pm$316.76 & - & - & -\\
        & Total & 3185.02$\pm$3.89 & 158.32$\pm$9.18 & -27.89$\pm$2.09 & -3886.99$\pm$103.87 & - & - & -\\
        
        \hline

        & 1 & 19615.98$\pm$0.02 & 8.08$\pm$0.04 & -26.92$\pm$0.43 & -231.62$\pm$15.89 & 106.71$\pm$12.10 & 7.81$\pm$0.82 & 14.22$\pm$1.49\\
        NVSS\,J134035+444817$^{\star}$ & 2 & 19601.39$\pm$0.01 & 2.18$\pm$0.03 & -14.99$\pm$0.31 & -34.75$\pm$2.48 & 45.46$\pm$3.72 & 0.98$\pm$0.09 & 1.79$\pm$0.16\\
        & Total & 19615.98$\pm$0.02 & 8.09$\pm$0.04 & -26.92$\pm$0.44 & -266.37$\pm$18.20 & 106.71$\pm$12.10 & 8.80$\pm$0.37 & 16.01$\pm$0.66\\

        \hline

        & 1 & 28554.18$\pm$0.13 & 13.34$\pm$0.30 & -112.87$\pm$2.68 & -1603.41$\pm$67.99 & 15.90$\pm$0.32 & 2.21$\pm$0.09 & 4.02$\pm$0.16\\
        4C\,+57.23$^{\diamond}$ & 2 & 28586.70$\pm$0.56 & 29.81$\pm$1.44 & -38.34$\pm$1.51 & -1216.82$\pm$82.54 & 5.12$\pm$0.19 & 1.61$\pm$0.11 & 2.94$\pm$0.20\\
        & Total & 28554.27$\pm$0.31 & 13.69$\pm$0.30 & -114.30$\pm$2.50 & -2820.23$\pm$87.46 & 16.12$\pm$0.28 & 3.83$\pm$0.10 & 6.96$\pm$0.19\\
        
        \hline
        
        & 1 & 1793.54$\pm$0.92 & 36.34$\pm$4.63 & -20.39$\pm$3.83 & -788.65$\pm$179.96 & 5.85$\pm$1.13 & 2.25$\pm$0.52 & 4.09$\pm$0.94\\
        & 2 & 1835.52$\pm$8.77 & 88.28$\pm$12.65 & -14.86$\pm$1.23 & -1396.91$\pm$233.01 & 4.23$\pm$0.35 & 3.95$\pm$0.66 & 7.20$\pm$1.20\\
        & 3 & 1750.99$\pm$1.92 & 13.34$\pm$5.18 & -5.16$\pm$1.72 & -73.19$\pm$37.49 & 1.45$\pm$0.49 & 0.21$\pm$0.11 & 0.37$\pm$0.19\\
        NVSS\,J141314-031227$^{\star}$ & 4 & 1967.18$\pm$0.69 & 8.50$\pm$1.66 & -12.06$\pm$1.88 & -109.09$\pm$27.39 & 3.42$\pm$0.54 & 0.31$\pm$0.08 & 0.56$\pm$0.14\\
        & 5 & 1988.74$\pm$0.85 & 21.19$\pm$2.21 & -15.80$\pm$1.22 & -356.36$\pm$46.93 & 4.51$\pm$0.35 & 1.01$\pm$0.13 & 1.84$\pm$0.24\\
        & Total & 1796.39$\pm$4.96 & 70.81$\pm$8.32 & -28.66$\pm$2.05 & -2724.20$\pm$72.15 & 8.33$\pm$0.61 & 7.78$\pm$0.13 & 14.16$\pm$0.24\\

        \hline

        & 1 & 73947.99$\pm$0.15 & 8.43$\pm$0.36 & -131.97$\pm$7.15 & -1184.52$\pm$81.60 & 10.21$\pm$0.56 & 0.90$\pm$0.06 & 1.64$\pm$0.11\\
        & 2 & 73959.82$\pm$0.08 & 14.00$\pm$0.43 & -366.30$\pm$24.56 & -5459.84$\pm$402.28 & 31.40$\pm$2.41 & 4.47$\pm$0.35 & 8.13$\pm$0.64\\
        NVSS\,J141558+132024$^{\diamond}$ & 3 & 73961.54$\pm$0.32 & 25.42$\pm$0.68 & -261.89$\pm$24.03 & -7087.14$\pm$677.90 & 21.40$\pm$2.16 & 5.61$\pm$0.57 & 10.21$\pm$1.04\\
        & 4 & 73992.74$\pm$2.18 & 101.72$\pm$3.44 & -23.72$\pm$1.09 & -2568.21$\pm$146.61 & 1.76$\pm$0.08 & 1.90$\pm$0.10 & 3.46$\pm$0.19\\
        & Total & 73960.11$\pm$0.52 & 21.20$\pm$1.01 & -643.71$\pm$24.47 & -16299.72$\pm$254.45 & 64.17$\pm$3.13 & 14.31$\pm$0.01 & 26.05$\pm$0.02\\ 

        \hline

        & 1 & 10205.76$\pm$0.85 & 6.17$\pm$2.77 & -5.63$\pm$1.91 & -36.97$\pm$20.80 & 1.13$\pm$0.39 & 0.07$\pm$0.04 & 0.13$\pm$0.08\\
        NVSS\,J160332+171158$^{\star}$ & 2 & 10223.57$\pm$1.08 & 28.93$\pm$1.89 & -18.00$\pm$1.16 & -554.32$\pm$51.22 & 3.65$\pm$0.23 & 1.12$\pm$0.10 & 2.04$\pm$0.19\\
        & 3 & 10219.03$\pm$1.00 & 4.77$\pm$2.95 & -4.26$\pm$2.02 & -21.64$\pm$16.88 & 0.85$\pm$0.41 & 0.04$\pm$0.03 & 0.08$\pm$0.06\\
        & Total & 10219.49$\pm$1.07 & 31.60$\pm$1.98 & -21.19$\pm$1.25 & -612.93$\pm$11.69 & 4.31$\pm$0.25 & 1.24$\pm$0.02 & 2.25$\pm$0.03\\
        
        \hline
        
        \end{tabular}
\end{table*}

\begin{table*}
	\fontsize{6.8}{8}\selectfont
    \centering
	\caption{Same as Table ~\ref{known_absorption_table_1}, but for new detected \hi absorption shown in this paper.}
	\label{new_absorption_table_1}
	\begin{tabular}{|c|c|c|c|c|c|c|c|c|}
	\hline
	Radio Source & Comp & cz$_{\rm peak}$ & FWHM & $S_{\hi, \rm peak}$ & $\int S_{\hi}dv$ & $\tau_{\rm peak}\times10^{2}$ & $\int\tau dv$ & $N_{\hi}/T_{s}$\\
		 & & (\kms) & (\kms) & (mJy) & (mJy\kms) &  & (\kms) & (10$^{18}$cm$^{-2}$K$^{-1}$)\\

        \hline
        
        & 1 & 79032.90$\pm$0.02 & 3.12$\pm$0.10 & -205.40$\pm$12.02 & -682.79$\pm$45.31 & 13.21$\pm$0.79 & 0.43$\pm$0.03 & 0.78$\pm$0.05\\
        & 2 & 79035.75$\pm$0.29 & 6.32$\pm$0.42 & -97.46$\pm$6.76 & -655.78$\pm$63.53 & 6.05$\pm$0.42 & 0.40$\pm$0.04 & 0.73$\pm$0.07\\
        NVSS\,J004219+570836$^{\diamond}$ & 3 & 79039.81$\pm$1.04 & 17.03$\pm$1.29 & -14.83$\pm$2.45 & -268.81$\pm$48.97 & 0.90$\pm$0.15 & 0.16$\pm$0.03 & 0.30$\pm$0.05\\
        & Total & 79033.10$\pm$0.30 & 4.24$\pm$0.43 & -272.44$\pm$9.07 & -1607.39$\pm$27.56 & 17.92$\pm$0.58 & 1.02$\pm$0.01 & 1.85$\pm$0.01\\

        \hline
        
        & 1 & 76385.99$\pm$0.21 & 13.53$\pm$0.78 & -13.41$\pm$0.96 & -193.19$\pm$17.82 & 25.95$\pm$2.14 & 3.60$\pm$0.35 & 6.55$\pm$0.64\\
        NVSS\,J011322+251852$^{\diamond}$ & 2 & 76391.39$\pm$0.71 & 34.84$\pm$1.68 & -10.26$\pm$0.96 & -380.63$\pm$40.21 & 19.22$\pm$2.00 & 6.93$\pm$0.78 & 12.61$\pm$1.42\\
        & 3 & 76424.11$\pm$0.51 & 7.69$\pm$1.32 & -3.23$\pm$0.45 & -26.47$\pm$5.85 & 5.67$\pm$0.81 & 0.46$\pm$0.10 & 0.84$\pm$0.19\\
        & Total & 76386.52$\pm$0.54 & 19.39$\pm$1.37 & -23.08$\pm$0.99 & -600.29$\pm$9.92 & 49.96$\pm$2.84 & 11.73$\pm$0.11 & 21.36$\pm$0.21\\

        \hline

        & 1 & 75179.93$\pm$0.83 & 24.42$\pm$3.37 & -9.90$\pm$2.28 & -257.45$\pm$69.07 & 3.14$\pm$0.73 & 0.81$\pm$0.22 & 1.48$\pm$0.40\\
         4C\,+26.07$^{\otimes}$ & 2 & 75198.50$\pm$0.36 & 12.38$\pm$0.83 & -11.89$\pm$0.88 & -156.68$\pm$15.72 & 3.78$\pm$0.28 & 0.50$\pm$0.05 & 0.90$\pm$0.09\\
        & 3 & 75180.84$\pm$0.83 & 51.24$\pm$3.68 & -10.81$\pm$2.34 & -589.36$\pm$134.82 & 3.43$\pm$0.75 & 1.86$\pm$0.43 & 3.39$\pm$0.78\\
        & Total & 75196.81$\pm$0.75 & 41.88$\pm$3.16 & -22.17$\pm$2.12 & -1003.49$\pm$17.59 & 7.16$\pm$0.70 & 3.21$\pm$0.02 & 5.84$\pm$0.04\\

        \hline
        
         & 1 & 87558.50$\pm$10.44 & 82.80$\pm$9.40 & -12.84$\pm$2.85 & -1131.77$\pm$282.60 & 21.68$\pm$5.36 & 18.50$\pm$4.91 & 33.68$\pm$8.94\\
        NVSS\,J033529+195621$^{\otimes}$ & 2 & 87611.86$\pm$8.98 & 73.22$\pm$7.36 & -10.53$\pm$3.80 & -820.99$\pm$307.78 & 17.42$\pm$6.86 & 13.23$\pm$5.26 & 24.08$\pm$9.57\\
        & Total & 87583.92$\pm$0.73 & 112.37$\pm$1.21 & -16.92$\pm$0.33 & -1952.77$\pm$42.44 & 29.68$\pm$0.56 & 32.97$\pm$0.40 & 60.01$\pm$0.74\\

        \hline

        & 1 & 8938.21$\pm$1.41 & 56.91$\pm$3.70 & -12.51$\pm$0.95 & -758.19$\pm$77.19 & 29.64$\pm$3.08 & 17.18$\pm$2.06 & 31.28$\pm$3.74\\
        NVSS\,J040845+001306$^{\star}$ & 2 & 8928.80$\pm$0.29 & 4.46$\pm$0.75 & -16.28$\pm$2.29 & -77.24$\pm$16.97 & 40.61$\pm$7.43 & 1.81$\pm$0.44 & 3.30$\pm$0.79\\
        & 3 & 8937.80$\pm$0.87 & 5.21$\pm$2.28 & -5.91$\pm$2.15 & -32.79$\pm$18.68 & 12.91$\pm$5.05 & 0.70$\pm$0.41 & 1.28$\pm$0.75\\
        & Total & 8928.84$\pm$1.29 & 16.08$\pm$3.39 & -27.89$\pm$1.18 & -868.22$\pm$21.12 & 84.77$\pm$8.44 & 20.65$\pm$0.25 & 37.59$\pm$0.46\\

        \hline

        & 1 & 14282.00$\pm$1.48 & 48.46$\pm$2.28 & -7.20$\pm$0.38 & -371.21$\pm$26.20 & 19.96$\pm$1.14 & 10.00$\pm$0.73 & 18.20$\pm$1.33\\
        NVSS\,J055437+271126$^{\otimes}$ & 2 & 14340.71$\pm$1.41 & 63.72$\pm$4.38 & -9.34$\pm$0.21 & -633.51$\pm$45.93 & 26.77$\pm$0.65 & 17.45$\pm$1.27 & 31.76$\pm$2.31\\
        & 3 & 14378.90$\pm$0.97 & 21.13$\pm$3.60 & -2.68$\pm$0.46 & -60.30$\pm$14.57 & 6.98$\pm$1.24 & 1.55$\pm$0.38 & 2.83$\pm$0.69\\
        & Total & 14339.09$\pm$1.41 & 121.95$\pm$3.60 & -9.48$\pm$0.28 & -1065.02$\pm$17.54 & 27.22$\pm$0.90 & 29.66$\pm$0.21 & 53.98$\pm$0.39\\      
        
        \hline

        NVSS\,J073755+264652$^{\star}$ & 1 & 71621.72$\pm$0.14 & 13.92$\pm$0.34 & -39.50$\pm$1.03 & -585.12$\pm$23.57 & 79.70$\pm$4.17 & 10.47$\pm$0.41 & 19.06$\pm$0.75\\
        
        \hline
        
        & 1 & 27414.93$\pm$2.68 & 47.72$\pm$5.52 & -34.91$\pm$2.15 & -1773.50$\pm$232.98 & 34.73$\pm$3.47 & 16.76$\pm$2.61 & 30.50$\pm$4.74\\
	NVSS\,J080101-075121$^{\otimes}$ & 2 & 27442.87$\pm$0.71 & 14.52$\pm$2.41 & -38.38$\pm$5.56 & -593.23$\pm$130.65 & 38.94$\pm$7.40 & 5.68$\pm$1.41 & 10.34$\pm$2.56\\
	& Total & 27441.92$\pm$2.19 & 50.79$\pm$4.74 & -52.31$\pm$3.05 & -2366.73$\pm$43.17 & 57.90$\pm$6.37 & 23.34$\pm$0.95 & 42.48$\pm$1.73\\

        \hline

        & 1 & 99217.81$\pm$0.81 & 13.33$\pm$1.69 & -19.65$\pm$1.62 & -279.02$\pm$53.09 & 33.55$\pm$0.28 & 0.47$\pm$0.09 & 0.86$\pm$0.16\\
        NVSS\,J092351+281527$^{\diamond}$ & 2 & 99210.55$\pm$0.30 & 3.04$\pm$0.88 & -16.62$\pm$4.02 & -53.73$\pm$21.15 & 2.83$\pm$0.69 & 0.09$\pm$0.04 & 0.17$\pm$0.06\\
        & Total & 99210.75$\pm$0.73 & 14.23$\pm$1.56 & -25.46$\pm$2.02 & -332.75$\pm$38.66 & 43.67$\pm$3.49 & 0.57$\pm$0.06 & 1.03$\pm$0.11\\

        \hline

        NVSS\,J093150+254034$^{\diamond}$ & 1 & 80339.26$\pm$0.07 & 2.51$\pm$0.16 & -60.43$\pm$3.39 & -161.71$\pm$5.46 & 41.30$\pm$0.03 & 1.04$\pm$0.17 & 1.89$\pm$0.32\\
        
	\hline

        & 1 & 39901.98$\pm$1.34 & 20.74$\pm$2.82 & -16.93$\pm$4.29 & -373.70$\pm$107.63 & 1.20$\pm$0.31 & 0.27$\pm$0.08 & 0.48$\pm$0.14\\
        NVSS\,J094208+135152$^{\diamond}$ & 2 & 39903.49$\pm$0.31 & 5.87$\pm$0.72 & -63.76$\pm$5.00 & -398.34$\pm$58.13 & 4.61$\pm$0.36 & 0.29$\pm$0.04 & 0.52$\pm$0.08\\
        & 3 & 39909.17$\pm$0.83 & 4.77$\pm$1.51 & -20.71$\pm$4.99 & -105.11$\pm$41.78 & 1.48$\pm$0.36 & 0.07$\pm$0.03 & 0.14$\pm$0.05\\
        & Total & 39903.51$\pm$0.81 & 8.60$\pm$1.71 & -80.85$\pm$4.82 & -877.16$\pm$14.22 & 5.89$\pm$0.35 & 0.63$\pm$0.02 & 1.15$\pm$0.03\\

        \hline

        NVSS\,J095058+375758$^{\star}$ & 1 & 12202.08$\pm$0.23 & 35.23$\pm$0.54 & -46.07$\pm$0.94 & -1727.84$\pm$30.54 & 86.14$\pm$0.04 & 28.38$\pm$0.37 & 51.65$\pm$0.67\\
        
        \hline
        
        & 1 & 95828.71$\pm$0.05 & 7.69$\pm$0.21 & -61.20$\pm$2.26 & -500.94$\pm$23.38 & 23.14$\pm$0.63 & 1.25$\pm$0.06 & 2.27$\pm$0.10\\
        NVSS\,J095812+112643$^{\otimes}$ & 2 & 95827.53$\pm$0.22 & 20.26$\pm$0.78 & -26.57$\pm$2.16 & -572.89$\pm$51.57 & 15.60$\pm$0.57 & 1.38$\pm$0.12 & 2.52$\pm$0.23\\
        & Total & 95828.64$\pm$0.14 & 9.54$\pm$0.52 & -87.53$\pm$2.49 & -1073.82$\pm$17.73 & 6.48$\pm$0.53 & 2.71$\pm$0.01 & 4.94$\pm$0.01\\
        
        \hline
        
        & 1 & 67051.73$\pm$0.81 & 11.56$\pm$1.93 & -9.97$\pm$1.43 & -122.67$\pm$27.03 & 12.97$\pm$2.04 & 1.57$\pm$0.36 & 2.85$\pm$0.65\\
        & 2 & 67079.17$\pm$0.34 & 6.50$\pm$1.02 & -21.57$\pm$2.92 & -149.13$\pm$30.94 & 30.53$\pm$4.97 & 2.02$\pm$0.45 & 3.67$\pm$0.81\\
        & 3 & 67088.58$\pm$1.56 & 22.18$\pm$2.53 & -24.61$\pm$1.52 & -581.01$\pm$75.42 & 35.68$\pm$3.00 & 7.99$\pm$1.11 & 14.55$\pm$2.03\\
        NVSS\,J100755+405519$^{\otimes}$ & 4 & 67105.29$\pm$1.45 & 9.34$\pm$2.31 & -32.31$\pm$10.31 & -321.04$\pm$129.62 & 50.10$\pm$20.86 & 4.62$\pm$2.13 & 8.41$\pm$3.87\\
        & 5 & 67116.37$\pm$0.79 & 11.59$\pm$4.45 & -60.95$\pm$4.58 & -751.91$\pm$294.11 & 136.05$\pm$23.77 & 13.64$\pm$5.57 & 24.83$\pm$10.14\\
        & 6 & 67126.05$\pm$0.84 & 8.54$\pm$1.40 & -58.59$\pm$15.48 & -532.55$\pm$165.50 & 125.38$\pm$66.67 & 9.42$\pm$4.66 & 17.15$\pm$8.49\\
        & 7 & 67137.91$\pm$0.43 & 10.01$\pm$0.92 & -35.58$\pm$1.68 & -379.02$\pm$39.34 & 56.91$\pm$4.42 & 5.57$\pm$0.65 & 10.14$\pm$1.18\\
        & Total & 67125.02$\pm$0.96 & 34.06$\pm$1.48 & -69.65$\pm$1.08 & -2837.33$\pm$44.29 & 189.42$\pm$52.69 & 51.86$\pm$0.37 & 94.39$\pm$0.68\\

        \hline

        NVSS\,J104941+133255$^{\diamond}$ & 1 & 73958.31$\pm$0.11 & 5.45$\pm$0.25 & -57.07$\pm$2.43 & -331.45$\pm$24.54 & 49.53$\pm$3.27 & 2.67$\pm$0.25 & 4.86$\pm$0.46\\
        
        \hline
        
        & 1 & 71474.54$\pm$0.10 & 3.47$\pm$0.37 & -149.76$\pm$27.20 & -553.11$\pm$116.71 & 9.51$\pm$1.81 & 0.35$\pm$0.07 & 0.63$\pm$0.14\\
        NVSS\,J115948+582020$^{\diamond}$ & 2 & 71475.90$\pm$0.21 & 12.51$\pm$0.17 & -603.46$\pm$51.30 & -8034.81$\pm$691.79 & 45.48$\pm$4.83 & 5.66$\pm$0.57 & 10.30$\pm$1.03\\
        & 3 & 71478.65$\pm$0.33 & 7.84$\pm$0.83 & -302.32$\pm$56.23 & -2521.73$\pm$539.16 & 20.22$\pm$4.16 & 1.64$\pm$0.37 & 2.98$\pm$0.67\\
        & Total & 71475.70$\pm$0.23 & 10.95$\pm$0.33 & -917.04$\pm$52.53 & -11109.65$\pm$175.56 & 81.04$\pm$6.94 & 8.69$\pm$0.01 & 15.81$\pm$0.02\\

        \hline

        NVSS\,J162549+402921$^{\star}$ & 1 & 8776.56$\pm$0.75 & 17.51$\pm$1.76 & -12.43$\pm$1.10 & -231.80$\pm$35.15 & 15.93$\pm$1.66 & 2.90$\pm$0.78 & 5.28$\pm$1.41\\

        \hline
        
        & 1 & 42738.30$\pm$1.96 & 113.76$\pm$4.43 & -55.02$\pm$5.83 & -6662.34$\pm$754.05 & 9.79$\pm$1.08 & 11.69$\pm$1.35 & 21.28$\pm$2.46\\
        NVSS\,J225900+274356$^{\star}$ & 2 & 42754.27$\pm$1.21 & 61.95$\pm$4.29 & -39.57$\pm$5.96 & -2609.05$\pm$433.04 & 6.94$\pm$1.08 & 4.53$\pm$0.76 & 8.25$\pm$1.39\\
        & 3 & 42850.82$\pm$5.61 & 336.03$\pm$7.55 & -20.34$\pm$0.70 & -7276.34$\pm$303.37 & 3.51$\pm$0.11 & 12.49$\pm$0.49 & 22.73$\pm$0.89\\
        & Total & 42750.67$\pm$3.44 & 100.47$\pm$5.78 & -108.34$\pm$3.91 & -16547.73$\pm$285.55 & 20.30$\pm$0.74 & 29.58$\pm$0.19 & 53.83$\pm$0.35\\
        
        \hline
	\end{tabular}
\end{table*}

\section*{Acknowledgements}
This work made use of the data from FAST (Five-hundred-meter Aperture Spherical radio Telescope)(https://cstr.cn/31116.02.FAST). FAST is a Chinese national mega-science facility, operated by National Astronomical Observatories, Chinese Academy of Sciences. This work is supported by the National SKA Program of China (Nos. 2022SKA0110100 and 2022SKA0110101), the NSFC International (Regional) Cooperation and Exchange Project (No. 12361141814), and NSFC grants (Nos. 11973047, 12203061, 12303004). WH acknowledges support from the South African Radio Astronomy Observatory and National Research Foundation (Grant No. 84156). The authors thank Sne$\rm\check{z}$ana Stanimirovi$\rm\acute{c}$, James R. Allison, and Renzhi Su for helpful discussion. 

%%%%%%%%%%%%%%%%%%%%%%%%%%%%%%%%%%%%%%%%%%%%%%%%%%
\section*{Data Availability}

The radio data analyzed in this work can be accessed by sending a request to the FAST Data Centre or the corresponding authors of this paper.

\appendix

\section{Gaussian Model Selection}

When fitting models, adding more parameters can increase the maximum likelihood, but this often leads to overfitting. To address this, the Bayesian Information Criterion (BIC; \citealt{1978AnSta...6..461S,24ce203a-855a-3aa9-952f-976d23b28943}) is used in statistics for model selection among a finite set of models. BIC introduces a penalty term for the number of parameters to prevent overfitting. Its main objective is to identify the best model by balancing the goodness of fit with model complexity. In general, models with lower BIC values are preferred. The BIC is formally defined as:
\begin{eqnarray}
    \mathrm{BIC} = k\mathrm{ln}(n) -2\mathrm{ln}(\hat{L}),
    \label{BIC}
\end{eqnarray}
where $k$ is the number of parameters estimated by the model, $n$ is the number of data points, and $\hat{L}$ is the maximum value of the likelihood function for the model. For each \hi absorption, we fitted the spectra from follow-up observations using models with varying numbers of Gaussian functions and calculated the BIC for each. The model with the lowest BIC was chosen for the final \hi absorption profile fitting and the estimation of physical parameters.

\begin{figure*}[hbt!]
    \centering
    \includegraphics[width=0.32\textwidth]{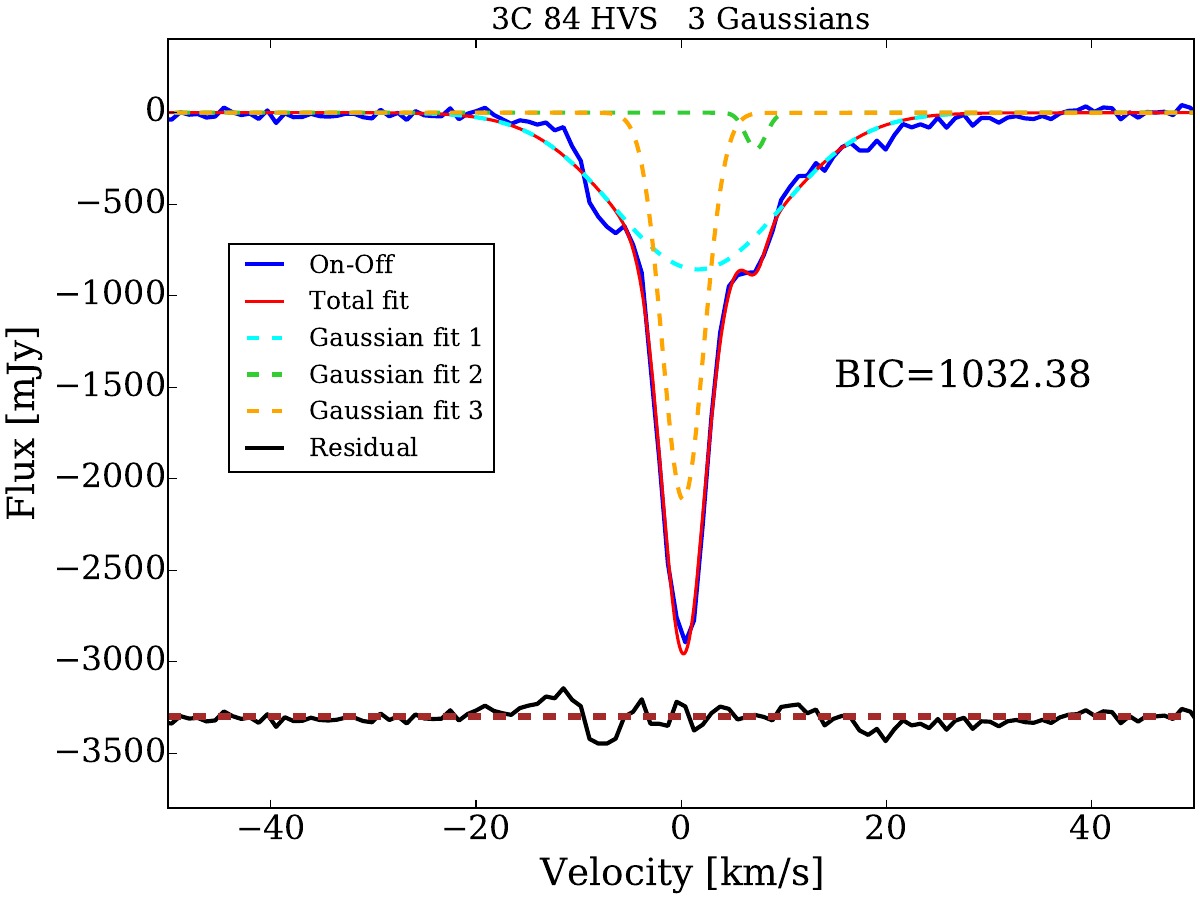}
    \includegraphics[width=0.32\textwidth]{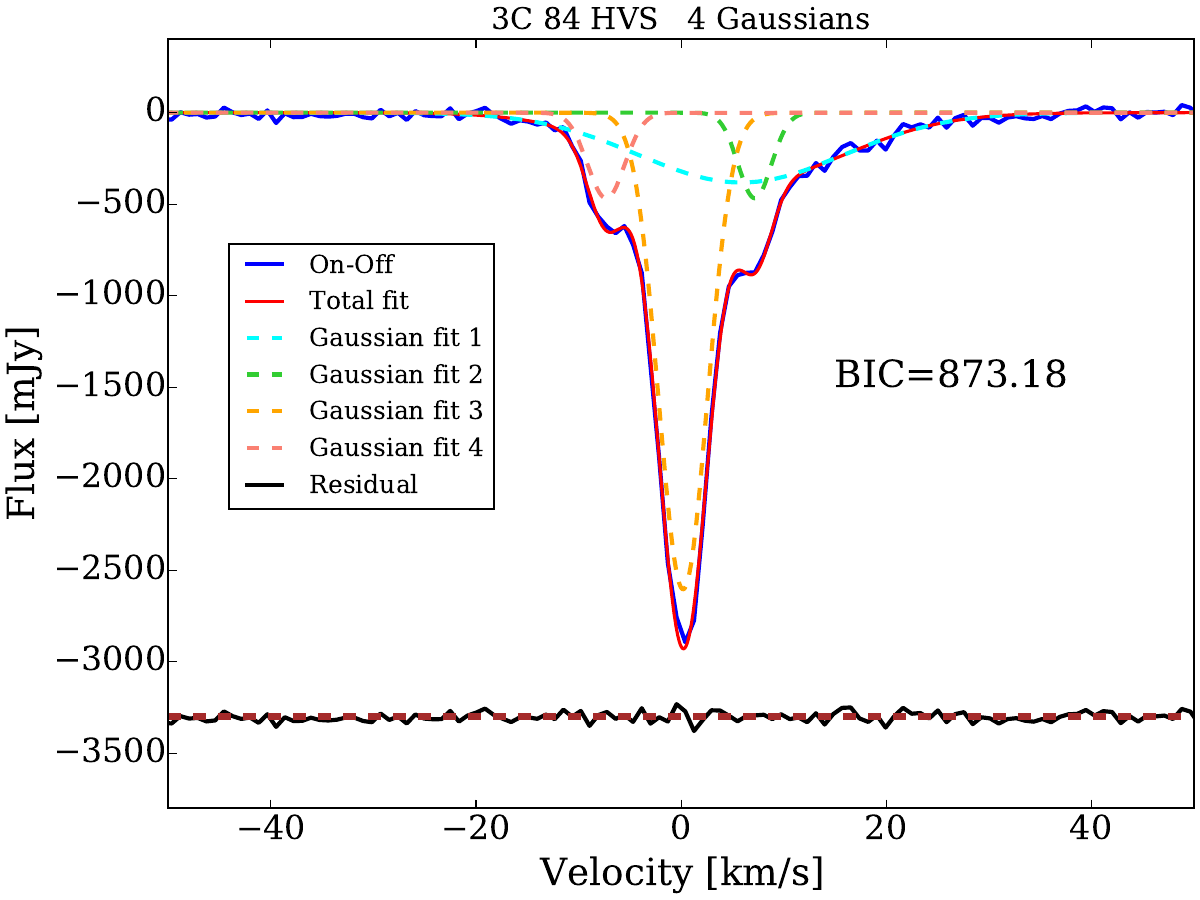}
    \includegraphics[width=0.32\textwidth]{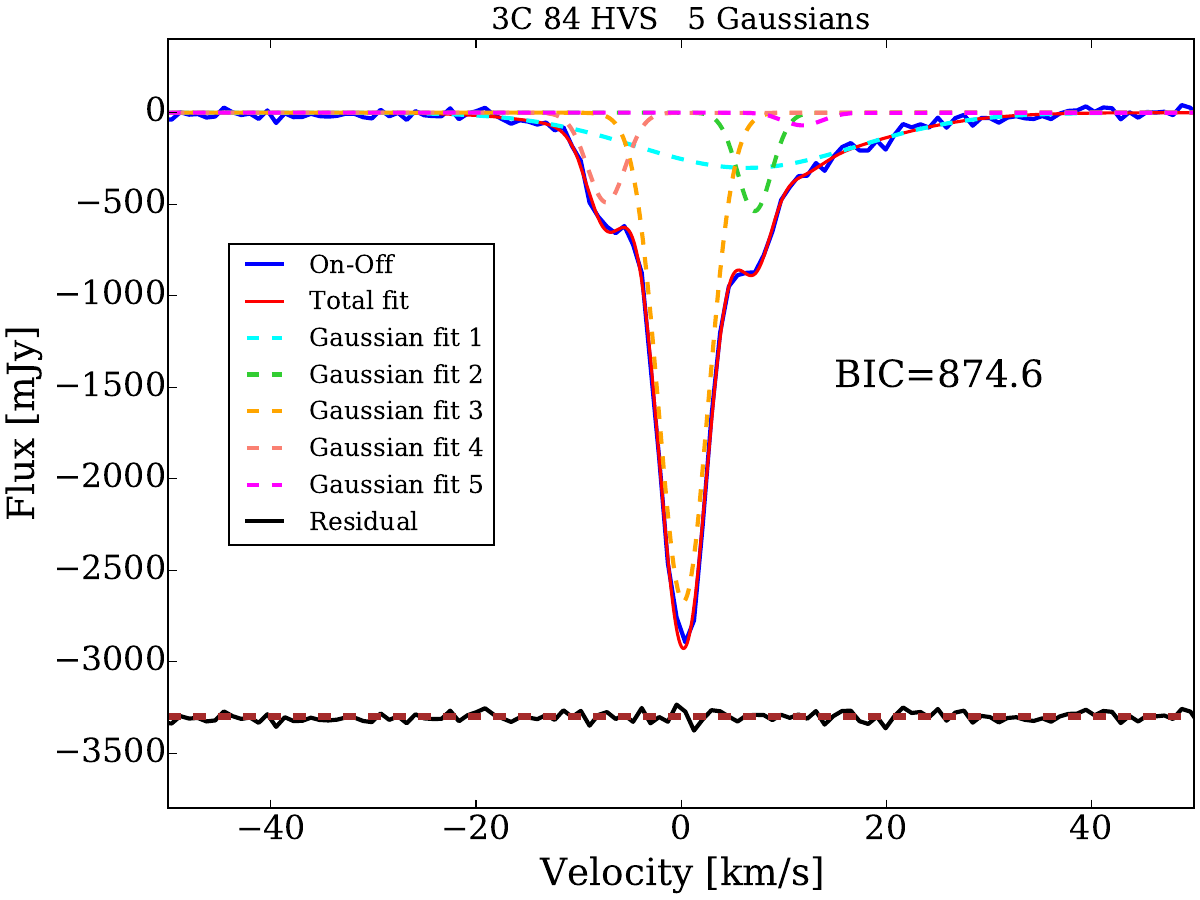}
    \caption{The fitting results for the \hi absorption of the HVS in the 3C\,84 system using three, four, and five Gaussian models are displayed in the left, center, and right panels, respectively. The BIC values for each model are also provided in the corresponding panels.}
    \label{3C_84_HVS_fit_BIC}
\end{figure*}

Figure~\ref{3C_84_HVS_fit_BIC} presents the fitting results for the \hi absorption of the HVS in the 3C\,84 system using three, four, and five Gaussian models. The fit with three Gaussian functions produces residual spectra with significant fluctuations, indicating that this model is insufficient to accurately capture the \hi absorption profile. Although the five-Gaussian model results in residual spectra with minimal fluctuations, the fifth Gaussian component (represented by the magenta dashed line) appears unnecessary, with no clear physical interpretation, suggesting overfitting. On the other hand, the four-Gaussian model also yields a flat residual spectrum with minimal fluctuations and achieves the lowest BIC value, making it the optimal model we adopted for the \hi absorption in the HVS from the 3C\,84 system.

\bibliography{HIabs}{}

\begin{thebibliography}{}
\expandafter\ifx\csname natexlab\endcsname\relax\def\natexlab#1{#1}\fi
\providecommand{\url}[1]{\href{#1}{#1}}
\providecommand{\dodoi}[1]{doi:~\href{http://doi.org/#1}{\nolinkurl{#1}}}
\providecommand{\doeprint}[1]{\href{http://ascl.net/#1}{\nolinkurl{http://ascl.net/#1}}}
\providecommand{\doarXiv}[1]{\href{https://arxiv.org/abs/#1}{\nolinkurl{https://arxiv.org/abs/#1}}}

\bibitem[{{Aditya}(2019)}]{2019MNRAS.482.5597A}
{Aditya}, J.~N.~H.~S. 2019, \mnras, 482, 5597, \dodoi{10.1093/mnras/sty3062}

\bibitem[{{Aditya} {et~al.}(2021){Aditya}, {Jorgenson}, {Joshi}, {Singh}, {An}, \& {Chandola}}]{2021MNRAS.500..998A}
{Aditya}, J.~N.~H.~S., {Jorgenson}, R., {Joshi}, V., {et~al.} 2021, \mnras, 500, 998, \dodoi{10.1093/mnras/staa3306}

\bibitem[{{Aditya} \& {Kanekar}(2018{\natexlab{a}})}]{2018MNRAS.473...59A}
{Aditya}, J.~N.~H.~S., \& {Kanekar}, N. 2018{\natexlab{a}}, \mnras, 473, 59, \dodoi{10.1093/mnras/stx2325}

\bibitem[{{Aditya} \& {Kanekar}(2018{\natexlab{b}})}]{2018MNRAS.481.1578A}
---. 2018{\natexlab{b}}, \mnras, 481, 1578, \dodoi{10.1093/mnras/sty2184}

\bibitem[{{Aditya} {et~al.}(2016){Aditya}, {Kanekar}, \& {Kurapati}}]{2016MNRAS.455.4000A}
{Aditya}, J.~N.~H.~S., {Kanekar}, N., \& {Kurapati}, S. 2016, \mnras, 455, 4000, \dodoi{10.1093/mnras/stv2563}

\bibitem[{{Aditya} {et~al.}(2024){Aditya}, {Yoon}, {Allison}, {An}, {Chhetri}, {Curran}, {Darling}, {Emig}, {Glowacki}, {Kerrison}, {Koribalski}, {Mahony}, {Moss}, {Morgan}, {Sadler}, {Soria}, {Su}, {Weng}, \& {Whiting}}]{2024MNRAS.527.8511A}
{Aditya}, J.~N.~H.~S., {Yoon}, H., {Allison}, J.~R., {et~al.} 2024, \mnras, 527, 8511, \dodoi{10.1093/mnras/stad3722}

\bibitem[{{Allison}(2021)}]{2021MNRAS.503..985A}
{Allison}, J.~R. 2021, \mnras, 503, 985, \dodoi{10.1093/mnras/stab518}

\bibitem[{{Allison} {et~al.}(2016){Allison}, {Zwaan}, {Duchesne}, \& {Curran}}]{2016MNRAS.462.1341A}
{Allison}, J.~R., {Zwaan}, M.~A., {Duchesne}, S.~W., \& {Curran}, S.~J. 2016, \mnras, 462, 1341, \dodoi{10.1093/mnras/stw1722}

\bibitem[{{Allison} {et~al.}(2012){Allison}, {Curran}, {Emonts}, {Ger{\'e}b}, {Mahony}, {Reeves}, {Sadler}, {Tanna}, {Whiting}, \& {Zwaan}}]{2012MNRAS.423.2601A}
{Allison}, J.~R., {Curran}, S.~J., {Emonts}, B.~H.~C., {et~al.} 2012, \mnras, 423, 2601, \dodoi{10.1111/j.1365-2966.2012.21062.x}

\bibitem[{{Allison} {et~al.}(2015){Allison}, {Sadler}, {Moss}, {Whiting}, {Hunstead}, {Pracy}, {Curran}, {Croom}, {Glowacki}, {Morganti}, {Shabala}, {Zwaan}, {Allen}, {Amy}, {Axtens}, {Ball}, {Bannister}, {Barker}, {Bell}, {Bock}, {Bolton}, {Bowen}, {Boyle}, {Braun}, {Broadhurst}, {Brodrick}, {Brothers}, {Brown}, {Bunton}, {Cantrall}, {Chapman}, {Cheng}, {Chippendale}, {Chung}, {Cooray}, {Cornwell}, {DeBoer}, {Diamond}, {Edwards}, {Ekers}, {Feain}, {Ferris}, {Forsyth}, {Gough}, {Grancea}, {Gupta}, {Guzman}, {Hampson}, {Harvey-Smith}, {Haskins}, {Hay}, {Hayman}, {Heywood}, {Hotan}, {Hoyle}, {Humphreys}, {Indermuehle}, {Jacka}, {Jackson}, {Jackson}, {Jeganathan}, {Johnston}, {Joseph}, {Kendall}, {Kesteven}, {Kiraly}, {Koribalski}, {Leach}, {Lenc}, {Lensson}, {Mackay}, {Macleod}, {Marquarding}, {Marvil}, {McClure-Griffiths}, {McConnell}, {Mirtschin}, {Norris}, {Neuhold}, {Ng}, {O'Sullivan}, {Pathikulangara}, {Pearce}, {Phillips}, {Popping}, {Qiao}, {Reynolds}, {Roberts}, {Sault}, {Schinckel}, {Serra}, {Shaw},
  {Shields}, {Shimwell}, {Storey}, {Sweetnam}, {Troup}, {Turner}, {Tuthill}, {Tzioumis}, {Voronkov}, {Westmeier}, \& {Wilson}}]{2015MNRAS.453.1249A}
{Allison}, J.~R., {Sadler}, E.~M., {Moss}, V.~A., {et~al.} 2015, \mnras, 453, 1249, \dodoi{10.1093/mnras/stv1532}

\bibitem[{{Allison} {et~al.}(2020){Allison}, {Sadler}, {Bellstedt}, {Davies}, {Driver}, {Ellison}, {Huynh}, {Kapi{\'n}ska}, {Mahony}, {Moss}, {Robotham}, {Whiting}, {Curran}, {Darling}, {Hotan}, {Hunstead}, {Koribalski}, {Lagos}, {Pettini}, {Pimbblet}, \& {Voronkov}}]{2020MNRAS.494.3627A}
{Allison}, J.~R., {Sadler}, E.~M., {Bellstedt}, S., {et~al.} 2020, \mnras, 494, 3627, \dodoi{10.1093/mnras/staa949}

\bibitem[{{Allison} {et~al.}(2022){Allison}, {Sadler}, {Amaral}, {An}, {Curran}, {Darling}, {Edge}, {Ellison}, {Emig}, {Gaensler}, {Garratt-Smithson}, {Glowacki}, {Grasha}, {Koribalski}, {Lagos}, {Lah}, {Mahony}, {Mao}, {Morganti}, {Moss}, {Pettini}, {Pimbblet}, {Power}, {Salas}, {Staveley-Smith}, {Whiting}, {Wong}, {Yoon}, {Zheng}, \& {Zwaan}}]{2022PASA...39...10A}
{Allison}, J.~R., {Sadler}, E.~M., {Amaral}, A.~D., {et~al.} 2022, \pasa, 39, e010, \dodoi{10.1017/pasa.2022.3}

\bibitem[{{Augarde} \& {Lequeux}(1985)}]{1985A&A...147..273A}
{Augarde}, R., \& {Lequeux}, J. 1985, \aap, 147, 273

\bibitem[{{Baan} \& {Haschick}(1990)}]{1990ApJ...364...65B}
{Baan}, W.~A., \& {Haschick}, A. 1990, \apj, 364, 65, \dodoi{10.1086/169385}

\bibitem[{{Becker} {et~al.}(1995){Becker}, {White}, \& {Helfand}}]{1995ApJ...450..559B}
{Becker}, R.~H., {White}, R.~L., \& {Helfand}, D.~J. 1995, \apj, 450, 559, \dodoi{10.1086/176166}

\bibitem[{{Beichman} {et~al.}(1981){Beichman}, {Neugebauer}, {Soifer}, {Wootten}, {Roellig}, \& {Harvey}}]{1981Natur.293..711B}
{Beichman}, C.~A., {Neugebauer}, G., {Soifer}, B.~T., {et~al.} 1981, \nat, 293, 711, \dodoi{10.1038/293711a0}

\bibitem[{{Biggs} {et~al.}(2016){Biggs}, {Zwaan}, {Hatziminaoglou}, {P{\'e}roux}, \& {Liske}}]{2016MNRAS.462.2819B}
{Biggs}, A.~D., {Zwaan}, M.~A., {Hatziminaoglou}, E., {P{\'e}roux}, C., \& {Liske}, J. 2016, \mnras, 462, 2819, \dodoi{10.1093/mnras/stw1786}

\bibitem[{{Bordoloi} {et~al.}(2022){Bordoloi}, {O'Meara}, {Sharon}, {Rigby}, {Cooke}, {Shaban}, {Matuszewski}, {Rizzi}, {Doppmann}, {Martin}, {Moore}, {Morrissey}, \& {Neill}}]{Bordoloi2022}
{Bordoloi}, R., {O'Meara}, J.~M., {Sharon}, K., {et~al.} 2022, \nat, 606, 59, \dodoi{10.1038/s41586-022-04616-1}

\bibitem[{{Bregman} {et~al.}(1981){Bregman}, {Lebofsky}, {Aller}, {Rieke}, {Aller}, {Hodge}, {Glassgold}, \& {Huggins}}]{1981Natur.293..714B}
{Bregman}, J.~N., {Lebofsky}, M.~J., {Aller}, M.~F., {et~al.} 1981, \nat, 293, 714, \dodoi{10.1038/293714a0}

\bibitem[{{Carilli} {et~al.}(1992){Carilli}, {Perlman}, \& {Stocke}}]{1992ApJ...400L..13C}
{Carilli}, C.~L., {Perlman}, E.~S., \& {Stocke}, J.~T. 1992, \apjl, 400, L13, \dodoi{10.1086/186637}

\bibitem[{{Casoli} {et~al.}(1989){Casoli}, {Combes}, {Augarde}, {Figon}, \& {Martin}}]{1989A&A...224...31C}
{Casoli}, F., {Combes}, F., {Augarde}, R., {Figon}, P., \& {Martin}, J.~M. 1989, \aap, 224, 31

\bibitem[{{Caulet} {et~al.}(1992){Caulet}, {Woodgate}, {Brown}, {Gull}, {Hintzen}, {Lowenthal}, {Oliversen}, \& {Ziegler}}]{1992ApJ...388..301C}
{Caulet}, A., {Woodgate}, B.~E., {Brown}, L.~W., {et~al.} 1992, \apj, 388, 301, \dodoi{10.1086/171153}

\bibitem[{{Chambers} {et~al.}(2016){Chambers}, {Magnier}, {Metcalfe}, {Flewelling}, {Huber}, {Waters}, {Denneau}, {Draper}, {Farrow}, {Finkbeiner}, {Holmberg}, {Koppenhoefer}, {Price}, {Rest}, {Saglia}, {Schlafly}, {Smartt}, {Sweeney}, {Wainscoat}, {Burgett}, {Chastel}, {Grav}, {Heasley}, {Hodapp}, {Jedicke}, {Kaiser}, {Kudritzki}, {Luppino}, {Lupton}, {Monet}, {Morgan}, {Onaka}, {Shiao}, {Stubbs}, {Tonry}, {White}, {Ba{\~n}ados}, {Bell}, {Bender}, {Bernard}, {Boegner}, {Boffi}, {Botticella}, {Calamida}, {Casertano}, {Chen}, {Chen}, {Cole}, {Deacon}, {Frenk}, {Fitzsimmons}, {Gezari}, {Gibbs}, {Goessl}, {Goggia}, {Gourgue}, {Goldman}, {Grant}, {Grebel}, {Hambly}, {Hasinger}, {Heavens}, {Heckman}, {Henderson}, {Henning}, {Holman}, {Hopp}, {Ip}, {Isani}, {Jackson}, {Keyes}, {Koekemoer}, {Kotak}, {Le}, {Liska}, {Long}, {Lucey}, {Liu}, {Martin}, {Masci}, {McLean}, {Mindel}, {Misra}, {Morganson}, {Murphy}, {Obaika}, {Narayan}, {Nieto-Santisteban}, {Norberg}, {Peacock}, {Pier}, {Postman}, {Primak}, {Rae}, {Rai},
  {Riess}, {Riffeser}, {Rix}, {R{\"o}ser}, {Russel}, {Rutz}, {Schilbach}, {Schultz}, {Scolnic}, {Strolger}, {Szalay}, {Seitz}, {Small}, {Smith}, {Soderblom}, {Taylor}, {Thomson}, {Taylor}, {Thakar}, {Thiel}, {Thilker}, {Unger}, {Urata}, {Valenti}, {Wagner}, {Walder}, {Walter}, {Watters}, {Werner}, {Wood-Vasey}, \& {Wyse}}]{2016arXiv161205560C}
{Chambers}, K.~C., {Magnier}, E.~A., {Metcalfe}, N., {et~al.} 2016, arXiv e-prints, arXiv:1612.05560, \dodoi{10.48550/arXiv.1612.05560}

\bibitem[{{Chandola} {et~al.}(2013){Chandola}, {Gupta}, \& {Saikia}}]{2013MNRAS.429.2380C}
{Chandola}, Y., {Gupta}, N., \& {Saikia}, D.~J. 2013, \mnras, 429, 2380, \dodoi{10.1093/mnras/sts499}

\bibitem[{{Chandola} \& {Saikia}(2017)}]{2017MNRAS.465..997C}
{Chandola}, Y., \& {Saikia}, D.~J. 2017, \mnras, 465, 997, \dodoi{10.1093/mnras/stw2705}

\bibitem[{{Chandola} {et~al.}(2020){Chandola}, {Saikia}, \& {Li}}]{2020MNRAS.494.5161C}
{Chandola}, Y., {Saikia}, D.~J., \& {Li}, D. 2020, \mnras, 494, 5161, \dodoi{10.1093/mnras/staa1029}

\bibitem[{{Chandola} {et~al.}(2024){Chandola}, {Saikia}, {Ma}, {Zheng}, {Tsai}, {Li}, {Tramonte}, \& {Pan}}]{2024ApJ...973...48C}
{Chandola}, Y., {Saikia}, D.~J., {Ma}, Y.-Z., {et~al.} 2024, \apj, 973, 48, \dodoi{10.3847/1538-4357/ad5d5c}

\bibitem[{{Chandola} {et~al.}(2011){Chandola}, {Sirothia}, \& {Saikia}}]{2011MNRAS.418.1787C}
{Chandola}, Y., {Sirothia}, S.~K., \& {Saikia}, D.~J. 2011, \mnras, 418, 1787, \dodoi{10.1111/j.1365-2966.2011.19607.x}

\bibitem[{{Cohen} {et~al.}(2007){Cohen}, {Lane}, {Cotton}, {Kassim}, {Lazio}, {Perley}, {Condon}, \& {Erickson}}]{2007AJ....134.1245C}
{Cohen}, A.~S., {Lane}, W.~M., {Cotton}, W.~D., {et~al.} 2007, \aj, 134, 1245, \dodoi{10.1086/520719}

\bibitem[{{Combes} {et~al.}(2023){Combes}, {Gupta}, {Muller}, {Balashev}, {Deka}, {Emig}, {Kl{\"o}ckner}, {Klutse}, {Knowles}, {Mohapatra}, {Momjian}, {Noterdaeme}, {Petitjean}, {Salas}, {Srianand}, \& {Wagenveld}}]{2023A&A...671A..43C}
{Combes}, F., {Gupta}, N., {Muller}, S., {et~al.} 2023, \aap, 671, A43, \dodoi{10.1051/0004-6361/202245482}

\bibitem[{{Curran}(2017)}]{2017A&A...606A..56C}
{Curran}, S.~J. 2017, \aap, 606, A56, \dodoi{10.1051/0004-6361/201731666}

\bibitem[{{Curran} {et~al.}(2016){Curran}, {Duchesne}, {Divoli}, \& {Allison}}]{2016MNRAS.462.4197C}
{Curran}, S.~J., {Duchesne}, S.~W., {Divoli}, A., \& {Allison}, J.~R. 2016, \mnras, 462, 4197, \dodoi{10.1093/mnras/stw1938}

\bibitem[{{Curran} {et~al.}(2017){Curran}, {Hunstead}, {Johnston}, {Whiting}, {Sadler}, {Allison}, \& {Bignell}}]{2017MNRAS.470.4600C}
{Curran}, S.~J., {Hunstead}, R.~W., {Johnston}, H.~M., {et~al.} 2017, \mnras, 470, 4600, \dodoi{10.1093/mnras/stx1572}

\bibitem[{{Curran} {et~al.}(2011){Curran}, {Whiting}, {Murphy}, {Webb}, {Bignell}, {Polatidis}, {Wiklind}, {Francis}, \& {Langston}}]{2011MNRAS.413.1165C}
{Curran}, S.~J., {Whiting}, M.~T., {Murphy}, M.~T., {et~al.} 2011, \mnras, 413, 1165, \dodoi{10.1111/j.1365-2966.2011.18209.x}

\bibitem[{{Darling}(2012)}]{2012ApJ...761L..26D}
{Darling}, J. 2012, \apjl, 761, L26, \dodoi{10.1088/2041-8205/761/2/L26}

\bibitem[{{de Vries} {et~al.}(2009){de Vries}, {Snellen}, {Schilizzi}, {Mack}, \& {Kaiser}}]{2009A&A...498..641D}
{de Vries}, N., {Snellen}, I.~A.~G., {Schilizzi}, R.~T., {Mack}, K.~H., \& {Kaiser}, C.~R. 2009, \aap, 498, 641, \dodoi{10.1051/0004-6361/200811145}

\bibitem[{{De Young} {et~al.}(1973){De Young}, {Roberts}, \& {Saslaw}}]{1973ApJ...185..809D}
{De Young}, D.~S., {Roberts}, M.~S., \& {Saslaw}, W.~C. 1973, \apj, 185, 809, \dodoi{10.1086/152456}

\bibitem[{{Deka} {et~al.}(2024{\natexlab{a}}){Deka}, {Gupta}, {Chen}, {Johnson}, {Noterdaeme}, {Combes}, {Boettcher}, {Balashev}, {Emig}, {J{\'o}zsa}, {Kl{\"o}ckner}, {Krogager}, {Momjian}, {Petitjean}, {Rudie}, {Wagenveld}, \& {Zahedy}}]{2024A&A...687A..50D}
{Deka}, P.~P., {Gupta}, N., {Chen}, H.~W., {et~al.} 2024{\natexlab{a}}, \aap, 687, A50, \dodoi{10.1051/0004-6361/202348464}

\bibitem[{{Deka} {et~al.}(2024{\natexlab{b}}){Deka}, {Gupta}, {Jagannathan}, {Sekhar}, {Momjian}, {Bhatnagar}, {Wagenveld}, {Kl{\"o}ckner}, {Jose}, {Balashev}, {Combes}, {Hilton}, {Borgaonkar}, {Chatterjee}, {Emig}, {Gaunekar}, {J{\'o}zsa}, {Klutse}, {Knowles}, {Krogager}, {Mohapatra}, {Moodley}, {Muller}, {Noterdaeme}, {Petitjean}, {Salas}, \& {Sikhosana}}]{2024ApJS..270...33D}
{Deka}, P.~P., {Gupta}, N., {Jagannathan}, P., {et~al.} 2024{\natexlab{b}}, \apjs, 270, 33, \dodoi{10.3847/1538-4365/acf7b9}

\bibitem[{{Dickey}(1986)}]{1986ApJ...300..190D}
{Dickey}, J.~M. 1986, \apj, 300, 190, \dodoi{10.1086/163793}

\bibitem[{{Dickey}(1997)}]{1997AJ....113.1939D}
---. 1997, \aj, 113, 1939, \dodoi{10.1086/118408}

\bibitem[{{Dickey} \& {Benson}(1982)}]{1982AJ.....87..278D}
{Dickey}, J.~M., \& {Benson}, J.~M. 1982, \aj, 87, 278, \dodoi{10.1086/113103}

\bibitem[{{Dickey} {et~al.}(2013){Dickey}, {McClure-Griffiths}, {Gibson}, {G{\'o}mez}, {Imai}, {Jones}, {Stanimirovi{\'c}}, {Van Loon}, {Walsh}, {Alberdi}, {Anglada}, {Uscanga}, {Arce}, {Bailey}, {Begum}, {Wakker}, {Bekhti}, {Kalberla}, {Winkel}, {Bekki}, {For}, {Staveley-Smith}, {Westmeier}, {Burton}, {Cunningham}, {Dawson}, {Ellingsen}, {Diamond}, {Green}, {Hill}, {Koribalski}, {McConnell}, {Rathborne}, {Voronkov}, {Douglas}, {English}, {Ford}, {Lockman}, {Foster}, {Gomez}, {Green}, {Bland-Hawthorn}, {Gulyaev}, {Hoare}, {Joncas}, {Kang}, {Kerton}, {Koo}, {Leahy}, {Lo}, {Migenes}, {Nakashima}, {Zhang}, {Nidever}, {Peek}, {Tafoya}, {Tian}, \& {Wu}}]{2013PASA...30....3D}
{Dickey}, J.~M., {McClure-Griffiths}, N., {Gibson}, S.~J., {et~al.} 2013, \pasa, 30, e003, \dodoi{10.1017/pasa.2012.003}

\bibitem[{{Dickey} {et~al.}(2022){Dickey}, {Dempsey}, {Pingel}, {McClure-Griffiths}, {Jameson}, {Dawson}, {D{\'e}nes}, {Clark}, {Joncas}, {Leahy}, {Lee}, {Miville-Desch{\^e}nes}, {Stanimirovi{\'c}}, {Tremblay}, \& {van Loon}}]{2022ApJ...926..186D}
{Dickey}, J.~M., {Dempsey}, J.~M., {Pingel}, N.~M., {et~al.} 2022, \apj, 926, 186, \dodoi{10.3847/1538-4357/ac3a89}

\bibitem[{{Douglas} {et~al.}(1996){Douglas}, {Bash}, {Bozyan}, {Torrence}, \& {Wolfe}}]{1996AJ....111.1945D}
{Douglas}, J.~N., {Bash}, F.~N., {Bozyan}, F.~A., {Torrence}, G.~W., \& {Wolfe}, C. 1996, \aj, 111, 1945, \dodoi{10.1086/117932}

\bibitem[{Dunning {et~al.}(2017)Dunning, Bowen, Castillo, Chung, Doherty, George, Hayman, Jeganathan, Kanoniuk, Mackay, Reilly, Roush, Smart, Shaw, Smith, Tzioumis, \& Venables}]{8105012}
Dunning, A., Bowen, M., Castillo, S., {et~al.} 2017, in 2017 XXXIInd General Assembly and Scientific Symposium of the International Union of Radio Science (URSI GASS), 1--4, \dodoi{10.23919/URSIGASS.2017.8105012}

\bibitem[{{Dutta}(2019)}]{2019JApA...40...41D}
{Dutta}, R. 2019, Journal of Astrophysics and Astronomy, 40, 41, \dodoi{10.1007/s12036-019-9610-5}

\bibitem[{{Dutta} {et~al.}(2018){Dutta}, {Srianand}, \& {Gupta}}]{2018MNRAS.480..947D}
{Dutta}, R., {Srianand}, R., \& {Gupta}, N. 2018, \mnras, 480, 947, \dodoi{10.1093/mnras/sty1872}

\bibitem[{{Emonts} {et~al.}(2010){Emonts}, {Morganti}, {Struve}, {Oosterloo}, {van Moorsel}, {Tadhunter}, {van der Hulst}, {Brogt}, {Holt}, \& {Mirabal}}]{2010MNRAS.406..987E}
{Emonts}, B.~H.~C., {Morganti}, R., {Struve}, C., {et~al.} 2010, \mnras, 406, 987, \dodoi{10.1111/j.1365-2966.2010.16706.x}

\bibitem[{{Fischer} {et~al.}(1983){Fischer}, {Simon}, {Benson}, \& {Solomon}}]{1983ApJ...273L..27F}
{Fischer}, J., {Simon}, M., {Benson}, J., \& {Solomon}, P.~M. 1983, \apjl, 273, L27, \dodoi{10.1086/184123}

\bibitem[{{Fritz} {et~al.}(2006){Fritz}, {Franceschini}, \& {Hatziminaoglou}}]{2006MNRAS.366..767F}
{Fritz}, J., {Franceschini}, A., \& {Hatziminaoglou}, E. 2006, \mnras, 366, 767, \dodoi{10.1111/j.1365-2966.2006.09866.x}

\bibitem[{{Gallimore} {et~al.}(1999){Gallimore}, {Baum}, {O'Dea}, {Pedlar}, \& {Brinks}}]{1999ApJ...524..684G}
{Gallimore}, J.~F., {Baum}, S.~A., {O'Dea}, C.~P., {Pedlar}, A., \& {Brinks}, E. 1999, \apj, 524, 684, \dodoi{10.1086/307853}

\bibitem[{{Gehrz} {et~al.}(1983){Gehrz}, {Sramek}, \& {Weedman}}]{1983ApJ...267..551G}
{Gehrz}, R.~D., {Sramek}, R.~A., \& {Weedman}, D.~W. 1983, \apj, 267, 551, \dodoi{10.1086/160892}

\bibitem[{{Ger{\'e}b} {et~al.}(2015){Ger{\'e}b}, {Maccagni}, {Morganti}, \& {Oosterloo}}]{2015A&A...575A..44G}
{Ger{\'e}b}, K., {Maccagni}, F.~M., {Morganti}, R., \& {Oosterloo}, T.~A. 2015, \aap, 575, A44, \dodoi{10.1051/0004-6361/201424655}

\bibitem[{{Ger{\'e}b} {et~al.}(2014){Ger{\'e}b}, {Morganti}, \& {Oosterloo}}]{2014A&A...569A..35G}
{Ger{\'e}b}, K., {Morganti}, R., \& {Oosterloo}, T.~A. 2014, \aap, 569, A35, \dodoi{10.1051/0004-6361/201423999}

\bibitem[{{Giovannini} {et~al.}(2001){Giovannini}, {Cotton}, {Feretti}, {Lara}, \& {Venturi}}]{2001ApJ...552..508G}
{Giovannini}, G., {Cotton}, W.~D., {Feretti}, L., {Lara}, L., \& {Venturi}, T. 2001, \apj, 552, 508, \dodoi{10.1086/320581}

\bibitem[{{Giroletti} {et~al.}(2003){Giroletti}, {Giovannini}, {Taylor}, {Conway}, {Lara}, \& {Venturi}}]{2003A&A...399..889G}
{Giroletti}, M., {Giovannini}, G., {Taylor}, G.~B., {et~al.} 2003, \aap, 399, 889, \dodoi{10.1051/0004-6361:20021821}

\bibitem[{{Grasha} {et~al.}(2019){Grasha}, {Darling}, {Bolatto}, {Leroy}, \& {Stocke}}]{2019ApJS..245....3G}
{Grasha}, K., {Darling}, J., {Bolatto}, A., {Leroy}, A.~K., \& {Stocke}, J.~T. 2019, \apjs, 245, 3, \dodoi{10.3847/1538-4365/ab4906}

\bibitem[{{Gupta} {et~al.}(2006){Gupta}, {Salter}, {Saikia}, {Ghosh}, \& {Jeyakumar}}]{2006MNRAS.373..972G}
{Gupta}, N., {Salter}, C.~J., {Saikia}, D.~J., {Ghosh}, T., \& {Jeyakumar}, S. 2006, \mnras, 373, 972, \dodoi{10.1111/j.1365-2966.2006.11064.x}

\bibitem[{{Gupta} {et~al.}(2012){Gupta}, {Srianand}, {Petitjean}, {Bergeron}, {Noterdaeme}, \& {Muzahid}}]{2012A&A...544A..21G}
{Gupta}, N., {Srianand}, R., {Petitjean}, P., {et~al.} 2012, \aap, 544, A21, \dodoi{10.1051/0004-6361/201219159}

\bibitem[{{Gupta} {et~al.}(2009){Gupta}, {Srianand}, {Petitjean}, {Noterdaeme}, \& {Saikia}}]{2009MNRAS.398..201G}
{Gupta}, N., {Srianand}, R., {Petitjean}, P., {Noterdaeme}, P., \& {Saikia}, D.~J. 2009, \mnras, 398, 201, \dodoi{10.1111/j.1365-2966.2009.14933.x}

\bibitem[{{Gupta} {et~al.}(2021){Gupta}, {Jagannathan}, {Srianand}, {Bhatnagar}, {Noterdaeme}, {Combes}, {Petitjean}, {Jose}, {Pandey}, {Kaski}, {Baker}, {Balashev}, {Boettcher}, {Chen}, {Cress}, {Dutta}, {Goedhart}, {Heald}, {J{\'o}zsa}, {Kamau}, {Kamphuis}, {Kerp}, {Kl{\"o}ckner}, {Knowles}, {Krishnan}, {Krogager}, {Kulkarni}, {Momjian}, {Moodley}, {Passmoor}, {Schr{\"o}eder}, {Sekhar}, {Sikhosana}, {Wagenveld}, \& {Wong}}]{2021ApJ...907...11G}
{Gupta}, N., {Jagannathan}, P., {Srianand}, R., {et~al.} 2021, \apj, 907, 11, \dodoi{10.3847/1538-4357/abcb85}

\bibitem[{{Healey} {et~al.}(2007){Healey}, {Romani}, {Taylor}, {Sadler}, {Ricci}, {Murphy}, {Ulvestad}, \& {Winn}}]{2007ApJS..171...61H}
{Healey}, S.~E., {Romani}, R.~W., {Taylor}, G.~B., {et~al.} 2007, \apjs, 171, 61, \dodoi{10.1086/513742}

\bibitem[{{Heiles} \& {Troland}(2003)}]{2003ApJ...586.1067H}
{Heiles}, C., \& {Troland}, T.~H. 2003, \apj, 586, 1067, \dodoi{10.1086/367828}

\bibitem[{{Helmboldt} {et~al.}(2007){Helmboldt}, {Taylor}, {Tremblay}, {Fassnacht}, {Walker}, {Myers}, {Sjouwerman}, {Pearson}, {Readhead}, {Weintraub}, {Gehrels}, {Romani}, {Healey}, {Michelson}, {Blandford}, \& {Cotter}}]{2007ApJ...658..203H}
{Helmboldt}, J.~F., {Taylor}, G.~B., {Tremblay}, S., {et~al.} 2007, \apj, 658, 203, \dodoi{10.1086/511005}

\bibitem[{{Helou} {et~al.}(1991){Helou}, {Madore}, {Schmitz}, {Bicay}, {Wu}, \& {Bennett}}]{1991ASSL..171...89H}
{Helou}, G., {Madore}, B.~F., {Schmitz}, M., {et~al.} 1991, {The NASA/IPAC extragalactic database.}, ed. M.~A. {Albrecht} \& D.~{Egret}, Vol. 171, 89--106, \dodoi{10.1007/978-94-011-3250-3\_10}

\bibitem[{{Hu} {et~al.}(2020){Hu}, {Catinella}, {Cortese}, {Staveley-Smith}, {Lagos}, {Chauhan}, {Oosterloo}, \& {Chen}}]{2020MNRAS.493.1587H}
{Hu}, W., {Catinella}, B., {Cortese}, L., {et~al.} 2020, \mnras, 493, 1587, \dodoi{10.1093/mnras/staa257}

\bibitem[{{Hu} {et~al.}(2021){Hu}, {Cortese}, {Staveley-Smith}, {Catinella}, {Chauhan}, {Lagos}, {Oosterloo}, \& {Chen}}]{2021MNRAS.507.5580H}
{Hu}, W., {Cortese}, L., {Staveley-Smith}, L., {et~al.} 2021, \mnras, 507, 5580, \dodoi{10.1093/mnras/stab2431}

\bibitem[{{Hu} {et~al.}(2019){Hu}, {Hoppmann}, {Staveley-Smith}, {Ger{\'e}b}, {Oosterloo}, {Morganti}, {Catinella}, {Cortese}, {Lagos}, \& {Meyer}}]{2019MNRAS.489.1619H}
{Hu}, W., {Hoppmann}, L., {Staveley-Smith}, L., {et~al.} 2019, \mnras, 489, 1619, \dodoi{10.1093/mnras/stz2038}

\bibitem[{{Hu} {et~al.}(2023){Hu}, {Wang}, {Li}, {Xu}, {Yang}, {Lagache}, {Pen}, {Zheng}, {Shu}, {Zheng}, {Li}, {Ching}, \& {Chen}}]{2023A&A...675A..40H}
{Hu}, W., {Wang}, Y., {Li}, Y., {et~al.} 2023, \aap, 675, A40, \dodoi{10.1051/0004-6361/202245549}

\bibitem[{{Jaffe}(1990)}]{1990A&A...240..254J}
{Jaffe}, W. 1990, \aap, 240, 254

\bibitem[{{Jiang} {et~al.}(2020){Jiang}, {Tang}, {Hou}, {Liu}, {Kr{\v{c}}o}, {Qian}, {Sun}, {Ching}, {Liu}, {Duan}, {Yue}, {Gan}, {Yao}, {Li}, {Pan}, {Yu}, {Liu}, {Li}, {Peng}, {Yan}, \& {FAST Collaboration}}]{2020RAA....20...64J}
{Jiang}, P., {Tang}, N.-Y., {Hou}, L.-G., {et~al.} 2020, Research in Astronomy and Astrophysics, 20, 064, \dodoi{10.1088/1674-4527/20/5/64}

\bibitem[{{Jiang} {et~al.}(2024){Jiang}, {Chen}, {Gan}, {Sun}, {Zhu}, {Li}, {Zhu}, {Wu}, {Chen}, {Zhang}, \& {An}}]{2024AstTI...1...84J}
{Jiang}, P., {Chen}, R., {Gan}, H., {et~al.} 2024, Astronomical Techniques and Instruments, 1, 84, \dodoi{10.61977/ati2024012}

\bibitem[{{Jiao} {et~al.}(2020){Jiao}, {Zhang}, {Zhang}, {Yu}, {Zhu}, \& {Li}}]{2020JCAP...01..054J}
{Jiao}, K., {Zhang}, J.-C., {Zhang}, T.-J., {et~al.} 2020, \jcap, 2020, 054, \dodoi{10.1088/1475-7516/2020/01/054}

\bibitem[{{Jing} {et~al.}(2024){Jing}, {Wang}, {Xu}, {Liu}, {Chen}, {Liang}, {Xu}, {Cao}, {Wang}, {Hu}, {Zhang}, {Guo}, {Gao}, {Ai}, {Gan}, {Gao}, {Han}, {Hou}, {Hou}, {Jiang}, {Kong}, {Li}, {Liu}, {Shao}, {Pan}, {Pan}, {Qian}, {Sun}, {Tang}, {Yang}, {Zhang}, {Zhang}, \& {Zhu}}]{2024SCPMA..6759514J}
{Jing}, Y., {Wang}, J., {Xu}, C., {et~al.} 2024, Science China Physics, Mechanics, and Astronomy, 67, 259514, \dodoi{10.1007/s11433-023-2333-8}

\bibitem[{{Kanekar} {et~al.}(2009){Kanekar}, {Prochaska}, {Ellison}, \& {Chengalur}}]{2009MNRAS.396..385K}
{Kanekar}, N., {Prochaska}, J.~X., {Ellison}, S.~L., \& {Chengalur}, J.~N. 2009, \mnras, 396, 385, \dodoi{10.1111/j.1365-2966.2009.14661.x}

\bibitem[{{Kanekar} {et~al.}(2014){Kanekar}, {Prochaska}, {Smette}, {Ellison}, {Ryan-Weber}, {Momjian}, {Briggs}, {Lane}, {Chengalur}, {Delafosse}, {Grave}, {Jacobsen}, \& {de Bruyn}}]{2014MNRAS.438.2131K}
{Kanekar}, N., {Prochaska}, J.~X., {Smette}, A., {et~al.} 2014, \mnras, 438, 2131, \dodoi{10.1093/mnras/stt2338}

\bibitem[{Kang {et~al.}(2024)Kang, Lu, Zhang, \& Zhu}]{Kang_2024}
Kang, J., Lu, C.-Z., Zhang, T.-J., \& Zhu, M. 2024, Research in Astronomy and Astrophysics, 24, 075002, \dodoi{10.1088/1674-4527/ad48d1}

\bibitem[{Kass \& Raftery(1995)}]{24ce203a-855a-3aa9-952f-976d23b28943}
Kass, R.~E., \& Raftery, A.~E. 1995, Journal of the American Statistical Association, 90, 773.
\newblock \url{http://www.jstor.org/stable/2291091}

\bibitem[{{Kloeckner} {et~al.}(2015){Kloeckner}, {Obreschkow}, {Martins}, {Raccanelli}, {Champion}, {Roy}, {Lobanov}, {Wagner}, \& {Keller}}]{2015aska.confE..27K}
{Kloeckner}, H.~R., {Obreschkow}, D., {Martins}, C., {et~al.} 2015, in Advancing Astrophysics with the Square Kilometre Array (AASKA14), 27.
\newblock \doarXiv{1501.03822}

\bibitem[{{Lacy} {et~al.}(2020){Lacy}, {Baum}, {Chandler}, {Chatterjee}, {Clarke}, {Deustua}, {English}, {Farnes}, {Gaensler}, {Gugliucci}, {Hallinan}, {Kent}, {Kimball}, {Law}, {Lazio}, {Marvil}, {Mao}, {Medlin}, {Mooley}, {Murphy}, {Myers}, {Osten}, {Richards}, {Rosolowsky}, {Rudnick}, {Schinzel}, {Sivakoff}, {Sjouwerman}, {Taylor}, {White}, {Wrobel}, {Andernach}, {Beasley}, {Berger}, {Bhatnager}, {Birkinshaw}, {Bower}, {Brandt}, {Brown}, {Burke-Spolaor}, {Butler}, {Comerford}, {Demorest}, {Fu}, {Giacintucci}, {Golap}, {G{\"u}th}, {Hales}, {Hiriart}, {Hodge}, {Horesh}, {Ivezi{\'c}}, {Jarvis}, {Kamble}, {Kassim}, {Liu}, {Loinard}, {Lyons}, {Masters}, {Mezcua}, {Moellenbrock}, {Mroczkowski}, {Nyland}, {O'Dea}, {O'Sullivan}, {Peters}, {Radford}, {Rao}, {Robnett}, {Salcido}, {Shen}, {Sobotka}, {Witz}, {Vaccari}, {van Weeren}, {Vargas}, {Williams}, \& {Yoon}}]{2020PASP..132c5001L}
{Lacy}, M., {Baum}, S.~A., {Chandler}, C.~J., {et~al.} 2020, \pasp, 132, 035001, \dodoi{10.1088/1538-3873/ab63eb}

\bibitem[{{Li} {et~al.}(2018){Li}, {Wang}, {Qian}, {Krco}, {Jiang}, {Yue}, {Jin}, {Zhu}, {Pan}, {Nan}, \& {Dunning}}]{2018IMMag..19..112L}
{Li}, D., {Wang}, P., {Qian}, L., {et~al.} 2018, IEEE Microwave Magazine, 19, 112, \dodoi{10.1109/MMM.2018.2802178}

\bibitem[{{Li} {et~al.}(2023){Li}, {Wang}, {Deng}, {Yang}, {Hu}, {Liu}, {Zhao}, {Zuo}, {Shu}, {Li}, {Timbie}, {Ansari}, {Perdereau}, {Stebbins}, {Wolz}, {Wu}, {Zhang}, \& {Chen}}]{2023ApJ...954..139L}
{Li}, Y., {Wang}, Y., {Deng}, F., {et~al.} 2023, \apj, 954, 139, \dodoi{10.3847/1538-4357/ace896}

\bibitem[{{Lu} {et~al.}(2023){Lu}, {Zhang}, \& {Zhang}}]{2023MNRAS.521.3150L}
{Lu}, C.-Z., {Zhang}, T., \& {Zhang}, T.-J. 2023, \mnras, 521, 3150, \dodoi{10.1093/mnras/stad761}

\bibitem[{{Ma} {et~al.}(2019){Ma}, {Xu}, {Zhu}, {Hu}, {Li}, {Shan}, {Zhu}, {Gu}, {Li}, {Liu}, \& {Wu}}]{2019ApJS..240...34M}
{Ma}, Z., {Xu}, H., {Zhu}, J., {et~al.} 2019, \apjs, 240, 34, \dodoi{10.3847/1538-4365/aaf9a2}

\bibitem[{{Maccagni} {et~al.}(2017){Maccagni}, {Morganti}, {Oosterloo}, {Ger{\'e}b}, \& {Maddox}}]{2017A&A...604A..43M}
{Maccagni}, F.~M., {Morganti}, R., {Oosterloo}, T.~A., {Ger{\'e}b}, K., \& {Maddox}, N. 2017, \aap, 604, A43, \dodoi{10.1051/0004-6361/201730563}

\bibitem[{{Mahony} {et~al.}(2022){Mahony}, {Allison}, {Sadler}, {Ellison}, {Mao}, {Morganti}, {Moss}, {Seta}, {Tadhunter}, {Weng}, {Whiting}, {Yoon}, {Bell}, {Bunton}, {Harvey-Smith}, {Kimball}, {Koribalski}, \& {Voronkov}}]{2022MNRAS.509.1690M}
{Mahony}, E.~K., {Allison}, J.~R., {Sadler}, E.~M., {et~al.} 2022, \mnras, 509, 1690, \dodoi{10.1093/mnras/stab3041}

\bibitem[{{Maina} {et~al.}(2022){Maina}, {Mohapatra}, {J{\'o}zsa}, {Gupta}, {Combes}, {Deka}, {Wagenveld}, {Srianand}, {Balashev}, {Chen}, {Krogager}, {Momjian}, {Noterdaeme}, \& {Petitjean}}]{2022MNRAS.516.2050M}
{Maina}, E.~K., {Mohapatra}, A., {J{\'o}zsa}, G.~I.~G., {et~al.} 2022, \mnras, 516, 2050, \dodoi{10.1093/mnras/stac1752}

\bibitem[{{Martin} {et~al.}(2005){Martin}, {Fanson}, {Schiminovich}, {Morrissey}, {Friedman}, {Barlow}, {Conrow}, {Grange}, {Jelinsky}, {Milliard}, {Siegmund}, {Bianchi}, {Byun}, {Donas}, {Forster}, {Heckman}, {Lee}, {Madore}, {Malina}, {Neff}, {Rich}, {Small}, {Surber}, {Szalay}, {Welsh}, \& {Wyder}}]{2005ApJ...619L...1M}
{Martin}, D.~C., {Fanson}, J., {Schiminovich}, D., {et~al.} 2005, \apjl, 619, L1, \dodoi{10.1086/426387}

\bibitem[{{McCarthy} {et~al.}(1997){McCarthy}, {Miley}, {de Koff}, {Baum}, {Sparks}, {Golombek}, {Biretta}, \& {Macchetto}}]{1997ApJS..112..415M}
{McCarthy}, P.~J., {Miley}, G.~K., {de Koff}, S., {et~al.} 1997, \apjs, 112, 415, \dodoi{10.1086/313035}

\bibitem[{{McHardy} {et~al.}(1991){McHardy}, {Abraham}, {Crawford}, {Ulrich}, {Mock}, \& {Vanderspeck}}]{1991MNRAS.249..742M}
{McHardy}, I.~M., {Abraham}, R.~G., {Crawford}, C.~S., {et~al.} 1991, \mnras, 249, 742, \dodoi{10.1093/mnras/249.4.742}

\bibitem[{Momjian {et~al.}(2002)Momjian, Romney, \& Troland}]{Momjian_2002}
Momjian, E., Romney, J.~D., \& Troland, T.~H. 2002, The Astrophysical Journal, 566, 195, \dodoi{10.1086/337993}

\bibitem[{{Morganti} \& {Oosterloo}(2018)}]{2018A&ARv..26....4M}
{Morganti}, R., \& {Oosterloo}, T. 2018, \aapr, 26, 4, \dodoi{10.1007/s00159-018-0109-x}

\bibitem[{{Morganti} {et~al.}(2023){Morganti}, {Murthy}, {Oosterloo}, {Blanchard}, {Cook}, {Paragi}, {Orienti}, {Nagai}, \& {Schulz}}]{2023A&A...678A..42M}
{Morganti}, R., {Murthy}, S., {Oosterloo}, T., {et~al.} 2023, \aap, 678, A42, \dodoi{10.1051/0004-6361/202347117}

\bibitem[{{Murray} {et~al.}(2018){Murray}, {Stanimirovi{\'c}}, {Goss}, {Heiles}, {Dickey}, {Babler}, \& {Kim}}]{2018ApJS..238...14M}
{Murray}, C.~E., {Stanimirovi{\'c}}, S., {Goss}, W.~M., {et~al.} 2018, \apjs, 238, 14, \dodoi{10.3847/1538-4365/aad81a}

\bibitem[{{Murthy} {et~al.}(2022){Murthy}, {Morganti}, {Kanekar}, \& {Oosterloo}}]{2022A&A...659A.185M}
{Murthy}, S., {Morganti}, R., {Kanekar}, N., \& {Oosterloo}, T. 2022, \aap, 659, A185, \dodoi{10.1051/0004-6361/202142550}

\bibitem[{{Murthy} {et~al.}(2021){Murthy}, {Morganti}, {Oosterloo}, \& {Maccagni}}]{2021A&A...654A..94M}
{Murthy}, S., {Morganti}, R., {Oosterloo}, T., \& {Maccagni}, F.~M. 2021, \aap, 654, A94, \dodoi{10.1051/0004-6361/202141566}

\bibitem[{{Myers} {et~al.}(2015){Myers}, {Palanque-Delabrouille}, {Prakash}, {P{\^a}ris}, {Yeche}, {Dawson}, {Bovy}, {Lang}, {Schlegel}, {Newman}, {Petitjean}, {Kneib}, {Laurent}, {Percival}, {Ross}, {Seo}, {Tinker}, {Armengaud}, {Brownstein}, {Burtin}, {Cai}, {Comparat}, {Kasliwal}, {Kulkarni}, {Laher}, {Levitan}, {McBride}, {McGreer}, {Miller}, {Nugent}, {Ofek}, {Rossi}, {Ruan}, {Schneider}, {Sesar}, {Streblyanska}, \& {Surace}}]{2015ApJS..221...27M}
{Myers}, A.~D., {Palanque-Delabrouille}, N., {Prakash}, A., {et~al.} 2015, \apjs, 221, 27, \dodoi{10.1088/0067-0049/221/2/27}

\bibitem[{{Nagar} {et~al.}(2002){Nagar}, {Oliva}, {Marconi}, \& {Maiolino}}]{2002A&A...391L..21N}
{Nagar}, N.~M., {Oliva}, E., {Marconi}, A., \& {Maiolino}, R. 2002, \aap, 391, L21, \dodoi{10.1051/0004-6361:20021039}

\bibitem[{{Nair} \& {Abraham}(2010)}]{2010ApJS..186..427N}
{Nair}, P.~B., \& {Abraham}, R.~G. 2010, \apjs, 186, 427, \dodoi{10.1088/0067-0049/186/2/427}

\bibitem[{{Nan} {et~al.}(2011){Nan}, {Li}, {Jin}, {Wang}, {Zhu}, {Zhu}, {Zhang}, {Yue}, \& {Qian}}]{2011IJMPD..20..989N}
{Nan}, R., {Li}, D., {Jin}, C., {et~al.} 2011, International Journal of Modern Physics D, 20, 989, \dodoi{10.1142/S0218271811019335}

\bibitem[{{O'Dea} {et~al.}(1987){O'Dea}, {Sarazin}, \& {Owen}}]{1987ApJ...316..113O}
{O'Dea}, C.~P., {Sarazin}, C.~L., \& {Owen}, F.~N. 1987, \apj, 316, 113, \dodoi{10.1086/165183}

\bibitem[{{Oh} {et~al.}(2018){Oh}, {Koss}, {Markwardt}, {Schawinski}, {Baumgartner}, {Barthelmy}, {Cenko}, {Gehrels}, {Mushotzky}, {Petulante}, {Ricci}, {Lien}, \& {Trakhtenbrot}}]{2018ApJS..235....4O}
{Oh}, K., {Koss}, M., {Markwardt}, C.~B., {et~al.} 2018, \apjs, 235, 4, \dodoi{10.3847/1538-4365/aaa7fd}

\bibitem[{{Ostorero} {et~al.}(2017){Ostorero}, {Morganti}, {Diaferio}, {Siemiginowska}, {Stawarz}, {Moderski}, \& {Labiano}}]{2017ApJ...849...34O}
{Ostorero}, L., {Morganti}, R., {Diaferio}, A., {et~al.} 2017, \apj, 849, 34, \dodoi{10.3847/1538-4357/aa8ef6}

\bibitem[{{Perlman} {et~al.}(1996){Perlman}, {Carilli}, {Stocke}, \& {Conway}}]{1996AJ....111.1839P}
{Perlman}, E.~S., {Carilli}, C.~L., {Stocke}, J.~T., \& {Conway}, J. 1996, \aj, 111, 1839, \dodoi{10.1086/117922}

\bibitem[{{Perlman} {et~al.}(2002){Perlman}, {Stocke}, {Carilli}, {Sugiho}, {Tashiro}, {Madejski}, {Wang}, \& {Conway}}]{2002AJ....124.2401P}
{Perlman}, E.~S., {Stocke}, J.~T., {Carilli}, C.~L., {et~al.} 2002, \aj, 124, 2401, \dodoi{10.1086/344109}

\bibitem[{{Perlman} {et~al.}(2001){Perlman}, {Stocke}, {Conway}, \& {Reynolds}}]{2001AJ....122..536P}
{Perlman}, E.~S., {Stocke}, J.~T., {Conway}, J., \& {Reynolds}, C. 2001, \aj, 122, 536, \dodoi{10.1086/321149}

\bibitem[{{Rahmani} {et~al.}(2012){Rahmani}, {Srianand}, {Gupta}, {Petitjean}, {Noterdaeme}, \& {V{\'a}squez}}]{2012MNRAS.425..556R}
{Rahmani}, H., {Srianand}, R., {Gupta}, N., {et~al.} 2012, \mnras, 425, 556, \dodoi{10.1111/j.1365-2966.2012.21503.x}

\bibitem[{{Readhead} {et~al.}(2021){Readhead}, {Ravi}, {Liodakis}, {Lister}, {Singh}, {Aller}, {Blandford}, {Browne}, {Gorjian}, {Grainge}, {Gurwell}, {Hodges}, {Hovatta}, {Kiehlmann}, {L{\"a}hteenm{\"a}ki}, {Mcaloone}, {Max-Moerbeck}, {Pavlidou}, {Pearson}, {Peirson}, {Perlman}, {Reeves}, {Soifer}, {Taylor}, {Tornikoski}, {Vedantham}, {Werner}, {Wilkinson}, \& {Zensus}}]{2021ApJ...907...61R}
{Readhead}, A.~C.~S., {Ravi}, V., {Liodakis}, I., {et~al.} 2021, \apj, 907, 61, \dodoi{10.3847/1538-4357/abd08c}

\bibitem[{{Roger} {et~al.}(1978){Roger}, {Caswell}, {Murray}, {Cole}, \& {Cooke}}]{1978MNRAS.182..209R}
{Roger}, R.~S., {Caswell}, J.~L., {Murray}, J.~D., {Cole}, D.~J., \& {Cooke}, D.~J. 1978, \mnras, 182, 209, \dodoi{10.1093/mnras/182.2.209}

\bibitem[{{Rubin}(1978)}]{1978ApJ...224L..55R}
{Rubin}, V.~C. 1978, \apjl, 224, L55, \dodoi{10.1086/182758}

\bibitem[{{Sadler} {et~al.}(2020){Sadler}, {Moss}, {Allison}, {Mahony}, {Whiting}, {Johnston}, {Ellison}, {Lagos}, \& {Koribalski}}]{2020MNRAS.499.4293S}
{Sadler}, E.~M., {Moss}, V.~A., {Allison}, J.~R., {et~al.} 2020, \mnras, 499, 4293, \dodoi{10.1093/mnras/staa2390}

\bibitem[{{Saintonge}(2007)}]{2007AJ....133.2087S}
{Saintonge}, A. 2007, \aj, 133, 2087, \dodoi{10.1086/513515}

\bibitem[{{Saraf} {et~al.}(2023){Saraf}, {Wong}, {Cortese}, \& {Koribalski}}]{2023MNRAS.519.4128S}
{Saraf}, M., {Wong}, O.~I., {Cortese}, L., \& {Koribalski}, B.~S. 2023, \mnras, 519, 4128, \dodoi{10.1093/mnras/stac3695}

\bibitem[{{Schwarz}(1978)}]{1978AnSta...6..461S}
{Schwarz}, G. 1978, Annals of Statistics, 6, 461

\bibitem[{{Skrutskie} {et~al.}(2006){Skrutskie}, {Cutri}, {Stiening}, {Weinberg}, {Schneider}, {Carpenter}, {Beichman}, {Capps}, {Chester}, {Elias}, {Huchra}, {Liebert}, {Lonsdale}, {Monet}, {Price}, {Seitzer}, {Jarrett}, {Kirkpatrick}, {Gizis}, {Howard}, {Evans}, {Fowler}, {Fullmer}, {Hurt}, {Light}, {Kopan}, {Marsh}, {McCallon}, {Tam}, {Van Dyk}, \& {Wheelock}}]{2006AJ....131.1163S}
{Skrutskie}, M.~F., {Cutri}, R.~M., {Stiening}, R., {et~al.} 2006, \aj, 131, 1163, \dodoi{10.1086/498708}

\bibitem[{{Srianand} {et~al.}(2022){Srianand}, {Gupta}, {Petitjean}, {Momjian}, {Balashev}, {Combes}, {Chen}, {Krogager}, {Noterdaeme}, {Rahmani}, {Baker}, {Emig}, {J{\'o}zsa}, {Kloeckner}, \& {Moodley}}]{2022MNRAS.516.1339S}
{Srianand}, R., {Gupta}, N., {Petitjean}, P., {et~al.} 2022, \mnras, 516, 1339, \dodoi{10.1093/mnras/stac1877}

\bibitem[{{Stern} {et~al.}(2012){Stern}, {Assef}, {Benford}, {Blain}, {Cutri}, {Dey}, {Eisenhardt}, {Griffith}, {Jarrett}, {Lake}, {Masci}, {Petty}, {Stanford}, {Tsai}, {Wright}, {Yan}, {Harrison}, \& {Madsen}}]{2012ApJ...753...30S}
{Stern}, D., {Assef}, R.~J., {Benford}, D.~J., {et~al.} 2012, \apj, 753, 30, \dodoi{10.1088/0004-637X/753/1/30}

\bibitem[{{Struve} \& {Conway}(2012)}]{2012A&A...546A..22S}
{Struve}, C., \& {Conway}, J.~E. 2012, \aap, 546, A22, \dodoi{10.1051/0004-6361/201218768}

\bibitem[{{Su} {et~al.}(2022){Su}, {Sadler}, {Allison}, {Mahony}, {Moss}, {Whiting}, {Yoon}, {Aditya}, {Bellstedt}, {Robotham}, {Garratt-Smithson}, {Gu}, {Koribalski}, {Soria}, \& {Weng}}]{2022MNRAS.516.2947S}
{Su}, R., {Sadler}, E.~M., {Allison}, J.~R., {et~al.} 2022, \mnras, 516, 2947, \dodoi{10.1093/mnras/stac2257}

\bibitem[{{Thuan} \& {Wadiak}(1982)}]{1982ApJ...252..125T}
{Thuan}, T.~X., \& {Wadiak}, E.~J. 1982, \apj, 252, 125, \dodoi{10.1086/159539}

\bibitem[{{van Gorkom} {et~al.}(1989){van Gorkom}, {Knapp}, {Ekers}, {Ekers}, {Laing}, \& {Polk}}]{1989AJ.....97..708V}
{van Gorkom}, J.~H., {Knapp}, G.~R., {Ekers}, R.~D., {et~al.} 1989, \aj, 97, 708, \dodoi{10.1086/115016}

\bibitem[{{Varenius} {et~al.}(2017){Varenius}, {Costagliola}, {Kl{\"o}ckner}, {Aalto}, {Spoon}, {Mart{\'\i}-Vidal}, {Conway}, {Privon}, \& {K{\"o}nig}}]{2017A&A...607A..43V}
{Varenius}, E., {Costagliola}, F., {Kl{\"o}ckner}, H.~R., {et~al.} 2017, \aap, 607, A43, \dodoi{10.1051/0004-6361/201629819}

\bibitem[{{Vermeulen} {et~al.}(2003){Vermeulen}, {Pihlstr{\"o}m}, {Tschager}, {de Vries}, {Conway}, {Barthel}, {Baum}, {Braun}, {Bremer}, {Miley}, {O'Dea}, {R{\"o}ttgering}, {Schilizzi}, {Snellen}, \& {Taylor}}]{2003A&A...404..861V}
{Vermeulen}, R.~C., {Pihlstr{\"o}m}, Y.~M., {Tschager}, W., {et~al.} 2003, \aap, 404, 861, \dodoi{10.1051/0004-6361:20030468}

\bibitem[{{Weedman}(1972)}]{1972ApJ...171....5W}
{Weedman}, D.~W. 1972, \apj, 171, 5, \dodoi{10.1086/151250}

\bibitem[{{Weng} {et~al.}(2022){Weng}, {Sadler}, {Foster}, {P{\'e}roux}, {Mahony}, {Allison}, {Moss}, {Su}, {Whiting}, \& {Yoon}}]{2022MNRAS.512.3638W}
{Weng}, S., {Sadler}, E.~M., {Foster}, C., {et~al.} 2022, \mnras, 512, 3638, \dodoi{10.1093/mnras/stac747}

\bibitem[{{Wilson} {et~al.}(1976){Wilson}, {Penston}, {Fosbury}, \& {Boksenberg}}]{1976MNRAS.177..673W}
{Wilson}, A.~S., {Penston}, M.~V., {Fosbury}, R.~A.~E., \& {Boksenberg}, A. 1976, \mnras, 177, 673, \dodoi{10.1093/mnras/177.3.673}

\bibitem[{{Wolfe} {et~al.}(2005){Wolfe}, {Gawiser}, \& {Prochaska}}]{2005ARA&A..43..861W}
{Wolfe}, A.~M., {Gawiser}, E., \& {Prochaska}, J.~X. 2005, \araa, 43, 861, \dodoi{10.1146/annurev.astro.42.053102.133950}

\bibitem[{{Wolfire} {et~al.}(2003){Wolfire}, {McKee}, {Hollenbach}, \& {Tielens}}]{2003ApJ...587..278W}
{Wolfire}, M.~G., {McKee}, C.~F., {Hollenbach}, D., \& {Tielens}, A.~G.~G.~M. 2003, \apj, 587, 278, \dodoi{10.1086/368016}

\bibitem[{{Wright} {et~al.}(2010){Wright}, {Eisenhardt}, {Mainzer}, {Ressler}, {Cutri}, {Jarrett}, {Kirkpatrick}, {Padgett}, {McMillan}, {Skrutskie}, {Stanford}, {Cohen}, {Walker}, {Mather}, {Leisawitz}, {Gautier}, {McLean}, {Benford}, {Lonsdale}, {Blain}, {Mendez}, {Irace}, {Duval}, {Liu}, {Royer}, {Heinrichsen}, {Howard}, {Shannon}, {Kendall}, {Walsh}, {Larsen}, {Cardon}, {Schick}, {Schwalm}, {Abid}, {Fabinsky}, {Naes}, \& {Tsai}}]{2010AJ....140.1868W}
{Wright}, E.~L., {Eisenhardt}, P. R.~M., {Mainzer}, A.~K., {et~al.} 2010, \aj, 140, 1868, \dodoi{10.1088/0004-6256/140/6/1868}

\bibitem[{{Yan} {et~al.}(2016){Yan}, {Stocke}, {Darling}, {Momjian}, {Sharma}, \& {Kanekar}}]{2016AJ....151...74Y}
{Yan}, T., {Stocke}, J.~T., {Darling}, J., {et~al.} 2016, \aj, 151, 74, \dodoi{10.3847/0004-6256/151/3/74}

\bibitem[{{York} {et~al.}(2000){York}, {Adelman}, {Anderson}, {Anderson}, {Annis}, {Bahcall}, {Bakken}, {Barkhouser}, {Bastian}, {Berman}, {Boroski}, {Bracker}, {Briegel}, {Briggs}, {Brinkmann}, {Brunner}, {Burles}, {Carey}, {Carr}, {Castander}, {Chen}, {Colestock}, {Connolly}, {Crocker}, {Csabai}, {Czarapata}, {Davis}, {Doi}, {Dombeck}, {Eisenstein}, {Ellman}, {Elms}, {Evans}, {Fan}, {Federwitz}, {Fiscelli}, {Friedman}, {Frieman}, {Fukugita}, {Gillespie}, {Gunn}, {Gurbani}, {de Haas}, {Haldeman}, {Harris}, {Hayes}, {Heckman}, {Hennessy}, {Hindsley}, {Holm}, {Holmgren}, {Huang}, {Hull}, {Husby}, {Ichikawa}, {Ichikawa}, {Ivezi{\'c}}, {Kent}, {Kim}, {Kinney}, {Klaene}, {Kleinman}, {Kleinman}, {Knapp}, {Korienek}, {Kron}, {Kunszt}, {Lamb}, {Lee}, {Leger}, {Limmongkol}, {Lindenmeyer}, {Long}, {Loomis}, {Loveday}, {Lucinio}, {Lupton}, {MacKinnon}, {Mannery}, {Mantsch}, {Margon}, {McGehee}, {McKay}, {Meiksin}, {Merelli}, {Monet}, {Munn}, {Narayanan}, {Nash}, {Neilsen}, {Neswold}, {Newberg}, {Nichol}, {Nicinski},
  {Nonino}, {Okada}, {Okamura}, {Ostriker}, {Owen}, {Pauls}, {Peoples}, {Peterson}, {Petravick}, {Pier}, {Pope}, {Pordes}, {Prosapio}, {Rechenmacher}, {Quinn}, {Richards}, {Richmond}, {Rivetta}, {Rockosi}, {Ruthmansdorfer}, {Sandford}, {Schlegel}, {Schneider}, {Sekiguchi}, {Sergey}, {Shimasaku}, {Siegmund}, {Smee}, {Smith}, {Snedden}, {Stone}, {Stoughton}, {Strauss}, {Stubbs}, {SubbaRao}, {Szalay}, {Szapudi}, {Szokoly}, {Thakar}, {Tremonti}, {Tucker}, {Uomoto}, {Vanden Berk}, {Vogeley}, {Waddell}, {Wang}, {Watanabe}, {Weinberg}, {Yanny}, {Yasuda}, \& {SDSS Collaboration}}]{2000AJ....120.1579Y}
{York}, D.~G., {Adelman}, J., {Anderson}, John~E., J., {et~al.} 2000, \aj, 120, 1579, \dodoi{10.1086/301513}

\bibitem[{{Yu} {et~al.}(2014){Yu}, {Zhang}, \& {Pen}}]{2014PhRvL.113d1303Y}
{Yu}, H.-R., {Zhang}, T.-J., \& {Pen}, U.-L. 2014, \prl, 113, 041303, \dodoi{10.1103/PhysRevLett.113.041303}

\bibitem[{{Yu} {et~al.}(2023){Yu}, {Fang}, {Wang}, \& {Wu}}]{2023ApJ...952..144Y}
{Yu}, Q., {Fang}, T., {Wang}, J., \& {Wu}, J. 2023, \apj, 952, 144, \dodoi{10.3847/1538-4357/acdb76}

\bibitem[{{Zhang} {et~al.}(2021){Zhang}, {Zhu}, {Wu}, {Yu}, {Jiang}, {Yue}, {Huang}, \& {Hao}}]{2021MNRAS.503.5385Z}
{Zhang}, B., {Zhu}, M., {Wu}, Z.-Z., {et~al.} 2021, \mnras, 503, 5385, \dodoi{10.1093/mnras/stab754}

\bibitem[{{Zhang} {et~al.}(2024){Zhang}, {Zhu}, {Jiang}, {Cheng}, {Wang}, {Wang}, {Xu}, {Liu}, {Yu}, {Qian}, {Yu}, {Ai}, {Jing}, {Xu}, {Liu}, {Guan}, {Sun}, {Yang}, {Huang}, {Hao}, \& {FAST Collaboration}}]{2024SCPMA..6719511Z}
{Zhang}, C.-P., {Zhu}, M., {Jiang}, P., {et~al.} 2024, Science China Physics, Mechanics, and Astronomy, 67, 219511, \dodoi{10.1007/s11433-023-2219-7}

\bibitem[{{Zheng} {et~al.}(2020){Zheng}, {Li}, {Sadler}, {Allison}, \& {Tang}}]{2020MNRAS.499.3085Z}
{Zheng}, Z., {Li}, D., {Sadler}, E.~M., {Allison}, J.~R., \& {Tang}, N. 2020, \mnras, 499, 3085, \dodoi{10.1093/mnras/staa3033}

\bibitem[{{Zovaro} {et~al.}(2019){Zovaro}, {Sharp}, {Nesvadba}, {Bicknell}, {Mukherjee}, {Wagner}, {Groves}, \& {Krishna}}]{2019MNRAS.484.3393Z}
{Zovaro}, H. R.~M., {Sharp}, R., {Nesvadba}, N. P.~H., {et~al.} 2019, \mnras, 484, 3393, \dodoi{10.1093/mnras/stz233}

\bibitem[{{Zwaan} {et~al.}(2015){Zwaan}, {Liske}, {P{\'e}roux}, {Murphy}, {Bouch{\'e}}, {Curran}, \& {Biggs}}]{2015MNRAS.453.1268Z}
{Zwaan}, M.~A., {Liske}, J., {P{\'e}roux}, C., {et~al.} 2015, \mnras, 453, 1268, \dodoi{10.1093/mnras/stv1717}

\end{thebibliography}
\bibliographystyle{aasjournal}



\end{document}